\documentclass[12pt,a4paper]{book}
\pdfoutput=1
\usepackage[table]{xcolor} 

\usepackage{geometry}
\usepackage{emptypage}

\usepackage[pdftex,pdfpagelabels,bookmarks,hyperindex,hyperfigures,hyperfootnotes=false]{hyperref}
\hypersetup{
    colorlinks = true,
    citecolor = blue,
    linkcolor = blue
}
\usepackage[figure]{hypcap} 
\usepackage{apacite}
\usepackage{color} 
\usepackage{tcolorbox}		
\definecolor{myblue}{rgb}{0, 0.6, 1.0}
\definecolor{mybrown}{rgb}{0.8, 0.8, 0.6}
\usepackage{amsmath}
\usepackage{amssymb}
\usepackage{setspace}
\usepackage{courier}
\usepackage{pdfpages}
\usepackage{longtable}
\usepackage{ctable}
\usepackage{textcomp}
\usepackage{enumitem}

\usepackage[titletoc]{appendix}

\usepackage[normalem]{ulem}

\begin{document}

\onehalfspacing

%
%
%
%
%
%
%
%
%
%
%
%

\frontmatter


\thispagestyle{empty}

\begin{center}

\hrulefill

$\;$

$\;$

{\Huge Modelling Early Transitions Toward Autonomous Protocells}

$\;$

\hrulefill

$\;$

{\Large Benjamin John Shirt-Ediss}

\begin{table}[h]
\centering
\begin{tabular}{c}
\includegraphics[width=7cm]{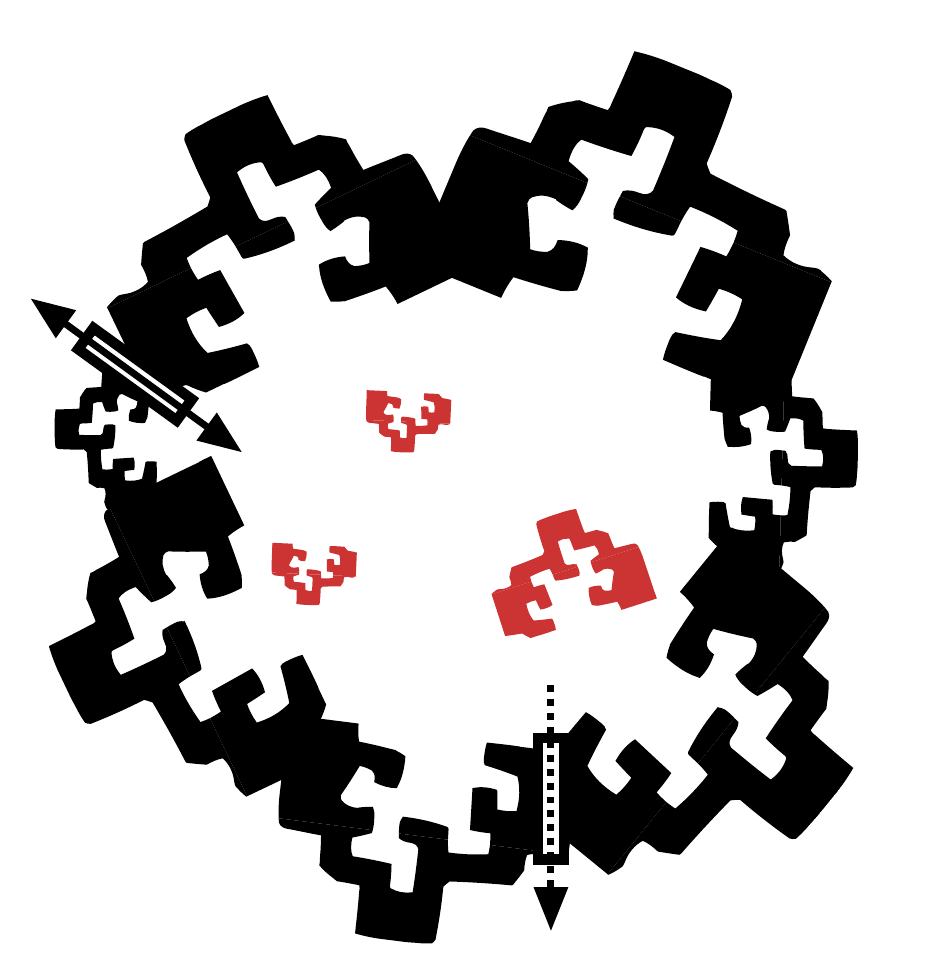}
\end{tabular}
\end{table}

{\large Thesis Submitted for the Degree of Doctor of Philosophy}

{\large Within the PhD Program {\em Filosof{\'i}a, Ciencia y Valores}}

{\large UPV / EHU}

$\;$

{\large January 2016}

Defended February 9th 2016 (Distinction). Version 1.0

$\;$

{\large \textbf{Thesis Supervisors:}}

\end{center}

\begin{table}[h]
\centering
\begin{tabular}{cc}
{\large Dr. Kepa Ruiz-Mirazo} & {\large Prof. Ricard V. Sol{\'e}} \\
$\;$ & $\;$ \\
\includegraphics[width=5cm]{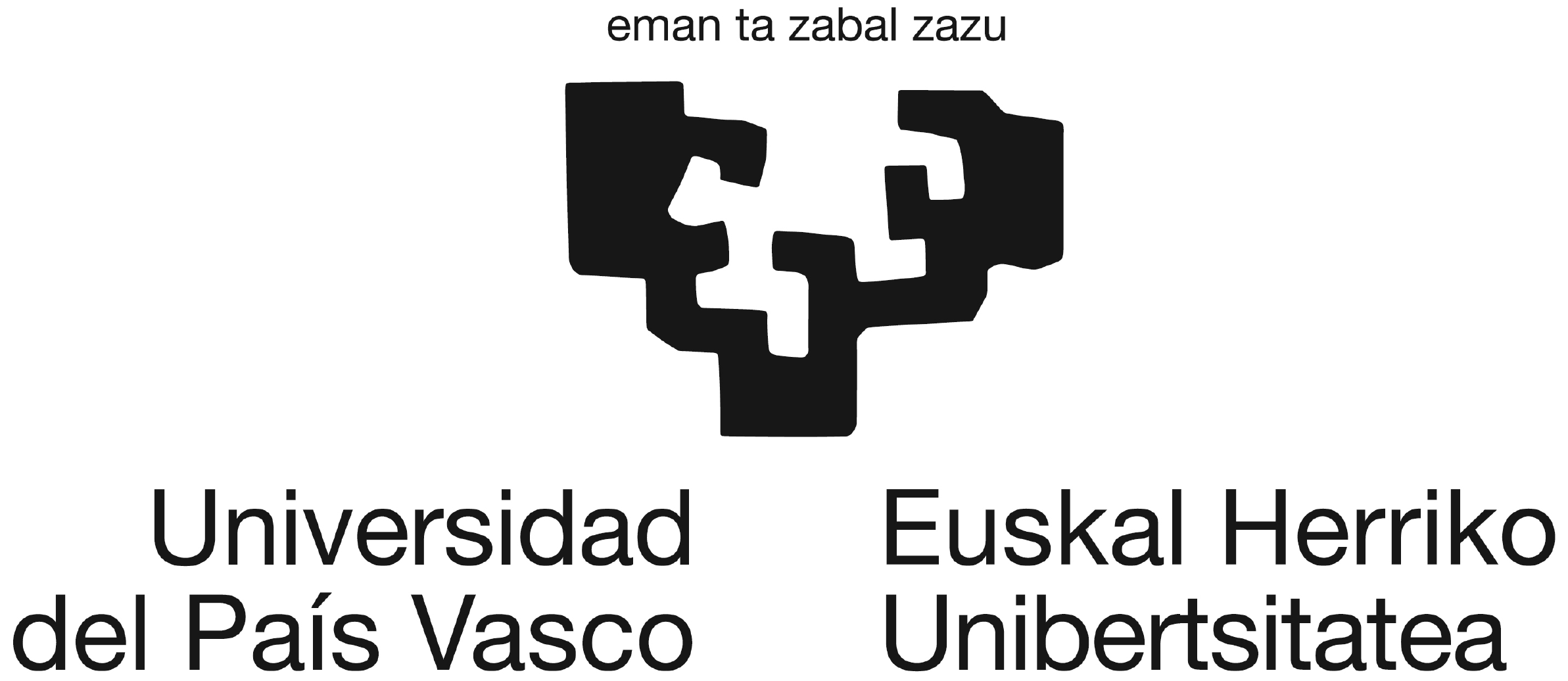} & \includegraphics[width=5.2cm]{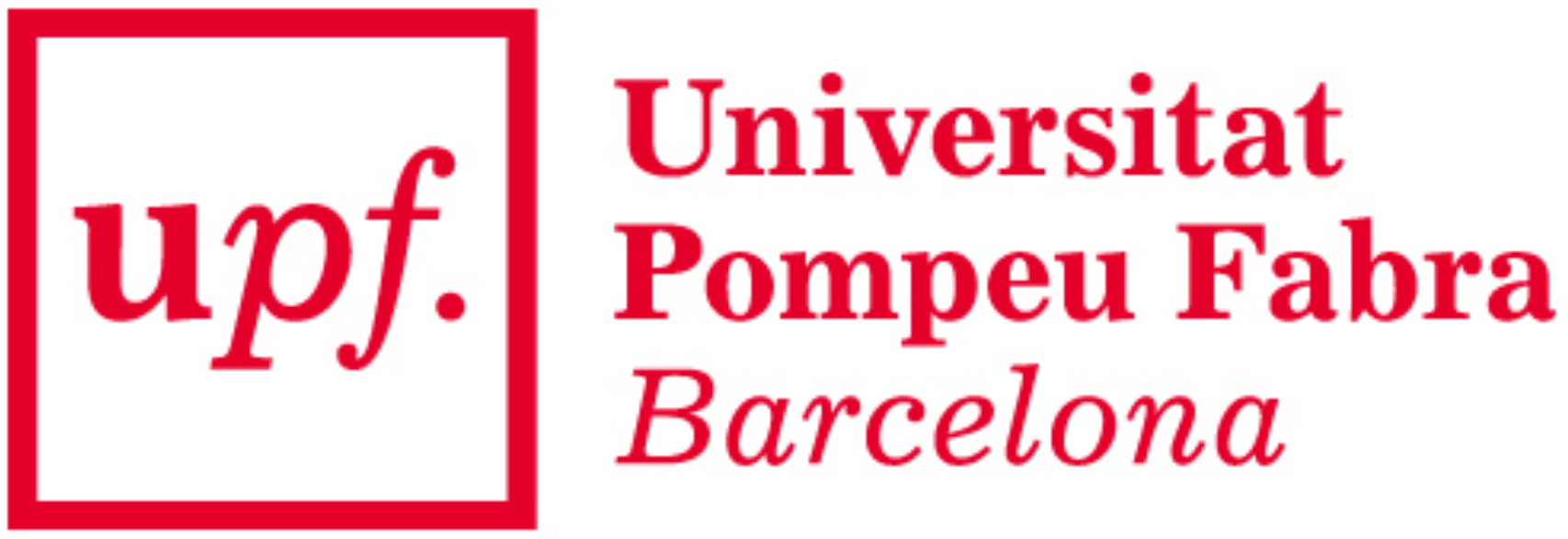}
\end{tabular}
\end{table}
\clearpage



\thispagestyle{empty}

\begin{centering}

{\Huge \textbf{Licence}}

$\;$

$\;$

$\;$

\includegraphics[width=4.5cm]{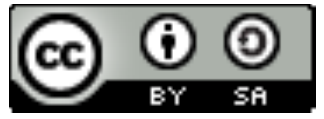}

$\;$

$\;$

$\;$

\end{centering}

For purposes of reproduction, reuse and modification, the text of this  manuscript and all original figures contained within are licensed under a \href{http://creativecommons.org/licenses/by-sa/4.0/}{Creative Commons Attribution-ShareAlike 4.0 International License}

$\;$

The original figures in this thesis are: 

Fig 1.1, Fig 2.1, Fig 2.2(a,b), Fig 2.3, Fig 2.4, Fig 2.5(b), Fig 3.2, Box 2, Fig 3.4, Fig 4.1(a,b,d,e), Fig 4.2, Fig 4.3, Fig 4.4, Fig 4.6, Fig 4.7, Fig 4.8(b), Fig 5.1, Fig A.1, Fig A.2.

$\;$

The other figures in this thesis have been reproduced from various sources and carry their own individual licences (thus, they are NOT included in the above Creative Commons licence). Permission to reproduce copyrighted figures in this thesis has been sought from the appropriate bodies. See \hyperref[chapter:copyright]{ Copyright Permissions for Reproduced Figures } at the end of this document.

\clearpage


\thispagestyle{empty}
\null\vspace{\stretch {1}}
\begin{flushright}
\begin{quotation}

``The mere formulation of a problem is often far more essential than its solution, which may be merely a matter of mathematical or experimental skill. To raise new questions, new possibilities, to regard old problems from a new angle requires creative imagination and marks real advances in science.''

\hfill Albert Einstein
\end{quotation}
\end{flushright}
\vspace{\stretch{2}}\null
\clearpage


\chapter*{Abstract}

\singlespacing

This thesis broadly concerns the origins of life problem, pursuing a joint approach that combines general philosophical/conceptual reflection on the problem along with more detailed and formal scientific modelling work oriented in the conceptual perspective developed.

The central subject matter addressed by this thesis is the emergence and maintenance of compartmentalised chemistries as precursors of more complex systems with a proper cellular organization. Whereas an evolutionary conception of life dominates prebiotic chemistry research and overflows into the protocells field, this thesis defends that the `autonomous systems perspective' of living phenomena is a suitable -- arguably {\em the most} suitable -- conceptual framework to serve as a backdrop for protocell research. The autonomy approach allows a careful and thorough reformulation of the origins of cellular life problem as the problem of how integrated autopoietic chemical organisation, present in all full-fledged cells, originated and developed from more simple far-from-equilibrium chemical aggregate systems. 

Moving away from the original highly abstract formulation of the theory of autopoiesis, this work demonstrates that a heuristic concept like autonomy can be converted into a set of concrete questions to be addressed through accurate scientific means, both experimentally and theoretically. A semi-empirical approach to modelling protocell systems is adopted to investigate the initial complementary relationships that could have been established between chemical reaction networks and self-assembling compartments. In particular, results of this thesis highlight that effects of water osmosis could have been highly influential on the dynamic capabilities of early protocells.

The overall message of this dissertation for origin-of-life researchers is to call attention to a set of commonly overlooked transitions in the development of increasingly complex material systems: namely those concerned with the complementary relationship between selectively permeable membranes and chemical reaction networks. This is especially important in terms of identifying and implementing systems with a relatively complex architecture of interactions among their molecular and supra-molecular components, which could be regarded as intermediate steps toward the immensely complex organisation of a minimal living cell. Although part of that huge gap may need to be covered by an evolutionary process ruled by natural selection, other central parts of the puzzle will require discovering complex molecular mixtures that engage in sufficiently robust, dynamic interactions leading to entities resembling cells.

\onehalfspacing


\tableofcontents

%
%
%
%
%
%
%
%
%
%
%
%


\mainmatter

{
\chapter*{Introductory Outline}
\addcontentsline{toc}{chapter}{Introductory Outline}
\renewcommand{\chaptermark}[1]{\markboth{}{\sffamily #1}}
\chaptermark{Introductory Outline}

The transition from non-living to living matter remains a major blindspot in science. Current research lines into the origins of life, whether based on single replicator molecules or on the newly emerging field of protocells, are typically oriented towards an {\em evolutionary} conception of what life entails. Under this conception, of prime interest is how inorganic molecules and chemical assemblies started to replicate, engage in selection, and increase in complexity, ultimately toward the biochemical complexity of living cells. This doctoral thesis tackles the problem of the origins of cellular life from an alternate conceptual starting point, from the {\em autonomous systems perspective}, based on the conceptual theory of Biological Autonomy. From this systems-theoretic perspective of cellular organisation, the fundamental transitions in the origins of cellular life problem become radically reformulated as how chemical compartment systems, i.e. protocells, started to develop {\em integrated} chemical infrastructures in order to stay far-from-equilibrium, and then how these integrated infrastructures further developed into the autopoietic organisation displayed by full-fledged biological cells. Autonomy defines a qualitatively different -- and it is argued wider and more appropriate -- global framework in which to place scientific research on protocells.

This thesis is a work located in the feedback loop between philosophical reflection on the origins of life problem, and scientific modelling of protocells. It demonstrates how the interaction of philosophy and science can bring about fruitful results and novel insights on a complex problem like the transition from physics and chemistry towards biology. On a conceptual level, this work explains the autonomy conception of life in detail, and follows through the (radical) implications that this organisational, systems view has for reformulating the major transitions in protocell evolution. On the scientific level, this thesis constructs realistic ``semi-empirical'' protocell models to contribute toward an initial stage in modelling autonomous protocells. Throughout the work, the conceptual and scientific levels are interwoven and reinforce each other: the conceptual analysis is essential for highlighting which type of scientific questions are relevant to focus on, and the scientific work carried out makes possible the further specification of overlooked open questions in the field of prebiotic protocell research.

\section*{Main Objectives}
\addcontentsline{toc}{section}{Main Objectives}

The main objectives of this thesis are:

\begin{enumerate}

\item To explain how the biological autonomy perspective has strong implications for the origins of cellular life, detailing the different set of scientific questions and major transitions that this perspective illuminates, as compared to standard evolutionary views.

\item To start investigating the physicochemical basis of biological autonomy in protocells by identifying a current gap of knowledge in protocell research programs: namely, the rigorous modelling of basic chemistry interacting with dynamic lipid compartments. To investigate this area as a first, necessary transition toward basic autonomous protocells. 

\item To develop semi-empirical protocell models which contribute new insights to the general question of how chemical reactions started to couple with dynamic lipid compartments, before self-producing, metabolic cells came to stage. 

\item To carefully examine and explain the main implications of the above modelling work for the general conceptual framework of autonomy, as applied to this prebiotic context. 

\item To locate and explain future challenges faced by the semi-empirical modelling approach, proposing strategies for further advancing this research line toward modelling autonomous protocells.

\end{enumerate}

Below follows a chapter summary of this thesis, highlighting how the main ideas and conceptual strand run throughout the work. Chapters \ref{chapter:1}-\ref{chapter:3} serve as an extended introduction and literature review. These initial chapters lay essential groundwork and take the time to properly develop the implications of the autonomy perspective for protocell research before narrowing down and identifying the sub-problem that is addressed by the scientific protocell modelling contributions of this thesis.

\section*{Thesis Summary}
\addcontentsline{toc}{section}{Thesis Summary}

This thesis begins in Chapter \ref{chapter:1} by providing a primer on protocells. In the origins of cellular life problem, protocells (very rudimentary physicochemical systems based on self-assembled compartments) are increasingly being perceived as providing a vital bridge between chemistry/physics and biological cellular systems. Many general reasons in favour of self-assembled compartments from an early stage in the origins of life can be cited. These reasons range from compartments being effective `localisers' of molecular populations (enabling chemical evolution), to compartments being necessary scaffolds to set up the correct conditions to host complex sequences of linked chemical reactions.

However, whilst a useful {\em vehicle} for explaining abiogenesis, protocells are concept-neutral, and do not by themselves define a general research program toward the origins of cellular life. In practice, protocells are used in scientific research programs that adhere to wider {\em general conceptions} of what the phenomenon of `life' itself entails. Different researchers hold (explicitly or implicitly) different conceptual orientations on life, and these orientations percolate into which protocell experiments they perform, and how they interpret their results as relevant.

Therefore, a more conceptual, philosophical reflection on the phenomena of what constitutes `life' is extremely relevant for protocell research, and origins of life research in general. While a universally accepted ``bright line'' definition separating living from non-living systems continues to be elusive, researchers in origins of life, artificial life and biology have meanwhile partitioned into two broad conceptual camps. The main purpose of Chapter \ref{chapter:2} is to explain, in detail, the central tenets held by each of these camps. The dominant camp at present, the {\em evolutionary} view of life, is based on a diachronic or `across generations' perspective of living systems, whereby life is seen to be manifest in chemical systems which can reproduce, proliferate and proceed through chemical evolution to higher levels of complexity. This perspective stems from the extension of evolution by natural selection to units which are much more simple than whole living systems, and underlies the RNA World and Ribocell protocell research projects. The other, marginalised camp is the {\em autonomy} view of life, which instead sees living systems from a synchronic or `in time' perspective, focussing on the fundamental autopoietic organisation of components and processes that allow cells to function as far-from-equilibrium systems in the here-and-now.

Even if prebiotic research is currently heavily embedded in an evolutionary conception of life, this thesis argues that the marginalised autonomy perspective is actually the most general and most appropriate conceptual framework for protocell research into the origins of cellular life problem. A critical blindspot of the evolutionary perspective is that, by perceiving life as primarily manifest `across time', it lacks (or finds no use for) a rigorous account of the physicochemical organisation of individual cellular systems. The evolutionary approach implicitly assumes that biological cells are simply template-directed chemical networks in `lipid bags', and this weak notion of cellularity correspondingly translates into an origins of life research program focussed on the increase in complexity of chemical networks, where protocells play a side part as useful `chemical containers' facilitating this process. 

The autonomy perspective, on the other hand, instills a deep systemic appreciation of the way molecules have to organise in space and time in order to form a functional far-from-equilibrium biological cell. This rigorous organisational view of cellularity translates into a protocell-centric research program into the origins of cellular life that aims to unravel how the chemical organisation of protocells became more integrated and complex over time, ultimately transforming into the self-fabricating, autopoietic organisation that biological cells possess. A key point is that the autonomy view makes explicit a critical {\em integration problem} in origins of life, typically overlooked by evolutionary approaches, and an autonomy-led protocell research program confronts this problem. Biological cells are only functional systems by virtue of the fact that they closely couple and coordinate compartment, metabolism and template-information subsystems. The same logic would apply equally to the earlier, more rudimentary protocell stages preceding full-fledged cells: in order to be viable, functional far-from-equilibrium systems, able to further increase in complexity, these protocells would also require some type of integrated chemical organisation able to overcome the key problems associated with cellularity -- such as the problems of selective permeability, osmotic water flow, and the harnessing and distribution of free energy resources. Autonomy implies that protocell research programs into the origins of cellular life should deal with cellular organisation and integration issues from an {\em early stage}, for a late-stage amalgamation of independently developed template, membrane and metabolic systems would be unlikely to result in functional cells. 

Chapter \ref{chapter:3} details how a research program into autonomous protocells can be realised, bridging the scientific and conceptual levels of the first two chapters. The chapter begins by discussing reasons why, although important, the theory of Biological Autonomy is difficult to convert into quantitative models. Computational and experimental approaches to implement minimal autopoietic systems {\em in-silico} and {\em in-vitro} are then reviewed as rudimentary attempts to model autonomous systems, along with their associated limitations. A hybrid ``semi-empirical'' approach is put forward as a theoretical yet physically-grounded route to properly address the problem of how autonomy originated and developed in protocells. A research program toward basic autonomous protocells is explained that involves semi-empirical modelling, and the aim of the program is to investigate how the metabolic and membrane systems of protocells co-evolved to result in integrated, functional protocell units. Recent semi-empirical protocell schemes based on a kinetic model of a lipid vesicle are reviewed. 

The concluding part of Chapter \ref{chapter:3} identifies the more specific sub-area where this thesis makes its scientific contributions: the realistic modelling of far-from-equilibrium chemistry in dynamic lipid compartments. This sub-area involves modelling early protocell `reactors' that hypothetically preceded self-producing protocells. These `reactors' would not have possessed the capacity to fabricate complex molecular components like lipids or peptides, but they could nevertheless already have started to demonstrate emergent and biologically relevant non-linear behaviours. 

Chapter \ref{chapter:4} provides a non-technical overview of the scientific contributions made by this thesis toward the realistic modelling of chemistry in dynamic lipid compartments. Four pieces of scientific work are reviewed. The scientific work includes an improved theoretical model of membrane lipid exchange kinetics, validated against experiments, for the semi-empirical vesicle model introduced in Chapter \ref{chapter:3}. On the alternate subject of protocell metabolism, a different model is used to demonstrate how osmotic water flow across the permeable membranes of early protocells could have been an important factor in creating complex reaction dynamics and complex protocell dynamics. A new general systems principle called `osmotic coupling' is proposed, which is applicable to all types of metabolism in early protocells. 

The discussion in Chapter \ref{chapter:5} re-assumes more of a global perspective, recapitulating how the autonomous systems view creates a valid conceptual backdrop for protocell research, and how the prevalent evolutionary view, which by seeing life `across time' rather than `in time', misses the important integration problem that all cellular systems must accomplish. A tentative set of major transitions in protocell development from an autonomy perspective are outlined, and the scientific results of Chapter \ref{chapter:4} are summarised and related to an early protocell stage on this set of transitions. Limitations of research, future challenges to the semi-empirical modelling approach, and future directions are considered. 

}

\chapter[\texorpdfstring{Protocells: In the Twilight Zone Between Non-Living and Living \\ Matter}{}]{Protocells: In the Twilight Zone Between Non-Living and Living Matter}
\chaptermark{Protocells Primer}
\label{chapter:1}

%
%
%
\sectionmark{Introduction}

The biological cell is the minimal physicochemical structure which is unanimously agreed to be alive by human observers, and forms the fundamental building block of all living systems on earth \shortcite{Brenner2010}. Cells can exist in colonies or cohorts of individuals, as tightly coupled heterogeneous assemblies composing multicellular organisms, and as free-living individuals. In fact, most cells can exist in the free-living state, given suitable environmental conditions. On this logic, asking how life originated is equivalent to asking how the first cells originated.

During abiogenesis, the 500 million year process dating from the stabilisation of the early earth ($\approx$4.0 billion years ago) up until the appearance of the first fossil evidence for cyanobacteria ($\approx$3.5 billion years ago), the spontaneous emergence of a single biological cell, in a single instant in time, is a possibility that can be ruled out with certainty. Current cells, even in their simplest prokaryote forms, are bewilderingly complex biochemical nano-machines, both in terms of their component parts list, and in terms of the spatial and temporal organisation of these components. The spontaneous assembly of a cell would not only demand the right mix of complex macromolecules to be present in the prebiotic environment, but these components would also have to localise and assemble in a precise order to create a far-from equilibrium organism as opposed to a complicated non-functional conglomerate. Therefore, it is legitimate to reason that a series of more basic `infra-biological' structures pre-dated the first full-fledged cells, and via a series of fundamental transitions, evolved into them. 

The exact identity of these infra-biological structures continues to be matter of dividing opinion, the debate revolving largely around which criteria are considered essential for life. However, one pertinent generalisation that can be made about contemporary cells is that they all depend critically on the existence of a surrounding lipid membrane \shortcite{morowitz1992}.\footnote{Although some authors dispute the primary role of a membrane for effective cell function. \shortciteA{pollackCellsGels} for example argues that the functioning of a cell is better explained by the fact that the cytoplasm is a {\em gel}.} At some point during abiogenesis, life became membrane bound: the question is {\em when} and {\em how} this happened.

The field of protocells works on the assumption that {\em compartmentalisation} of chemical constituents (which may be by a lipid membrane or by other means) was a fundamental step realised {\em very early} in the origins of life. As such, the field of protocells investigates rudimentary cell-like structures as candidates leading up to the emergence of the first cell \shortcite{rasmussen2009}. 

One recent definition\footnote{Other definitions biased more toward evolutionary capabilities have also been proposed. A protocell is: ``a simple cell-like entity (with a compartment and genetic material) capable of self-replication, metabolism, and Darwinian evolution'' \shortcite{Blain2014}; ``an entity thermodynamically separated from the environment and able to replicate using available nutrient molecules and energy sources.'' \shortcite[p103]{morowitz1992}.} of a protocell in the {\em Encyclopedia of Astrobiology} is:

\begin{quotation}

any experimental or theoretical model that involves a self-assembled compartment (typically a supramolecular structure, like a lipid vesicle) linked to chemical processes taking place around or within it, aimed at explaining how more complex biological cells or alternative forms of cellular organization may come about.

\hfill {\small \shortcite{ruizMirazoProtocellDefn}}

\end{quotation}

\begin{figure}
\begin{center}
\includegraphics[width=15.5cm]{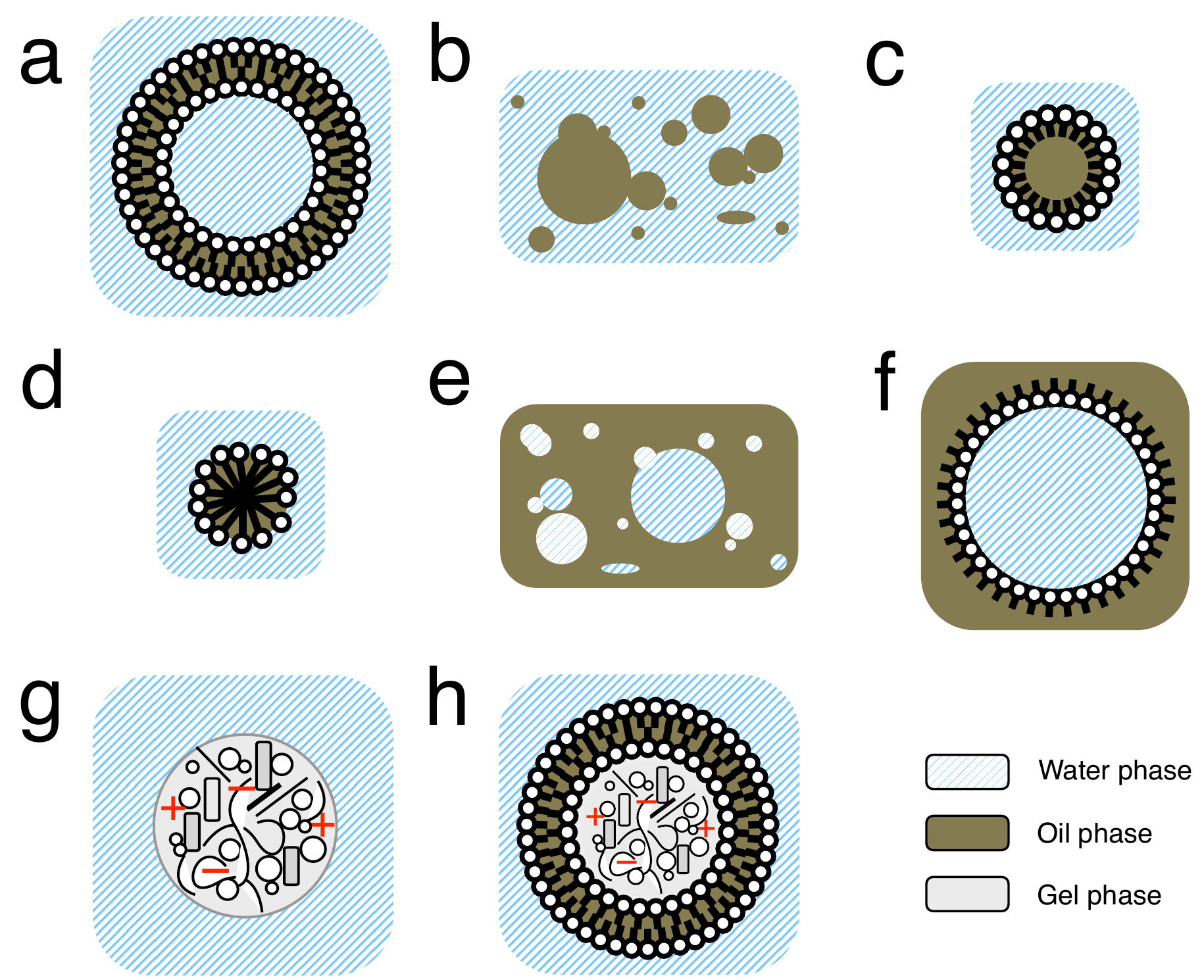}
\end{center}
\caption{
{\textbf{Protocell Architectures.}} (a) Unilamellar vesicle. (b) Oil droplets in water. (c) Surfactant covered oil droplet in water. (d) Micelle. (e) Water droplets in oil. (f) Surfactant covered water droplet in oil. (g) Coacervate. (h) Vesicle enclosing coacervate.}
\label{fig:ch1_protocell_architectures}
\end{figure}

Protocells are therefore the hypothesised prototypes of the first cells on earth. They are soft, mutable supramolecular structures held together by the weak forces of self-assembly (see Section \ref{sec:4_1_1}), just as the membranes of contemporary cells are, and they partially implement some of the essential functionalities observed in cells.\footnote{A chemical system occupying a micro-metre sized rock pore is also `compartmentalised', but it is not strictly a protocell.} These functionalities can include the maintenance of a spatial identity over time, adaptive response to environmental changes, the metabolic use of environmental resources to sustain, grow and reproduce, the excretion of waste, and the use of a genetic apparatus to coordinate operations, grant heredity and enable evolution. Figure \ref{fig:ch1_protocell_architectures} depicts a variety of self-assembled compartment structures currently being explored as protocell candidates.

Apart from a membrane being crucial, extant cellular life also relies on a metabolic system and a template-based genetic system for operation. This triad of subsystems \shortcite{Ganti2003a,szathmary2005} is useful for categorising experimental and theoretical approaches to protocells too \shortcite{rasmussen2009b,Sole2007,RuizMirazo2014}. By definition, all protocells have some type of self-assembled compartment $C$, which may or may not be a semi-permeable membrane, in addition to a metabolism $M$ and/or molecular template copying system $T$.  The valid permutations of this trio are $T+C$, $M+C$ and $T+M+C$, with the latter signifying an advanced protocell design. By itself, an empty self-assembled compartment $C$ is not strictly a protocell, but is rather an equilibrium supramolecular structure. Protocells have the extra requirement that self-assembly needs to be coupled to an energy and matter flow in some way, i.e. protocells operate {\em far-from-equilibrium}. 

In summary, the field of protocells represents an attempt at an extreme form of reverse engineering of the cell. Not only do we want to know the fundamental principles underlying the operation of modern living cells, but we also want to know the history of prototypes, re-designs and failures which incrementally lead up to such a complicated final product becoming possible and arising on the early earth. Protocell research approaches the full-fledged living cell from the `bottom up', with the ultimate goal of creating synthetic cellular systems in the lab by starting from inert molecules and molecular assemblies which are combined in some careful predefined sequence.\footnote{perhaps relying on in-vitro evolution for some of the stages of preparation.} On the way to achieving this grand objective, protocells have immediate technological applications such as drug delivery capsules or as nano-sized chemical reactors. Protocells are also starting to be used in synthetic biology as agents to interact with living cell populations, steering their collective behaviour without the need for genetic modification of the individual living cells \shortcite{Lentini2014,Gardner2009}.

The alternative to the `bottom up' approach of protocells is the `top down' approach whereby existing cells are artificially simplified to successively lower levels of complexity e.g. by reducing the size of their genome. Whilst this reverse process does give useful information about the minimum complexity threshold required by extant cellular life and it does provide a potential reference for the final steps of biogenesis, it does not represent a paradigm to understand the emergence of cellular life in general. Typically, modern cells cannot tolerate much simplification without becoming completely non-functional, and so ancient cellular prototypes must be made by synthesis, not by decomposition.

The remainder of this factual introduction to protocells is organised as follows. First, a brief diversion is taken to explain in more detail how a remote historical event like the origins of life can be investigated using the scientific method. Following that, other more traditional (and still ongoing) non-protocell approaches to the origin of life are outlined, to give a wider perspective. Then, Section \ref{sec:1_3} considers why protocells have risen to prominence and lists general reasons why self-assembled compartments feature in modern theories of abiogenesis. Section \ref{sec:1_4} outlines general routes to constructing protocells in the laboratory, and Section \ref{sec:1_5} concludes by noting that a conception of life is critical in determining the emphasis and direction of all protocell research programs. This leads onto more conceptual discussions of life in Chapter \ref{chapter:2}.

%
%
%

\section[A Synthetic Approach Makes Origins of Life Scientifically Accessible]{A Synthetic Approach Makes Origins of Life Scientifically Accessible%
\sectionmark{A Synthetic Approach to Origins}}
\sectionmark{A Synthetic Approach to Origins}
\label{sec:1_1}

Making a full historical reconstruction of the exact sequence of physico-chemical transitions which lead to the emergence of cellular-based life on earth is impossible. This information is forever lost. The origin of life process not only happened at an {\em extremely} remote period in the past, making prebiotic conditions difficult to be rigorously established, but also the appearance of life erased any direct evidence of the initial developmental trajectory of pre-life. As Morowitz points out, origin events generally prohibit historical investigation because ``Each origin radically transforms the system and changes the rules of operation.'' \shortcite[p14]{morowitz1992}. The origin of the universe and the origin of human culture are two other examples. Once life appeared, it radically changed the boundary conditions away from those in early prebiotic times: it oxygenated the atmosphere, and it established a biosphere pervaded by hierarchical ecosystems. Now, all life invariably comes from pre-existing life, and well-adapted living organisms preclude the re-emergence of simpler, more fragile forms of `protolife', even if such protolife were able to appear under the changed environmental conditions. 

As a result, a huge chasm in complexity separates biological life and the closest `near-life' systems, which are arguably self-organising systems termed {\em dissipative structures} \shortcite{Nicolis1989}. Dissipative structures, like living organisms, are physical systems that degrade their environment to dynamically maintain their own ordered existence, but in nature they only constitute relatively simple examples such as hurricanes, candle flames, oscillating chemical reactions and convection patterns in liquids (e.g. Rayleigh-Bernard cells).\footnote{See \shortciteA{Mossio2010} for a discussion of the difference between inert systems with only {\em physical} self-organisation, and organisms with {\em biological} self-organisation. }

Fortunately, a remaining (and scientifically accessible) option exists for the exploration of origins: a {\em synthetic} approach to abiogenesis (e.g. see \shortciteNP{Attwater2014}). This approach involves the bottom-up synthesis of life and ``life as it could be'' in the laboratory. Using educated guesses about approximate prebiotic conditions (e.g. the energy sources, molecules and prevailing conditions present: see \shortciteNP{deamer2011,Deamer1997}), often combined with generalisations about current living cells (e.g. all rely heavily on protein catalysts synthesised by nucleic acids), diverse experiments are performed to uncover what could be the most relevant physical effects and transitions leading from prebiotic chemistry to the first cellular-based life. A synthetic approach to the origins of life therefore simplifies abiogenesis to only those physical effects and transitions most essential, and gives us a way to provide an account of origins by constructing a {\em story based on educated guesses} involving the sequence of these transitions. Lab experiments will continue to shed more light on what phenomena were possible, and which are difficult to achieve to narrow space of possibilities for origins. 

A synthetic approach has the advantage of being able to disregard the complicated contingent aspects of abiogenesis. Indeed, much of what happened during the history of abiogenesis -- the exact details of all the chemical and physico-chemical structures existing, their spatial distributions and their myriad interactions -- is probably not necessary to know in detail in order to have a comprehensive picture of origins. Most chemical phenomena taking place would have been fleeting and completely irrelevant to the emergence of life, {\em but a select few phenomena would have been fundamental}. These latter phenomena are those searched for by a synthetic experimental approach to origins of life.

Before general arguments are presented for protocells, the next section briefly reviews different approaches to origins of life and the traditional theories.

%
%
%
\section[Non-Protocell Approaches to Origins of Life]{Non-Protocell Approaches to Origins of Life%
\sectionmark{Non-Protocell Approaches}}
\sectionmark{Non-Protocell Approaches}
\label{sec:1_2}

The origins of life field is indeed a broad sphere of enquiry, with many avenues currently open and active. Lines of investigation can be approximately divided into (at least) three broad categories, but these categories all remain interrelated to some extent. In order to set protocells into a wider context, these categories are very briefly reviewed below, with more emphasis on the third category, to which protocells belong.

Ever since the landmark Miller-Urey experiment \shortcite{Miller1953,Lazcano2003}, a traditional category in origins research has been to conduct experiments to find out how the individual building blocks of life could have been synthesised on the prebiotic earth in conditions where efficient and specific protein catalysts did not yet exist. Under `plausible' prebiotic assumptions, these  experiments aim to uncover potential chemical synthesis pathways to monomers like amino acids, nucleotides and lipids, and then to their functional polymerised forms (e.g. proteins and nucleic acids). The field of abiotic synthesis has been comprehensively reviewed elsewhere \shortcite[Section 2 therein]{RuizMirazo2014}, but for some brief examples, it includes the work of \shortciteA{cairnsSmith1985} who cited that clay crystals could have been pivotal helpers in the organised assembly of complex organic molecules, and it also includes more recent work by the Sutherland group who have uncovered possible abiotic synthesis pathways for C and U nucleotides \shortcite{Powner2009} and for the precursors of amino acids and lipids \shortcite{Patel2015}. Finally, the possibility that essential material was delivered to the early earth by meteorites in an already synthesised form also falls into this first category of research.

A second category of origins research tries to narrow down the geographical locations suitable for the genesis of life on the early earth. Alternative `cradles of life' are sought that could have provided suitable bioenergetic and nutrient conditions to kick-start life. Scenarios recently argued for have included deep sea hydrothermal vents \shortcite{Martin2008,Lane2012}, inland geothermal fields \shortcite{Mulkidjanian2012} and scenarios allowing the construction of natural `fuel cells' \shortcite{Barge2014}.

The third category in origins research aims to tackle the overarching question of how a general infrastructure became established in prebiotic chemistry, and what the identity of that infrastructure was, such that the emergence of full-fledged living systems became eventually possible. Recognising that genetic and metabolic systems are both essential parts of all current living organisms, this category traditionally divided into two fiercely defended schools of thought: ``genes preceded metabolism'' (exemplified by \shortciteNP{Anet2004})  and ``metabolism preceded genes'' (exemplified by \shortciteNP{Shapiro2000}). These approaches existed before compartments or membranes were conceded as relevant in the overall scheme of origins, and research down these lines is still ongoing today, with neither of the two schools having yet a clear advantage over the other \shortcite{Fry2011,Lazcano2010}.

The genes-first school is based on an evolutionary view of life (see Section \ref{sec:2_1}) and holds that the prebiotic environment was populated by self-replicating template molecules, most likely RNA, which somehow arose spontaneously in fairly high concentrations. This approach criticises the metabolism-first school (discussed below) along these lines: Metabolic cycles in living cells are very complicated sequences of reactions only possible because of the presence of very specific protein catalysts. Such advanced catalysts were not available on the early earth, and in their absence, it is extremely unlikely that long sequences of reactions could have become organised spontaneously \shortcite{Orgel2008}. Also, generally only hypothetical models and not experimental evidence support the metabolism-first approach.

The genes-first view holds that replicators competed in different scenarios and evolved longer lengths over time, acquiring novel functions such as the ability to catalyse their own formation, the ability to catalyse complex sequences of simpler chemical reactions, or the ability to produce proteins, some of which could act as the first enzymes. The RNA World hypothesis sits at the centre of the genes-first approach and remains under active development as the mainstream theory in origins of life (see Section \ref{sec:2_1}). 

Conversely, the metabolism-first school is based more on a thermodynamic view of life. It postulates that multi-step chemical networks and cycles existed on the prebiotic earth, where reactions either proceeded spontaneously or were catalysed by e.g. various surfaces. This approach criticises the genes-first school along these lines: RNA is a very specific and complicated macromolecule, and even though it could be conceivably synthesised in the prebiotic environment,  nature is indiscriminate, and does not have a special bias for creating RNA molecules. In composing RNA by non-enzymatic means, interfering factors would mean that the majority of molecules produced would be non-functional \shortcite{Shapiro2007}. Also, heredity can be implemented without information-storing molecules, for example, information exists in the concentrations of individual compounds in a chemically reproducing network.

The metabolism-first approach holds that chemical networks, when driven by an energy flow, created localised regions of increased order that consisted of small abundant molecules at first, but later developing so that complex macromolecules began to be synthesised. Nucleotides, the monomers of nucleic acids, eventually appeared on the scene to fulfil roles like catalysis and energy storage, before ultimately being used to synthesise RNA. Stuart Kauffman's autocatalytic set model \shortcite{Kauffman1986,Farmer1986} 
is often cited as holding a central position within the metabolism first approach. It suggests a way in which complex polymers can be built from a collective network of cleavage and condensation reactions, rather than by template base-pairing. It is a model which continues to attract attention (e.g. \shortciteNP{Hordijk2013,Hordijk2010}). Freeman Dyson's model for how disordered molecular populations jumped into an ordered `metabolic' state also falls into the metabolism-first camp \shortcite{Dyson1999}, as does the hypothesis of \shortciteA{Wachtershauser1988}  who proposed that the surface of pyrites could ignite basic metabolic cycles, acting as a basic catalyst and forming a ``pioneer organism''.

Whereas both the gene-first and metabolism-first schools seek to explain the emergence of a complex metabolic system directed by enzymes synthesised by nucleotides, most modern theories on the origins of life now also see {\em compartmentalisation} as a key phenomenon. The next section reviews general arguments for protocells: why compartmentalisation of chemical systems is now broadly thought of as an essential step in abiogenesis.

%
%
%
%
%
\begin{tcolorbox}[colback=mybrown!50!white,colframe=mybrown!75!black,title=Box 1: Focus on Lipid Vesicles as Protocell Compartments]

\begin{center}
\includegraphics[width=14.5cm]{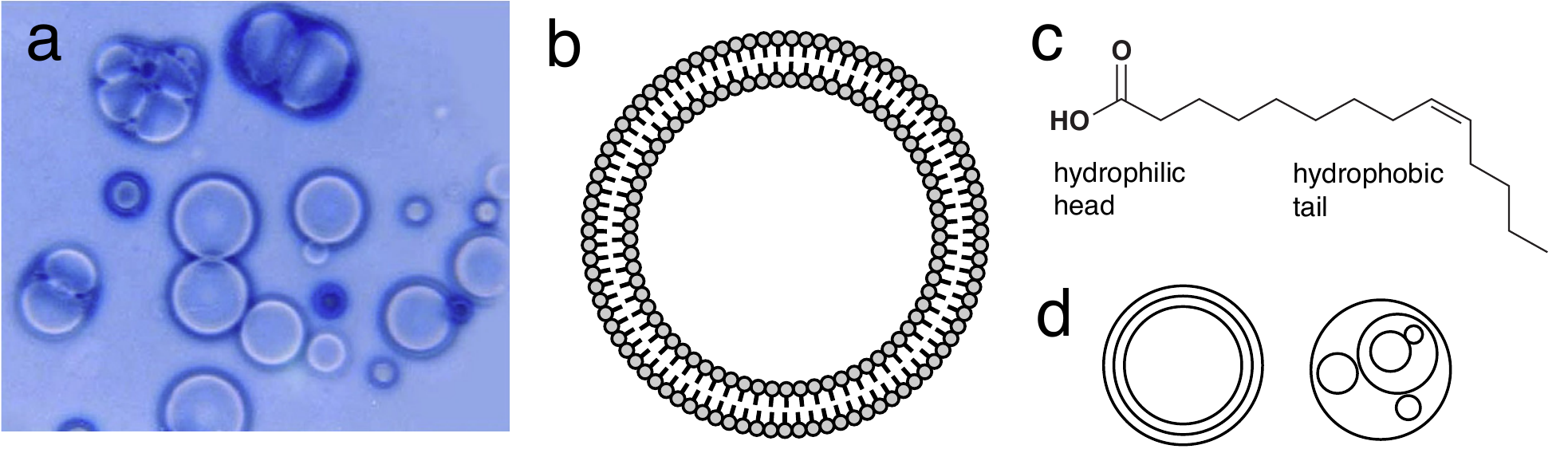}
\end{center}

\emph{Above:} (a) Vesicles formed from lipids extracted from the Murchison meteorite (\shortciteNP{Pohorille2009}: see \hyperref[chapter:copyright]{ View Image Copyright Permissions.}). The largest spherical vesicles pictured are around 20$\mu$m in diameter. (b) Vesicle cross section cartoon, showing the bilayer arrangement of lipids. (c) Myristoleic acid, a simple type of lipid amphiphile able to form vesicles. (d) Multilamellar and nested vesicles.

\noindent\rule[0.5ex]{\linewidth}{1pt}

\singlespacing

\begin{itemize}
\item A lipid vesicle is a soft topologically closed supramolecular structure which self-assembles as a two-layer lipid membrane (or as a set of two-layer lipid membranes) in aqueous solution, keeping an aqueous volume within it. Given the correct conditions, these bilayer membranes form with the lipid heads contacting the inner out outer water phases, and with the lipid tails clustering together inside the hydrophobic centre of the bilayer (see Section \ref{sec:4_1_1} for more details).

\item A variety of different lipid types can constitute vesicles, and vesicles can be made from one single type of lipid or from diverse mixtures of lipids (as the membranes of all living cells are). In origins of life studies, the lipids used for protocell models tend to be very simple lipids called `fatty acids' because it is not unreasonable that these lipids would have been environmentally available on the early earth \shortcite{Rushdi2001, Dworkin2001,RuizMirazo2014}.

\item Lipid vesicles can contain nested, smaller vesicles, and/or multiple bilayer membranes stacked next to each other. This latter case is called a `multilamellar' vesicle. Vesicles forming naturally tend to be very heterogeneous in size and multilamellarity, but some measurements and calculations are much easier to carry out with more homogeneous populations. Therefore, experimentalists tend to prefer working with `unilamellar' vesicles, which consist of a single bilayer and have a more standard size. Unilamellar vesicles can be prepared by extruding a solution of fatty acids through a filter of small holes.
\end{itemize}

\onehalfspacing

\end{tcolorbox}
%
%
%
%
%
%

%
%
%
\section[Relevance of Self-Assembled Compartments in the Origins of Cellular Life]{Relevance of Self-Assembled Compartments in the Origins of Cellular Life%
\sectionmark{Relevance of Self-Assembled Compartments}}
\sectionmark{Relevance of Self-Assembled Compartments}
\label{sec:1_3}

The construction of rudimentary cell-like chemical systems is a scientific tradition stretching back, at least, to the latter half of the 19$^\text{th}$ century \shortcite{hanczyc2009}. Russian biochemist Aleksandr Oparin, investigating chemistry in coacervates in the 1950's and 1960's, was one of the first people to implement a systematic research program into protocells. Within the last 15 years, the field of protocells has attracted more sustained interest, and the current pace of research is accelerating exponentially.

So why is the subject of protocells now a hot topic? what reasons exist for considering self-assembled compartments as essential from an early stage in origins of life? Table \ref{table:ch1_compartment_reasons} below lists a range of diverse factors that have come to light, broadly supporting the relevance of compartmentalised protocells during abiogenesis. Some researchers hold self-assembled compartments as necessary for concentrating important populations of molecules, keeping them safe from environmental parasites, and allowing their reliable transmission to offspring. Alternatively, others see compartments as playing more of an active role in the `proto-metabolism' of an emerging cellular system, for example, helping to provide energy resources, catalyse reactions, and encourage otherwise unfavourable chemical synthesis. Both of these viewpoints are represented in Table \ref{table:ch1_compartment_reasons}.

In current research, the most popular experimental approach to protocells is to use membranous lipid vesicles as candidate compartments \shortcite{Chen2010}. For this reason, a short primer on lipid vesicles is given in Box 1. Other compartment media are also explored under the banner of `protocells', such as lipid micelles, ``membrane-free'' compartments like microemulsions (consisting of minute water droplets in oil, and the reverse), coacervates, droplets/coacervates covered in surfactants, gels and various inorganic membranes, to name but a few (see Fig. \ref{fig:ch1_protocell_architectures} and also \shortciteNP{Li2014} for a round up of some interesting non-lipid approaches to protocells). Table \ref{table:ch1_compartment_reasons} is biased toward phenomena enabled by lipid vesicles, but most of the points raised can generalise to different types of self-assembled compartment. Therefore, points in the table are labelled as `V' if they apply specifically to vesicles, `C' if they apply specifically to non-vesicle compartments and `V,C' if they should be applicable to all types of self-assembled micro-compartment.

\clearpage

%
%
%
%
%
%
{\rowcolors{2}{gray!10}{white}
\bgroup
\def\arraystretch{2}
\begin{longtable}{|p{0.5cm} p{13cm}|p{1cm}|} 
\specialrule{.3em}{0em}{0em}
\rowcolor{gray!40}
\multicolumn{3}{|p{15.35cm}|}{Self-assembled compartments (C), and especially lipid vesicles (V), are generally deemed relevant in the origins of life for the following reasons:} \\
\hline
1 & \textbf{Compartments, including lipid vesicles, readily self-assemble in aqueous solution} (given appropriate conditions). & V,C \\
2 & \textbf{Meteorites similar to those that bombarded the early earth have been found to contain lipids that self-assemble into vesicles} e.g. the Murchison meteorite \shortcite{Deamer1997}. & V \\
3 & \textbf{Vesicles obey the principle of continuity}. All extant cells use bilayer phospholipid membranes as a central part of their compartmentalisation (in conjunction with other structures, like the cell wall), and so it is logical to assume pre-life also used some type of lipid membrane. Additionally, giant unilamellar vesicles self-assemble in the size range of modern bacteria. & V \\
4 & \textbf{Vesicles are generally permeable to a variety of small molecules, when composed of simple lipids}, meaning that the first protocells could have had a `heterotrophic' lifestyle, absorbing the majority of their required nutrients from the environment \shortcite{Deamer2008,Mansy2010}. & V \\
5 & \textbf{Compartments enable an `isolation effect'}, keeping an internal chemistry protected from environmental parasites and inhibitory factors \shortcite{Ichihashi2014}. By the same token, \textbf{compartments entrap key molecules}. Macromolecules made inside compartments would not be membrane permeable, and would become concentrated inside. & V,C \\
6 & \textbf{Compartments represent units of selection} that co-localise populations of distinct molecules e.g. sequence-based biopolymers, or more general compositional information \shortcite{Segre2000}. Also, being held together by weak bonds, self-assembled compartments can crucially \textbf{grow by addition of components, and divide}, transferring all or part of their contents to their offspring. & V,C \\
7 & \textbf{Compartmentalisation allows the existence of protocell assemblies or communities} where symbiotic processes and different levels of selection operate \shortcite{Stano2014,Stano2007} & V,C \\
8 & \textbf{Vesicles (and micelles) can proliferate in an exponentially increasing, autocatalytic way}, given the correct conditions \shortcite{Stano2010}. & V \\
9 & \textbf{Compartments, particularly vesicles, create a phase separation in an otherwise homogeneous medium} \shortcite{Morowitz1981,morowitz1992}. This permits differences to exist between the inside and outside of the system, like concentration gradients, pH differences (e.g. \shortciteNP{Chen2004b}), and oxidation-reduction differences. These differences could be used as \textbf{energy storage mechanisms} eventually supporting endothermic transformations and against gradient active solute transport (as critically utilised in all extant cells; \shortciteNP{Skulachev1992}). & V,C \\
10 & \textbf{Lipid membranes, especially bilayers, can act as promoters and regulators to a variety of chemical reactions} \shortcite{Walde2014}. For example, lipid membranes can promote polymerisation reactions, which put together building block monomers into larger macromolecules. & V \\
11 & \textbf{Compartments, particularly vesicles, provide a scaffold of diverse reaction environments in close quarters}, e.g. water and oily phase reactions, and surface chemistry. These different types of chemistry can interface directly with each other and promote reactions that are otherwise not possible \shortcite{monnard2011}. & V,C \\
12 & \textbf{Vesicles provide a scaffold in which molecular nano-machines can be anchored} such as protein catalysts, receptors, ion pumps, mechanosensitive channels, flagella motors, etc. These machineries grant new possibilities to the unit as a whole. & V \\
13 & \textbf{Compartments can help generate phenotypic diversity.} The fusion of two protocell aggregates, each with different initial functions could, for example, lead to a new protocell aggregate inheriting all the components of the parents and combining them into a new organisation, giving a new overall function: i.e. they can support `aggregation generated novelty' \shortcite{rasmussen2009a}. & V,C \\
14 & \textbf{Self-assembled compartments typically contain micro-sized volumes making encapsulated reactions more favourable} \shortcite{Fallah-Araghi2014, Matsuura2012}, particularly when the micro-environment has molecular crowding \shortcite{Sokolova2013,Ellis2001}. & V,C \\
15 & \textbf{Compartmentalisation enables motility of a whole, collective, chemical system} into environmental regions with different conditions, resources, and competitors \shortcite{Hanczyc2011,Gutierrez2014}. & V,C \\
16 & \textbf{Vesicles could have been favourable for the development of multi-step and multi-stable chemical systems.} Vesicles can host nested vesicles inside, giving rise to heterogeneous compartmented reaction spaces. In this way, different spatial regions of a vesicle can perform different, but linked chemical tasks \shortcite{Elani2014} and potentially, reaction systems with more stable states can be generated \shortcite{Harrington2013}. & V \\
\specialrule{.3em}{0em}{0em}
\caption{\textbf{General Reasons for the Relevance of Self-Assembled Compartments in the Origins of Life.}}
\label{table:ch1_compartment_reasons}
\end{longtable}
\egroup
}
%
%
%
%
%
%


Although Table \ref{table:ch1_compartment_reasons} focusses on advantages of encapsulating chemical systems inside (or as part of) self-assembled enclosures, it should be noted that compartmentalisation is not without its corresponding difficulties. Vesicles in particular are only stable in a limited range of environmental conditions. Excessive temperature, extreme pH, high pressures, and the presence of salts (e.g. in oceans), are all factors negatively affecting the stability of vesicle bilayers \shortcite{Thomas2007,monnard2011}. Also, limited permeability of compartment membranes presents potential bottlenecks for an internal metabolism, as nutrient accessibility and dissipation of internal waste products are hindered. Vesicles, additionally, are susceptible to osmotic lysis or sudden burst, if the internal turgor pressure becomes too great (see later). 

Nevertheless, in spite of these potential problems, extant life organises all of its molecular components and processes around lipid compartments, controlling the spatial distribution of critical functional species, and avoiding their loss by diffusion. Proposing a prebiotic scenario in which compartmentalisation takes place relatively early provides room and time for chemical systems to develop these control mechanisms from the bottom-up.

%
%
%
\section{Basic and Hybrid Protocells}
\label{sec:1_4}

Current approaches to protocells can be roughly divided into two categories: basic protocells and hybrid protocells. Both approaches work fundamentally from the `bottom up', i.e. the idea is to put together isolated components in a lego-like way to form integrated systems with cell-like attributes. 

Basic protocells are focussed around simple molecular (often prebiotically plausible) components and simple chemical processes. Typically, these models are used to demonstrate important first-principles effects, like for example how protocell membrane morphology is related to factors such as osmotic pressure or the presence of internal macromolecules. Basic protocells are also used to demonstrate primitive pathways to well-established cellular functions, such as sustainable division or adaptive movement.

\begin{figure}
\begin{center}
\includegraphics[width=15.5cm]{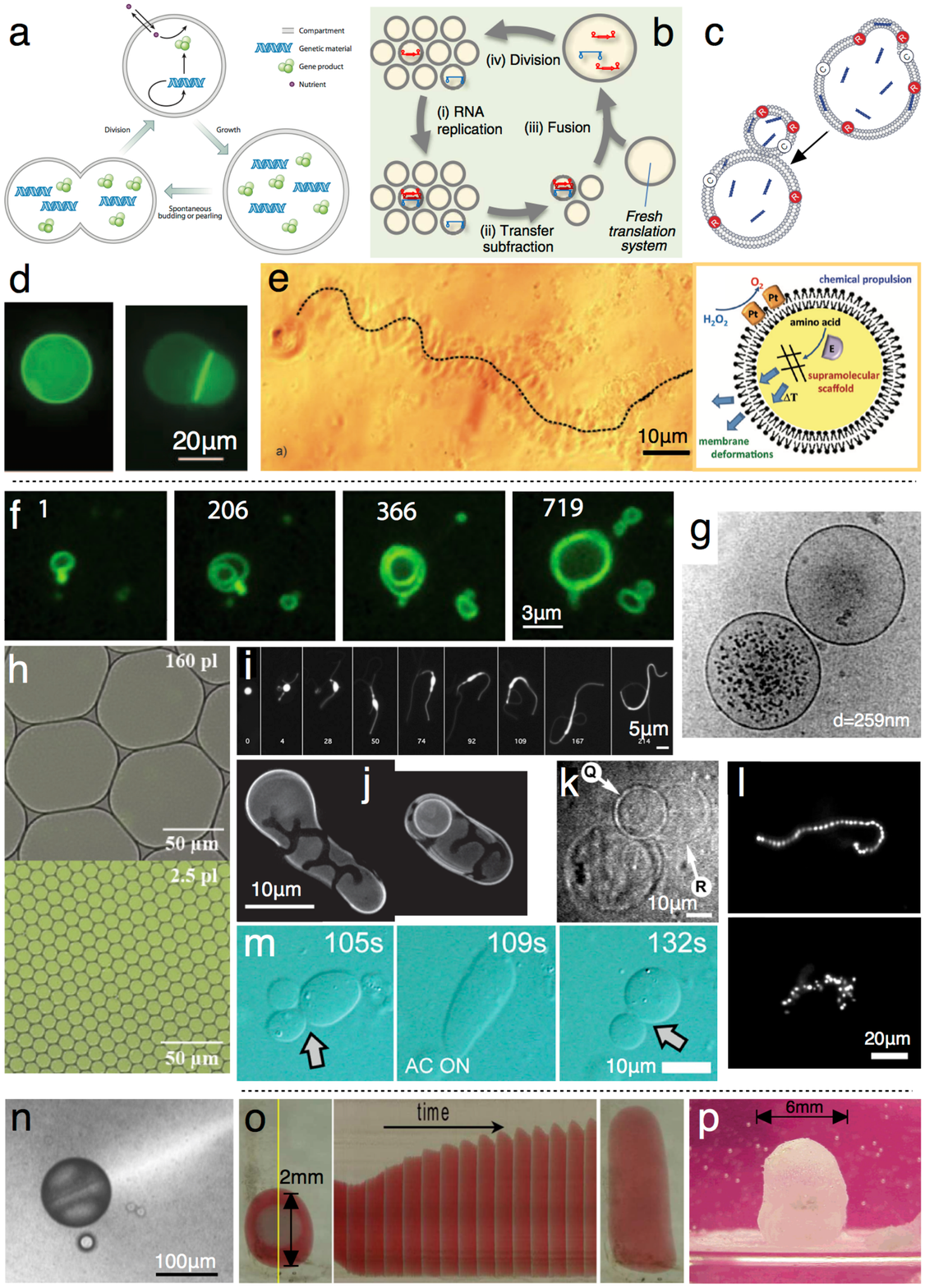}
\end{center}
\caption{
{\textbf{Diverse Experimental Approaches to Basic and Hybrid Protocells.}} See Table \ref{table:ch1_protocell_approaches} for legend and references. \hyperref[chapter:copyright]{ View Image Copyright Permissions.}
} 
\label{fig:ch1_protocell_approaches}
\end{figure}

Hybrid protocells, on the other hand, also incorporate biological machinery transplanted from existing cells. For example, in some cases, the whole DNA-RNA-protein transcription/translation system can be borrowed from organisms like {\em E. coli} and then encapsulated inside more primitive compartments. The use of hybrid protocells, although it may be `cheating' in a strict prebiotic sense, gives an opportunity to `jump ahead' to more advanced stages in the emergence of life, creating chimera systems that shed light on more developed properties of a cell, like the effect of micro-volume on protein synthesis, or the interaction of nucleic acids with lipid membranes. Figure \ref{fig:ch1_protocell_approaches} displays a diverse selection of basic and hybrid protocell approaches implemented to date in the lab.

%
%
%
%
%
%
%

%
%
%
{\rowcolors{2}{gray!10}{white}
\begin{table}
\centering
\bgroup
\def\arraystretch{2}
\scalebox{0.8}{\begin{tabular}{|p{0.5cm}p{14cm}p{4cm}|}
\specialrule{.3em}{0em}{0em}
\rowcolor{gray!40}
\multicolumn{3}{|c|}{\textbf{Hybrid Protocell Models ($\mu$m size scale)}} \\
\hline
(a) & Schematic for a minimal evolving RNA protocell, the central target of mainstream protocell research. & \shortciteNP{Blain2014,Szostak2001} \\
(b) & An RNA template system which evolves in water droplets over repeated generations. & \shortciteNP{Ichihashi2013} \\
(c) & Giant vesicles which reproduce themselves and amplify DNA. & \shortciteNP{Kurihara2011} \\
(d) & Cell-like bioreactor made from phospholipid vesicle encapsulating a cell-free expression system from {\em E. coli}. An internally synthesised $\alpha$-hemolysin pore embeds in the membrane and allows prolonged expression via increased nutrient accessibility. & \shortciteNP{Noireaux2004} \\
(e) & Chemically propelled phospholipid vesicle with cytoskeleton-like interior. & \shortciteNP{KrishnaKumar2011} \\
\hline
\rowcolor{gray!40}
\multicolumn{3}{|c|}{\textbf{Basic Protocell Models ($\mu$m size scale)}} \\
\hline
(f) & Phospholipid vesicles which grow and divide into daughter vesicles that continue to grow and divide. & \shortciteNP{Hardy2015} \\
(g) & `Super concentration' effect, whereby proteins congregate at high concentrations in certain newly formed vesicles, leaving others empty. & \shortciteNP{Luisi2010} \\
(h) & Droplets of micrometer size enhance internal chemical synthesis by surface chemistry. & \shortciteNP{Fallah-Araghi2014} \\
(i) & Protocell competition. Vesicles containing a phospholipid fraction steal lipid from surrounding vesicles and assume filamentous shapes that divide easily. & \shortciteNP{Budin2011} \\
(j) & Giant vesicles that undergo complex shape changes in osmotic gradients.  &  \shortciteNP{oglecka2012} \\
(k) & Protocell proliferation by synthesis of smaller vesicles inside giant vesicles. Smaller vesicles pass through the giant vesicle's membrane to exit. & \shortciteNP{Takakura2003} \\
(l) & Protocell division induced by intense illumination. & \shortciteNP{Zhu2012} \\
(m) & Giant phospholipid vesicles containing macromolecules spontaneously assume a budded shape, ready to divide. & \shortciteNP{Terasawa2012} \\
(n) & An oil droplet `intelligently' moves around its environment, due to an internal convective cycle driven by a chemical reaction at the droplet surface. & \shortciteNP{Hanczyc2014}; \shortciteNP{Hanczyc2011} \\
\hline
\rowcolor{gray!40}
\multicolumn{3}{|c|}{\textbf{Basic Protocell Models (mm size scale - visible to naked eye)}} \\
\hline
(o) & Growth of a droplet coupled to an internal Belouzov-Zhabotinsky reaction, a coupling between chemistry and mechanics. & \shortciteNP{szymanski2013} \\
(p) & Inorganic `cell' formed from pellet of calcium chloride in a solution of sodium carbonate. Semi-permeable membrane allows system to sustain itself far-from-equilibrium. & \shortciteNP{maselko2004} \\
\specialrule{.3em}{0em}{0em}
\end{tabular}}
\egroup
\caption{\textbf{Legend for Fig. \ref{fig:ch1_protocell_approaches}.}}
\label{table:ch1_protocell_approaches}
\end{table}
}

It is useful to note that the field of protocells is closely related to two underlying fields, also in their infancy, and which are also extremely relevant to origins research. The larger framework of which protocells are part is that of \emph{systems chemistry} \shortcite{RuizMirazo2014,kiedrowski2010}. Systems chemistry explores the \emph{dynamic properties} of complex chemical mixtures made of many different types of components engaging in myriad interactions. It can be seen as the `difficult cousin' of traditional chemistry, which instead focusses on deducing the \emph{structure} of individual types of molecules and on the study of isolated reactions in well-controlled conditions. Systems chemistry is the new discipline aiming to properly address the chemical infrastructures existing on the early prebiotic earth, and it is replacing the traditional metabolism-first vs. genes-first dichotomy with a more unified systems-oriented view \shortcite{DelaEscosura2015}. A sub-field of systems chemistry is that of \emph{structured media} \shortcite{Epstein2012,Showalter2015}. This sub-field investigates chemical reactions occurring in the presence of self-assembled structural entities, like microemulsions, droplet arrays and gels, paying attention to the new dynamical regimes which can result. For example, \shortciteA{Vanag2001} found that a standard reaction-diffusion chemical system has its pattern generation capabilities extended when it takes place in an oil populated by tiny water droplet compartments. 

Protocells are in turn a sub-field of structured media: the boundary between the two fields is blurry, but structured media tend to investigate reactions in solution pervaded with a multitude of simple self-assembled elements, whereas protocells focus more on individual instances of chemically active compartment systems, where the self-assembled compartment is more elaborate (e.g. a lipid bilayer, not just a micelle or phase difference). The emphasis in the protocells field is more on the behaviour of individual structures rather than on the patterning behaviour of an entire solution. Examples right on the boundary, which could either be classified as structured media or protocells,  include work done on the BZ oscillating reaction inside arrays of vesicles which signal to each other \shortcite{tomasi2014}, and the investigation of floating gel particles made from clay minerals as structures which can house and catalyse biochemical reactions \shortcite{Yang2013}.

%
%
%
\section[A Core Conception of Life Underlies All Protocell (and Origins) Research]{A Core Conception of Life Underlies All \\ Protocell (and Origins) Research%
\sectionmark{A Core Conception of Life Is Necessary}}
\sectionmark{A Core Conception of Life Is Necessary}
\label{sec:1_5}

In this short scientific introduction, we have been concerned with what protocells essentially are, how they are currently being investigated and several preliminary arguments for how they constitute a suitable vehicle for crossing the non-living and living worlds.

However, because the fundamental aim of protocells is to enlighten how biological cells came about, it follows that all protocell research programs are inevitably immersed in some general conception of what the phenomenon of life entails. By themselves, protocells are only like a blank piece of paper: they give a plausible object with which to construct a story about abiogenesis, but they do not dictate what that story should be. Rather, it is the core conception of life held knowingly or unconsciously by an investigator that determines the exact experiments she will involve the protocells in, and how she will interpret the results as being relevant to a larger picture. Without a conception of life, the ultimate goal of a protocell research program would be undefined.

Jack Szostak, a prominent figure in origins of life and protocell research, recently appeared to express a conflicting sentiment, stating ``Attempts to define life are irrelevant to scientific efforts to understand the origin of life.'' \shortcite{szostak2012a}. Szostak argued that a definition of life amounts to specifying a precise dividing line between chemistry and biology, whereas the emergence of life probably manifested as a temporally extended process, unfolding via a series of important transitions. Such transitions are cited as ``the true unknowns and subject of origin-of-life studies'', and it is argued that effort would be better spent in characterising the physical and chemical forces driving these transitions, rather than in defining life. This may well be a valid viewpoint, but what the article misses, however, is that even if an exact definition of life is irrelevant, a weaker {\em general conception} of the phenomenology of life is still required. It is this general conception which colours and biases what type of transitions will be searched for on the way toward a `living' cell. For example, by holding an evolutionary perspective, Szostak is inevitably biased toward searching for phenomena relevant to the evolutionary potential of protocells. 

Science is not performed in a conceptual vacuum, especially not origins of life studies. Instead of an ability to uncover an `objective truth', we are instead often constrained to find what we are looking for, as physiologist Claude Bernard once paraphrased ``The hypothesis ... had prepared my mind [and my predecessors'] for seeing things in a certain direction ... We had the fact under our eyes and did not see it because it conveyed nothing to our mind.'' (cited in \shortciteNP{gross1998}, p382). Chapter \ref{chapter:2} examines two different conceptual ways to characterise life, and argues that conceptualising life as autonomy, the way that has received by far the least attention, brings forward a relevant alternate research program in which protocells are the key players in the origin of life.

\chapter{Conceptualising Life: The Evolution--Autonomy Dichotomy}
\chaptermark{The Evolution--Autonomy Dichotomy}
\label{chapter:2}

%
%
%
\sectionmark{Defining Life}

If we observed a pink gooey blob on the surface of Mars, under what criteria would we classify it as alive or not?\footnote{A phrasing of `What is life?' by Inman Harvey.}

As human observers, we recognise earthly life without having to reference an explicit definition. In the here-and-now, we identify other living organisms heuristically. Given a split-second glimpse, or a freeze-frame photograph, we may suspect that a system is living if its physical embodiment has special superficial features such as intricate `organic-like' shapes, patterning, fluids, and soft interfaces (which all hint at a complex underlying biochemistry). Given a short observation time window, our confidence to attribute `life' is further increased if we can discern a state change in either the whole system and/or some of its internal/external parts, particularly if this state change seems asserted on behalf of the organism, rather than a direct result of prevailing environmental conditions. The most obvious form of state change perceptible by our human senses is movement, and thus the display of goal-directed movements (such as feeding, recoil, avoidance, hunting, etc.) from organic-like structures is typically how we identify life in brief encounters. Over longer time scales, we would also be inclined to identify a system as living if we witnessed it reproduce itself, or witnessed it carry out a gradual morphogenetic or metamorphic development process.

Our mental `checklist' approach may be sufficient to recognise living individuals in daily life, but it does not constitute a formal definition of the \emph{essential} qualitative difference between living and non-living systems.\footnote{Our heuristic approach is sometimes faulty in fact, and we attribute life when there only exists a superficial resemblance e.g. with life-like robots, or with a piece of dust twitching in the breeze.} A formal definition distilling the \emph{core logic} of living systems, if possible, would be much preferred to a list of heuristics, for it would not only allow earthly life to be recognised, but should also go some way to explaining how living systems universally generate their attributes from a deeper organisational scheme. Such a definition of life could help identify life in other parts of the universe, could establish a strict criterion on what could be considered alive in ethical debates, and, in the ideal circumstances, could act as a unifying principle to transform biology from an observational, descriptive science into an exact theoretical one. 

Life, however, has traditionally resisted a `water tight' consensus definition, despite a multitude of attempts throughout history (see \shortciteNP{Popa2004}). Whereas some systems easily yield to a characterisation of their core logic, for example a ``triangle'' is precisely ``a plane figure with three straight sides and three angles'', the logical scheme common to living systems has been much harder to pin down. Definitions often end up being too broad and abstract, admitting trivial cases of non-life, or too specific and narrow, excluding obvious cases of life. So, what type of problems stand in the way of and complicate a universal definition of life? A few key examples will be discussed below. 

A first problem is what status cells should take in a definition of life. The cell is the lowest complexity physical unit unanimously agreed to be alive, but does this necessarily mean that `defining life' should be equivalent to `defining cellular life'? In other words, are non-cellular forms of life possible? For example, are viruses alive? \shortciteA{Forterre2010} explains that, rather than being branded as a pure parasite, viruses can be seen through another lens as a strange type of organism that only only exists in cellular form {\em intermittently}. Viruses can be seen to propagate in intracellular space the form of non-metabolising virions, and then manifest their `organism' form once inside the cytoplasm of cells where they set up complex virion factories.\footnote{These virion factories replicate the genome of the virus and produce more virions to propagate the virus further.} Furthermore, some viruses can completely replace the genome of a host cell with their own (effectively giving a virus with the appearance of a cell), and other viruses can become `ill' when their viral factory is hijacked by yet further viruses. 

Cells and organisms themselves can also move into the grey area between the inert and living. For example, \shortciteA{Ganti2003a} highlights that dried seeds and frozen organisms are dormant configurations of matter not currently `alive', but with the potential for life, given the right conditions. Indeed, some microbial organisms perform the living-inert transition routinely, transiting into a non-metabolising state as a survival strategy whenever environmental conditions become too harsh ({\em cryptobiosis}, see \shortciteNP{Tsujimoto2016}). At the other end of the spectrum, simple chemical systems resulting from reactions and diffusions of molecules can display life-like properties, such as division \shortcite{Virgo2011}. These borderline cases mean that the set of life is at least a fuzzy set with different possible levels of membership.

Even disregarding borderline cases, `normal' metabolising life is also diverse and rife with counterexamples ready to challenge any comprehensive definition. For example, some cell types cannot reproduce (red blood cells have no nucleus at their final stage of development), some eukaryotes (called {\em coenocytes}) have division of their internal nuclei that is not accompanied by division of the cytoplasm, and some cells can even survive as just cytoplasm without a surrounding lipid membrane for transient periods \shortcite{Kim2001}.

Another issue, discussed further in this Chapter, is that living cells have both an individual and collective dimension to them. A cell exists as an individual in its own right, but at the same time it performs activities which transcend itself, such as dividing and passing on information in a heritable way such that an evolutionary lineage can be established (see the discussion of G{\'a}nti's life criteria below). In a definition, should life be defined at the individual or collective dimension, or both?

Finally, not helping matters is that the only example of life known about so far is terrestrial cellular-based life. This `sample size of one' problem means that it is generally not know to what extent the biochemistry of earthly living systems is necessary, and to what extent historical accidents have played a role in shaping their structure and organisation, for no other independent examples exist.\footnote{Some defend the possibility that there could exist undiscovered groups (or even whole ecosystems) of micro-organisms right here on earth that could employ a weird chemistry and hence constitute a qualitatively different `alien' example of life. This `shadow biosphere' hypothesis is discussed by \shortciteNP{Davies2009}.} Could life exist in different solvents, or use different molecules for catalysis or for information storage?

Precisely due to reasons like these, Harold Morowitz has referred to the definition of life as an `intellectual maze' \shortcite[p4]{morowitz1992}. Some people have seen little point in entering this maze because they believe life emerged along a continuum and thus a `bright line' cut-off between living and inert is meaningless (as in Section \ref{sec:1_5}). Others think that biological theory is not yet advanced enough to propose such a definition \shortcite{Cleland2002}. And, even if an impregnable definition of life did appear one day, some could argue that it may serve little practical purpose, for being able to {\em define} a phenomenon does not necessarily imply that a full or even partial {\em explanation} of that phenomena has been reached, only that some particular idiosyncrasy of it has been identified.\footnote{In order to be possible, a life definition would likely have to target one specific subgroup of life i.e. {\em actively metabolising} and {\em singular} cellular systems. Further, in order to be concise, a definition of life would undoubtedly have to be hierarchical, i.e. it would need to make use of specific terms loaded with meaning that require whole definition trees themselves (e.g. like the definition of detailed concepts in mathematics do).} In light of the these problems, it has been suggested that synthetic cellular life, if eventually created in the lab, may be best identified via a practical Turing-style imitation game scenario where living cells, not humans, are the judges \shortcite{Cronin2006}.

\begin{figure}[!b]
\begin{center}
\includegraphics[width=16.5cm]{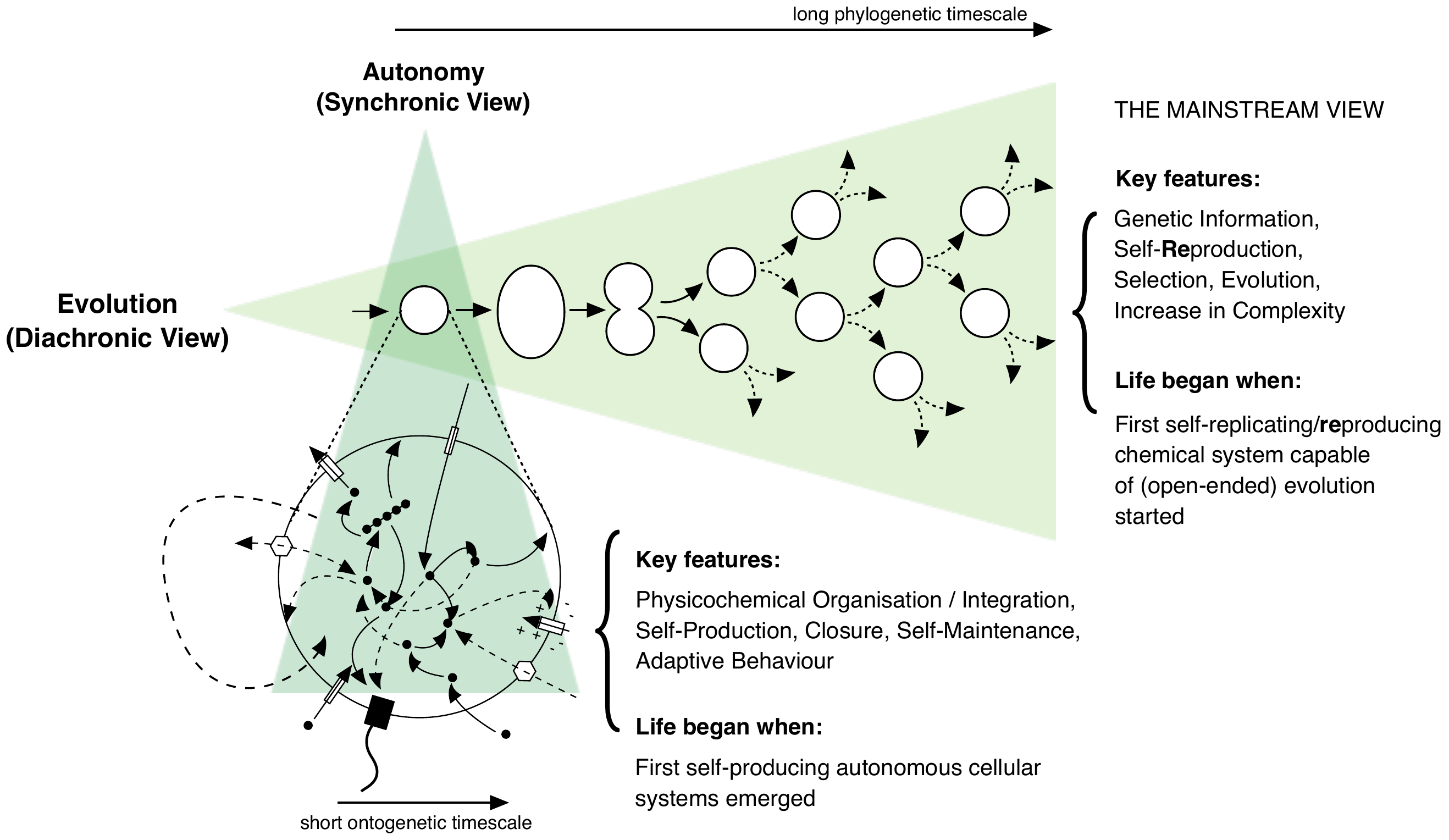}
\end{center}
\caption{
{\textbf{The Evolution-Autonomy Dichotomy.}}
}
\label{fig:ch2_dichotomy}
\end{figure}

Nevertheless, and as raised in Section \ref{sec:1_5}, the {\em general constellation} of ideas we perceive as meaningful and fruitful for investigating living systems is undoubtedly important when trying to synthesise life. Such a constellation sets a makeshift direction for enquiry, whose path can be subsequently modified as new discoveries are made. In the absence of a unanimously agreed definition of life, two distinct general conceptions of life have emerged in the literature (Fig, \ref{fig:ch2_dichotomy}). The first view is diachronic, or `across time'. It identifies life with the historical evolutionary process that gave rise to it, and views self-reproduction and evolution as being the primary properties of living systems enabling them to engage in population-level competition and selection. The second general conception of life is synchronic, or `in time'. It sees life as more connected with the organisational logic underpinning the operation of individual living systems over their short lifetimes, and holds that self-production, self-maintenance and adaptive behaviour are the primary properties of living systems.

These `evolution' and `autonomy' conceptions of life, as they will be called here, are two poignant aspects of biological cells, and cells typically embody {\em both} dimensions simultaneously.\footnote{\shortciteA{Ruiz-Mirazo2004} in fact combine both dimensions to attempt a universal definition of life: living beings, they say, are `autonomous systems with open-ended evolution capacities'.} However, most researchers in the fields of origins of life or artificial life end up gravitating toward one or the other of these conceptual poles, leaving a vacuous chasm in the middle. Therefore, the duo exist as almost opposing schools of investigation.\footnote{It is noticeable, for example, that past International and European conferences on Artificial Life have always had two seemingly orthogonal tracks: one focussed on population evolution, and the other on individual behaviour (see \shortciteNP{Aguilar2014}).} This artificial separation between the life conceptions is largely due to a difference in timescales, the role of the individual in each and hence the different theoretical tools and concepts employed in each. Researchers more biased toward an evolutionary perspective are concerned with the appearance of traits in populations of individuals over long trans-generational timescales, and the existence of these individuals is taken for granted. On the other hand, researchers biased toward the autonomy conception investigate the particular type of far-from-equilibrium physicochemical organisation that allows individual living systems to robustly distinguish themselves from their environment (and at the same time, respond adaptively to changes in it), over comparatively short time windows.

The `evolution' and `autonomy' schools of thought in fact actually closely mirror the {\em potential life criteria} and {\em absolute life criteria} classes, respectively, introduced by Tibor G{\'a}nti in the 1970's. As part of his Chemoton theory (see Section \ref{sec:2_2_3}), G{\'a}nti aimed to clarify how the units of evolution were related to the units of life, and made the pertinent point that living cells have attributes existing in two distinct classes. Cells have to perform essential activities in the here-and-now, in order to stay viable, and these he called these the {\em absolute criteria}. Cells need to (i) remain as distinct individuals, (ii) build their own structures by metabolising nutrients, (iii) demonstrate stability despite internal and environmental fluctuations, and (iv) must have their metabolism, development and evolution regulated by an information carrying subsystem. In addition to these essential prerequisites, G{\'a}nti noted, cells must perform activities which are not necessary for their instantaneous existence, but that are instead necessary for the long term continuance of the living world of which they are a part. These {\em potential criteria} require living systems to (i) be capable of growth and multiplication, (ii) have hereditary change with a capacity for a long term increase in complexity (`open-ended evolution', see later), and (iii) to be capable of dying so that their composing materials can be recycled and reused within the wider life system. G{\'a}nti argued that systems that only evolve tend to satisfy some of the potential criteria, whereas systems that live typically span and satisfy both sets of criteria, but minimally are required to satisfy the absolute criteria.

Hence, G{\'a}nti's view was that bare evolving systems, like mixtures of replicator molecules, were strictly not alive. Nevertheless, the mainstream trend is currently to identify life more with the `evolution' point of view and to marginalise or forget completely about the `autonomy' aspect. Although the autonomy conception has deeper historical roots, in the scientific era immediately proceeding Darwin, most people seem more comfortable thinking in a diachronic way, seeing self-reproduction and evolution as the primary defining features of life. This view transfers into the protocell world. Protocells are generally conceived as passive membranous containers whose function is to segregate and aid replicating genetic molecules.

The remainder of this Chapter reviews the `evolutionary' (Section \ref{sec:2_1}) and `autonomy' (Section \ref{sec:2_2}) perspectives in detail, hence looking a life through diachronic and synchronic lenses respectively. Even though current mainstream protocell research adopts the evolutionary view as its conceptual basis, the aim of this Chapter is to bring the autonomy perspective to the fore and explain how this marginalised conception of life defines a vital, but until now largely missing, research agenda for protocells (Section \ref{sec:2_3}).

%
%
%
\section[Diachronic Lens: Living Systems are Self-Reproducing Systems with Potential for Open-Ended Evolution]{Diachronic Lens: Living Systems are \\ Self-Reproducing Systems with Potential for \\ Open-Ended Evolution%
\sectionmark{The Evolutionary Perspective}}
\sectionmark{The Evolutionary Perspective}
\label{sec:2_1}

Darwin's theory of evolution by natural selection stands as one of the great unifying principles in all of biology. The theory outlines one important mechanism helping to build a picture of how cellular life, starting from a common ancestor (the so called `LUCA', around 3.5-3.8 billion years ago) subsequently diversified, resulting in the grand tree of life whose branches host the incredible biodiversity that we observe today (around 1.4-1.9 million species).\footnote{Evolution by natural selection `helps' to build the picture, because as \shortciteA{Gould1994} points out in his critical review, natural selection is only a principle to help explain local adaptations. Natural selection is {\em not} an engine inevitably driving towards a general increase in the complexity of species over time, although it is often depicted that way. Natural selection does not explain increases in complexity. \shortciteA{Szathmary1995} have suggested that increases in complexity are achieved through major evolutionary transitions.}

Evolution by natural selection comes about in a population of interacting individuals or `units', whenever those units (i) have phenotypic variations (varying morphological or behavioural characteristics) which in turn (ii) lead to differential fitness (different probabilities of survival and hence reproduction) and where (iii) the fitness of parents is somehow partially inherited by their offspring \shortcite{Lewontin1970}. Over time, individuals better adapted to coping with their environments survive longer and reproduce more (their successful characteristics perpetuated in their offspring), whereas less well adapted individuals are forced out of the population for indirect reasons (e.g. they ineffectively compete for limited food and/or space) or for direct reasons (e.g. they are actively preyed upon by higher fitness individuals).

From a historical viewpoint, a valid argument could be advanced that the ability for self-reproduction and a capacity for long term evolvability (or `open-ended evolution', see Section \ref{sec:2_1_2} below) are {\em necessary} characteristics of living systems. After all, organisms form part of a long-running evolutionary process, and it is this overarching process which connects them to both the past and the future.\footnote{Even though they are not capable of having offspring, sterile mules, grandmothers, red blood cells and neurones are isolated dead ends in an {\em overall} evolutionary process. These examples of non-reproducing living systems do not constitute complete counterexamples to evolution being important for life.}

However, the contemporary mainstream view leans toward the more extreme (and simplistic) position that the capacities for self-reproduction and evolvability are {\em sufficient} to define life.  ``Life is that which evolves'' is the bold opening statement of \shortciteA{Chen2012}, for example, and a similar sentiment is echoed in many current papers. \shortciteA{Trifonov2011} recently reported that the `average' definition of life, drawn from 123 definitions, would be (trivially) `Life is self-reproduction with variations'. Similarly a working definition of life referenced by many, and originally proposed by an internal NASA panel, is that `Life is a self-sustaining chemical system capable of undergoing Darwinian evolution' \shortcite{Joyce1994}, recently clarified by Gerald Joyce to mean `A chemical system capable of undergoing Darwinian evolution in a self-sustained manner.' \shortcite{Mullen2013}.\footnote{This clarification shows that the NASA definition is purely evolutionary. `Self-sustained' refers to evolution being self-sustained, and not a chemical organisation being self-sustained or self-maintained.} This mainstream position, as \shortciteA{szathmary2005} remark (and actually object against), is the same as saying that the units of life are the same as the units of evolution. This equality is convenient, since the units of evolution are relatively easy to define.

In fact, the mainstream evolutionary view of life often goes one step further, demoting or discarding completely the capacity for long term evolvability and promoting the more mechanistic and tangible aspect of self-reproduction. The reasoning is that self-reproduction at the individual level ultimately drives evolution at the population level, as the late Nobel Prize Cristian de Duve wrote in a {\em Nature} essay `The key notion in this theory [evolution] is reproduction. The rest follows obligatorily.' \shortcite[p581]{DeDuve2005}. Therefore, we arrive at the point where living systems are equated to be those which have growth, self-reproduction and proliferation as their basic characteristic. In this light, the single cell is effectively seen as (i.e. effectively {\em reduced} to) a self-reproducing system based on the replication of nucleic acids and the catalytic and functional abilities of proteins.

A brief clarification is necessary at this point with regards to the terminology being introduced. `Self-replication' is often used indiscriminately to refer to systems which bring about more of themselves, but in a strict sense, self-replication should only refer to systems which make {\em exact} copies of themselves (or an exact copy, within a small margin of error). When offspring resemble, but are far from being identical to their parents, `self-reproduction' should be used instead \shortcite{Dyson1999,Luisi2006,Szathmary1997}. Individual molecules can self-replicate, but cells always self-reproduce because the division process never creates perfectly faithful replicas: considerable stochastic differences always exist between parent and offspring.

Darwin himself did not write explicitly about the origin of life, but nor did he discount that a scientific exploration of origins was impossible \shortcite{Pereto2009}. Nevertheless, since the units of evolution are not necessarily restricted to the class of biological organisms\footnote{The triad of prerequisites is general, not restricted to organisms: `the principles can be applied equally to genes, organisms, populations, species, and at opposite ends of the scale, prebiotic molecules and ecosystems.' \shortcite[p2]{Lewontin1970}.}, a prevailing idea of the last century has been to extrapolate Darwinian evolution back towards the origin of life, proposing a precursor `chemical evolution' as the general mechanism to explain abiogenesis also. With life effectively conceptualised as a self-reproducing system, the emergence of life has been logically hypothesised as a {\em continuum from self-replication to self- reproduction}. At the beginning, individual replicator molecules (i.e. naked templates not requiring enzymes for self-replication) are considered to have evolved into more complicated chemical systems of replicators (templates catalysing each other, perhaps involving intermediate metabolic reactions too), which ultimately culminated in cellular-based reproducers, that contained replicating macromolecules (e.g. see \shortciteNP{Pross2011,Joyce2002}).

From this evolutionary perspective, the origin of life commenced when chemical evolution got started with the most basic kinds of molecular replicators.\footnote{It should be noted that evolution is often inextricably linked with template replicating molecules. But others have noted that (limited) heredity can also be ensured by systems without a template, such as in a `lipid world' of mixed composition vesicles \shortcite{Segre2000} or in compartmented autocatalytic sets with multiple `autocatalytic cores' \shortcite{Vasas2012}.} The popular RNA World hypothesis, championed by early origin of life theorist Leslie Orgel \shortcite{Orgel1973,Orgel2004}, proposes that the first self-replicating systems were based on RNA. Orgel stated ``It may be claimed, without too much exaggeration, that the problem of the origin of life is the problem of the origin of the RNA World, and that everything that followed is in the domain of natural selection.'' \shortcite[p100]{Orgel2004}. 

The logic for favouring a pre-living world of RNA replicators is the following. The metabolism of self-reproducing cells is an extremely complicated affair whereby polynucleotides specify the synthesis of amino acids into proteins, and at the same time proteins catalyse the formation of polynucleotides from nucleotides. The precise organisation of such a system would prohibit its sudden spontaneous emergence in prebiotic conditions, and so there should be some more minimal precursor system leading up to it. Of all the components involved, RNA is singled out as the most likely candidate to constitute a simple {\em one-molecule} precursor system because of its versatility:

\begin{enumerate}[label=\textbf{\arabic*})]
\item RNA plays a central role in current cell metabolism, acting as an information messenger sandwiched between the fixed and stable information archive of DNA, and the world of proteins, which put the informational content of the DNA into action.

\item RNA can weakly imitate the functional roles of both DNA and proteins: it is a nucleic acid like DNA and thus able to store information and potentially replicate by complementary base pairing. Also, at the same time, RNA can use its folded structure to catalyse a small but relatively diverse set of reactions.

\item RNA forms the catalytic site for peptide bond formation in cellular ribosomes, suggesting an ancient link between RNA and protein synthesis.
\end{enumerate}

Even though the origin of the first RNA molecules remains largely unclear (\shortciteNP{Shapiro2007,Shapiro2000} but see \shortciteNP{Powner2009}), the RNA World hypothesis is still actively developed \shortcite{Higgs2015,Pressman2015} and has historically spurred a number of developments. These include the theoretical concept of hypercycles, as a solution to the error-prone copying of long replicators \shortcite{Eigen1979}.\footnote{A hypercycle is a group of shorter templates that each self-replicate with high fidelity and help each other replicate (they catalyse each other in a closed cycle). The idea is that, rather than a long template holding all of the `information' and self-replicating unreliably, the same amount of `information' can be distributed amongst a mutually dependent group of short replicators that can each perform accurate self-replication.} Other theoreticians have also considered populations of replicators on two-dimensional surfaces \shortcite{Scheuring2003,Kamimura2014}, populations of replicators confined in small `honeycomb' enclosures in porous rock \shortcite{Branciamore2009} and populations of replicators inside compartments which divide \shortcite{Szathmary1987,Bianconi2013}. In general, however, the RNA World hypothesis has traditionally disregarded compartments, and has been silent on how replicators in free solution, or on surfaces, could have made the transition to organisationally complex cellular-based lifeforms.

%
%
%
%
%
%

\subsection[Experimental Mainstream in Protocell Research: The Ribocell]{Experimental Mainstream in Protocell Research: \\ The Ribocell}
\label{sec:2_1_1}

The current experimental mainstream in origins of life research inherits heavily from the conceptual landscape fashioned by evolutionary thinking and the RNA World, but in line with the current protocell line of research reviewed in Chapter \ref{chapter:1}, focusses more on the emergence of cellular, i.e. compartmentalised life. In addition to nucleic acid replicators, lipid compartmentalisation is considered as an important factor from the beginning, not just as a late innovation. Enclosing lipid vesicles are seen as key to helping the Darwinian evolution of an entrapped replicase because they spatially segregate replicator populations, forming units of selection, and allow successful mutant replicases to keep their selective advantage. The main protagonist Jack Szostak has written `Although a protocellular structure poses more problems initially, it is actually simpler to solve these problems up front rather than leave them till later when they could become completely intractable' \shortcite[pp1-2]{Szostak2012b}.

The mainstream project, seeded by the article `Synthesizing Life' in {\em Nature} over a decade ago \shortcite{Szostak2001} and followed up periodically since then \shortcite{Mansy2008,Ricardo2009,Blain2014}, has the long term aim to create a self-replicating nucleic acid protocell or `ribocell' (see Fig. \ref{fig:ch1_protocell_approaches}a).\footnote{Another notable attempt to synthesise a minimal type of evolving life was the Los Alamos Bug (\shortciteNP{Rasmussen2004}, popularised in {\em New Scientist}, February 2005), a proliferating lipid aggregate system that used a membrane embedded template molecule to synthesise membrane lipids using precursors and energy from light.} More specifically, the challenge is to establish the conditions under which a protocell compartment (a lipid vesicle) can grow and divide in synchronisation with the base pairing and strand separation of an entrapped RNA. The ribocell is thought to constitute an extremely minimal cellular-like reproducing system {\em within reach of complete laboratory synthesis}. Such a protocell would ideally divide into daughters, which then go on to do the same, hence establishing a primitive cell cycle. The argument is that once this cell cycle becomes established (and sustainable), then the chemical system has the golden ticket to start Darwinian evolution. From there, presumably, ``evolution is cleverer than you are'' (Orgel's second rule).\footnote{The long term evolutionary potential of the ribocell system is a problem presumably postponed until the first reproduction stage has been realised.}

Whilst the grand aim of the ribocell is to achieve the coordination of molecular genome replication and supramolecular compartment reproduction, current efforts are invested in tackling two pre-problems which need to be solved before the main problem can be properly attempted. The first problem is how non-enzymatic replication of the RNA genome comes about \shortcite{Szostak2012b}. The second problem, which is a {\em fundamental} problem in the origins of cellular life, is to establish how protocells started to divide in a semi-reliable way, given that they could not possibly possess the complex biochemical machinery (septal or `Z ring', see \shortciteNP{Weiss2004}) that modern cells use to accomplish this task. In this regard, the roles of simple physical forces and external stimuli are being investigated as primitive protocell division mechanisms \shortcite{Errington2013,Murtas2013,Svetina2009,Sole2009b,Hanczyc2004}. Another relevant aspect in this puzzle is that of {\em protocell competition}. Even if the ribocell could be made to do a lifecycle, there would be multiplication with heredity, but no selection as such. Therefore, how protocells might locally compete with each other is an area under active investigation \shortcite{Budin2011,Chen2004,shirtediss2014}, particularly when the genome inside the protocell could cause the compartmentalised system as a whole to become a better competitor \shortcite{Adamala2013}.

The most recent empirical advances on an integrated, self-replicating nucleic acid protocell have come in the form of partial solutions covering different parts of the grand scheme outlined above. \shortciteA{Ichihashi2013} report on a ``cell-like'' system based on micrometer water droplets in oil, whereby each droplet contains RNA templates which can replicate thanks to an added translation system. Over generations of manually fusing and dividing the droplets, they find that RNA replication (which has been introducing mutations in the RNAs) generally increases in efficiency and parasitic side reactions tend to decrease in efficiency. Therefore this system, whilst requiring significant human intervention, constitutes an empirical example of chemical evolution in cell-like environments. Working with lipid-based protocells, \shortciteA{Terasawa2012} have reported that deflated GUV vesicles containing an inert polymer in their aqueous pool (mimicking a macromolecule, like a nucleic acid) can readily assume a budded shape and divide. They postulate that a simple physical link could exist between the presence of genetic material and protocell compartment division, mediated by the effect of excluded volume. \shortciteA{Kurihara2011} have reported on another relationship apparently holding between macromolecules and compartment division: DNA macromolecules can speed up the spontaneous growth and division of lipid vesicles when lipid precursor is added because they embed in the lipid membrane and cause local deformations. Finally, \shortciteA{Hardy2015} have recently reported a vesicle system that can keep reproducing indefinitely (but lacking encapsulated templates).

Being well-defined, the `synchronisation problem' of simultaneous genome replication and compartment reproduction has also attracted theoretical modelling and analysis. By constructing minimal linear models of growing protocell systems composed of two variables, genetic material $X$ and compartment material $C$, \shortciteA{Carletti2008} have been able to prove analytically that such models will converge to constant division times. Furthermore, they demonstrate constant division times result regardless of whether the genetic material is anchored in the membrane, or is floating inside the enclosed aqueous volume of the protocell. Using a more sophisticated model with physically realistic parameters, \shortciteA{Mavelli2012} has performed stochastic simulations of the ribocell, mimicking the original \shortciteA{Szostak2001} idea in-silico in order to investigate parameter regions leading to synchronisation and factors introduced by fluctuations. Finally, \shortciteA{Mavelli2013} have recently focussed in detail on the compartment reproduction half of the problem. Adopting a more general perspective, they have derived a formula specifying what relationships must hold between different parameters in a realistic protocell model - regardless of whether the metabolism is template controlled or not - such that a sustainable division cycle becomes established, creating the conditions for population evolution.

%
%
%
%
%
%

\subsection{Von Neumann's Logical Organisation for Self-Replicating Machine Automata}
\label{sec:2_1_2}

Hungarian mathematician John von Neumann started an intriguing body of work generally related to the topic of systems that self-replicate/self-reproduce, well worth including as part of this section. Von Neumann worked quite independently from the fields of origin of life and definition of life\footnote{von Neumann did not comment on how his logical definition of a self-replicating machine was appropriate for a definition of life. Maturana and Varela, on the other hand (see Section \ref{sec:2_2_2}), made explicit that their logical definition of a self-producing `autopoietic' machine was a necessary and sufficient condition for life.} and he was enthralled by one abstract problem concerning the limiting capabilities of automata which can construct other automata. Von Neumann was interested in the answer to the general question ``Can the construction of automata by automata progress from simpler types to increasingly complicated types?'' \shortcite[p92, Question E]{VonNeumann1966}. According to \shortciteA{McMullin2000}, by posing such a question, von Neumann was actually pursuing the important problem of open-ended evolution and the minimum complexity machine that could achieve it.

A unit of selection with the capacity for open-ended evolution is able to evolve, but crucially is also able to {\em sustain evolutionary change in the long term}, without reaching a pre-determined upper limit of organisational complexity \shortcite{Ruiz-Mirazo2008b}.\footnote{For an example, \shortciteA{Dawkins1986} touched on the idea of open-ended evolution when discussing the `evolution of evolvability'. Here, he postulated that some types of organisms may not only be good at {\em surviving}, but also good at actually {\em evolving}. He noted that certain genotype encodings of his computational `biomorph' creatures are more pregnant with possibility when it comes to evolution e.g. the ability to specify body segments in a genotype leads to a multitude of forms not possible without segmentation.}  Biological systems are systems capable of evolving in an open-ended way, and it is this non-asymptotic feature of their construction which has permitted the emergence of the `endless forms' of life, including complex multi-cellular beings. If biological systems had possessed no open-ended capacity, evolution may have produced just well-adapted microbial communities on Earth (or, simpler, chemical assemblies) and then reached an impassable ceiling of complexity, a plateau. However, for its subtlety, the capacity for open-endedness is an often forgotten ingredient in evolutionary definitions of life.

\begin{figure}
\begin{center}
\includegraphics[width=16cm]{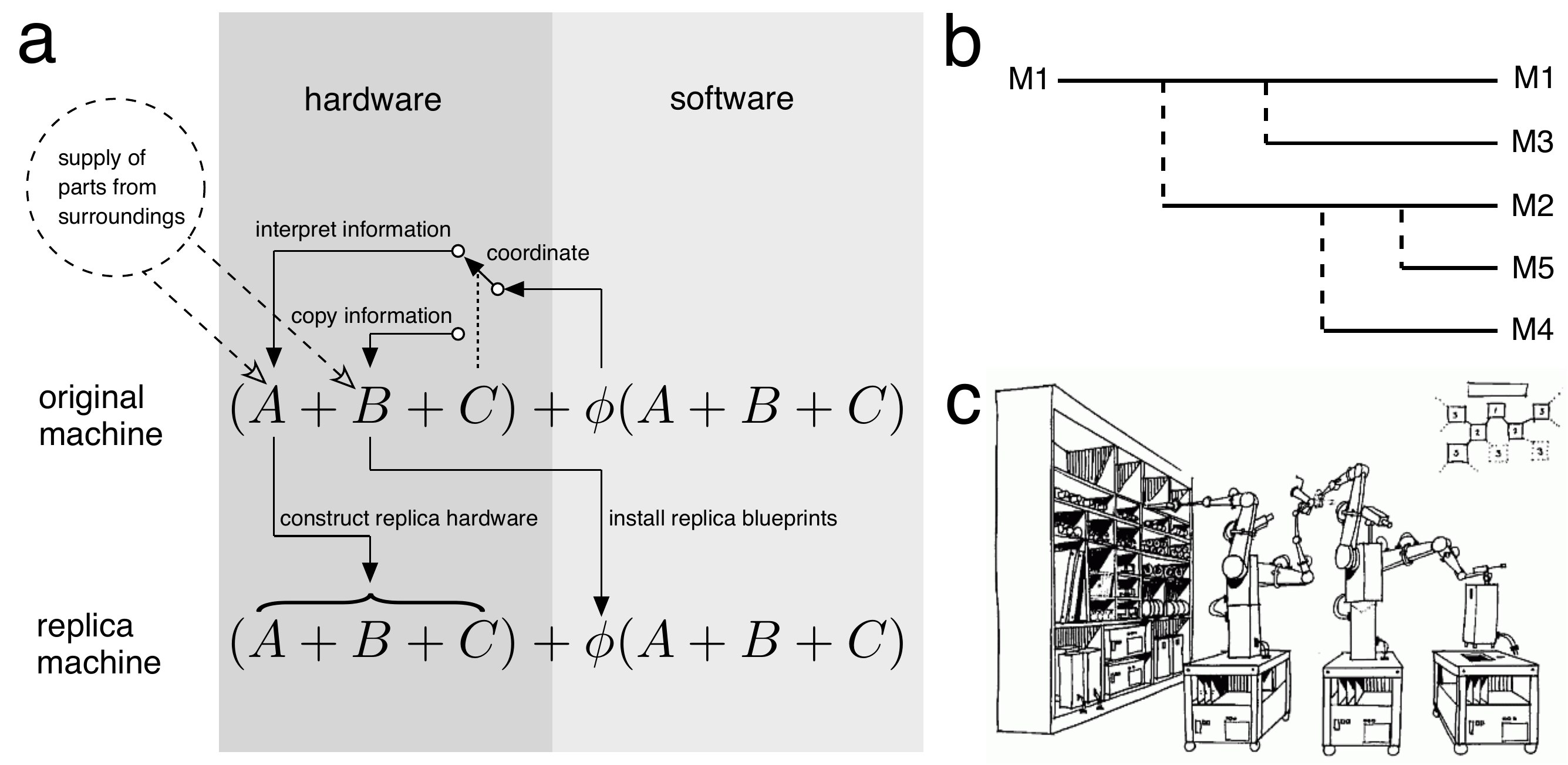}
\end{center}
\caption{
{\textbf{Self-Replicating Machines.}} (a) Von Neumann's logic of self-replication (see text for explanation). (b) Lineage resulting from self-replicating machines of the type Von Neumann envisaged. The parent machine constructs, from scratch, an independent copy of itself in the environment, and {\em both} machines survive to further replicate. (c) A hypothetical physical system implementing Von Neumann's logic: a series of self-replicating robots in a NASA cartoon (see \protect\shortciteNP[p45]{freitas2004}). Self-replicating robots are seen as a key solution to the economical colonisation of distant planets.
}
\label{fig:ch2_von_neumann_machine}
\end{figure}

Von Neumann's `Question E' cited above is interesting because, in some evolutionary lineages, there has been an increase in the complexity of biological molecular machines over time, whereas the type of electromechanical machines that we currently build through engineering generally involve a large {\em degeneration} of complexity at each stage of production. To construct a machine of a given complexity with our current technology, an even more complicated machine is required, and ultimately humans are required to construct the most complicated machines.\footnote{In fact, humans are often required as an essential part of many complex production processes, such as in microelectronics, or car production. Machines rarely independently create others. The machine currently coming closest is probably the Rep-Rap 3d printer: version 1 can print 50\% of its plastic parts (but not metal parts, nor electronics). Virtual physics worlds, such as cellular automata (CA), are more forgiving than the real world, and self-replicating `machines' of grid cells can be realised. Von Neumann himself demonstrated a CA design requiring 29 states for each cell. Since then, simpler designs like the self-replicating loop by Christopher Langton have been proposed.} In manufacturing, at the intersection between a degeneration and a growth of complexity lies self-replication, whereby a machine creates another in the precise image of itself. Therefore, leading up to Question E above, von Neumann posed a pre-problem: ``Can any automata construct other automata that are exactly like it? Can it be made, in addition, to perform further tasks, e.g. also construct certain other, prescribed automata?'' \shortcite[p92, Question D]{VonNeumann1966}. It was in answer to this question that von Neumann devised the essential logic of self-replication (Fig. \ref{fig:ch2_von_neumann_machine}a).

%

\begin{figure}
\begin{center}
\includegraphics[width=16cm]{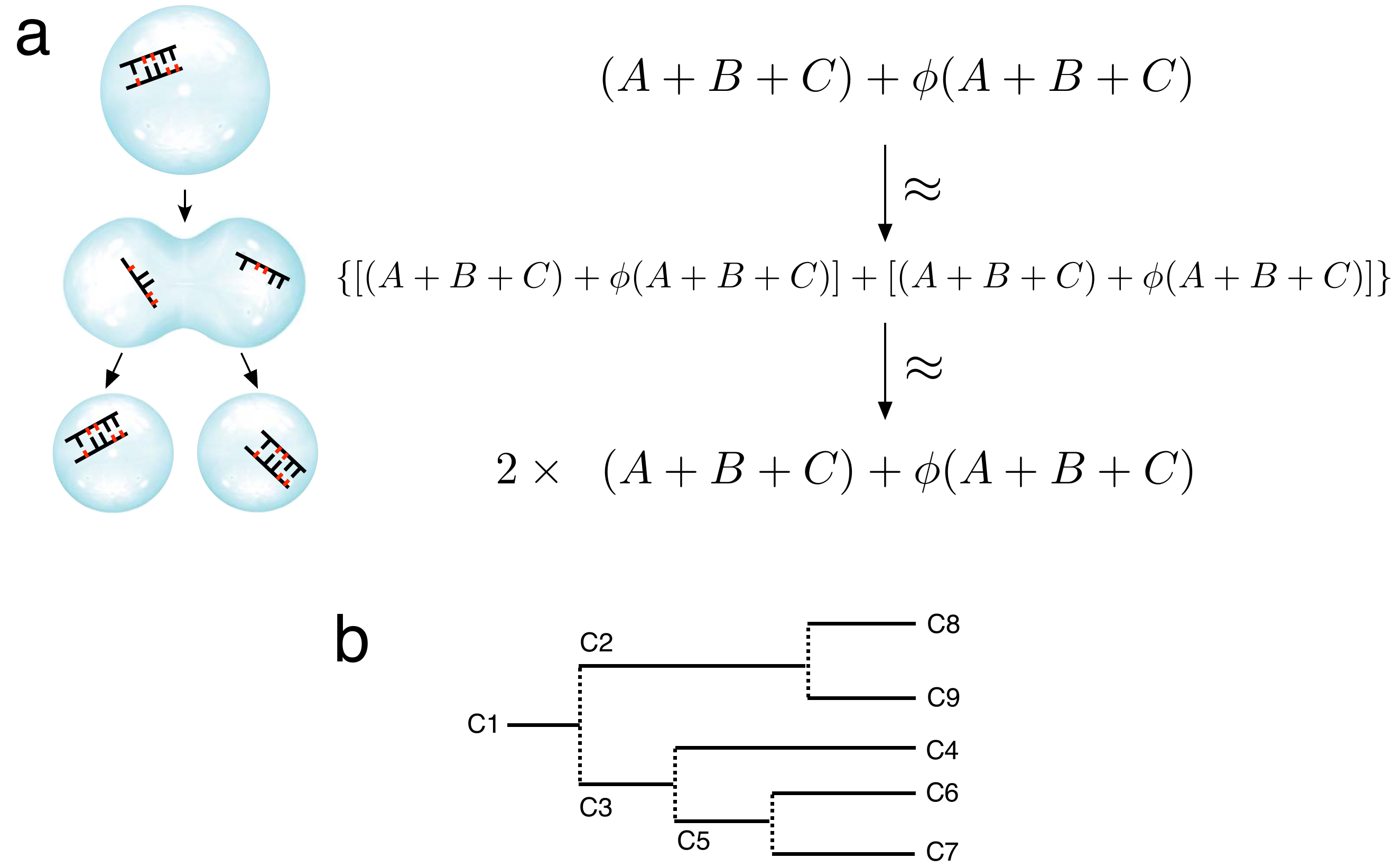}
\end{center}
\caption{
{\textbf{Mitotic Cell Division in Von Neumann Terms.}} (a) The cell grows into a conglomerate machine consisting of two joined original machines, and then divides when both are sufficiently complete. (b) The lineage of dividing cells is different to that of replicating mechanical machines (Fig. \ref{fig:ch2_von_neumann_machine}b). At each stage, the parent is lost, and two new daughters are formed, which may each have stochastic differences to the parent and to each other. Cell shapes adapted from \protect\shortciteA{Sole2007}.
}
\label{fig:ch2_von_neumann_cell}
\end{figure}

According to von Neumann, a self-replicating system must consist of four logical parts: a constructor A, a duplicator B, a controller C and blueprint instructions $\Phi$(A+B+C) that specify how to construct the machine of parts A, B and C from available resources. In the original machine, the constructor A {\em interprets} stored blueprint instructions $\Phi$ to construct a physical replica of itself by utilising material from the surroundings. The duplicator B in the original machine also {\em copies} the blueprint instructions into the replica without interpreting them. The controller C ensures that the constructor and duplicator are well-coordinated as the replica is formed, so that it eventually ends up with identical hardware and blueprints to the original machine. Without a constructor A, a machine cannot synthesise anything physically new, rendering self-replication impossible; missing blueprint duplicator B, a machine can synthesise a replica, but the replica will not have operating instructions, and thus will be forever static; lacking a controller C implies no coordination to the overall self-replication of material and information. It should be noted that the instructions $\Phi$ can make the machine carry out other useful behaviours unrelated to self-replication. Self-replication may be a crucial but infrequent aspect of the machine.

The self-replication logic proposed by von Neumann maps directly into the scenario of a chain of self-replicating robots which construct themselves from a parts store (Fig. \ref{fig:ch2_von_neumann_machine}c). A pre-existing initial robot fetches parts from a specially arranged parts store and uses them to construct the body of a second robot. On completing the construction task, the controller of the original robot then copies the software program into the second robot. On completion, the second robot is switched on, and both work together to construct the third robot, and so on.\footnote{Alternatively, the second robot could go and locate another parts store to construct the third robot from there.}

It is an interesting exercise to see how far von Neumann's self-replicating logic applies to mitotic cell division. In mitotic division, the cell does not build a copy of itself in the environment {\em de novo}. Rather, the cell takes in nutrients, grows, and then radically re-organises its internal structure, duplicating its essential components at each pole before dividing in the middle. Figure \ref{fig:ch2_von_neumann_cell} shows how von Neumann's replication logic could approximately apply to this process. Approximate signs on the arrows acknowledge that cell division is actually a case of {\em self-reproduction} and not exact {\em self-replication} (i.e. the daughter cells are not identical to the parent, nor to each other). In the cell, the information $\Phi$ corresponds to the DNA, the constructor A corresponds to the general system of cell metabolism, the duplicator B corresponds to the spindle apparatus and the controller C consists of the complex interactions in which the cell segregates its components, and then constricts the membrane. Mitotic cell division always results in the loss of the parent cell, leading to a different type of lineage than does machine replication (compare Fig. \ref{fig:ch2_von_neumann_cell}b with Fig. \ref{fig:ch2_von_neumann_machine}b).

For self-replicators of lower complexity (e.g. template molecules, or protocells), the essential elements of von Neumann's replicator logic often meld together. For example, for template molecules, construction, duplication and control all seem to be entailed by the process of complementary base pairing. Likewise, in early protocells, the information instructing the system how to divide is somehow embedded in the physical components themselves and their interaction with outside influences, rather than being explicitly encoded in a molecular medium. The same applies to G{\'a}nti's chemoton (see Section \ref{sec:2_2_3} later); it proliferates and is program controlled, but the program does not contain an independent description of how to build a new microsphere from scratch.

%
%
%
\section[Synchronic Lens: Living Systems are Autonomous Agents]{Synchronic Lens: Living Systems are \\ Autonomous Agents%
\sectionmark{The Autonomy Perspective}}
\sectionmark{The Autonomy Perspective}
\label{sec:2_2}

Even though living systems can be legitimately described as self-reproducing systems which evolve, this does not fully meet our daily experience of them. Indeed, when confronted with the pink gooey blob on Mars, applying the criteria of self-reproduction and evolution would actually seem like very indirect and distant ways to establish whether it was alive or not. Rather, in the here-and-now, what immediately strikes us about individual organisms is their apparent {\em purposefulness}, a property typically absent from other types of physicochemical system. During their ontogenetic lifetimes, in between bouts of self-reproduction, organisms act conspicuously as {\em autonomous agents}, forever reaching out to manipulate their world in the apparent pursuit of their own agenda and goals. 

Bacteria are the often cited example of biological autonomous agents, performing actions which tend to benefit their own survival and eventual reproduction. As Pross writes, ``bacteria when placed in a glucose solution gradient `swim' upstream to take advantage of the higher concentration of nutrient available there. Or, if glucose, the cell's primary energy source, is replaced by lactose, then the cell synthesises the enzyme necessary to break down the complex sugar into its constituent simple sugars, glucose and galactose.'' \shortcite[p725]{Pross2008}.\footnote{This is the lac-operon mechanism.} Indeed, prokaryotes are impressive metabolisers, able to switch their metabolic configuration to extract resources from almost anywhere. Apart from a flexible metabolic capacity, other remarkable cases of adaptive behaviour are also routinely demonstrated in individual cells and single-celled organisms. For example, individual cells of the red alga {\em Antithamnion} can show meticulously coordinated self-repair when its cell wall is torn into two pieces by a fine needle, and the gel-like single-celled amoeba can adjust its rate of reproduction to match food supply, use different engulfing tactics to suit different types of microscopic prey, and in some species, can even construct an intricate protective shell from inert materials found in the environment \shortcite{Ford2009}. In another example amongst many, {\em E. coli} have been shown to have abilities of environmental anticipation embedded into their metabolic networks \shortcite{Mitchell2009}.

In the intellectual landscape before Darwin, the purposeful or so called `teleological' character of organisms was the original hallmark used to differentiate living systems from the non-living. The question was, how could such teleology exist in a universe governed by objective natural law? How could there be {\em doings}, and not merely {\em happenings}? Early philosophers such as Aristotle, Descartes and then later Kant invested much thought on how teleology was manifest in the living, and to what extent organisms could be viewed as mere machines (see \shortciteNP[Section 1 therein]{Bedau2010}). In particular, Kant termed organisms {\em natural purposes} and hinted that their teleology may be brought about through self-organisation of matter, seeding the embryo ideas of self-organisation and emergence at the same time \shortcite{Weber2002}.

An autonomous agent can be roughly defined as a system (i) with some capacity for self-generating the rules that determine its behaviour, and that (ii) directs this behaviour toward ongoing activity beneficial for the survival of the system over the short or long term (like eating, hunting, hibernating etc.). Put more bluntly, autonomous agents are some special arrangement of matter that appears to be `selfish' \shortcite{Kauffman2000}. In a different phrasing, \shortciteA{Boden2008} writes that ``autonomy is self-determination: the ability to do what one does independently, without being forced so to do by some outside power...an individual's autonomy is the greater, the more its behaviour is directed by self-generated (and idiosyncratic) inner mechanisms'' (pp1-2). So autonomous agents are not mere slaves to their environment. Another key characteristic of autonomous agents, highlighted by \shortciteA{Collier2008}, is that they are often masters of {\em anticipation}, able to project probable (favourable or unfavourable) future states from present conditions and past history and then use this information to determine present dynamics accordingly. In such a way, the behaviour of autonomous agents is not directly tied to perception, but also guided by history and past associations made. Table \ref{table:ch2_autonomous_vs_other} summarises some of the striking differences existing between autonomous agents and other physicochemical systems.

{\rowcolors{2}{gray!10}{white}
\begin{table}
\centering
\bgroup
\def\arraystretch{2}
\begin{tabular}{|p{7.5cm}|p{7.5cm}|}
\showrowcolors
\specialrule{.3em}{0em}{0em}
\rowcolor{gray!40}
\textbf{Autonomous Agents} & \textbf{Other Physicochemical Systems} \\
\rowcolor{gray!40}
Paramecium cell, multicellular organism... & Orbiting planet, mountain, a dead organism... \\
\hline
Assert actions on their own behalf & Buffeted and determined by external forces \\
Can show different behaviours in the same conditions or the same behaviours in different conditions$^*$ & Same inputs generally give same predictable outputs \\ 
Actively enquire, constantly explore new possibilities & Respond only when prompted, do not expand their possibilities of interaction \\
Seemingly preoccupied with their own well being and existence & Indifferent to their own existence \\
{\em Follow only natural law}, but this is not the right level of description to capture their behaviour & {\em Follow only natural law}, and their behaviour can be described by its application \\
\specialrule{.3em}{0em}{0em}
\end{tabular}
\egroup
\caption{{\textbf{The Teleological Character of Autonomous Agents.}} The statement marked $^*$ is due to \protect\shortciteA[p60]{Suzuki2009}.}
\label{table:ch2_autonomous_vs_other}
\end{table}
}

Efforts to precisely define autonomy, and give a quantitative measure allowing different systems to be directly compared, have been harder to develop. Existing attempts have tended to equate autonomy with self-determination (as echoed by Boden above), granting higher autonomy values to systems that are more determined by their own processes rather than by external influences, i.e. autonomy is equivalent to the statistical {\em independence} of an agent from its environment \shortcite{Seth2010,Bertschinger2008}. However, as \shortciteA{Barandiaran2009} articulate, although this method has some merit, autonomous agents occupy a conceptually more tricky position than this.

For example, sometimes an agent will let its behaviour be largely determined by the environment and will take advantage of or `surf' prevailing conditions (e.g. a bird gliding on air currents), only to minimally intervene at key moments, changing its relationship to the environment in very subtle ways (e.g. a wing movement), such that a goal is achieved in the long run (e.g. intercontinental migration). In this context, statistical analysis has a difficult time in distinguishing the agent as the overall source of its own activity when the interventions it makes are so small, and the role played by the environmental forces is so large.

Evidently `autonomy', like `life', is a slippery concept of which we have some intuition, but which becomes difficult to formalise.\footnote{`Intelligence' or `complex system' are other major concepts we talk of routinely without possessing agreed and comprehensive definitions for.} However, some progress can be made into the type of physicochemical organisation which lies at the root of - and ultimately gives rise to - autonomous behaviour in living systems. This is discussed below for the remainder of Section \ref{sec:2_2}.

%
%
%
%
%
%

\subsection[The Behavioural and Constitutive Dimensions of Autonomous Agents]{The Behavioural and Constitutive Dimensions \\ of Autonomous Agents}
\label{sec:2_2_1}

The class of autonomous agents is evidently wider than all biological organisms, for on the rough definition given above, it would also encompass (at least) the artificial agents created by man, such as learning Khepera robots, Google self-driving cars and the Mars curiosity lander module, to name a few. Within this wide class, two qualitatively different types of autonomy can be usefully distinguished (see \shortciteNP{Froese2007,Moreno2008}).

All autonomous agents, regardless of whether embodied as exquisitely organised collections of biomolecules (living systems) or as coarse assemblies of metal, plastics and semiconductors (robots), implement a sense of {\em behavioural  autonomy}. To some extent, all regulate their coupling with the environment in an adaptive way in order to carry out their own agenda. Behavioural autonomy is generally termed `agency'.

In addition, the subset of biological systems also implement a much stronger form of autonomy, called {\em constitutive autonomy} or `biological autonomy', as used by \shortciteA{Moreno2015} and originally by \shortciteA{varela1979}. Living organisms not only regulate their interactions with the environment, but they also actively {\em distinguish} themselves from the environment at the same time: that is, they build and maintain their own bodies. This is an absolute necessity, since most of the complex molecules composing cells have extremely short half lives as compared to the lifetime of the organism, and therefore demand constant re-synthesis. Cells thus exist at the molecular level as remarkable `fluid machines', organised such that part of the operation of the machine is to continually replace the decaying or diffusing components of which the machine itself is made (self-maintenance). Thus, whilst a cell appears to persist as a unity on the macro-level, this persistence is only of {\em form} and not of {\em matter}: the microscopic structure is actually a constant turnover or flow of different molecules. This operative scenario stands in stark contrast to (current) man-made machines. Man-made, or so called `heteronomous' machines - all robots included - come pre-embodied as a linkage of coarse-grain components that simply persist over time. The fuel input to such machines is not used to construct or repair part of the structure, but is rather just used to move the pre-existing structure, with only the fuel itself undergoing chemical transformation.

The idea of constitutive autonomy is an important one, for it means that {\em the activity of a cell, is coincident with, and not detached from, its very being}. As \shortciteA{Kampis1995} phrases, ``In cells, the action results in its own blueprint'' (p96). In other words, what a cell is materially, and what it does behaviourally, are deeply and intimately connected.\footnote{For higher, multicellular organisms, the behaviour of the organism may be more decoupled from its material realisation, e.g. determined more by the action of a neural system.} The way cells continually build their structure, and synthesise new components in response to environmental changes is, naturally, deeply intertwined with the way they interact with their environments. For cells, the constitutive and behavioural aspects of autonomy are two inseparable sides of the same coin.\footnote{However, theoretical cell models nearly always presume the existence of the cellular body, and just focus on the dynamics of behaviour, or of internal chemical reactions. Few models explicitly tackle the issue of body construction, although this was the vision for modelling efforts when the European Conferences for Artificial Life were founded \shortcite{varelaBourgine1992}.}

Of crucial importance here is to realise that the specific type of autonomy possessed by biological systems goes all the way down to the root of the existence of the system itself. This `radical embodiment' of biological systems creates a unique state of affairs, a new paradigm. On the one hand, all man-made machines, including even robots that demonstrate some degree of behavioural autonomy, all bear the indelible mark of an external human designer. A robot may be able to learn new associations, but this capacity is somehow primed and requested by the external designer, rather than being of {\em intrinsic value} to the robot itself.\footnote{But, it is an open question, if beyond a threshold complexity of `wiring', intelligence could emerge in a robot based on a non-biological substrate (however, such a robot built at the macro-scale level would not be able to synthesise its own components, and so would be unable to grow or self-repair).}, The designer is the one who points out what a machine should find relevant about the world it inhabits. Thus, the function attained by the machine holds relevance for the designer but is alien to the machine itself: the machine materially persists regardless of whether performing its intended function or not.\footnote{Ray Bradbury's collection of short stories {\em The Martian Chronicles} \shortcite{Bradbury1950} has several poignant sci-fi examples of intelligent robotic devices carrying out an essentially meaningless existence after humans have departed. In one case, after the wipe out of earthly civilisation following nuclear war, intelligent houses continue to tidy away dust, wash dishes and make breakfast as usual, but for no purpose. In another case, a captain's android family go on mindlessly in the same routine after his death.} On the other hand, by continually synthesising their own embodiment, biological systems cannot be reduced to mere artefacts that simply extend the autonomy of a human designer, like machines can. Instead, there becomes a case to talk about the presence of `genuine' autonomy in biological systems, a concern about the world that is not externally imposed. In a biological system, say a cell, what is `good' and `bad' for the system, the so called {\em norms} of operation, are not decided a priori by an external authority, but instead spontaneously arise from the very organisation and ongoing dynamics of the system itself (see \shortciteNP{Barandiaran2014}). A cell determines what is relevant for itself and behaves in a way that generally ensures its continued material existence. Likewise, the distinguishable components of a self-generated biological system acquire specific roles or {\em functions}, not because these were stipulated by an outside agent, but because the components each assist, in some specific capacity, the maintenance of the whole system. As \shortciteA{Rosen1991} explains, the function of a component in a self-maintaining system can be ascertained by removing that component, and observing the behaviour of the whole system in its absence. From this perspective, the function of mechano-sensitive channels in bacteria is to help the organism survive large and sudden osmotic shocks.

The idea of constitutive autonomy (self-maintenance) stems from and is articulated in more detail by the theory of {\em autopoiesis}, reviewed in the following subsection.

%
%
%
%
%
%

\subsection[Maturana and Varela's Logical Organisation for the Living Cell: Autopoiesis]{Maturana and Varela's Logical Organisation for the \\Living Cell: Autopoiesis}
\label{sec:2_2_2}

The theory of autopoiesis was introduced in the 1970's by Chilean biologists Humberto Maturana and Francisco Varela as an attempt to describe the abstract logical organisation of the cell and to provide an essentialist definition of life \shortcite{autopoiesis_cog1980,Luisi2003,Fleischaker1988}. Although the main ideas behind autopoiesis can be traced back (at least) to Immanuel Kant and then identified in the writings of numerous scientists and philosophers since (including Claude Bernard, Hans Jonas and Aleksandr Oparin, for example\footnote{The reader should refer to \shortciteNP{Moreno2015} for a general account of the historical development of autopoietic ideas.}), Maturana and Varela were the first to take these ideas, shape them into a concise definition of life, and then rigorously pursue what such a definition implied for the autonomy of living systems. Indeed, it was this extra step, this connection with autonomy, that set it apart from other theories.

Autopoiesis is a theory situated within the field of {\em relational biology} (see \shortciteNP[Ch5]{Rosen1991}). Relational biology considers biological systems from the point of view of the abstract relations or logical organisation that connect the components together into a functioning {\em whole}, and forgets about the exact biochemical nature of these components. This top-down description thus captures the `forest' at the expense of blurring the `trees', and represents the exact opposite of the traditional analytical approach to biology. In the analytical or reductive approach, the workings of individual components or sub-systems are characterised in detail (the trees), but yet this information cannot be pieced together to form a fully integrated account of the organism (the forest).\footnote{Mathematical physicist Nicholas Rashevsky moved into relational biology in the 1950's, later in his career (and invented the discipline), after working for many years on detailed (and pioneering) analytical models of isolated aspects of organisms.} The two approaches are complementary, but require different branches of mathematics to express. Unconventional for some, relational biology is not concerned with time nor with material structure, nor even states, but is instead concerned with invariant organisation of processes, usually expressed as graphs, hypersets or category theory.

Autopoietic theory itself can be seen to arise from a number of basic and quite undeniable observations that can be made about a living prokaryote cell. A prokaryote is:

\begin{enumerate}[label=\textbf{\arabic*})]
\item a system localised in space, distinguishable from the environment by the person observing it
\item an open system far-from-equilibrium, absorbing nutrients and excreting waste
\item internally manufacturing the majority of components it requires and of which it is made, including highly complex macromolecules (DNA, RNA, proteins etc.) -  from simpler and less diverse materials in the nutrient medium (e.g. glucose and various salts).\footnote{The fact that a cell is able to synthesise all of its components is evidenced by cell division. Cell division can only result in two daughters resembling the parent cell only if a copy can be made of everything not immediately available from the environment. DNA could be argued to copy itself, but the fact is, it still requires enzymes from the cell to unwind its double helix, and help pairing of nucleotides to the separated strands making two new identical DNA molecules.}
\subitem \textbf{3a}) including its own boundary i.e. a surrounding semi-permeable phospholipid membrane which selectively absorbs nutrients and expels waste.
\item effective only as an organised whole. A prokaryote put through a blender results in a system of molecules with the same chemical composition as the original intact cell, but it is simply an inert paste, not a living organism.
\end{enumerate}

To formulate autopoiesis, Maturana and Varela asked the following pertinent question: what does a cellular machine {\em essentially} have to be in order to generate the basic observations above? 

By `to be', they were not searching for a detailed biochemical and physical description of a particular species of bacteria. Rather, they were interested in trying to define life in general and so they were seeking to identify the abstract relational logic a bacterium was implementing. As such, they focussed on how the cell must be globally organised as a series of processes\footnote{a ``process transforms something into something else (a chemical reaction transforms its reactants into its products; a transport process transforms the spatial distribution of a substance; friction transforms kinetic energy into heat)'' \shortcite{Virgo2011}.}, such that the system was overall a {\em self-producing} and {\em self-bounding} machine in physical space. They called this type of machine an `autopoietic' machine (greek: {\em auto}=self, {\em poiesis}=creating/producing), defined as follows: 

\begin{quotation}

An autopoietic machine is a machine organized (defined as a unity) as a network of processes of production (transformation and destruction) of components that produces the components which:

(i) through their interactions and transformations continuously regenerate and realize the network of processes (relations) that produced them; and

(ii) constitute it (the machine) as a concrete unity in the space in which they (the components) exist by specifying the topological domain of its realization as such a network.

\hfill {\small Maturana and Varela, De Maquinas y Seres Vivos (1972)}

\hfill {\small Reprinted in \shortcite[pp78-79]{autopoiesis_cog1980}}

\end{quotation}

Phrased in more colloquial language, an autopoietic machine is a peculiar type of ``biochemical factory that autonomously fabricates itself'' \shortcite{hofmeyr2007}.

Maturana and Varela hold that all cells are autopoietic machines.\footnote{Autopoiesis was developed in the context of the single cell. Multicellular living systems are also self-producing systems, but how the `first order' autopoiesis of the cell extends to such `second order' structures is less clear and not a well developed part of the theory.}$^{,}$\footnote{Stating that all cells are autopoietic machines is, on its own, equivalent to saying that autopoiesis is {\em necessary} for life and leaves room for life entailing something more besides (such as the capacity for open-ended evolution for example). However, Maturana and Varela also adopted the stronger position, that autopoiesis was {\em sufficient} to decisively demarcate living systems from non-living systems.} At the level of the whole cell system, there exists web of interrelated processes (e.g. catalysed reactions, molecular diffusions, energy transduction mechanisms, active transport mechanisms, etc.) whose operation (i.e. cell component dynamics) ensures the re-creation of the very same process network, and at the same time builds an enclosing supra-molecular membrane structure in physical space to give the process network an identity distinct from the surrounding milieu. 

On a first reading, autopoietic machines and Von Neumann's self-replicating machines (Fig. \ref{fig:ch2_von_neumann_machine}) seem to be almost equivalent ideas. After all, both are machines able to fabricate themselves. However, the ideas are quite distinct. Von Neumann's replicators fabricate {\em an independent replica copy} of themselves in the environment, and once assembled, a machine has a static enduring embodiment. Autopoietic machines, on the other hand, are organised to keep {\em re-fabricating their own embodiment} at each moment in time; they have a dynamic structure constantly in molecular turnover where most components and assemblies quickly decay and need continual re-synthesis.

An alternative way to see autopoietic machines, is as an extreme type of homeostatic device:

\begin{quotation}

an autopoietic machine is a homeostatic (or rather relations-static) system which has its own organization (defining network of relations) as the fundamental variable which it maintains constant.

\hfill {\small \shortcite[p79]{autopoiesis_cog1980}}

\end{quotation}

Whilst man-made homeostatic devices (Watt's governors, thermostats or switched-mode power supplies, etc.) have fixed evolution equations to maintain just one essential variable within bounds (fuel flow, temperature and voltage respectively), the `set point' of an autopoietic machine is its autopoietic organisation, i.e. the topology of relations allowing the system to exist in the first place. Cells maintain their basic autopoietic organisation at all times, despite the fact that the microscopic structure of the system is in a continual state of molecular turnover (the cell could be growing, moving, changing shape etc.). Losing the autopoietic organisation implies disintegration, according to Maturana and Varela.

\subsubsection{Autopoiesis as Organisational Closure}

The definition of autopoiesis can be understood as a certain type of {\em closure} (see \shortciteNP{Mossio2013}), called {\em organisational closure}. Organisms maintain their organisational closure invariant as their molecular structure changes. It is useful at this stage to make a brief detour to develop an intuition of what closure actually means. This can be done by considering a simple example of closure in a catalysed chemical reaction network. 

In chemical reaction systems, catalyst species help accelerate the spontaneous rate of chemical reactions, sometimes by many orders of magnitude, by lowering the activation energy barrier of a reaction, providing an alternate, easier route from reactants to products. In this process the catalyst is not consumed or transformed itself.\footnote{ In a reaction, the catalyst does disappear temporarily -- for example, forming a brief intermediate called an enzyme-substrate complex -- but it is recreated in the final reaction step. Also, a catalyst need not necessarily be a single molecule, it can be a surface, or the overall action of a chemical cycle for example.}

A collectively autocatalytic set is ``a subset of molecules and reactions where each molecule is created by at least one reaction from this set, and each reaction is catalysed by at least one molecule from the set'' \shortcite{Kauffman1986,hordijk2004}. The crucial point is that each reaction in the autocatalytic set can only take place when all of the other reactions in the set are also taking place, because each reaction requires catalysts/products made by the other reactions (no molecule catalyses its own formation). This `collective existence' property of the reaction set is called {\em catalytic closure}.

\begin{figure}
\begin{center}
\includegraphics[width=16cm]{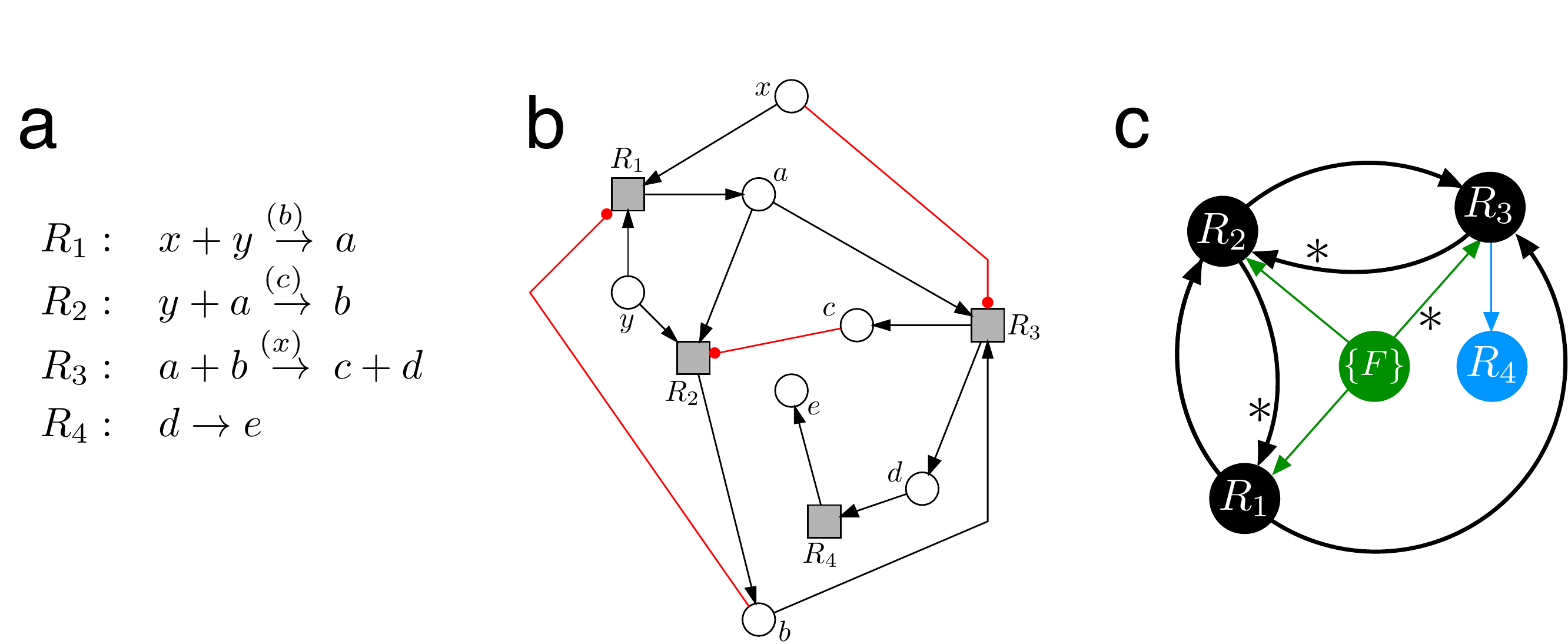}
\end{center}
\caption{
{\textbf{Catalytic Closure.}} An autocatalytic set of reactions expressed (a) as a series of catalysed chemical transformations, (b) as a petri net (bipartite graph) where circles represent chemical species types, boxes represent reactions and red lines denote that a certain species catalyses (but is not consumed by) a reaction, and (c) as a closed set of mutually-enabling processes. See text for discussion.
}
\label{fig:ch2_autocatalytic_set}
\end{figure}

Figure \ref{fig:ch2_autocatalytic_set} shows a hypothetical autocatalytic set of reactions in three different representations. Representations (a) and (b) are equivalent, with the latter petri net graph giving a more intuitive picture of the flow and roles of chemical species. Representation (c) shows the same autocatalytic set in a slightly more abstract way; as a series of connected {\em processes}. An arrow from one process $a$ to another process $b$, indicates that $a$ enables the occurrence of $b$ in some way, and arrows marked with a star represent that $a$ enables $b$ specifically because $a$ catalyses $b$.

All four reactions feature in this process representation, as they can be legitimately described as processes of molecular transformation. Also, a fifth process $\{F\}$ is present, and refers to the continuous pumping of a high energy food set of molecules ($x$ and $y$) as being necessary to enable reactions $R1$, $R2$ and $R3$ (and indirectly $R4$). Another process omitted from the diagram could be the `containment process' ensuring that all reactions take place in sufficient proximity such that their products are accessible by the other reactions: this would have an `enables' arrow to all four reactions.

From the process graph in Fig. \ref{fig:ch2_autocatalytic_set}c, it is easiest to observe the feature of (catalytic) closure, highlighted with thick black arrows. If reaction $R1$ is taken out of the system indefinitely, reactions $R2$ and $R3$ can no longer take place, as $R1$ was supplying the reactant a needed by both of them. Likewise, if reaction $R2$ is stopped permanently then $R1$ cannot take place as it's catalyst $b$ is no longer produced and $R3$ also stops because $b$ was required as a reactant. Finally, knocking out $R3$ will first stop $R2$ because catalyst $c$ is no longer produced, and this in turn will stop the manufacture of catalyst $b$, halting $R1$.

Thus, given that a continual supply of food resources is available (i.e. process $\{F\}$ is `on'), the closed set of reactions $\{R1, R2, R3\}$ in a sense form a `strongly symbiotic' group, producing the molecular and catalytic requirements of each other, in a self-sustained manner. In the long term, either all reactions $\{R1, R2, R3\}$ will be taking place together, or none of them will be taking place. Reaction $R4$ is not in the catalytically closed network, because it simply feeds off product $d$ made by reaction $R3$ (a `parasite' or `side reaction'). Permanently disabling $R4$ has no effect on the continuance of $\{R1, R2, R3\}$.\footnote{The minimal example presented here is engineered to convey the general idea of catalytic closure. Recent formal treatment of the problem \shortcite{hordijk2004,Hordijk2013} has addressed detecting autocatalytic sets in large randomly generated reaction systems. Typically, large AC sets exist (maxRAFs), but these may be further composed of sub-networks which by themselves are autocatalytic (subRAFs). In turn, subRAFs may be further decomposed until minimal irreducible self-sustaining sets are reached (irrRAFs). Therefore, in large systems, it is generally not the case that permanently disabling one reaction leads to the disappearance of the entire autocatalytic set.}

Now returning to autopoiesis, \shortciteA[Ch7 therein]{varela1979} stated that autopoietic machines implement organisational closure. He meant by this that, because an autopoietic machine continually constructs its own material embodiment from ambient nutrients, then every process in the system necessarily has to be made possible from the actions of other processes in the system, and thus the group of processes constituting the system are {\em only viable collectively}, just like the closed set of reactions $\{R1, R2, R3\}$ in our autocatalytic example. That is, if an autopoietic machine is taken apart or severely damaged, its multimolecular, macromolecular and supramolecular components will disintegrate quickly. An autopoietic machine can be depicted as a process closure diagram, similar to Fig. \ref{fig:ch2_autocatalytic_set}c. Autopoietic machines however implement a specific type of process closure, because one extra requirement is that some of the processes must build a membrane and constitute the system as a unity in space (part (ii) of the definition of autopoiesis).

\begin{figure}
\begin{center}
\includegraphics[width=16cm]{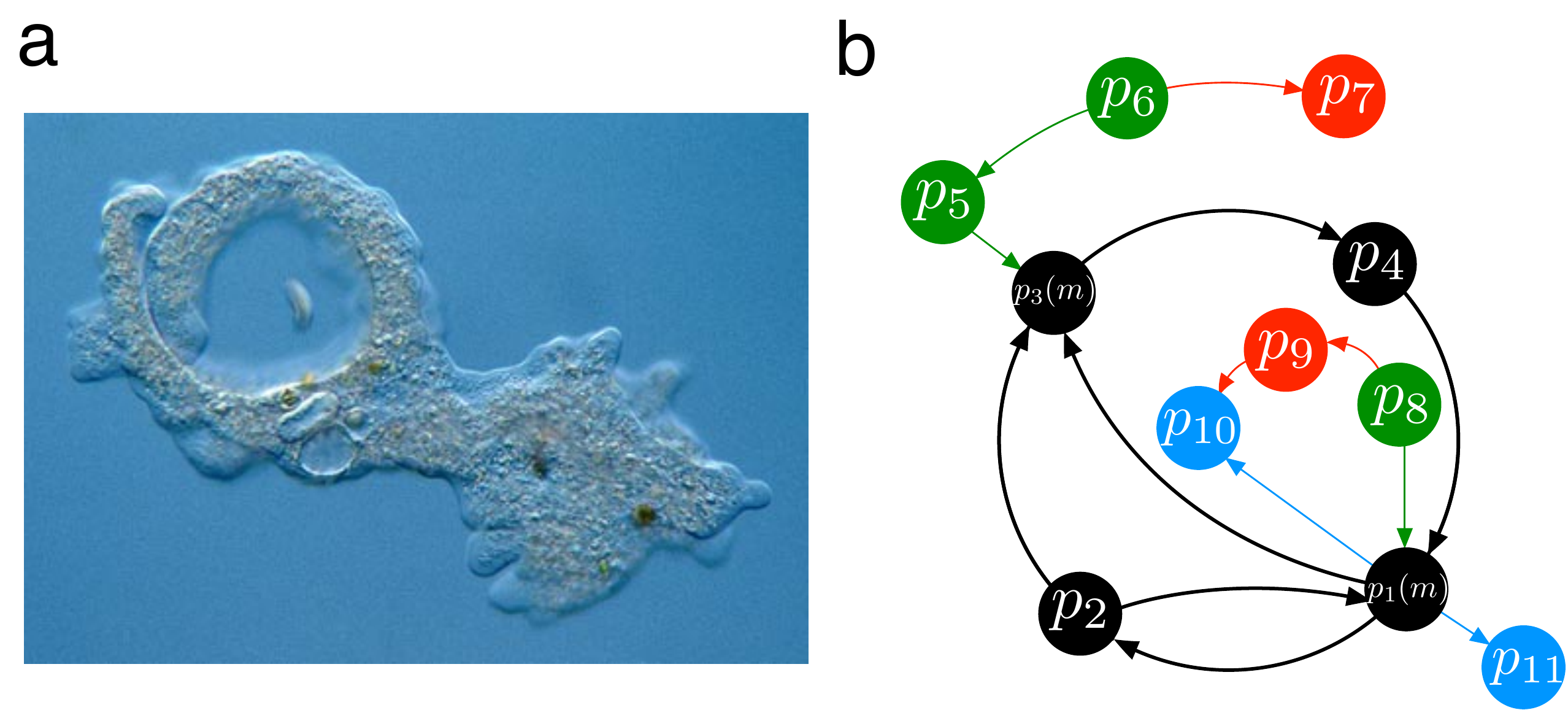}
\end{center}
\caption{
{\textbf{Autopoiesis as Organisational Closure.}} A cellular organism, such as an amoeba (a) can be thought of as a special type of physico-chemical system that exists because of a mutually supporting set of processes that form a closed set (b). Closed set pictured in black. See text for discussion.
}
\label{fig:ch2_op_closure}
\end{figure}

Figure \ref{fig:ch2_op_closure} shows a single-celled amoeba both in physical space as we observe it, and the autopoietic organisation giving rise to such a complex physicochemical system represented (very) abstractly as a hypothetical graph of closed process dependencies. Processes marked with $(m)$ are processes concerned with the production of a semi-permeable membrane, or processes which are made possible by the fact that a semi-permeable membrane is produced, such as passive or active molecular diffusions across the membrane boundary. The closed network of processes $\{p1(m), p2, p3(m), p4\}$, marked in black, has the collective action to realise the same network over time, whilst also materially constructing the amoeba in space. Dissecting the amoeba, disrupting any one of the processes $\{p1(m), p2, p3(m), p4\}$ will eventually cause all of the other processes in the closed network to stop, leading to the irrecoverable disintegration of the amoeba.\footnote{ Although there could be some redundancy: if some processes are stopped, organisational closure could still be ensured by other processes filling their role.} Green process circles denote events whose operation continues regardless of whether or not the amoeba exists, but whose presence is required to create a niche in which the amoeba can survive (e.g. by providing nutrients, or sunlight if the organism is autotrophic). Blue process circles are events that the amoeba enables through its operation, but are of no further consequence to the amoeba (like the excretion of waste products). Red circles are processes entirely independent of the amoeba, and their existence, or non-existence, is inconsequential for the bug.

Processes and their closure in Fig. \ref{fig:ch2_op_closure}b are at a higher level of abstraction than those in Fig. \ref{fig:ch2_autocatalytic_set}c. In the latter case, processes corresponded to catalysed chemical reactions, and each process had its enabling conditions met quite simply: i.e. when the required reactant and catalyst molecules were present in the (well-stirred) reaction solution. The autopoietic machine of Fig. \ref{fig:ch2_op_closure}, however, has a much richer set of physicochemical phenomena simultaneously happening, not just catalysed reactions, but processes such as solvent volume changes, the maintenance of molecular crowding, the use of pH gradients to drive endergonic reactions, and the formation of supramolecular structures. This dense web of entangled processes matter for the continuation of the whole machine. As such, processes in Fig. \ref{fig:ch2_op_closure}b generally exist on different levels of abstraction, run over different timescales, and are generally only made possible by other processes constructing complicated material constraints (for example, a supra-molecular membrane must be constructed for an ion-pumping process to become possible). For this reason, processes in Fig. \ref{fig:ch2_op_closure}b are harder to pinpoint exactly and are labelled more anonymously as ``p''.\footnote{The difference in complexity between a cellular organism represented as a closed set of processes, and an autocatalytic reaction net represented in the same manner can be seen in a different way. An autocatalytic set can bootstrap itself into existence from a food set, if some of the initial reactions can proceed spontaneously at a slow rate, or if some members of the food set catalyse the initial reactions. However, an autopoietic machine cannot spontaneously emerge from a nutrient medium. Rather, it has to come from an existing autopoietic machine, in an unbroken lineage. The processes in the machine result from the build-up of complicated material constraints. As such, the machine can self-maintain from a previous functional state (and go on to divide, continuing the lineage), but its complexity prohibits a full re-emergence of the system from the nutrients alone.
}

Employing an organisational closure diagram to represent an autopoietic system is beneficial in the following ways:

\begin{enumerate}[label=\textbf{\arabic*})]
\item The diagram gives a method to be specific about the closed set of processes and map their relationships. This is one level of abstraction less than autopoiesis, which does not detail particular process relationships but simply requires that overall, the process network should continuously re-create itself.

\item The diagram makes clear that the closed set of processes constituting the autopoietic system need not all necessarily take place {\em within} the physical membrane of the system. In other words, the  `frontier' of the autopoietic system is not the same as the membrane boundary \shortcite{Virgo2011}. Indeed, vital processes in the organisational closure may take place across or even outside the compartment boundary. An organism may fashion its environment into an extension of itself, a niche necessary for its own self-maintenance (a spider building a web, for example). Thus, it is more correct to consider the full extent of an autopoietic system to include part of the environment as well.

\item The diagram makes clear that organisational closure is a non-trivial concept. Processes can have different types of enabling relationships. One process may be a precondition for the existence of another, or it may simply just modify its rate. Alternatively, there could be complicated enabling conditions for a process, where different subsets of incoming arrows would equally serve to enable it \shortcite{DiPaolo2014}.

\end{enumerate}

%
%
%
%
%
%

\subsection[Other Relational Biology Perspectives Similar to Autopoiesis]{Other Relational Biology Perspectives Similar to \\Autopoiesis}
\label{sec:2_2_3}

Autopoiesis is not unique in providing a theory of the logical organisation of biological systems. Other researchers have made different but closely related attempts, and four of these are reviewed below. The four attempts focussed on, excepting Kauffman, don't tend to relate the logical organisation of a living system with its {\em autonomy} like autopoiesis crucially does, but nevertheless, they usefully illuminate the ideas in autopoiesis from different angles and often in a less abstract language. In this way, they are useful to include in this discussion to form a solid appreciation of what autopoiesis means.\footnote{Another cross-board comparison of organisational approaches to life, with an emphasis on M-R systems, can be found in \shortciteA{Letelier2011}.} After the related theories have been reviewed, Section \ref{sec:2_3} then explains how an autonomous systems perspective of life converts into a protocell research program.

\subsubsection{Kauffman's Collectively Autocatalytic Sets and Work-Constraint Cycle}

Autocatalytic sets were first proposed by Stuart Kauffman in 1986 as an alternative paradigm to resolve both the emergence of complex polymers and the appearance of connected chemical networks in the origin of life \shortcite{Kauffman1986,Farmer1986}. Already briefly reviewed in Section \ref{sec:2_2_2}, the idea of an autocatalytic set is a hypothetical network of (two or more) ongoing reactions that act to mutually catalyse each other and sustain the reaction network as a whole. This is a different idea to isolated catalysis, where the product of one reaction catalyses another reaction, and also a different idea to isolated autocatalysis, where a single reaction speeds itself up by making more of the catalytic factor which is driving the reaction. Rather, catalysis is achieved collectively: alone, the reactions would not proceed, but as a network with catalytic closure, they can.

As well as having implications in the emergence of life\footnote{Briefly, based on the theory of phase transitions in random graphs, Kauffman and colleagues argued that the sudden emergence of a large autocatalytic web of reactions would be inevitable in prebiotic chemistry, given a sufficient diversity of smaller molecules that could weakly catalyse the formation of other molecules, including larger ones. The notion of an autocatalytic set is a general one, not particularly restricted to certain molecule types, and such autocatalytic sets were envisaged at different levels: between organic molecules and unspecific catalysts, to explain how the first connected metabolisms emerged, or between peptides and peptide catalysts to explain the appearance of complex peptides.}, catalytic closure can be seen as a cornerstone feature of all extant cellular life. In one sense, cellular metabolism can be viewed as a complex autocatalytic set. Certainly, in order to function at all, protein enzymes are required to selectively catalyse reactions that would otherwise proceed extremely slowly, and the enzymes themselves are produced as part of the whole metabolic network. In this regard, an autocatalytic set seems to realise part (i) of the definition of autopoiesis: it forms a network (of chemical reactions) whose operation (production of products and catalysts) regenerates that same network (makes the catalysed chemical reactions possible). From simple mycoplasma, to all instances of free living cells, Kauffman remarked ``The system as a whole is collectively autocatalytic. Every molecular species has its formation catalysed by some molecular species in the system, or else is supplied exogenously as food'' (cited in \shortciteNP[p377]{Bedau2010}).

In his subsequent work, {\em Investigations} \shortcite{Kauffman2000} and later, Kauffman appeared to accept that autocatalytic chemical networks were a crucial part of cellular operation, but not the whole story. Efficient channelling and synchronisation of reaction pathways by highly specific enzymes is indeed a feature of cell metabolism because without the influence of enzymes making specific reactions occur rapidly in certain sequences, side reactions would ruin the organisation of metabolism. But, another important feature of metabolism is that cells coax non-spontaneous (endergonic) synthesis to occur overall by cleverly linking spontaneous (exergonic) reactions, via various mechanisms. Living cells don't obtain all the high energy and structurally complex organic molecules they require by simply absorbing them from the environment. Rather, they internally synthesise the majority of these molecules by tapping an energy flow to piece together smaller molecules into larger ones. Cells perform biosynthesis by storing some of the energy released from the oxidation of food (spontaneous reactions) in energy carrier molecules (ATP being the traditional example) that diffuse throughout the cell. Such energy carriers then `power' non-spontaneous processes to convert smaller molecules into products with higher energy \shortcite{Alberts2002}. Sometimes, energy carriers enable not only non-spontaneous chemical transformations, but also non-spontaneous membrane transport processes as well, like the pumping of ions against a gradient. However, autocatalytic sets do not deal explicitly with the logistics of endergonic-exergonic reaction couplings, and contain no notion of a surrounding membrane, nor any concept of spatial organisation in fact.\footnote{Also, whether non-catalysed reactions are permitted by the definition is unclear.}

Therefore, perhaps sensing a deeper problem, Kauffman started down the line of developing a more abstract, more encompassing approach to investigate the nature of autonomous agents in general, ``that mysterious concatenation of matter, energy, information and something more that we call life'' \shortcite[p47]{Kauffman2000}. Inspired by the example of a bacterium swimming up a glucose gradient to `get food', he asked {\em what must a physical system be to constitute an autonomous agent?} and came up with a tentative answer: an autonomous agent is a self-replicating system that is able to perform at least one thermodynamic work cycle \shortcite{Kauffman2003}. 

The perspective on the logic of autonomous agents that Kauffman arrived at in his book {\em Investigations} was similar in character to autopoiesis, but instead of being purely relational, made more use of thermodynamic concepts. Rather than talking about abstract `component production', he instead looked at the problem of cellular organisation through the lens of work and energy. Any system able to do useful work has to set up some kind of constraints on a raw energy flow, to channel this flow toward some specific means (Kauffman followed \shortciteA{Atkins1984} in seeing work as the ``constrained release of energy into fewer degrees of freedom''). Kauffman said that the work performed by cells was indeed due to constraints existing on energy flows, but these constraints were mostly not fixed, instead they were actually constructed or manipulated by the cell itself as it performed work. So, the idea of a work-constraints (W-C) cycle was born: in cells, work is made possible by a web of constraints, and this web of constraints exists largely because work is done.\footnote{ Cells also directly use work, for movement, etc.} This circular thinking has the same flavour as autocatalytic sets, but is pitched at a higher level of abstraction: instead of a network of reactions being maintained (that produce species, enabling those same reactions), a network of constraints is maintained (that enable work, remaking those same constraints).

Helping to understand the concept better, the following hypothetical example of propagating work and constraint construction in the cell was given by \shortciteA[p101]{Kauffman2000}. Firstly, lipid synthesis for the membrane takes work to accomplish i.e. spontaneous catabolic reactions releasing energy from food need to have this energy channelled into high energy carriers, which then permit the non-spontaneous biosynthesis reactions making the lipids, to proceed. Once produced, the lipids then incorporate into and maintain the bilayer membrane, and this supramolecular structure provides an oily internal environment favourable to some reactions, because it changes the translational, vibrational and rotational motions of the reactant molecules. In other words, the bilayer has modified the boundary conditions of those reactions (modified the {\em constraint} acting upon them), so that they now take place. Products of such reactions may go on to do more work and modify more constraints, like diffusing across the cell, giving up vibrational energy to do work opening an ion channel, and so on.

Therefore, apart from the ubiquitous use of specific catalysts to channel reaction pathways, cells are generally also in the business of coupling spontaneous processes to non-spontaneous ones in complex webs to synthesise components (and also capitalise on the affordances offered by self-assembly e.g. protein folding and spatial membrane formation), such that they can exist as complex non-equilibrium systems. We have no theory, Kauffman claims, that allows us to capture this kind of organisation where work propagates, makes constraints, and these constraints enable further work to be done.

\begin{figure}
\begin{center}
\includegraphics[width=16cm]{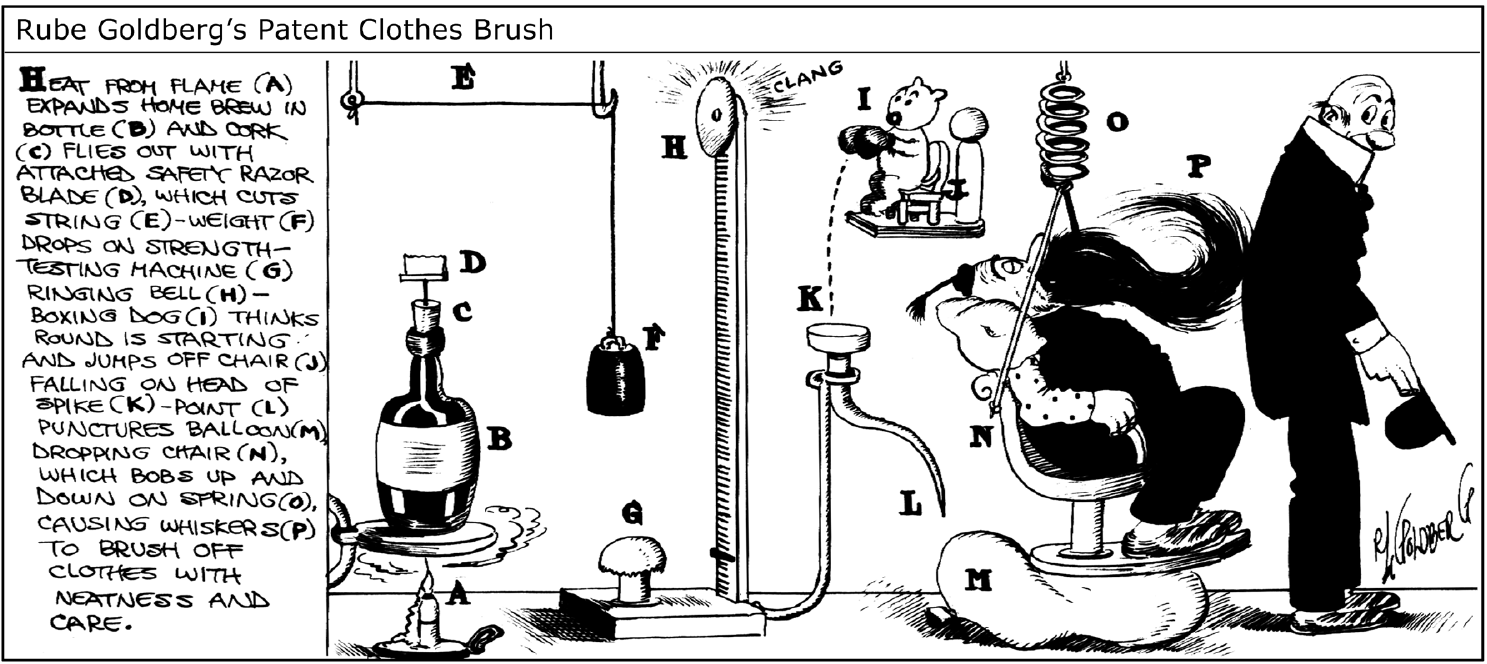}
\end{center}
\caption{
{\textbf{A Rube Goldberg Machine}} is a comical device, but also serves well in a more scholarly sense to  convey the concept of propagating work, as discussed by Kauffman (see text). \footnotesize{Artwork Copyright \textcopyright$\;$and TM Rube Goldberg Inc. All Rights Reserved. RUBE GOLDBERG \textregistered$\;$ is a registered trademark of Rube Goldberg Inc. All materials used with permission. rubegoldberg.com.}
}
\label{fig:ch2_rube_goldberg}
\end{figure}

A useful metaphor employed by Kauffman to illustrate the concept of {\em propagating work} are machines of the Rube Goldberg type (Fig. \ref{fig:ch2_rube_goldberg}). These machines consist of a series of stages where in each stage some kind of potential energy is converted to work, and the work done triggers the next stage. The machine is able to operate, because the constraints (i.e. the positioning of the apparatus) have been carefully set up by a human beforehand, and work propagates through the chain only once. An engine is a machine at a higher level of organisation, able to bend a linear chain of work tasks into a cycle, arriving back at its original configuration, because some of the energy released during the spontaneous steps (gas explosion in the cylinder head) is used to drive the non-spontaneous steps necessary to reset the cycle (pumping of fuel, re-compression of cylinder). At a yet further level of organisation lies a cell, able to perform continuous work like an engine does, but additionally some of the work it continually produces is used for its own means to actually remake the cellular structure (the crucial constraints) in the process.\footnote{Rube Goldberg machines and engines can also dynamically build constraints, but to a much more limited extent than a cell does. (For example, a step in a Rube Goldberg machine might spray a wall white, and this act enables a subsequent step to reflect a beam of light from that wall.)}

Departing from autopoiesis, Kauffman cites `closure' as having being achieved when the cell divides `making a rough copy of itself'. For him, reproduction is a vital aspect of an autonomous agent, and the final outcome of the web of work and constraints.\footnote{but some cells, like heart and brain cells do not divide after their initial development phase, yet still constitute bonafide autonomous living agents.} Autopoietic theory, on the other hand, is typically dismissive of cell division as a central feature of life, and instead uses `closure' to mean the closure of processes allowing the continual re-synthesis of the same system (because for Maturana \& Varela, robust self-maintenance logically preceded the ability to divide).

\subsubsection{Rosen's (M,R) Systems}

In the late 1950's, and preceding the notion of autocatalytic sets, relational biologist Robert Rosen developed a mathematical framework called (M,R) or metabolism-repair systems aimed at capturing the essential logic of life (later summarised in \shortciteA{Rosen1991}; see \shortciteA{Kampis1995} or \shortciteA{Cornish-Bowden2007} for introductory overviews). Rosen observed that living organisms possess two distinct mechanisms, one for the actual functioning, and another for the repair of the functional part. Human-engineered machines, on the other hand, are different: they are typically produced to execute a specific function, and have no (or extremely limited) means for self-repair, instead relying on an external agency for this task. Indeed, as \shortciteA{Letelier2006} remark, a human-made machine even has difficulty providing information to an external agency about the operational status of all of its components, let alone having the capacity to repair them.

Rosen observed that every living cell is based on a metabolic network that relies on catalysts for its operation, but rather than being given from the outside, these catalysts are synthesised by the metabolic network itself. Furthermore, Rosen observed that these catalysts (or as he saw them, mathematical mappings from sets of reactants to sets of products) had a tendency to become degraded by chemical transformations and/or diluted by system growth, lessening their concentration, and so would continually need to be `repaired' by some mechanism. This set up the following conundrum that Rosen tried to resolve with the mathematical machinery of functions and category theory: 

{\em Catalysts (M) are required for the successful operation of a metabolic network, but each catalyst inherently degrades and thus is required to be continually `repaired' by a subnetwork of reactions (R). In turn, the subnetworks repairing each catalyst also rely on catalysts which also need to be repaired, and so on.}

Only two escape routes exist out of this puzzle; either an infinite regress (not a valid solution), or catalytic closure, where it is the collective action of the {\em whole} metabolic network that constitutes the repair mechanism for the repair mechanisms. Rosen's work was an abstract - some would argue obscure - mathematical attempt to resolve how catalytic closure could be realised by a metabolic network. Rosen additionally required the stronger condition that this catalytic closure should only be realisable in a single, unique way (so called `organisational invariance' or `invertibility'). Essentially, Rosen was seeking solutions to the equation

\begin{equation*}
f=f(f)
\end{equation*}

which states that metabolism is effectively a mapping function $f$, which acts on itself, to produce itself \shortcite{Cornish-Bowden2007}.

At their core, both (M,R) systems and autopoiesis were theories emphasising that the underlying {\em cause} for the existence of an organism is a closed network of relations, and these relations remain invariant despite the flowing (far-from-equilibrium) material structure of the system. Rosen expressed this idea by saying that organisms were `closed to efficient causation' but open to material causation. Both autopoiesis and (M,R) were formulated in abstract terms, at the systems level, in a true relational biology spirit without any mention of material implementation nor thermodynamic considerations. However, crucial differences also exist between the two theories. One difference is that (M,R) systems only appeared to focus on the replacement of catalysts, not {\em all} cellular structures and components as autopoiesis did. Another obvious difference is that no notion of a membrane entered into (M,R) systems. Other comparisons have been made by \shortcite{Letelier2003}.

Although effort has been made to make Rosen's mathematical approach a little more understandable \shortcite{Letelier2006,Wolkenhauer2001}, it is fair to say that his approach to biological organisation remains the one most shrouded in mystery and uncertainty. To the extent that (M,R) systems have been reviewed above, it seems unclear how they are qualitatively different to autocatalytic sets. Autocatalytic sets seem to trivially solve the issue of a catalysed reaction network in which all catalysts are produced by the network itself, and thus the recursive problem of `what repairs the repair system' attacked by Rosen seems irrelevant. (M,R) systems without the `organisational invariance' property seem to be identical to autocatalytic sets which do not have catalysts supplied in the food set\footnote{For example, the (M,R) metabolic system analysed in \shortciteA{Piedrafita2010} could equally be labelled as an `autocatalytic set that has none of the catalysts supplied in the food set'.}, and no concrete chemical examples of (M,R) systems with the elusive `organisational invariance' property seem to exist. One extra puzzle is why Rosen himself never discussed worked examples of (M,R) systems using abstract chemical notation.

\subsubsection{G{\'a}nti's Chemoton}

\begin{figure}
\begin{center}
\includegraphics[width=15cm]{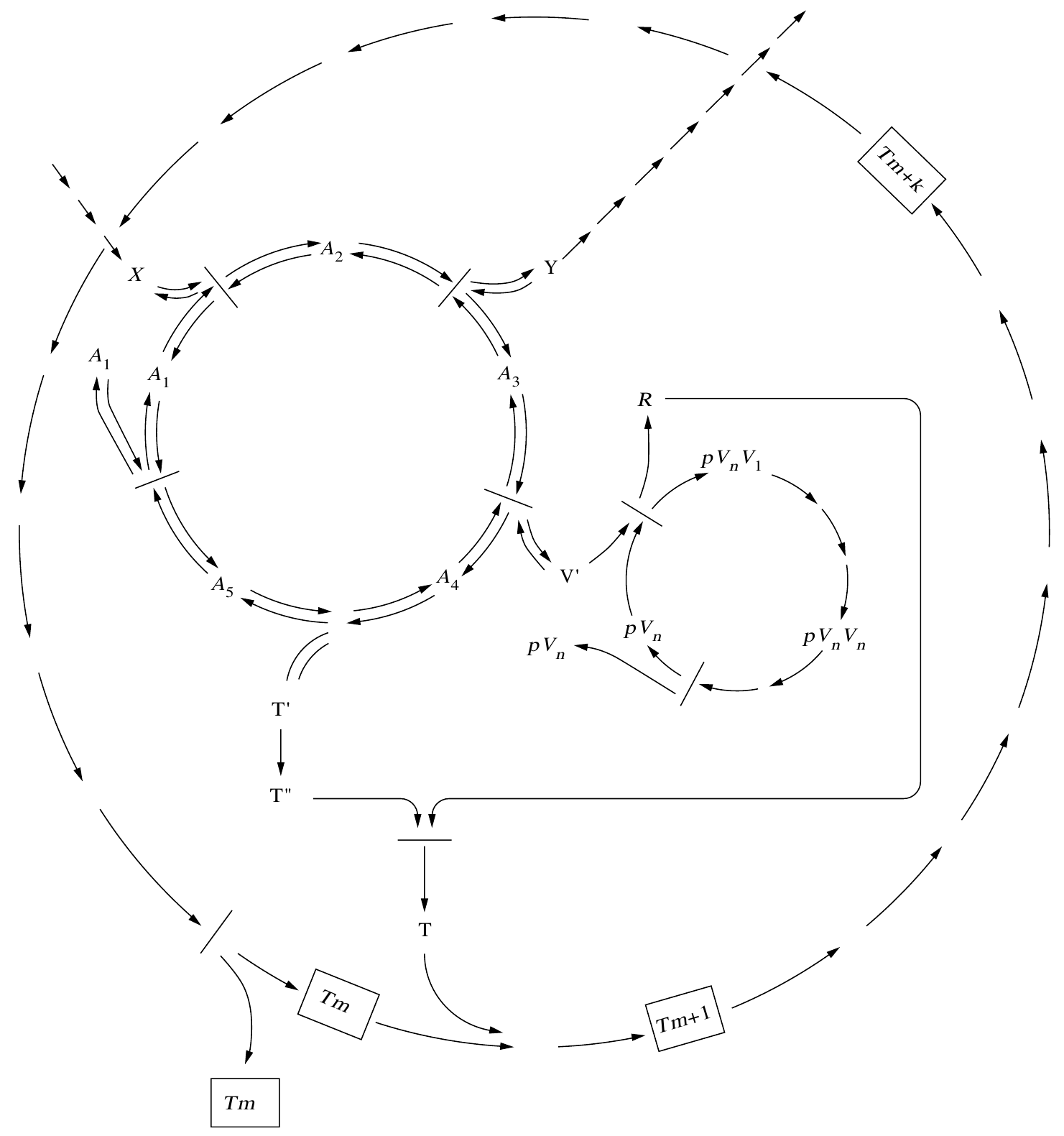}
\end{center}
\caption{
{\textbf{The Chemoton.}} The logical coupling of a metabolic, genetic and membrane system, claimed by Tibor G{\'a}nti to minimally implement a program controlled proliferating microsphere system. On this diagram, it should be noted that the outer circle corresponds to actual {\em physical} organisation of matter, i.e. a self-assembled physical membrane in space, whereas the inner circles correspond only to relational organisations of processes, i.e. the chemical cycles don't actually occupy fixed spatial positions, able to physically enclose other cycles: instead, reactions happen at all points in the well-stirred interior. Reprinted from \protect\shortcite{ganti1997}.
}
\label{fig:ch2_chemoton1997}
\end{figure}

One intellectual contribution made by Tibor G{\'a}nti to clarify attempts at a definition of life were his absolute and potential lists of life criteria, already recited in the introduction of this chapter. However, these life criteria formed just the starting point of G{\'a}nti's research program. The centrepiece was his Chemoton model, a kind of minimal blueprint for a proliferating, instruction controlled and membrane-bound chemical system that satisfied all the life criteria, and one which G{\'a}nti claimed captured the core organisational logic of all living cells (\shortciteNP{Ganti2003a} see  Fig. \ref{fig:ch2_chemoton1997}). G{\'a}nti developed the chemoton to fill the role of the most fundamental and indivisible unit of biology, in the hope of transforming the science into a more exact and well grounded discipline like physics or chemistry.

The chemoton model can be seen as a culmination of some key observations made by G{\'a}nti. First of all, he observed an abstract property shared by all continuously operating machines, including all lifeforms: all must set up constrained pathways on the flow of energy between a high energy and low energy source, transforming part of this energy into useful work (an observation not dissimilar to Kauffman's, above). Secondly, he noted that biological systems are unlike any technological machines that we can currently produce, in that they manipulate {\em chemical} energy by {\em chemical} means. Biological cells are `ticking chemical machines', and because all parts are implemented at the size scale of molecular and supramolecular chemistry, this endows they with unique abilities, such as the ability to synthesise all of their own components for repair, growth and eventual reproduction. Hence, he narrowed to domain of living systems to that of fluid automata, remarking that ``in looking for the fundamental principle of life, we have to examine the nature and organisation of `constrained paths' involving chemical changes capable of work performance, regulation and control in {\em solutions}'' (\shortciteNP[p74]{Ganti2003a}, emphasis added. See also \shortciteNP{Ganti2002}). Finally, G{\'a}nti observed that it would make sense to base a characterisation of life's organisational  logic at the level of the simplest cells, the prokaryotes. He observed that all prokaryotes function by coupling three distinct subsystems: a metabolic system converting energy and building structures, a genetic information system that controls this metabolic system and encodes heritable traits, and finally a membrane system keeping the whole operation together and facilitating system reproduction.\footnote{As remarked in Chapter \ref{chapter:1}, this triad is currently used to guide protocell research, and back in 1971, G{\'a}nti was quite unique amongst his peers (excepting Maturana and Varela) in suggesting a bounding membrane as one of the necessary components for life.} Viruses are ruled out by this triad classification, because they lack metabolism. Single replicating molecules are also ruled out as `living' too. For G{\'a}nti, life was not embodied in single molecules, but was rather a {\em systemic} property based on the tight integration of the latter three subsystems.

With the chemoton model, G{\'a}nti essentially tried to solve the problem, to a first approximation, of how a metabolic, genetic and membrane chemical system must be {\em stoichiometrically coupled together}, in order to form the most basic instance of a proliferating chemical microsphere system, whereby microsphere reproduction was program controlled and sustainable.\footnote{ \shortciteA{Morowitz1988} also described a logic for a proliferating minimum protocell system, based on realistic prebiotic considerations. Their hypothetical system did not have template program control, but could still proliferate by capturing light energy via pigment molecules in the membrane. This energy was stored in electrochemical ion gradients across the membrane, which, in turn, could enable internal lipid synthesis from a precursor, growing and proliferating the system.} Referring to Fig. \ref{fig:ch2_chemoton1997}, the chemoton functions as follows. The autocatalytic metabolic cycle in the chemoton transforms high energy nutrient $X$ into low energy waste product $Y$, and as it turns, produces template monomers $V'$ and membrane precursor $T'$. The template monomers $V'$ accumulate inside the system until a threshold concentration is reached. Once the threshold concentration is superseded, the template molecule $pVn$ spontaneously `unzips' into two complementary base strands.\footnote{Although the problem of non-enzymatic replication of nucleic acids is actually far from being solved \shortcite{Szostak2012b}.} At this moment, monomers $V'$ start pairing to the base strands, constructing two new whole template molecules, whilst also producing a byproduct $R$ in the process. This byproduct combines with the membrane precursor $T'$, creating new amphiphilic molecules for the membrane, which grows, deforming in shape until division is achieved. Therefore, the membrane grows in bursts, synchronised with the replication of the template molecule inside the chemoton.\footnote{G{\'a}nti did not indicate how reliable spatial segregation of the newly formed template molecules into the daughter chemotons was achieved. This could also be argued as very relevant to the logic of proliferating life.} The system is program controlled by the template, in the minimal sense that the {\em length} of the template (the template, after all, has no sequence data) affects the reproduction time and thus the dynamics of the cell cycle of the chemoton.\footnote{ In order for a pattern to count as `information', G{\'a}nti required that another system existed capable of reading and interpreting the pattern. The chemoton system read and interpreted the length of the template to control the dynamics of the division cycle. } Crucially, all three subsystems are mutually dependent on each other, and the program controlled reproduction of the whole `super system' is only possible when all three are present and functioning. Notably, G{\'a}nti argued that correct functioning could only be obtained if each of the three cycles was autocatalytic.\footnote{ The membrane growth process can be regarded as autocatalytic because the membrane absorbs membrane monomers at a rate proportional to its surface area, accelerating its growth as it grows larger. In the personal view of the present author, the claim that all cycles require to be autocatalytic for correct functioning is not immediately obvious. }

To summarise, the chemoton system used chemical cycles to channel an energy difference existing between nutrients and waste into the construction of more program controlled micro-sized `ticking' machines. Undoubtedly, the chemoton is a {\em false} model of cellular life, for indeed no life form either directly implements, nor could be artificially constructed using the exact chemical sequences proposed by G{\'a}nti. However, G{\'a}nti's objective was to propose a false model which was nevertheless instructive or heuristic in understanding something about the essence of living organisation. G{\'a}nti's engineering background gave him a practical approach to the problem of `what is life?', absent from the other more abstract relational biology approaches discussed in this subsection. The emblematic schematic of the chemoton represents a visual thinking point that has directly inspired a number of protocell simulation studies (e.g. \shortciteNP{fernando2004,Munteanu2006,Mavelli2007,Ruiz-Mirazo2008,VanSegbroeck2009,zachar2011}).

Similar to an autopoietic system, a chemoton operates in physical space, synthesising its own components, including a bounding membrane that enables the internal processes to keep running. Departing from autopoiesis (and the other perspectives discussed here on biological organisation), the chemoton also includes a heritable `genetic information' system as an essential participant in the organisational logic of life. The chemoton is constructed to address G{\'a}nti's absolute {\em and} potential life criteria, whereas autopoiesis only targets the absolute criteria. In fact, the chemoton does not operate in a stationary self-maintaining state, as an autopoietic system is envisaged to do. Rather, to avoid an osmotic burst eventually caused by the exponential accumulation of impermeable internal components, a functioning chemoton {\em must} continuously synthesise membrane material and then divide. On this difference, G{\'a}nti remarked that ``living systems are fundamentally growing (accumulating) systems, in which more matter enters than leaves. A growing system cannot be in a stationary state'' \shortcite[p73]{Ganti2003a}. Differing from autocatalytic sets and M-R systems, the chemoton circuit diagram contains no chemical paths accelerated by catalysts. Only spontaneous reactions are included. Rather than being essential for life, G{\'a}nti argued that catalysts only speed up reactions that can already occur spontaneously and don't influence the stoichiometry of the overall chemical change between reactants and products. Stoichiometric coupling and not kinetics was the level of analysis that interested him. He stated that ``The truly fundamental elementary units of biology should not be sought in the enzymatic regulation, but in the system regulated by the enzymes'' \shortcite[p63]{Ganti2003a}.

\subsubsection{Constructive Dynamical Systems and Component Systems}\label{ch2_constructive_dynamical_systems}

Through a series of inspiring papers in the mid 90's, Walter Fontana and Leo Buss enquired into the shortcomings of the traditional dynamical systems perspective as being able to represent biological organisation. They argued that the way Newton taught us to formulate and solve dynamics problems was correct, but did not reach far enough to describe the processes and material transformations happening in whole organisms. What was needed, they said, was a new theory of {\em constructive dynamical systems} \shortcite{fontanaBuss1996,fontanaBuss1994}.

Their essential point was as follows. Traditional dynamical models can only capture change in the system they model by value changes in scalar state variables. From the outset, the modeller is charged with identifying {\em all} of the relevant variables and their coupling for the {\em entire lifetime} of the system being modelled, such that the state space is of fixed size and mathematical tools such as stability, sensitivity and bifurcation analysis can be applied. However, this setup is quite brittle, for it cannot cope with more profound changes in the system being modelled, such as the creation of qualitatively new objects or new constraints (boundary conditions) which emerge through the action of the system dynamics.\footnote{It is true that some finite amount of structural change in a system can be accommodated by using a traditional dynamical system. For example, varying parameters can be handled by making them state variables. Or, changing degrees of freedom can be accommodated by starting with a high dimensional state space and specifying dormant state variables that become active when necessary. However, most systems with emergent constraints can outgrow a fixed description.} Fontana and Buss called this is the `object problem' of dynamical systems \shortcite{fontanaBuss1996}.\footnote{Advances have been made in the analysis of hybrid dynamical systems that combine continuous and discontinuous state changes \shortcite{Goebel2009}. But, such hybrid systems still have a fixed equation structure, and thus still face the object problem.} The problem arises because converting a real world system into a traditional dynamical model always loses a handle on the actual objects which are interacting; these objects are instead dissolved into scalar state variables. In the final dynamical equations, there exists no mechanism and no information with which to construct new evolution equations if the underlying objects represented by the system change, i.e. when the current state variables fall into critical ranges or ratios. Indeed, this is by design, as any dynamical model `freezes' the problem domain, and it is assumed all that is relevant about the problem has been included in the model. 

However, the construction/destruction of objects and the appearance/disappearance of constraints is a crucial feature in the biochemical dynamics of living systems.\footnote{Also, constraint construction via dynamics is also feature of the operation of social insect systems, i.e. in the concept of {\em stigmergy} \shortcite{Bonabeau1999} whereby individual agents lay down structural constraints in the environment as a cue to later guide the behaviour of other passing agents. By this process, wasps are able to indirectly communicate and build a honeycomb, and ants are e.g. able to organise collective cemeteries for the dead.} In response to perturbations, cells can synthesise new components and radically change the properties of their embodiment, effectively becoming a dynamical system with different relevant state variables and  evolution equations. Cells can do more than simply change their state. Therefore, Fontana and Buss said that a new type of dynamical model would be required to fully capture biological organisation: one that can move in an {\em object space} (a structural space of evolution equations), as well as in the standard phase space (a space of states possible with the current set of evolution equations). This new type of dynamical system would not only flow in phase space, but could also change the dimensionality and flow features of the phase space, as it evolved along the state trajectory.

As a first example of a dynamical system able to deal with the object problem, Fontana and Buss proposed a model artificial chemistry able to execute reactions over a growing variety of chemical species (where molecules were implemented as expressions in lambda calculus). This constituted a rudimentary example of a dynamical system able to modify its object structure (the types of species present: novel ones could appear) as well as changing its state (the concentration of each species present). Appendix \ref{appendix:A} of this thesis presents a different approach to simulate general constructive dynamical systems by using a framework based on petri nets.

From the dynamical systems viewpoint offered by Fontana and Buss, a self-maintaining system is one whose equation structure is not fixed a priori, but results as a stabilised aspect of the component dynamics it causes (i.e. the system occupies a stable attractor in {\em object space}). This notion of biological organisation can be seen as a more formal statement of autopoiesis. It differs only in that there is no explicit mention of a bounding membrane.

Another proposal congruent to constructive dynamical systems is {\em component systems} \shortcite{Kampis1991}. Component systems are made of interacting components able to create an open-ended variety of new components, or destroy existing ones. Therefore, they also constitute systems able to build their own internal constraints and modify their governing dynamical description. Kampis holds the view that component systems are only computable over the short term when their dynamical description remains static. In the longer term, he says, they are not computable because their changing description prohibits them from being mapped to any single simulation algorithm.

%
%
%
\section[Autonomy is a Missing Conceptual Backdrop for Origins of Life Protocell Research]{Autonomy is a Missing Conceptual Backdrop \\ for Origins of Life Protocell Research%
\sectionmark{Autonomy in Protocell Research}}
\sectionmark{Autonomy in Protocell Research}
\label{sec:2_3}

At this point, it is useful to recapitulate what has been said in this chapter so far. Firstly, the introduction revealed that two distinct general conceptions of life are followed by (and indeed partition most) scientists and philosophers, in lieu of a unanimously agreed definition of life. The pair were the `evolution' and `autonomy' conceptions of life. 

Under the popular `evolution' (or diachronic) conception of life, {\em a cell is primarily considered as a self-reproducing entity that evolves}. This view holds that the most significant feature of living systems is their genetic information content (DNA) and the replication and passing on of this information content to future generations. Through a pure evolutionary lens, the origin of life is straightforwardly conceived as a continuum between replicating `informational' molecules at one end and more complex reproducing living cells (that embed replicating informational molecules) at the other end. The broad challenge of the evolutionary-based origins of life research program is to decipher how the first self-replicating RNA molecules appeared and then how they evolved, via natural selection at the level of chemical assemblies, into membrane bound DNA-RNA-protein architectures (``cells''). The current mainstream line of investigation in protocell research is located firmly within the evolutionary view. Szostak's ribocell thesis (Section \ref{sec:2_1_1}) posits that a major transition in prebiotic chemical evolution was the insertion of RNA molecules into lipid compartments: this duo opened the floodgates of evolution (leading onto the development of internal metabolism, translation machinery, energetics etc.) as soon as the replication of the RNA and the division of the compartment became synchronised. The long term goal of the ribocell project is therefore to achieve a minimal synthetic system that solves this synchronisation problem.

Conversely, under the marginalised `autonomy' (or synchronic) conception of life, {\em a cell is primarily considered as an autonomous system that, through its organisation, self-produces and self-maintains its own localised material existence in space}. This view holds that the most significant feature of living systems is their system-level `autopoietic' organisation that allows for their continued existence and produces their outward purposive behaviour (as reviewed in Section \ref{sec:2_2_2}). The systems-oriented autonomy perspective does not exclude that genetic molecules are important, but they are not the key issue (see below). 

As has been discussed, the evolutionary perspective is already heavily embedded in research programs into origins of life. The questions that the remainder of this section attempts to answer are: what alternative type of research program into origins does an `autonomy' conception of life propose? what are its main claims? and why is such a program vital to develop protocell research further?

A key point is that the autonomy conception of life instils a deep systemic appreciation of the way molecules have to organise in space and time in order to form a functional far-from-equilibrium biological cell. The autonomy approach to origins projects these same systems-level considerations back onto the protocells that proceeded the first cells. Although perhaps more difficult to articulate than the evolutionary view, the autonomous systems view is primarily concerned with explaining how protocells developed in {\em organisation} up to the extremely complex autopoietic organisation of full-fledged cells. The main question posed is along these lines: how did protocells transition from more rudimentary self-maintaining chemical reactors into chemical `agents' that are able to {\em fabricate} the majority of their components by controlling energy/matter flow through a boundary that, also, is of their own making?

The role of the protocell membrane is a crucial aspect in the autonomy view of origins. Rather than being regarded simply as a secondary `container' holding things together, under the autonomy account, the membrane is regarded as a key integrated part of the protocell system. The membrane is the primary interface that controls energy and matter exchanges between a protocell and its environment. Autonomy brings to the fore the fundamental issue of how protocells were able to progressively take control of their membranes (by effectively coupling internal metabolic networks with membrane production and trans-membrane processes, see Chapter \ref{chapter:3}), first turning into {\em basic autonomous agents} capable of robust far-from-equilibrium self-production in varied environments, and then into more complex autonomous agents with reliable division and heredity properties \shortcite{Ruiz-Mirazo2004BAS,krm_moreno2000}. The role of a self-assembled membrane in channelling matter and energy flow through a protocell system only becomes obvious from an autonomy perspective. Most evolutionary-based protocell research does also see membranes as relevant, but not in such a strong sense: here, membranes are simply passive structures that serve to segregate molecular populations.

In trying to grasp the tenets of an autonomous system account of protocells, it is revealing to compare it with the evolutionary ribocell approach. The ribocell approach postulates that `limping' protocells composed of naked RNA genes in compartments, with the help of fortuitous environmental conditions\footnote{such as the availability of nucleic acid precursors, shear forces to prompt compartment division and heat cycling to separate RNA strands.}, managed to get onto the runway of evolution by natural selection. Conveniently, evolution then solved all of the harder problems to make functioning, {\em fully integrated} cellular systems. How evolution actually achieved each stage of protocell integration is outside the scope of the ribocell project. On the other hand, the autonomous systems perspective {\em directly confronts} the hard issue of cellular integration. It tries to explicitly explain how protocells incrementally developed in order to solve the key problems of cellularity -- such as problems of selective permeability (adequate nutrient access and waste disposal from compartment), osmotic water flow, and the harnessing and distribution of free energy resources -- so that they could eventually exist as robust and adaptive far-from-equilibrium systems. Whilst postponed `for evolution to deal with' by ribocell project\footnote{Although under criticism here, it should be noted that the ribocell research program does not {\em contradict} the autonomous systems approach. It is simply more of a shortcut towards integrated protocells. The ribocell program does indeed see the linking of protocell metabolism and membrane as relevant (see for example \shortciteNP{Chen2004,Chen2004b,Adamala2013}). Additionally, the ribocell program is generating a wealth of excellent data about lipid vesicles (e.g. permeability measurements) and relevant physical effects (e.g. protocell competition) within the origins of life field.}, the systems integration problem is the {\em main} problem in the origins of cellular life and imperative to confront in detail if a comprehensive account of origins is to be reached. This is the reason why the autonomy perspective is the `big picture' framework vital for protocell research (even if the integration problem is a harder and more diffuse scientific challenge than the well-defined synchronisation problem of the ribocell).

Additionally, protocell {\em behaviour} (and how this stems from the chemical organisation of a protocell) is also naturally included under an autonomous systems account of protocell development. How protocells originated the ability to act on their own behalf as agents, to adapt to new or varied environments, or how they developed mechanisms to move in a more directed way (other than just by brownian motion or convection) are important lines of research that are not normally included under the evolutionary view. Recently, an increasing number of studies are acknowledging that a protocell that only replicates would be a poor mimic of cellular life. As \shortciteA{Forlin2012} state, ``life is not simply a machine that divides. Instead, life is integrated with its surroundings, both on a cellular and a chemical level.'' (p591). Therefore, there now exists a building inertia to investigate how the non-replicating' aspects of cellular life could have had precursors in protocells too \shortcite{Monnard2015,Forlin2012,DelBianco2012,Melkikh2012,Froese2014,Hanczyc2011,Mann2012}.

Finally, on the controversial issue of genetic material, the autonomous systems account favours the appearance of simpler self-maintaining protocell systems first, followed by protocells with template-controlled metabolisms later.\footnote{In terms of the Template (T), Metabolism (M) and Compartment (C) subsystems of protocells, the autonomy approach tends to favour the hypothesis that M+C $\rightarrow$ T+M+C (far-from equilibrium self-maintaining reactor systems later incorporated templates), and the ribocell approach supports the hypothesis T+C $\rightarrow$ T+M+C (close to equilibrium `gene bags' later evolved metabolism).} This is for the reason that protocells would first have to become units of a basic complexity, adequately controlling energy and matter flows, in order to be able to support a further increase of complexity brought about by templates. As \shortciteA[p249]{Ruiz-Mirazo2004BAS} explain: ``Before higher levels of complexity (based on macromolecular mechanisms, e.g., genetic or enzymatic mechanisms) are achieved, there has to be some self-constructing organization...through which the material and energetic problems associated to the actual capacity to generate that complexity are solved.''. Another, more basic reason for considering templates after metabolism is that metabolism may be necessary for assisting the fabrication and repair of the templates. Once templates had successfully become incorporated into metabolism, they would then open up new possibilities of metabolic control (e.g. enzymatic control of metabolic pathways) and would enable the synthesis of new structural elements not possible before. Additionally, templates would enter autonomous protocells into an evolutionary dimension, allowing for the reliable division and heredity of their complex organisations (discussed in \shortciteNP{Ruiz-Mirazo2012, Moreno2015}).

Now that a broad outline has been sketched of how the autonomous systems conception of life transfers into a protocells research agenda, Chapter \ref{chapter:3} goes into more detail about how the origin of autonomy in protocells can be modelled and defines the area in which this thesis makes its contributions. Later on, the discussion (Section \ref{sec:5_1}) elaborates more on the arguments made in this section.

\chapter[Autonomy in Protocells: A Semi-Empirical Approach]{\texorpdfstring{Autonomy in Protocells: \\ A Semi-Empirical Approach}{}}
\chaptermark{Autonomy: Semi-Empirical Modelling}
\label{chapter:3}

%
%
%

The biological autonomy conception of life was introduced in detail in Chapter \ref{chapter:2} as a constellation of ideas revolving around the abstract principle of autopoiesis, and the conclusion of Chapter \ref{chapter:2} started outlining how such an autonomous perspective can transfer into an important systems-oriented framework for research on protocells. This Chapter aims to bridge the conceptual and scientific worlds by making more precise how the origin of autonomy in protocells can be approached through a `semi-empirical' scientific research program based on theoretical modelling (Section \ref{sec:3_3}) and goes on to define the sub-area in which this thesis makes its primary contributions (Section \ref{sec:3_4}). To begin, however, it is explained why the theory of biological autonomy fundamentally requires parallel efforts in theoretical and empirical modelling in order to understand the origin of autonomy in protocells.

%
%
%
\section[Perspective: The Development of Biological Autonomy Theory]{Perspective: The Development of Biological \\ Autonomy Theory%
\sectionmark{Biological Autonomy}}
\sectionmark{Biological Autonomy}
\label{sec:3_1}

The concept of biological autonomy has undergone much development over the past 40 years. This can be attested by comparing Francisco Varela's original abstract exposition of it, closely tied to autopoiesis \shortcite{varela1979}, to the state of the art synthesis \shortcite{Moreno2015}, still conceptual and bearing the legacy of autopoiesis, but considerably expanded (notably, reconciling autonomy with evolution), and more closely linked to empirical examples.

Some of the key developments to biological autonomy will be briefly recapitulated here, with pointers to the relevant literature for the interested reader. To begin, one development has been the clarification of the notion of organisational closure (discussed in Chapter \ref{chapter:2}), and how it maybe more accurate to think of a closure of {\em constraints} operating in organisms, when speaking of an organisational closure \shortcite{Montevil2015,Mossio2010,Kauffman2000}.\footnote{`Constraints' are boundary conditions that emerge through the dynamics of the system itself. They are usually material configurations and in turn reduce the degrees of freedom of certain components in the system.} This work has helped sharpen what separates autonomous biological systems from less complex physical systems exhibiting closure, like the water cycle, or exhibiting self-organisation, like dissipative structures such as candle flames or hurricanes \shortcite{Mossio2010,Moreno2008}.

Other developments to biological autonomy have involved identifying important aspects that the theory of autopoiesis either missed out, or treated in an inappropriately trivial way \shortcite{Moreno2008}. The role of the cell membrane was one aspect not emphasised enough by autopoiesis. It has since been emphasised that whilst the membrane does indeed bound an autonomous agent in space, it more significantly serves as the interface through which the autonomous system controls its relations with the outside world, including the energy-matter flow through it \shortcite{Ruiz-Mirazo2004BAS,krm_moreno2000}. This shifted the perspective: autonomous systems regarded before as systems `resisting perturbations and maintaining their organisation' became more regarded as systems that necessarily interacted with the environment as agents as an essential part of their autonomy. Another important aspect neglected by autopoiesis was the historical dimension of autonomous agents, i.e. the relevance of genetic templates, populations and evolution. Some recent work has indicated that genetic templates, being {\em dynamically decoupled} from the self-maintaining organisation of an autonomous system are a relevant addition to the theory of biological autonomy.\footnote{Under this type of terminology, a `dynamically decoupled' component or subsystem is one that is ``sufficiently independent of the processes of material and energy flow that it can be varied without disrupting these basic processes, but still able to be linked to parts of the mechanism so as to be able to modulate their operations'' \shortcite{Ruiz-Mirazo2012}.} Such templates grant a biological system capacities of increased robustness (for example, by allowing the system to switch into qualitatively different modes of self-maintenance; \shortciteNP{Bich2015}), and they also open the way for reliable heredity and the eventual open-ended evolution of autonomous systems \shortcite{Ruiz-Mirazo2012,Ruiz-Mirazo2008b}.

Further clarifications to the biological autonomy framework have consisted in debates about about autopoiesis being necessary and sufficient (or just necessary) to capture the phenomena of cognition and life \shortcite{Bourgine2004,Bitbol2004,DiPaolo2005,Vakarelov2011}.\footnote{`Cognition' roughly means the operation of acquiring, processing and storing sensory inputs, and then using this information to guide successful future actions in the environment.} Whereas Maturana and Varela always held the position that cognition and life were entailed by the implications of autopoiesis, the consensus view emerging from these debates is that autopoiesis is only necessary and not sufficient for cognition and life. 

Yet other supplements to the theory of biological autonomy have dealt with making the notion of `norms' more concrete \shortcite{Barandiaran2014} and research on how autonomy could be extended beyond the paradigmatic case of the single cell towards communities or aggregates like biofilms or multicellular organisms \shortcite{Moreno2015}.

Finally, a very important contribution to the theory of biological autonomy has been the insistence that biological systems are {\em physical systems} \shortcite{Ruiz-Mirazo2004BAS,krm_moreno2000,Moreno1999,Kauffman2000,Fleischaker1990} and, therefore, should not be treated purely as abstract mathematical objects implementing closure (i.e. as autopoiesis regards them). Rather, organisms are constrained to further obey the laws of thermodynamics. This realisation has helped to make a list of basic ingredients necessary for achieving the synthesis of autonomous systems (e.g. semi-permeable membranes, catalysts and energy currencies are required\footnote{Semi-permeable membranes are necessary for any autopoietic system. If a membrane is fully permeable, then there is no distinction between cellular system and environment. If a membrane is completely impermeable, then the cell cytoplasm will settle to chemical equilibrium. Catalysts in turn are necessary for the temporal coordination of reaction pathways, and energy currencies are necessary to drive endothermic processes at different sites around the cell.}), and has also highlighted that all autonomous systems necessarily require to be thermodynamically open systems, continually traversed by a flow of matter and energy.

Therefore, the theory of biological autonomy has been quite fertile as regards to developments in the past 40 years. However, whilst undoubtedly valuable, these developments have not really ventured outside the domain of abstract relational biology. Instead of {\em modelling} cellular autonomy in a concrete way, they have been exercises in expanding, re-ordering and making more precise the {\em high-level characterisation} of biological systems.

\subsection{Why Biological Autonomy Needs Theoretical Models}
\label{sec:3_1_1}

Staying at the conceptual level is valid if the overall goal is to use biological autonomy as a tool to define life, or to articulate alternative general standpoints, for example as some authors have done in cognitive science.\footnote{In cognitive science, autopoiesis underlies new currents of thinking which links cognition in living systems with the fact that their bodies are self-individuating and `flowing' dynamic structures \shortcite{DiPaolo2014,VanDuijn2006}. This position stands in contrast to previous views that saw embodiment as irrelevant to cognition/intelligence, or only relevant in a limited role, i.e. as just an interface getting inputs from and delivering outputs to the environment).} However, if the aim is to (i) develop a grounded account of how autonomy is {\em actually realised} in specific instances of biological cells or (ii) to develop an understanding of how autonomy first originated in protocell systems (as is the aim in this thesis), then a level of analysis based at the overall organisational logic of autonomous systems is inappropriate. Rather, the level of analysis must be taken down towards the detailed molecular structure of particular cell or protocell systems and theoretical models need to be developed where time, states, frequencies, flow rates, spatial details, molecular details, molecular interactions etc. are all returned to the picture. An abstract theory of biological autonomy {\em is able} to discuss the generic organisation of autonomous systems at length, but lacking empirically-based models, it is prohibited from taking the further steps of being able to accurately explore system dynamics and make {\em predictions} about the behaviour of specific cells/protocells. Therefore, only theoretical/computational models, mathematical systems-level analysis and targeted experiments will potentially give insight into how autonomy is actually realised in living cells and how it arose in protocells.

The main problem with modelling biological autonomy is that the goal of autopoiesis (at the centre of the framework) was precisely the opposite: to {\em purposely} throw away all of the structural details of the cell in order to keep only the core organisational logic of the system. Autopoiesis and similar relational biology approaches like Kauffman's W-C cycle, or Rosen's M-R systems are hence - and unsurprisingly - not suggestive of immediate routes to low-level cell dynamics models.\footnote{In general, high-level theoretical frameworks give a hint of how a certain system operates overall, but no clue as to the behaviour of specific instances (i.e. they {\em underdetermine} the phenomena they relate to). In biological autonomy, autopoiesis broadly defines global cellular logic, but does not determine the complex dynamic behaviour of a particular cell. The theory of evolution by natural selection identifies the general requirements for an evolving system, but does not determine which lineages will develop in any particular population (or when increasing complexity will result). Likewise, in physics, the second law of thermodynamics states that entropy always increases in physical systems, but this does not determine the ordered phenomena demonstrated in certain far-from-equilibrium systems (dissipative systems).} In fact, mathematically speaking, relational models start from completely different primitives than dynamical models do, as \shortciteA{Kampis1995} has stated: ``[Relational biology gives] a different language talking about things, a language that does not translate to `ordinary' mathematics. In other words, there is no road whatsoever from relational models to differential equations or computer programs.'' (p95).

One way out of the dilemma of modelling autonomous systems is to start by constructing simple cell models that embody autopoiesis in the most reduced, minimal way possible. Such models are reviewed in Section \ref{sec:3_2} to follow. Before getting there, it is first worth commenting on the relationship between modelling biological autonomy and Systems Biology, since the latter discipline also takes an integrative approach to deal with the biological complexity of the cell.

\subsection{Systems Biology and Modelling Biological Autonomy}
\label{sec:3_1_2}

Systems biology, like biological autonomy, ultimately seeks an integrated systems account of living organisms: that is, how their structure and function emerges from a complex system of diverse components with various interdependencies. Whereas biological autonomy comes from the top-down, from a systems-theoretic account of cell organisation, systems biology works from the bottom-up, iteratively developing and refining precise quantitative models of organisms like {\em E. coli} or {\em Mycoplasma Pneumoniae} (for example, see \shortciteNP{Wodke2013,Guell2009,Kuhner2009}). As such, it would appear that systems biology {\em should be} ultimately guided by, and working towards, modelling the core concepts emerging from the theory of biological autonomy. However, even though certain eminent systems biologists have seen the work of Rosen, for example, as relevant to informing the field of systems biology \shortcite{Wolkenhauer2001}, and indeed the definition of ``Autopoietic Systems'' does exist in Springer's {\em Encyclopedia of Systems Biology} \shortcite{Bich2013}, in general, mainstream systems biology is largely ignorant of (or agnostic to) the more abstract system-theoretic approaches to living systems.

Why is this so? As \shortciteA{OMalley2005} explain, ``systems'' biology can be interpreted in a pragmatic way, or in a `hard line' systems-theoretic way. Most practitioners of systems biology are {\em pragmatists}. They are not so much concerned for abstract principles and laws of cellular organisation, but rather they perceive that an `understanding' of a living system is reached through a process of pooling all the empirical information sources available for a {\em particular} organism (e.g. genome, transcriptome, proteome, metabolism, spatial information etc.) and then creating a mathematical model that predicts some aspects of its behaviour. The fundamental question of how cellular behaviour is connected to the mode in which a cell fabricates its own embodiment -- the question that biological autonomy investigates -- is not of immediate relevance for a pragmatic systems biologist, whose principal task is instead to construct an effective model\footnote{ Systems biology models have traditionally focussed on simulating cells as `one pot' metabolic networks. Other forces and physical effects that come about through the embodiment of a chemical network in space are often disregarded (but this is changing).} of some sub-part of an {\em already} living, {\em already} autopoietic cellular system. It is perhaps appropriate that `cellular weather forecasting' has been used to refer to the practice of systems biology \shortcite{Wolkenhauer2005}.

Having made the critical observations above, it should be pointed out that systems biology does include a more conceptual/theoretical strand which seeks to identify more foundational principles defining the possible `design space' of all living systems. The beginnings of a theory of {\em biological robustness} \shortcite{Kitano2007,Kitano2004}, for example, is located within this strand, and as \shortciteA{Rosslenbroich2009} points out, this theoretical framework has certain parallels to biological autonomy. The theory of robustness generally tries to account for how organisms can survive as invariants units, maintaining their core functions in the face of constant internal changes and external environmental perturbations. Investigated under this approach are the mechanisms that an organism employs to maintain a stable overall presence, which can involve the organism switching into qualitatively different dynamical regimes, such as entering a dormant state under extreme dehydration. The idea of robustness is thus wider than that of homeostasis (which is just maintaining a set of dynamical variables within bounds), and could be seen under the light of biological autonomy as related to how organisational closure is maintained. However, systems biology seems to have been isolated from the systems-theoretic ideas and concepts in the biological autonomy literature, and hence ``curiously these two lines of discussion seem to have no connection to each other'' \shortcite[p625]{Rosslenbroich2009}.

The concept of biological autonomy, in fact, has not percolated significantly into biological science. One main reason could be that the framework has been perceived as too abstract: whilst it offers insightful semi-formal concepts about cellular organisation to `keep in the back of the mind', these concepts don't easily translate into low-level concrete models that can be related to real empirical examples of cells (as discussed above). As such, rather than specifically aiming to model biological autonomy, systems biology instead takes a low-level, practical, data-driven route to understanding the complexity of the cell. Nevertheless, biological autonomy still stands at the conceptual epicentre of systems biology, and its system-theoretic concepts will be necessary to incorporate eventually if cells are ever to be understood in their entirety.

%
%
%
\section[Dynamic Models Capturing Minimal Autopoiesis]{Dynamic Models Capturing Minimal Autopoiesis%
\sectionmark{Minimal Autopoiesis Models}}
\sectionmark{Minimal Autopoiesis Models}
\label{sec:3_2}

One start to modelling autonomy in cells or protocells could be argued to be present in minimal models of autopoiesis which reduce cellular autopoiesis down to its most nominal expression, and then try to implement this with dynamical components either {\em in-silico} inside the computer, or {\em in-vitro} in real chemistry. These models are critically reviewed below, leading on to Section \ref{sec:3_3} where a combined ``semi-empirical'' approach is put forward as a promising route for modelling basic autonomy in protocells.

%
%
%
%
%
%

\subsection[Computational Models of Autopoiesis in Artificial Chemistry]{Computational Models of Autopoiesis in \\ Artificial Chemistry}
\label{sec:3_2_1}

The field of `computational autopoiesis' \shortcite{McMullin2004} is a distinctive sub-field of Artificial Life, with the goal of simulating and formalising (minimal) autopoiesis in abstract `cell-like' models. Such models, typically implemented as 2D lattice artificial chemistries, embed the circular logic of autopoiesis by featuring a metabolism and membrane that co-support each other. The metabolism synthesises the membrane (that would decay otherwise) whilst the membrane encloses the metabolism and provides the correct conditions to sustain it (e.g. by limiting the diffusion of metabolites). In addition to being orders of magnitude more abstract, computational autopoiesis models depart from systems biology models of cells because they {\em explicitly} account for how the membrane of the cellular entity is produced and maintained as a whole. In other words, the cellular membrane is not treated as a pre-given boundary condition for the cell dynamics. Rather, the membrane is a boundary condition that emerges {\em through} the dynamics of the system components, and cannot be present without them.\footnote{`Boundary conditions' for a dynamical system are here meant in the sense of (i) {\em parameters} used in the evolution equations, or more generally (ii) factors affecting the algebraic structure of the evolution equations. This is usage is different to what `boundary conditions' refers to in boundary value problems in mathematics (i.e. solution of a differential equation, subject to extra constraints). Confusing matters further, `boundary' is sometimes used to refer to the cellular membrane!}

Models of computational autopoiesis bring autopoiesis out of the abstract relational biology domain, and provide an experimental test bed to start asking dynamical questions about organisational closure.\footnote{In a related field, the kinetic modelling of autocatalytic sets \shortcite{Filisetti2012} aims to explore catalytic closure in concrete instances of chemical reaction networks.} Such questions include: How does the cellular system respond to specific perturbations, of specific durations? What type of perturbations is the system robust to? If lesions are made to crucial pathways and/or some processes are temporarily stopped, can the system recover from this interference, or does its organisational closure break down permanently? To what extent is the agents behaviour (e.g. movement) related to its continuing metabolic self-maintenance? Generally, which are the open and unresolved issues in autopoietic theory requiring further formalisation?

\begin{figure}
\begin{center}
\includegraphics[width=15cm]{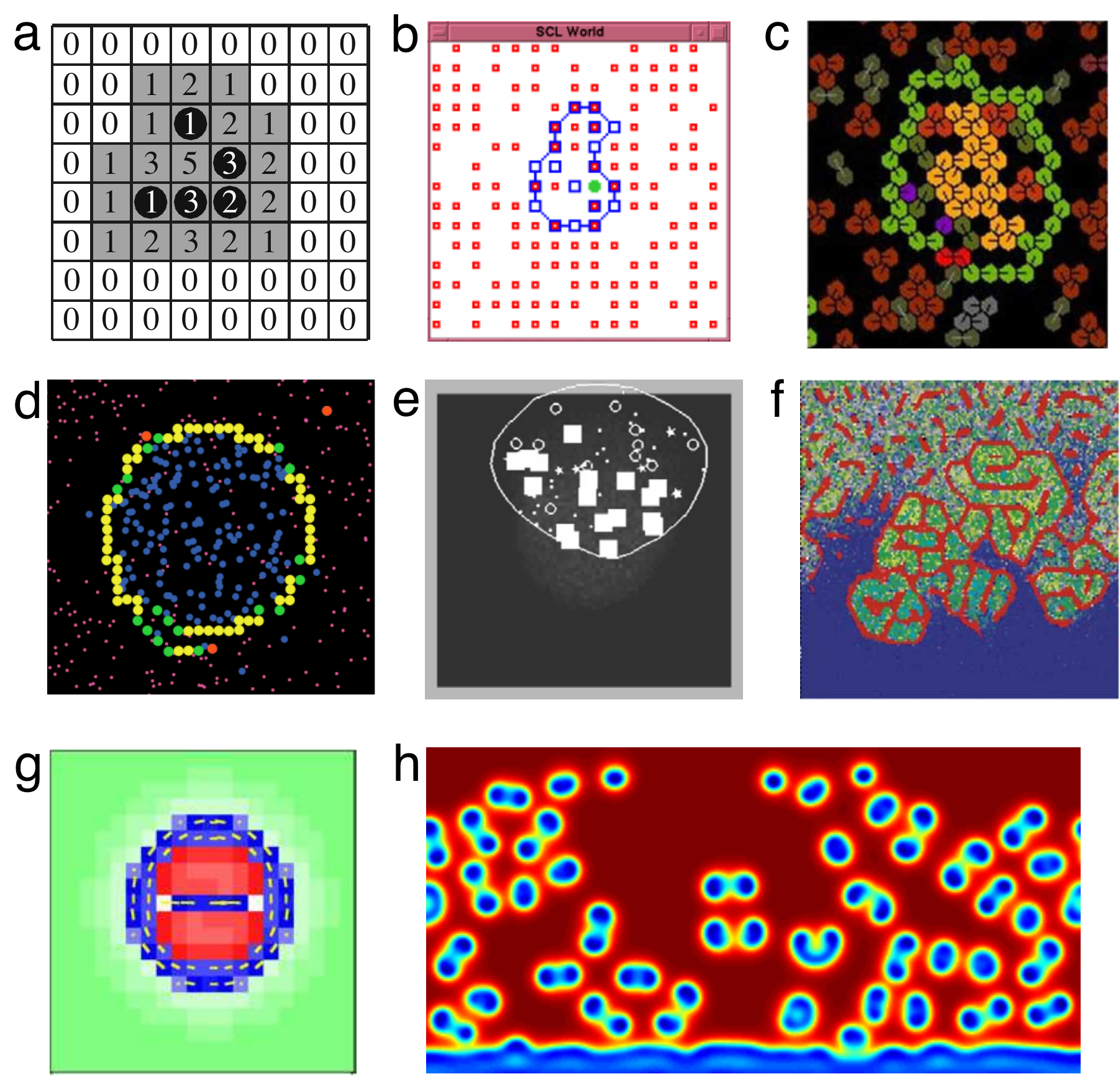}
\end{center}
\caption{
{\textbf{Example Models of Computational Autopoiesis.}} Moving from more abstract single molecule per lattice site models (a-e) toward more physically realistic coarse grain models (f-h), including models based on reaction-diffusion systems (g,h). \hyperref[chapter:copyright]{ View Image Copyright Permissions.}
}
\label{fig:ch3_comp_autopoiesis_models}
\end{figure}

Computational autopoiesis models span from extremely abstract models running on `toy' physics, at one end, to more elaborate models with physically justified component interactions at the other (Fig. \ref{fig:ch3_comp_autopoiesis_models}). To give an idea, a survey of notable models will be conducted below.

Perhaps at the most abstract end of the computational autopoiesis spectrum lies the analysis of recurrent spatiotemporal patterns in the Game of Life, such as gliders (\shortciteNP{Beer2004,Beer2014}; see Fig. \ref{fig:ch3_comp_autopoiesis_models}a) and blocks and blinkers \shortcite{Beer2015}. These special patterns can be seen as autopoietic systems because they are entities that produce and dynamically maintain their own bounded unity, albeit in a universe ruled by ``game-of-life physics''.  Although as far from living cells as one could imagine, the analysis of such mathematical patterns has helped sharpen some of the autopoietic concepts outlined by Maturana and Varela, such as the concept of a cognitive domain, and the dichotomy between {\em organisation} and {\em structure} in an autopoietic system.

Similarly abstract was the the original model of computational autopoiesis, presented as part of the first journal publication explaining the concept of autopoiesis (\shortciteNP{Varela1974}, corrected in \shortciteNP{McMullin1997}; see Fig. \ref{fig:ch3_comp_autopoiesis_models}b). In this model, three types of chemical component - substrates, catalysts and links - randomly floated around on a square lattice. The catalyst particles were able to convert substrates into links which, when adjacent, bonded together into larger membrane segments. Eventually, closed membranes would form around the catalyst particles, and, even though the membrane links would spontaneously decay, they would be continually re-synthesised by the entrapped catalyst. Thus, `autopoietic' self-maintaining entities were created.\footnote{Technically, the entrapped catalyst was not itself synthesised by the system, and so some disputed the `autopoietic' status of the cell-like entities formed.} Variants of the original 1974 model have since been implemented on hexagonal lattice grids (\shortciteNP{Sirmai2011,Sirmai2013}; see Fig. \ref{fig:ch3_comp_autopoiesis_models}c), using mixed lattice/off-lattice simulations (\shortciteNP{Wang2014}; see Fig. \ref{fig:ch3_comp_autopoiesis_models}d), and even just in one dimension \shortcite{Ono2000}.

One interesting extension to the \shortciteA{Varela1974} model, later made by \shortciteA{Suzuki2009}, was to modify the membrane link behaviour such that the autopoietic cell structure became {\em motile} and gradient-climbing. Motivated by the general autonomy-related question ``When does a chemical network bounded by a membrane become a cell that has its own intention?'' (p59), the latter authors conjectured that self-producing protocells could use their dynamic membrane shape to both sense environmental conditions and produce motor movement of the whole system (without the need for explicit sensors and effectors). Using a different cell-like model, (\shortciteNP{Egbert2009}; see Fig. \ref{fig:ch3_comp_autopoiesis_models}e) similarly explored how chemotactic movement and cell self-maintenance were related, and specifically the extent to which the latter determined the former. In the Egbert \& Di Paolo model, the membrane of the cell was represented as a flexible loop of connected springs and the whole cell translated to a new location determined by the overall direction of force produced by locally activated membrane cilia. In the Suzuki \& Ikegami model, movement was produced by the gradual amoeba-like flowing and migration of the membrane.

Moving onto more physically grounded models of computational autopoiesis, \shortciteA{Ono2005} presented a model to investigate general conditions for the emergence of membrane-bound protocells on a surface where there was a population of self-replicating catalysts that could also synthesise membrane molecules (Fig. \ref{fig:ch3_comp_autopoiesis_models}f). In this model, particles were designated as hydrophobic, hydrophilic or neutral, and they followed random walks across a lattice biased by interaction potentials between particles. Under the correct parameter regimes, the emergence of connected membrane compartments could be observed on the lattice, encapsulating the underlying chemistry and, in turn, produced and maintained by it (even when resource gradients were present). 

More recently, Agmon et al. (\shortciteNP{Agmon2014,Agmon2015}; see Fig. \ref{fig:ch3_comp_autopoiesis_models}g) have proposed a computational autopoiesis model based on a modified reaction-diffusion (RD) system. Reaction-diffusion systems simulate the time evolution of a smooth {\em concentration field} rather than the time evolution of single molecules. The RD system of Agmon et al. contains reacting and diffusing metabolic chemicals, and these can produce amphiphilic molecules for the membrane with special aggregation properties. The authors prime the two-dimensional model with an initial self-maintaining `protocell' configuration (a membrane surrounding a concentration of metabolic chemicals), and then quantify the robustness of the `protocell' configuration by systematically mapping how the configuration responds to localised perturbations of its structure.

Finally, it should be noted that the standard reaction-diffusion systems of physics have sometimes been considered as models of computational autopoiesis. Under certain parameter settings, RD systems can exhibit a stable regime of localised `spots' with differing chemical concentrations to the surrounding medium, both in still fluids \shortcite{Virgo2011} and in fluids with convection (\shortciteNP{Bartlett2015}; see Fig. \ref{fig:ch3_comp_autopoiesis_models}h). Even though such spots have no surrounding membrane surfactant layer, they are nevertheless distinguishable individuals that dynamically maintain their own continued existence. They can also move, responding to perturbations and following resource gradients. Therefore these `spot' patterns have been argued to fulfil the autopoietic definition.\footnote{\shortciteA{Popa2004} comments that dissipative structures lacking a topological membrane cannot store energy in alternative reserves, and also require to exploit their energy resource as is, with no energy transduction (p51).}

On reflection, the field of computational autopoiesis has been most useful for making more precise the abstract concepts of autopoiesis, and for outlining where grey areas still remain.\footnote{For example, through exploration with their model \shortciteA{Egbert2009} discuss the possibility that the network of process dependencies implementing organisational closure in an autopoietic system is not invariant, but can change over time, still retaining closure.} Additionally, computational autopoiesis models usually have a strong aesthetic appeal (i.e. they depict colourful cellular systems moving around and responding to environmental perturbations), and this quality helps them to capture attention and articulate the theory of autopoiesis to a wider audience. With regards to modelling autonomy, computational autopoiesis models make a rudimentary start: there is the notion of a minimal organisational closure wherein constraints emerge and support each other to create the cell entity (i.e. the membrane supports the metabolism which in turn supports the membrane).

However, the major shortcoming of computational autopoiesis models for investigating the origin of autonomy in protocells (or cellular autonomy) is that they bear only a very superficial resemblance to real protocells (or cells). Beyond a weak qualitative correspondence, it is difficult to map such models to empirical systems and experimental data. The computational models usually have an unrealistic size scale, an undefined timescale, and some even use implausible physics.\footnote{Most models are single-molecule implementations, meaning that the size scale of the membrane is unrealistically small with respect to the metabolites. Other implausible physical aspects have included membranes made from strong chemical bonds (whereas semi-permeable lipid membranes are actually held together only by weak forces) and waste particles magically disappearing in the cell interior. Also in all models described so far, there is no account of osmosis.} A main problem with computational autopoiesis models is identifying what role contingency plays. Often a circular collection of grid elements will be labelled as a `protocell' and then used as a vehicle to make claims about the origins of cellular life. But, lacking any empirical grounding, it remains unclear to what extent the behaviour and characteristics of the studied `protocell' model are dependent on the particular details of the computational implementation.

%
%
%
%
%
%

\subsection{Experimental Models of Autopoietic Lipid Aggregates}
\label{sec:3_2_2}

In past years, there has also been interest in implementing chemical aggregates that can be said to be `autopoietic' \shortcite{Stano2010}. The aim was to produce a synthetic chemical system that shared, albeit in a minimal form, the same fundamental chemical organisation implemented by all biological systems (thereby, arguably creating a type of `living' matter from inert components). Before reading the following experimental descriptions, the reader may benefit from reviewing Section \ref{sec:4_1_1}.

At the end of the 1980s, \shortciteA{Luisi1989} speculated that the implementation of minimal autopoietic systems in real chemical media was indeed feasible, and suggested reverse micelles as a candidate structure. They envisioned an organic (oily) solvent containing dissolved precursors and also tiny water droplets each wrapped by a single layer of surfactant (reverse micelles). The precursors would diffuse into the small watery core of the micelles and subsequently react to produce new amphiphiles for the micelle membrane. Each individual micelle would hence be autopoietic in the following sense: the molecular components forming the bounded micelle (a cluster of lipid amphiphiles) would provide the adequate conditions for a simple metabolic process (a single chemical reaction) that, in turn, would synthesise the components (lipid amphiphiles) for the bounded micelle. A secondary effect resulting from this process would be that the micelles would also grow and divide, self-reproducing at the same time. Hence, the micelles were said to carry out ``autopoietic self-reproduction''.\footnote{ In this low level of complexity, the autopoietic network is incredibly minimal. It could be argued that the most significant feature of the system is not the {\em self-maintenance} but the {\em self-reproduction} of the micelle aggregates. Some authors have focussed just on this latter aspect to develop thermodynamically grounded models of micelle reproduction life cycles \shortcite{fellermann2015,Fellermann2007a,Fellermann2007}.}

\begin{figure}
\begin{center}
\includegraphics[width=15cm]{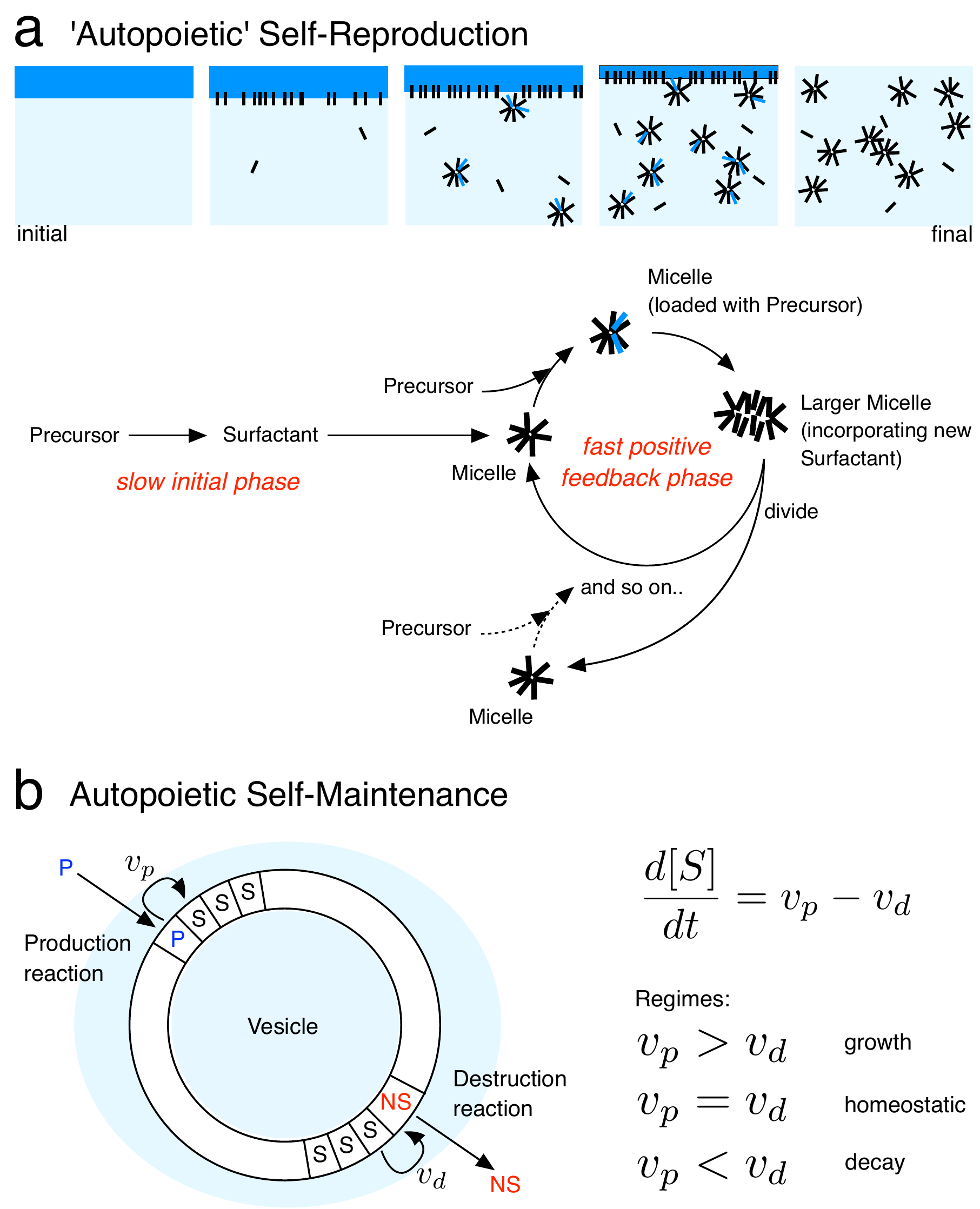}
\end{center}
\caption{
{\textbf{Minimal Autopoiesis {\em in vitro.}}} (a) `Autopoietic' self-reproduction of micelles. The self-reproduction of vesicles follows the similar logic. Autocatalytic self-reproduction of supramolecular lipid aggregates has been seen as an interesting alternative scenario to the self-replication of template molecules, and strengthens the role of compartment systems in origins of life. (b) Autopoietic self-maintenance of vesicles. In the homeostatic regime, vesicles can maintain a constant size - but with a dynamically changing structure - if the rate of surfactant production matches the rate of surfactant destruction.
}
\label{fig:ch3_chem_auto}
\end{figure}

The vision outlined by Luisi and Varela was empirically realised shortly afterward by Bachmann and colleagues, initially for reverse micelles in isooctane \shortcite{Bachmann1990}, and then for normal aqueous micelles \shortcite{Bachmann1992}. Two years later, autopoietic self-reproducing vesicles were realised by \shortciteA{Walde1994}. The essential logic of the latter two experimental studies, without going into too much chemical detail is depicted in Fig. \ref{fig:ch3_chem_auto}a and can be summarised as follows. Initially, a chemical mixture was prepared that consisted of two immiscible phases: an aqueous phase and oily phase made of lipid precursor (that floated on top, shown as dark blue in Fig. \ref{fig:ch3_chem_auto}a). Along the interface of these phases, the precursor would slowly be hydrolysed (e.g. split by accepting a water molecule) to form the lipid surfactant and a side product. The surfactant would slowly populate the interfacial area and exist in slowly increasing concentration in the water phase until such a time that the critical micelle concentration (or critical vesicle concentration) was reached (see Section \ref{sec:4_1_1}). At this point, micelles (or vesicles) would suddenly form spontaneously in the water phase and they would start to sequester some of the precursor inside their oily membranes. As a result, the contact area between the precursor and the water phase would start to increase significantly and a type of `physical autocatalysis' would start to take hold: the increased hydrolysis rate of precursor would create more micelles (or vesicles), creating more interfacial surface area and thereby increasing the hydrolysis rate of precursor yet further. This positive feedback effect is observed experimentally as a sharp increase in the concentration of micelles (or vesicles) after a long onset time, and ceases when all the precursor in the closed system has been used up, leaving a single phase suspension of micelles (or vesicles) in equilibrium. 

As a side note, \shortciteA{Bissette2014} have lately reported on an experimental scheme for autopoietic self-reproducing micelles closer in spirit to the original vision of Luisi \& Varela. In this scheme, the amphiphile for the micelle membrane is {\em synthesised} from two {\em simpler} molecules, rather than being created by the hydrolysation of a more complicated precursor which is more realistic from a prebiotic point of view.

In another recent experimental work, the idea of the autopoietic self-reproduction of vesicles has been extended to a system of {\em phospholipid} vesicles able to grow and divide in an indefinite cycle \shortcite{Hardy2015}. Phospholipids are more complicated two-tail lipids that form the basis of all modern cell membranes. For their synthesis to be achieved, a more elaborate chemistry, more like a minimal metabolism, had to be included onboard the vesicles. Vesicles had to include a specific catalyst that joined a precursor phospholipid to another molecule in order to synthesise the new phospholipids. This requirement meant that, in order for daughter vesicles to go on growing and re-producing themselves, the chemistry happening onboard the vesicles {\em also} required an extra reaction pathway to produce more of the catalyst. Thus, two reaction pathways took place on the vesicles (and three precursors were required in the external solution). Generally, this study serves as a neat illustration of an essential point: in order to be able to divide into functional progeny, complex protocell systems would have to possess chemical pathways to synthesise all of their key chemical components (e.g. have an autopoietic organisation). In fact, this experimental work is an implementation that comes very close to the original computational autopoiesis model \shortcite{Varela1974}. One difference is that, in the experimental system, all chemistry happens on or inside the vesicle membrane, not in the internal aqueous phase of the vesicles. In fact, the experimental system represents an advance over the computational model scenario, since the catalyst can be reproduced, permitting growing daughter cells.

Finally, another landmark study in the field of chemical autopoiesis is that of \shortciteA{Zepik2001}. Here, the authors experimentally demonstrated that aggregates could be made to simply self-maintain without necessarily entering into a self-reproductive regime (Fig. \ref{fig:ch3_chem_auto}b). Their experiment consisted of feeding a suspension of vesicles not only with a precursor that hydrolysed to give surfactant, but also with a second compound that had the opposite effect, causing the oxidation (the loss of electrons) of existing surfactants. When oxidised, the surfactants in the vesicle membranes transformed into a molecular form which did not form vesicles (marked `NS' in Fig. \ref{fig:ch3_chem_auto}b). Depending on the relative supply rates of the precursor and oxidising compound, they showed vesicles could be placed in a growing/proliferating regime (in the sense of Bachmann), in a homeostatic regime (closer to the basic idea of autopoiesis), or in a decaying regime.

It can be observed that the experimental route for demonstrating minimal chemical autopoiesis has a very different character from computational studies on abstract `cell-like' autopoietic systems. Real chemistry is difficult to control (specific aspects cannot be fine-tuned, and all interactions are frequently not known), and often the behaviour of aggregate chemistry at the nm/$\mu$m size scale cannot be observed directly. Instead nano or micro-scale mechanisms have to be carefully deduced by {\em indirect} means i.e. by making a suitable hypothesis consistent with population-level behaviour (although observation of single systems is now becoming possible with modern microscopy and imaging techniques). On the other hand, to counter these drawbacks, real chemical implementations of autopoiesis do not suffer the reality gap that plagues computational autopoiesis simulations.

Minimal chemical models of `aggregate autopoiesis' are relevant in the effort of modelling autonomous protocells, because they represent an empirical start in the direction of creating compartment-based systems where chemistry is coupled with self-assembly processes. The criticism of chemical autopoiesis is that it has not gone far enough to create {\em elaborate} self-maintaining chemical aggregates. In particular, the concept of `metabolism' was minimal and chemistry always took place on the aggregate surface, rather than being a complex aqueous-based chemistry inside a vesicle.\footnote{Recently, \shortciteA{Luisi2014} has issued a challenge for future chemical autopoiesis attempts: ``can one obtain such a Zepik system… with other kind of organic chemistry reactions, and possibly in a way that an internal mechanism -- more than reactions at the bilayer -- are involved?''}

%
%
%
\section[A Scientific Research Program Toward Basic Autonomous Protocells]{A Scientific Research Program Toward \\ Basic Autonomous Protocells%
\sectionmark{Autonomy Research Program}}
\sectionmark{Autonomy Research Program}
\label{sec:3_3}

In this thesis, the stages leading up to the first basic autonomous protocells are investigated by modelling the co-evolution of protocell metabolism and membrane through a `semi-empirical' approach. These terms will be described in due course. As compared to the minimal autopoiesis models of Section \ref{sec:3_2}, this approach entails constructing more chemically realistic (empirically based) computational models of early protocells and considers that the protocell metabolism, membrane and the coupling between metabolism and membrane can be much more elaborate. In particular, rather than the protocell membrane just serving to localise a metabolic network that in turn produces the membrane, the membrane is instead regarded as an active interface controlling energy and matter flow through the system (as explained in Section \ref{sec:2_3}) and various functional components for the membrane can be produced by the metabolism. Hence, an `autopoietic' systems perspective on protocell self-production and function is maintained, but the development of models is not constrained to the strict path of minimal autopoiesis.

The following section describes what a `co-evolution of protocell metabolism and membrane' approach entails, the rationale behind it, and why this approach defines a scientific research program into the origin of the first basic autonomous protocells. Then, Section \ref{sec:3_3_2} explains the `semi-empirical' modelling approach followed to construct different protocell membrane-metabolism couplings. Section \ref{sec:3_3_3} goes on to review two specific scenarios of membrane-metabolism co-evolution explored with semi-empirical modelling.

%
%
%
%
%

\subsection{Framework: Co-Evolution of Protocell Metabolism and Membrane}
\label{sec:3_3_1}

The early earth would have resembled a formidable cauldron; an expansive, heterogeneous and certainly {\em not} well-mixed chemical environment whose properties are nigh impossible to classify. Different chemical networks would have likely existed on catalytic surfaces, in micro-enclosures (e.g. in porous rocks), in water-oil emulsions, in spatially separated parts of solution, in the atmospheric gas phase, and in complicated structured media (arrays of interacting reaction zones, separated by, for example, self-assembled supramolecular structures), to name but a few scenarios.\footnote{Some constraints could be placed upon chemical possibilities if the geography, climate and chemical composition of the early earth was known, but this information is unavailable. Rather, educated guesses have to be relied upon, see \shortciteA{deamer2011}.} There would have been uncontrolled merging, mixing, and transfer between different chemical collectives and assemblies, driven by energy fluxes such as thermal and chemical gradients, lightening discharges, comet impacts, fluid convections, volcanic activity and solar radiation. In solution, other free-floating compartment systems apart from vesicles could have existed too, such as coacervates (blobs of macromolecules and other organic ingredients held together by electrostatic charges), and these could have also played a role in promoting certain kinds of reactions and in establishing the first protocells. Additionally, competition or symbiotic colony effects could have taken place in some communities of chemical aggregates.

As for the development of protocells, given the chaotic prebiotic chemistry conditions described above, it is certainly naive to assume that their lineage comprised a smooth set of transformations whereby free-floating equilibrium lipid vesicles gradually and elegantly became far-from-equilibrium biological cellular systems. As \shortciteA{Monnard2015} remark ``the appearance of protocells was not likely the consequence of a single system lineage that can be traced back to its roots on the early Earth'' (p1243). In the beginning, there could have been several very discontinuous transitions leading up to functional protocells, and indeed multiple scenarios can be envisaged for the creation of vesicle-based chemical systems. One hypothesis, called the `obcell', even advances that vesicles developed metabolic systems on the {\em outsides} of their membranes first, absorbing elements from a complex surrounding `protocytoplasm', before becoming inverted into protocells \shortcite{Griffiths2007,Blobel1980}.

Such a state of affairs might seem to present an intractable starting point for any investigation into the origin of autonomous protocells. It would seem that a multitude of unknown and contingent pathways could have lead to the first protocells. However, possibilities can be narrowed down. An important bottleneck can be identified that fixes a waypoint through which prebiotic protocell systems must have passed: the first functional or `basic autonomous' protocells with the ability to robustly maintain themselves far-from-equilibrium -- and with the ability to divide in a controlled way into equally functional progeny -- could only have existed as such if their metabolic and membrane systems were {\em tightly integrated}. This point is absolutely crucial.

Numerous times in prebiotic chemistry, there may have occurred the chance insertion of complex reaction systems, developed in different chemical contexts, into various self-assembled compartments (e.g. into the internal aqueous phase and/or membrane of lipid vesicles). However, such structures would have been unlikely to result in {\em  stable functional protocell systems} since the membrane would probably not possess the correct composition, nor contain the correct molecular machinery (selective channels, carriers, ion pumps, mechanisms to transduce external energy into a form usable by chemical reactions etc) in order to meet the permeability, catalytic and energy requirements to keep the metabolism running, and at the same time avoid e.g. osmotic burst. Equally, the metabolic processes would probably not be able to synthesise all of the key system components enabling division into equally functional offspring.

The bottleneck arising from the need of tight membrane-metabolism integration in protocells during origins of life is {\em useful} because it has implications for the stages immediately preceding the bottleneck.\footnote{However, the further back from this bottleneck one goes, the less clear the scenario becomes.} The bottleneck implies that a {\em co-evolution} of membrane and metabolism must have taken place over time in protocells, to ensure the high level of integration required to pass through it. Even if metabolism first developed directly on board protocell membranes, and not in their internal medium, then still an integration of membrane and metabolism must have occurred. Pure chance meetings of diverse chemical components would likely have produced complicated conglomerates, but it would not be the type of organised complexity necessary for functional protocells with the ability to proliferate into equally functional protocells. Therefore, from a certain point, protocell membrane and metabolism must have increased in complexity together, bootstrapping each other's functionality in a cyclic way so as to coordinate the dynamical behaviour of their respective components at each stage.

The co-evolution of metabolism and membrane implication is able to define a research program toward the first autonomous protocells. As \shortcite{Szathmary2007} articulates, the problem is to investigate how simple `leaky' protocells with a strong heterotrophic requirement were able to become progressively sequestered from their environments, turning into more impermeable and selective autotrophic systems that used sophisticated means, such as internal component synthesis, active transport and enzymatic metabolism, to stay far-from-equilibrium. 

%
%
%
%
%

\subsection{Methodology: Semi-Empirical Modelling}
\label{sec:3_3_2}

The co-evolution of protocell metabolism and membrane scenario introduced above is investigated in this thesis via a `semi-empirical' modelling approach. The semi-empirical approach aims to create a theoretical modelling paradigm that goes some way to bridging the gap between pure {\em in-silico} and pure {\em in-vitro} implementations of protocell systems, taking the best from both worlds. 

As discussed in Section \ref{sec:3_2_1}, pure in-silico cell or protocell models can be configured to an exact specification, directly monitored, and there is full knowledge of all mechanisms. But, often, they lack physical and chemical plausibility. Conversely, purely experimental protocell implementations (Section \ref{sec:3_2_2}) have the opposite problem; there is no reality gap, but it is extremely difficult to construct nano-sized chemical aggregates to an exact specification and often impossible to monitor their dynamic behaviour directly. The goal of the semi-empirical approach is to create theoretical protocell models able to incorporate realistic physical effects, empirical constants, size scales and time scales such that they can be directly interfaced to, and developed in parallel with, experimental research on lipid aggregates. Having a hybrid experimental-theoretical approach has considerable value. Realistic theoretical models can be suggestive of molecular mechanisms underlying experimentally observed phenomena, and models can further guide new experiments. Moreover, once validated experimentally, theoretical models can be used in an {\em extrapolative} way, to test more complex protocell scenarios that are difficult to construct directly in the lab. 

The {\em ideal} semi-empirical protocell model would have a full representation of space, be able to cover both short and long timescales, allow the inclusion of empirical data and constants (granting quantitative comparison against experiments) and would be amenable to full mathematical analysis to yield elegant and human-intuitive solutions. Of course, no theoretical model possesses all of these desirable qualities. Rather, the art is to reach a suitable compromise dependent on the context.

The semi-empirical protocell model used in the scientific work of this thesis is the kinetic vesicle model introduced by \shortciteA{Mavelli2010}\footnote{The vesicle model can be seen to have roots in earlier kinetics models designed to capture the mechanisms of autopoietic self-reproduction of micelles and vesicles (Section \ref{sec:3_2_2}), see \shortciteNP{Chizmadzhew1994, Mavelli1996,Mavelli2010a}.}, summarised in its most basic mathematical form in Box 2.\footnote{In Box 2, the model is stated in a deterministic formulation, in terms of particle numbers. It is possible to express the model in different ways. A deterministic formulation in terms of concentrations is given in \shortciteA{shirtediss2015} and the original stochastic formulation, in terms of particle numbers, is best summarised in \shortciteA{Mavelli2010,Mavelli2012}.} This model represents a good compromise of the factors mentioned above. The vesicle model is able to represent a protocell operating far-from-equilibrium, where the protocell metabolism and protocell membrane are both dynamic objects able to interact with each other. Many different membrane-metabolism couplings can be dynamically tested on the kinetic vesicle model, with the membrane able to admit different lipid compositions (e.g. changing its permeability and fluidity) and able to embed various hydrophobic molecules that further modify the membrane function (e.g. by making selective channels). Most implementations of the model so far have been similar to simulation models of G{\'a}nti's Chemoton (see Section \ref{sec:2_2_3}) but crucially differ in some important aspects. Most notably, osmotic water flow is rigorously included and the protocell has a variable volume interior aqueous pool. Volume is independent from the surface area of the bilayer membrane, which means that the protocell can assume different shapes, dependent on the volume-surface area relationship at any given time.\footnote{ Cell or protocell models that consider surface area and volume as independent properties are quite rare. Perhaps the closest current approach is the physicochemical cell model of \shortciteA{Surovstev2007}.} Also, passive diffusion of solutes by Fick's first law across the lipid bilayer membrane is included (an obvious feature, curiously absent in all Chemoton models so far). Moreover, the model incorporates realistic physical features such as accurate surface areas and volumes of lipids, molecule numbers, membrane thickness, membrane lipid exchange rates (to match the CVC - see Section \ref{sec:4_1_1}), concentration ranges and mass action kinetics of the internal chemistry. Based at the coarse grain level of {\em kinetics}, the model is able to directly incorporate rate constants from experimental papers (e.g. solute permeability constants) and therefore, the model timescale is generally well defined. The interested reader is directed to \shortciteA{Mavelli2014} for a recent review of applications. 

%
%

\singlespacing
\begin{tcolorbox}[colback=mybrown!50!white,colframe=mybrown!75!black,title=Box 2: Mavelli \& Ruiz-Mirazo Unilamellar Vesicle Model] 

\begin{center}
\includegraphics[width=14.5cm]{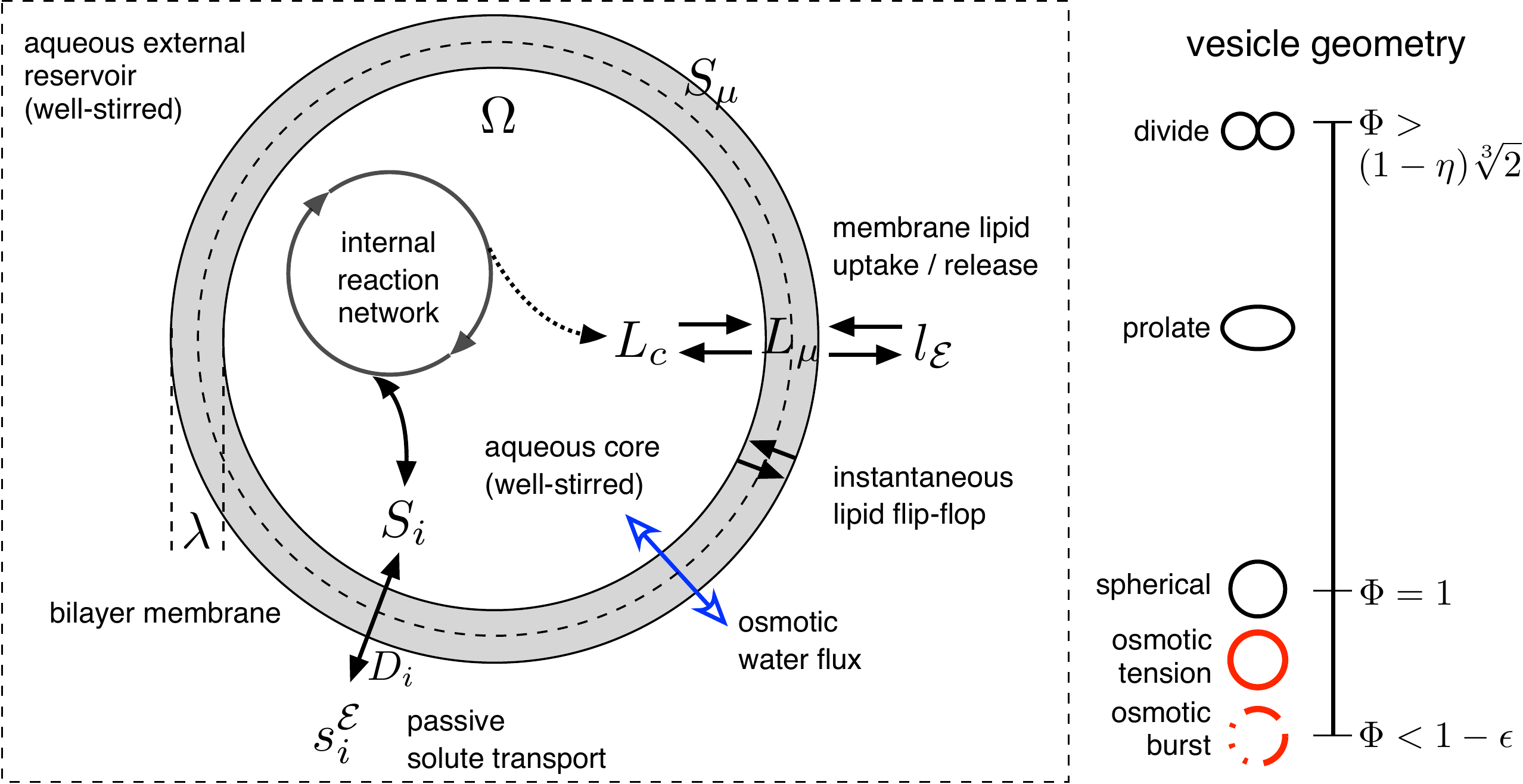}
\end{center}

\subsubsection*{Vesicle reduced surface (scalar value representing vesicle shape)}

\begin{equation}
\Phi = S_\mu / \sqrt[3]{36\pi(\Omega / N_A)^2}
\end{equation}

\subsubsection*{Lipid exchange kinetics for bilayer (of single lipid type)}

\begin{equation}
\frac{dL_c}{dt} = \Omega\textbf{r}_L({\vec s}) + k_{out}L_\mu - k_{in}S_\mu l_c \textbf{u}(\Phi)
\end{equation}

\begin{equation}
\frac{dL_\mu}{dt} = -2k_{out}L_\mu + k_{in}S_\mu(l_c + l_{\cal E})\textbf{u}(\Phi)
\end{equation}

\begin{equation}
S_\mu = \frac{1}{2}L_\mu\alpha
\end{equation}

\begin{equation}
\textbf{u}(\Phi) = 
\begin{cases}
    \text{exp}\left(\frac{1}{\Phi}-1\right) & \text{if } \Phi < 1\\
    1 & \text{otherwise}
\end{cases}
\end{equation}

\subsubsection*{Internal solute kinetics}

\begin{equation}
\frac{dS_i}{dt} = \Omega\textbf{r}_i({\vec s}) + S_\mu D_i\left(\frac{s_i^{\cal E} - s_i}{\lambda}\right)
\end{equation}

\subsubsection*{Internal aqueous volume changes}

\begin{equation}
\Omega = \frac{\sum_N S_j + B_T}{C_{\cal E}} \qquad \text{\footnotesize(if water flow considered instantaneous)}
\end{equation}

\begin{equation}
\frac{d\Omega}{dt} = N_Av_{aq}P_{aq}S_\mu\left(\sum_N s_j + b_T - C_{\cal E}\right) \qquad \text{\footnotesize(if water flow considered finite)}
\end{equation}

\end{tcolorbox}
\onehalfspacing

%
%
\begin{tcolorbox}[colback=mybrown!50!white,colframe=mybrown!75!black,title=Box 2 Continued: Mavelli \& Ruiz-Mirazo Unilamellar Vesicle Model] 

\subsubsection*{Description of model constants and symbols*}

\scalebox{0.95}{\begin{tabular}{p{0.8cm} p{8.7cm} p{2cm} p{2.5cm}}
\hiderowcolors
& {\footnotesize \textbf{Description}} & {\footnotesize \textbf{Unit}} & {\footnotesize \textbf{Value (typ)}} \\

$N_A$ 			& Avogadro's constant									& $\text{mol}^{-1}$		& {\scriptsize $6.02214129 \times 10^{23}$} \\
${\cal V}$ 		& Litre volume of vesicle aqueous pool 					& $\ell$ ($\text{dm}^3$)& $5.236 \times 10^{-19}$\\
$\Omega$ 		& Scaled vesicle volume: $\Omega=N_A{\cal V}$ 			& $\text{mol}^{-1} \text{dm}^3$	& \\
$S_\mu$ 		& Bilayer surface area (equal for both leaflets)		& $\text{dm}^2$ 		& $3.1416 \times 10^{-12}$\\
$\lambda$  		& Bilayer thickness										& dm  					& $4.0 \times 10^{-8}$\\
$\alpha$  		& Lipid $L$ head area 									& $\text{dm}^2$			& $3.0 \times 10^{-17}$ \\
$N$  			& Number of solute species types inside vesicle 		& 						& \\
$S_i$  			& Particle number of solute $i$ inside vesicle			& 						& \\
$s_i$  			& Concentration of solute $i$ inside vesicle 			& M						& up to mM \\
${\vec s}$		& Vector of all solute concentrations ${\vec s} = [s_1 \cdots s_N]$ & M 		& \\	
$\textbf{r}_i({\vec s})$  & MAK kinetics functions producing solute $i$ & $\text{M s}^{-1}$ 	& \\
$s_i^{\cal E}$  & Concentration of solute $i$ in environment 			& M						& up to mM \\
$D_i$  			& Diffusion constant for solute $i$ 					& {\footnotesize $\text{dm}^2 \text{s}^{-1} \text{mol}^{-1}$}			& $2.65 \times 10^8$ \\
$L_c$  			& Free lipid number inside vesicle 						& 						& $20$ \\
$l_c$  			& Free lipid concentration inside vesicle 				& M						& $5.0 \times 10^{-5}$ \\
$\textbf{r}_L({\vec s})$  & MAK kinetics functions producing lipid $L$ 	& $\text{M s}^{-1}$ 	& \\
$k_{out}$  		& Lipid desorption rate constant 						& $\text{s}^{-1}$ 		& $7.6 \times 10^{-2}$\\
$k_{in}$  		& Lipid absorption rate constant 						& {\footnotesize $\text{s}^{-1} \text{M}^{-1} \text{dm}^{-2}$}				& $7.6 \times 10^{19}$\\
$L_\mu$  		& Total lipid number in both bilayer leaflets 			& 						& $2 \times 10^5$  \\
$\textbf{u}(\Phi)$  & {\footnotesize Function accelerating lipid uptake when mem. tension} 		& & \\
$l_{\cal E}$  	& Free lipid concentration in environment 				& M						& $5.0 \times 10^{-5}$\\
$C_{\cal E}$  	& Total concentration of all species in environment 	& M						& \\
$B_T$  			& Number of buffer particles trapped inside vesicle 	& 						& $6.3 \times 10^4$ \\
$b_T$			& Concentration of buffer species inside vesicle 		& M 					& $0.2$ \\
$v_{aq}$  		& Molar volume of water at room temperature				& $\text{dm}^3 \text{mol}^{-1}$ & $1.8 \times 10^{-2}$ \\
$P_{aq}$  		& Macroscopic permeability constant for water 			& $\text{dm s}^{-1}$	& $1.0 \times 10^{-4}$ \\

\end{tabular}}

\;
* Values typical for a 100nm diameter unilamellar vesicle.

\end{tcolorbox}
%
%
%

%
%
%
%
%
%

\subsection[Semi-Empirical Models Toward Basic Autonomous Protocells]{Semi-Empirical Models Toward Basic Autonomous \\ Protocells}
\label{sec:3_3_3}

\begin{figure}
\begin{center}
\includegraphics[width=13cm]{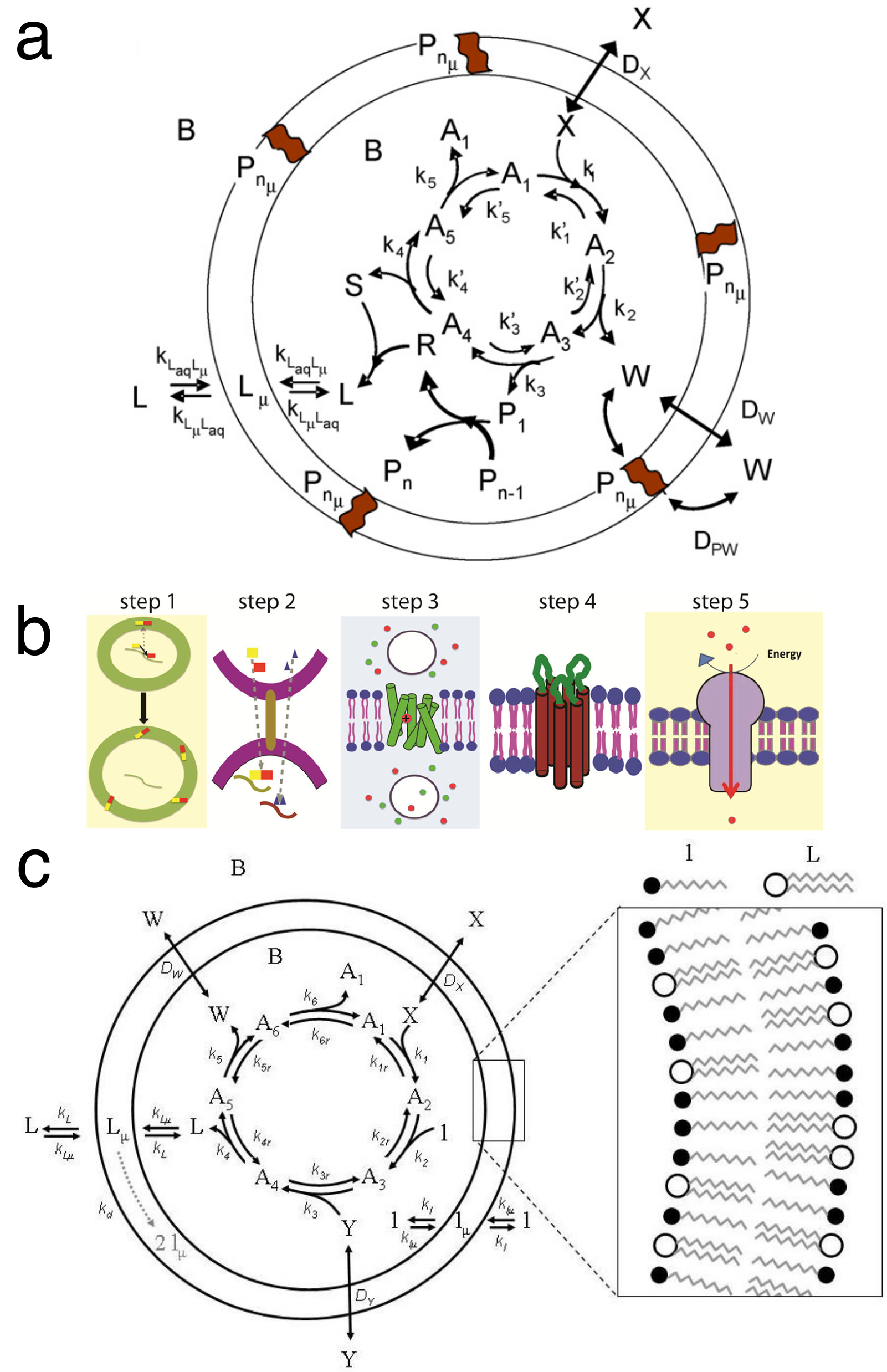}
\end{center}
\caption{
{\textbf{Semi-Empirical Protocell Models Toward Basic Autonomous Systems.}} (a) Lipid-Peptide protocell \protect\shortcite{Ruiz-Mirazo2008}. (b) Hypothesised evolutionary development of membrane protein channels, from \protect\shortciteA{Wilson2014}. (c) Lipid-producing protocell \protect\shortcite{Ruiz-Mirazo2011}. See text for discussion. \hyperref[chapter:copyright]{ View Image Copyright Permissions.}
}
\label{fig:ch3_krm_models}
\end{figure}

In the co-evolution of metabolism and membrane research program toward autonomous protocells, the vesicle model has so far been employed in two distinct ways (Fig. \ref{fig:ch3_krm_models}). 

The first scenario, depicted in Fig. \ref{fig:ch3_krm_models}a, has been to model a `lipid-peptide' protocell \shortcite{Ruiz-Mirazo2008}. In this scenario, an autocatalytic reaction cycle inside the protocell produced  peptide monomers that assembled into short peptides in the protocell interior (or inside the bilayer membrane itself), before orienting and spanning across the membrane where they functioned as rudimentary channels. Once in place, the peptide channels enhanced the permeability of certain solutes across the bilayer membrane -- including the waste made by the metabolic cycle -- and also contributed to the `elasticity' of the membrane, i.e. the range in which the membrane was stable. The logic was that by producing rudimentary transport channels for the membrane interface, the protocell system could gain some additional control over its own far-from-equilibrium maintenance and become more robust to changes in external conditions as a consequence. Preliminary results indicated that production of membrane channels could prevent an osmotic burst from certain initial conditions, but could also shorten the division time of the protocell system by making the membrane more rigid.

Related to the lipid-peptide protocell, Fig. \ref{fig:ch3_krm_models}b shows some steps for the evolution of protein channels in protocells hypothesised by one recent study (\shortciteNP{Wilson2014}; see also \shortciteNP{Pohorille2005}). The authors hypothesised that membrane channels co-evolved with protocell metabolism, initially starting from membrane-bound oligopeptides (Fig. \ref{fig:ch3_krm_models}b, steps 1 and 2), which then became ion conducting channels (step 3), turning into more rigid channels (step 4) and then, finally, channels capable of energy transduction and the utilisation of energy for active transport against concentration gradients (step 5). Contemporary biological cells critically rely on the latter to remain in far-from-equilibrium conditions.

The second scenario addressed with the semi-empirical vesicle model has been a `lipid-producing' protocell (Fig. \ref{fig:ch3_krm_models}c). The aim was to start investigating what advantages were brought to a protocell that could control its own membrane lipid composition (and hence permeability, fluidity, elasticity) by endogenously synthesising a more complex lipid type, like a two-tailed phospholipid \shortcite{Piedrafita2013,Ruiz-Mirazo2011,Piedrafita2011}. Internal synthesis of elaborate phospholipids to compose a surrounding membrane is a critical operation that all living cells perform. Given that primitive protocells would likely have self-assembled from simpler lipids types that were locally available, the transition from {\em self-assembled} to {\em self-produced} membranes (and its implications) is an important avenue to research in the development of protocells. 

In the model, the protocell reaction network consumed a simple single-tail lipid ($l$) already present in the protocell membrane (supplied by the outside medium), and by absorbing high energy precursors $X$ and $Y$, was able to modify this lipid into a two-tail lipid ($L$) whilst also producing waste products ($W$).\footnote{\shortciteA{Piedrafita2013} explored a similar model, but changed the internal metabolism from a G{\'a}nti autocatalytic cycle to an `M-R' system (that did not use the simple lipid type ($l$) as a precursor).} On incorporating the two-tail lipid, the membrane would become both less elastic, and less permeable to solutes (see below). Preliminary results indicated that, dependent on parameters, different asymptotic lipid-phospholipid membrane compositions could be achieved. Also, the observation was made that by requiring simple lipid ($l$), the metabolic reaction cycle naturally ran more slowly (hence producing less waste) in cases when the membrane composition was dominated by phospholipids and thus less permeable.

The `lipid-peptide' and `lipid-producing' protocell scenarios are models of how simple protocells began coupling internal reactions with trans-membrane processes or membrane composition respectively, in order to become more elaborate and more robust. In fact, the two scenarios are quite related. Protocells made of an increased phospholipid content would have faced a `sealing off' problem, like all living cells do. The close-packing and the long chain length of phospholipids grant liposomes with better stability characteristics\footnote{In terms of temperature range, pH range, and typically lower CVC value (see Section \ref{sec:4_1_1}).}, but at the same time make membranes that are quite impermeable to solutes. Cells solve the `sealing off' problem (and even take advantage of it\footnote{A closed membrane barrier with permeability controlled solely by transport mechanisms can allow selective and controlled molecular trans-membrane movement. The {\em cell} decides what is transported in and out, even against concentration gradients. On the other hand, a permeable membrane barrier that permits passive diffusions cannot be so selective, and it is actually the molecular properties of the {\em solutes} that determine what is transported in and out. }) by employing specialised machinery like channels to selectively move molecules across their phospholipid membranes. Hence, as protocell membranes became increasingly based on more elaborate lipid types like phospholipids, membrane protein machineries like channels would have likely co-evolved so that protocells could maintain their metabolic function \shortcite{Budin2011}.

One can imagine many immediate extensions to the `lipid-peptide' and `lipid-producing' protocell scenarios that would explore the co-evolution of the metabolism-membrane coupling further, hence moving the model closer toward basic autonomous protocells. For example, the two approaches could be combined to explore the dynamics of protocells that could synthesise impermeable `sealed' membranes that were nevertheless made selectively permeable through the parallel synthesis of peptide channels (a scenario closer to real biological membranes).  Also, a start in modelling the active transport of solutes across the protocell membrane could be made to see what general stability and control advantages the non-spontaneous molecular movement against concentration gradients could bring to the protocell system. Furthermore, a plethora of metabolic networks could be investigated beyond G{\'a}nti's autocatalytic cycles or Rosen's M-R systems. See Section \ref{sec:5_2_3} for a discussion of the challenges associated with extending the semi-empirical protocell research program.

The scientific modelling work carried out in this thesis actually constitutes a {\em prequel} rather than a {\em sequel} to the existing `lipid-peptide' and `lipid-producing' protocell schemes. As the following section explains, there exists rich protocell scenario to investigate {\em before} protocells started producing complex components such as lipids and membrane proteins.

%
%
%
\section[Before Basic Autonomous Protocells: Modelling the Early Interplay of Chemical Reactions and Dynamic Compartments]{Before Basic Autonomous Protocells: Modelling the Early Interplay of Chemical Reactions and Dynamic Compartments%
\sectionmark{Chemical Reactions in Dynamic Compartments}}
\sectionmark{Chemical Reactions in Dynamic Compartments}
\label{sec:3_4}

Both of the protocell models proposed as getting closer to basic autonomous protocells in the preceding Section \ref{sec:3_3_3} were `self-producing' protocells. Inside the lipid-peptide protocell, an autocatalytic reaction cycle produced peptide monomers that eventually assembled into membrane channels and also produced simple membrane lipids. Inside the lipid-producing protocell, a reaction cycle instead manufactured a complex two-tail lipid. Rather than extending these models further, a valid argument can be made for rigorously investigating {\em more simple} protocells instead, that did not synthesise their own membranes, nor complex membrane proteins.

Jumping straight into modelling lipid-synthesising protocells in fact brings some problems. A first issue is that, in cells, fatty acids and phospholipids are synthesised from {\em small molecule precursors} in a long sequence of linked reaction steps, where some steps require specific organic catalysts to proceed, or energy input from e.g. ATP in order to become spontaneous, or both factors. Realistically, lipid synthesis (fatty acids, or even more difficult, phospholipids) from small molecule precursors is {\em not} a trivial task, and would not be achievable by a simple chemical cycle. In early protocells that likely operated at a strongly reduced level of molecular and organisational complexity, it can be questioned whether `bottom up' lipid synthesis from small molecules was feasible, given the advanced catalytic and energetic requirements, or whether this was a relatively late addition. Incidentally, the chemical autopoiesis approaches in Section \ref{sec:3_3_2} did not solve the issue of prebiotic lipid synthesis, but rather used a shortcut to bypass this problem. This shortcut, available to human synthetic biologists but not necessarily to prebiotic protocells, involved supplying the self-sustaining/replicating system with precursors that were {\em already structurally very close} to being amphiphilic aggregate-forming molecules. These precursors typically required just a slight modification, often performed by just a single reaction, in order to make them amphiphilic.

A second issue with modelling internal lipid synthesis is that internal synthesis can have various biophysical effects on a vesicle. Some experimental studies with giant unilamellar vesicles have shown that in cases where lipid synthesis inside vesicles has been achieved artificially, the newly produced lipids don't always behave in an expected way i.e. they don't always grow the existing membrane uniformly. Rather, it has been observed that internal lipid synthesis can lead to multiple new vesicles self-assembling {\em inside} the water pool of the parent vesicle instead \shortcite{Takakura2003}.\footnote{Also, in \shortciteA{Hardy2015}, the phospholipid vesicle grows not by enlarging a single membrane, but by folding into a multilamellar vesicle.} Other have noted that asymmetries in area of the inner and outer leaflet of a vesicle bilayer -- as could take place when internally produced lipids associate with the inner leaflet -- can lead to vesicle morphology transformations \shortcite{Mui1995}. These findings point to the need for non-trivial theoretical models to accurately account for the effects of internal lipid synthesis.

In any case, an under-explored protocell area exists {\em before reaching} a scenario of protocells which endogenously produce complex components like lipids, or peptides: {\em  far-from-equilibrium chemistry in dynamic lipid compartments}. In this early protocell scenario, chemical networks could be envisaged to interact with semi-permeable membranes, not by producing them, but in more indirect ways, to form protocell aggregates or `reactors' with interesting non-linear dynamics, and possibly adaptive behaviours.

\begin{figure}
\begin{center}
\includegraphics[width=16cm]{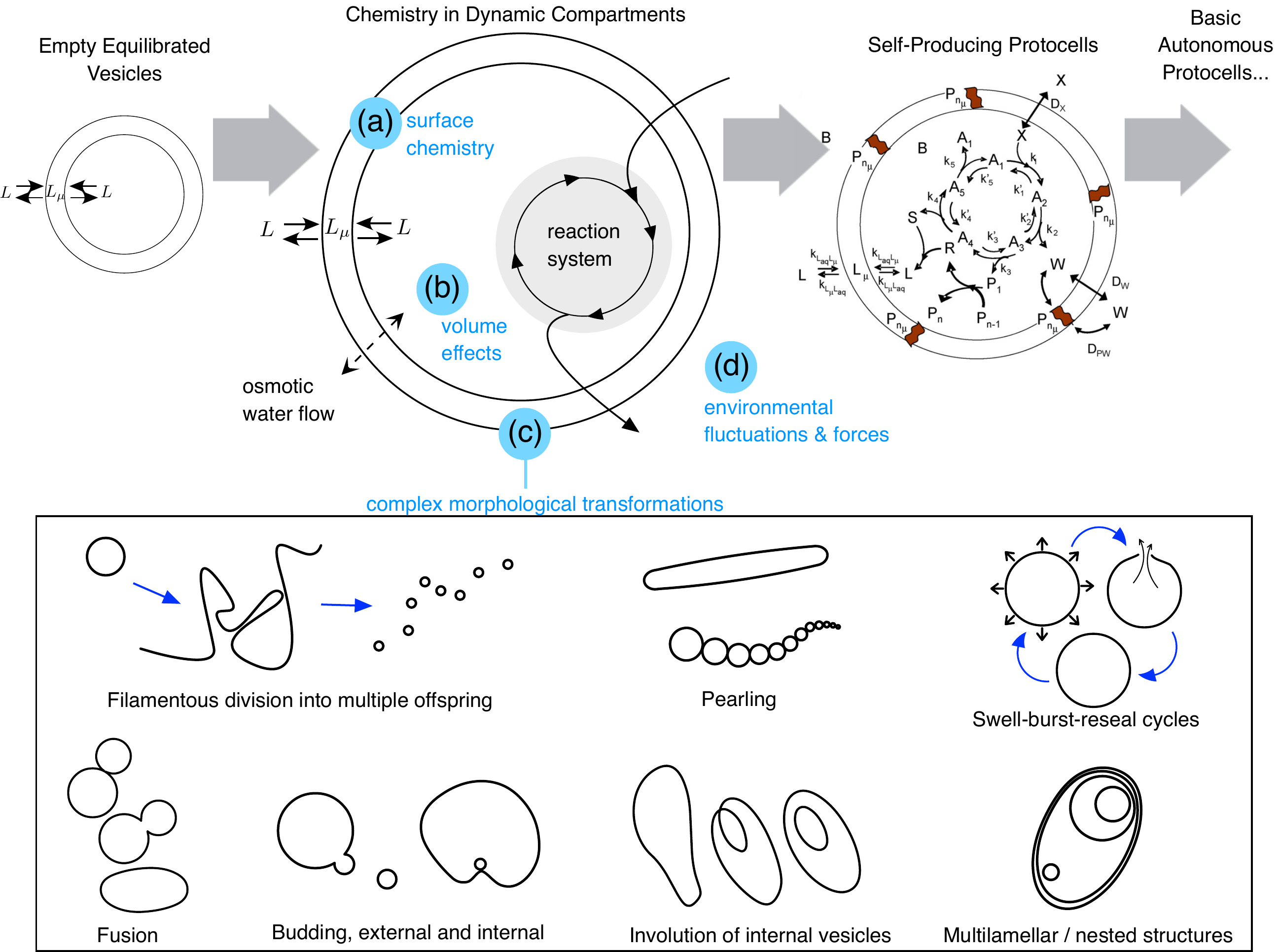}
\end{center}
\caption{
{\textbf{Chemistry in Dynamic Lipid Compartments.}} Before the development of protocells that internally synthesised complex components, a more simple protocell scenario can be postulated, consisting of chemistry interacting with dynamic lipid compartments. In this scenario, even though the protocell membrane is not produced by the internal  metabolism, non-linear protocell dynamics can still result from various factors (a)-(d). See text for discussion. 
}
\label{fig:ch3_dynamic_compartments}
\end{figure}

Figure \ref{fig:ch3_dynamic_compartments} gives four examples of factors that can potentially lead to interesting emergent chemical behaviour in a purely self-assembled protocell chemical reactor, based on a vesicle, that does not have lipid or peptide synthesis capability.

First of all, surface interactions can serve as an indirect coupling between metabolism and membrane (Fig. \ref{fig:ch3_dynamic_compartments}a). The interior surface of a vesicle (or alternatively the hydrophobic interior of its bilayer membrane) can provide an environment that favours certain types of chemical reactions. As such, a vesicle membrane is able to act as a promoter or regulator to certain reactions in the protocell metabolism \shortcite{Walde2014}. Moreover, surface chemistry has often been found to accelerate reactions inside small compartments with a high surface-to-volume ratio \shortcite{Fallah-Araghi2014,Sunami2010}.

Secondly, vesicles have a semi-permeable and elastic membrane enclosing an aqueous volume, and for these reasons, the internal aqueous volume can be significantly variable due to osmosis (Fig. \ref{fig:ch3_dynamic_compartments}b). This variation in the size of the molecular collision space can have various ramifications for the dynamics of an encapsulated chemistry (\shortciteNP{Lizana2008, Lizana2009}. See also Section \ref{sec:4_3_2} for a detailed discussion).

Thirdly, vesicles are not just spheres that divide. Rather, vesicles exhibit a rich variety of morphological transformations, triggered by changes in membrane composition, osmosis and external forces (Fig. \ref{fig:ch3_dynamic_compartments}c). For example, vesicles can grow into long filaments, and then divide \shortcite{Zhu2009}, or alternatively can reach a `pearling' instability whereby they rapidly convert into a beaded string \shortcite{Bar-Ziv1994}. Also, vesicles can undergo oscillatory dynamics when placed in a hypotonic (diluted) environment, swelling with water, and then transiently bursting to release contents before resealing -- iterating the cycle until solute gradients are equalised \shortcite{Oglecka2014,Mally2013,Popescu2008}. Vesicles can fuse together, mixing chemical contents and membrane composition, and increasing their surface-to-volume ratio \shortcite{Caschera2011}. Furthermore, vesicles can bud smaller vesicles both externally and internally \shortcite{oglecka2012}, and often exist in a range of multilamellar and nested structures (vesicles within vesicles, like Russian dolls) that provide an interesting compartmented scenario for reactions. 

Fourth and finally, dynamical effects can be provoked in vesicles by the presence of external forces, energy sources (e.g. \shortciteNP{Zhu2012}), and changing environmental concentrations which may also originate from neighbouring vesicles (Fig. \ref{fig:ch3_dynamic_compartments}d).

In summary, there exists a huge landscape of interesting possibilities for the interaction of chemical networks with the self-assembly physics of vesicles, well before vesicles would have started to produce important components internally. Even vesicle growth, which would seem to require an internal lipid production, could potentially have been achieved by other means, such as by vesicle fusion, lipid competition (see Section \ref{sec:4_2_1}) or by increased lipid uptake rate in osmotic stress conditions \shortcite{Mally2013,Chen2004}. Therefore, before jumping ahead to modelling self-producing protocell scenarios, a proper consolidation of more basic scenarios where protocell membrane and metabolism are more indirectly coupled is required.

The scientific contributions of this thesis, outlined in Chapter \ref{chapter:4} to follow, can be seen as contributions to the (wide) area of modelling chemistry in dynamic compartments.

\chapter[Overview of Scientific Contributions]{\texorpdfstring{Overview of \\ Scientific Contributions}{}}
\chaptermark{Scientific Contributions}
\label{chapter:4}

%
%
%

This chapter summarises, in a simplified and largely non-technical way, the main scientific contributions made by this thesis. Chapter \ref{chapter:5} to follow discusses in more detail how these results fit into the theme of modelling chemistry inside dynamic lipid compartments (the research area explained in Section \ref{sec:3_4}), and into the overarching problem of modelling the appearance of autonomy in protocells.

The published papers on which this chapter reports are included as Appendix \ref{appendix:D}. Material in Section \ref{sec:4_2} was published in {\em Scientific  Reports} as ``Modelling Lipid Competition Dynamics in Heterogeneous Protocell Populations'' \shortcite{shirtediss2014}. Material in Section \ref{sec:4_3} was published in {\em Life} as ``Emergent Chemical Behaviour in Variable-Volume Protocells'' \shortcite{shirtediss2015}. Material in Section \ref{sec:4_4} was published in the {\em Proceedings of the Twelfth European Conference of Artificial Life} as ``Steady state analysis of a vesicle bioreactor with mechanosensitive channels'' \shortcite{shirtediss2013}.

%
%
%
\section[Prelude: The Equilibrium Self-Assembly of Amphiphiles]{Prelude: The Equilibrium Self-Assembly of \\ Amphiphiles%
\sectionmark{Equilibrium Self-Assembly}}
\sectionmark{Equilibrium Self-Assembly}
\label{sec:4_1}

As described in Chapter \ref{chapter:1}, a protocell is a far-from-equilibrium chemical system embodied in a supra-molecular membrane structure which self-assembles through the action of weak forces. A good departure point on the path toward modelling protocells, like for modelling any phenomena, is to start simple: in this case, to develop intuition for how protocell compartments can self-assemble in an {\em equilibrium} setting without internal or external chemistry taking place. As \shortciteA{Sole2009} underlines, ``Self-assembly is an essential component in the path towards cellular systems.'' (p283).

In this regard, as a prelude to the scientific contributions to follow, the first scientific contribution presented below details a spatial lattice model developed to capture the self-assembly of amphiphile molecules, when these molecules form part of complex mixtures containing water and/or oil. Before the lattice model results are presented, Section \ref{sec:4_1_1} creates the context by recapitulating the basic principles underlying the self-assembly of lipids into micelles and vesicles.

%
%
%
%
%
%

\subsection{Basic Principles of Micelle and Vesicle Self-Assembly}
\label{sec:4_1_1}

Liquid water is a vital pre-requisite for life as we know it. One of the reasons why water is so important is due to the particular structure of water molecules: they are {\em polar} and attract each other.\footnote{Polar molecules have no overall charge, but their balance of charge throughout space is unequal and leads to localised regions of attraction and repulsion. Ionic molecules (ions) instead have an overall charge because they carry a surplus or absence of electrons.} In fact, water molecules engage in a highly dynamic network where hydrogen bonds between neighbouring molecules are continuously broken and re-established in another arrangement.\footnote{A hydrogen bond is the name given to the electrostatic attraction between polar molecules. Hydrogen bonds are stronger than Van der Waals forces, but weaker than covalent and ionic bonds.} This dynamic network of bonds leads to an important effect called the {\em hydrophobic effect}. In living systems, this effect drives the folding of proteins into functional three dimensional structures and is key to the formation of bilayer cell membranes.

\begin{figure}
\begin{center}
\includegraphics[width=14cm]{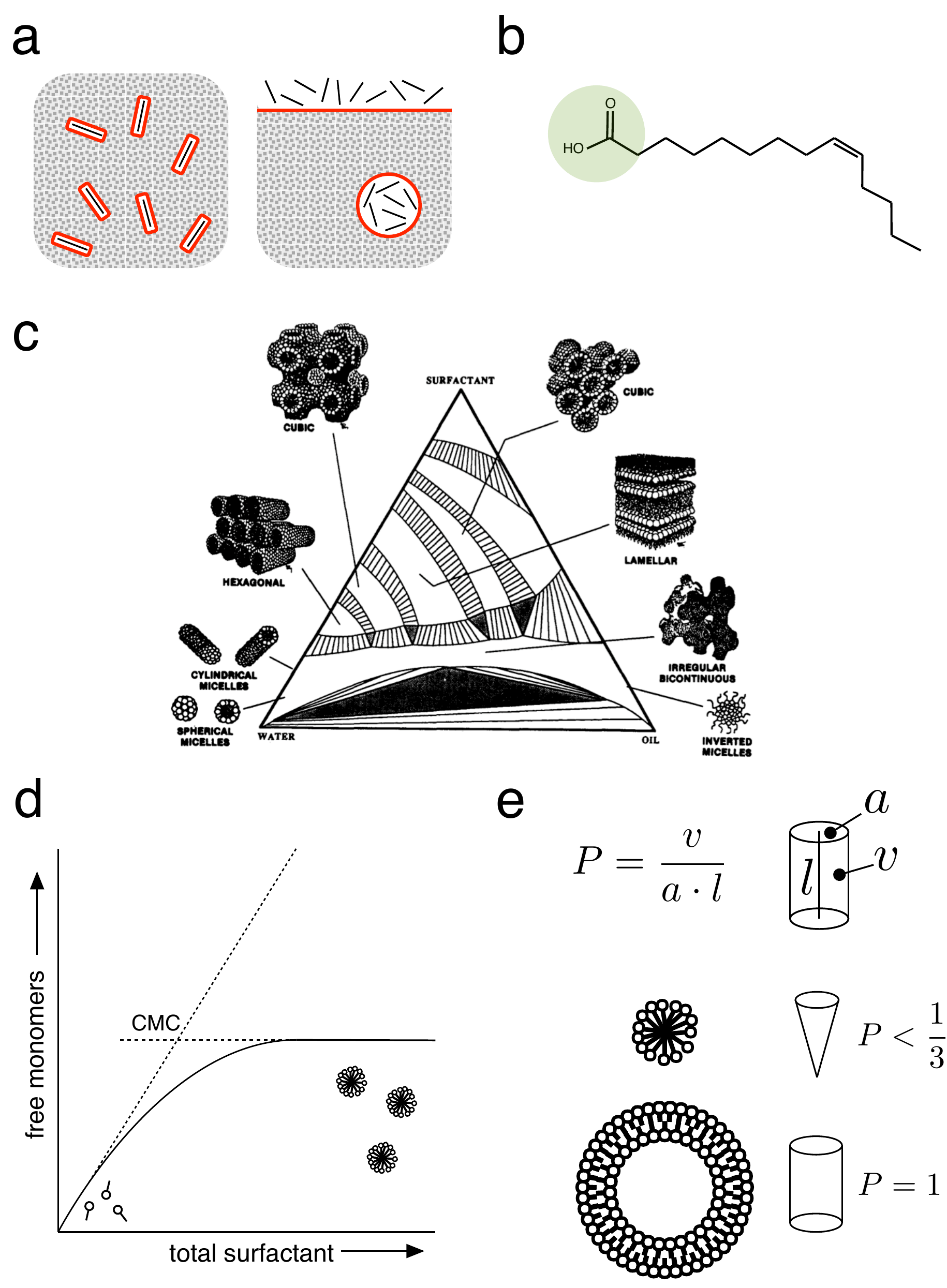}
\end{center}
\caption{
{\textbf{Fundamentals of Amphiphile Self-Assembly.}} (a) The hydrophobic effect. (b) A typical amphiphile (myristoleic acid) with polar head group (green) and non-polar hydrocarbon tail. (c) Mixtures of amphiphiles, water and oil give rise to a rich phase space of equilibrium structures \protect\shortcite{dawson1992}. (d) Self-assembly of micelles (or vesicles) only takes place after CMC (or CVC) surfactant concentration. (e) Effective amphiphile shape (packing parameter, $P$) determines aggregate type. See text for explanation. \hyperref[chapter:copyright]{ View Image Copyright Permissions.}
}
\label{fig:ch4_self_assembly}
\end{figure}

So, what is the hydrophobic effect? When molecules unable to form hydrogen bonds are put into water, like oily hydrocarbon chains for example, they restrict the hydrogen-bonding possibilities that the adjacent water molecules can engage in. This makes the bonding network more ordered as a whole. The hydrophobic effect is the name given to the water system striving to return to a more disordered (or maximal entropy) state again. It does this by aggregating non-polar molecules together, {\em minimising} their interface with the water hydrogen bonding network. This is the reason why oil and water don't mix. Figure \ref{fig:ch4_self_assembly}a illustrates the hydrophobic effect, using black rods to represent oil hydrocarbons and red lines to signify minimisation of the oil-water interface.

The hydrophobic effect becomes even more interesting when {\em amphiphiles} are added to water. Amphiphiles are molecules composed of two parts: a polar or ionic head part that can participate in the water hydrogen bonding network (`water loving' or {\em hydrophilic}), and a non-polar hydrocarbon tail that cannot (`water hating' or {\em hydrophobic}). An amphiphilic lipid molecule is drawn Fig. \ref{fig:ch4_self_assembly}b. In water, the dual hydrophilic-hydrophobic personality of amphiphiles (also called surfactants) means that they orient themselves such that the head parts contact the water interface whilst the tail parts are sequestered away together. This unique behaviour means that mixtures of amphiphiles, water and oil can display a huge diversity of self-assembled microstructures. Figure \ref{fig:ch4_self_assembly}c shows a so-called {\em phase diagram} that systematically maps how a three-part mixture (called a {\em ternary mixture}) of amphiphile, water and oil can transition between different regimes of self-assembly, dependent on the proportion of each of the three components in the mixture.\footnote{ The phase diagram is drawn as an equilateral triangle and uses a barycentric coordinate system. Each point on the triangle surface has the same total concentration of amphiphiles, water and oil; just their mix ratio is different. The vertices of the triangle represent 100\% concentration of amphiphile, water or oil respectively. Only a single amphiphile type is allowed. Cleverly, the area inside the triangle represents all the three-part (ternary) mixtures possible and the edges of the triangle represent all the two-part (binary) mixtures possible.} Although this phase diagram is only illustrative, it conveys well the rich variety of supramolecular structures that are able to exist in ternary mixtures under suitable conditions. Further, the diagram highlights that the transitions between different self-assembly regimes are often abrupt, and the transitions carve up the phase space in a non-trivial way. Phase diagrams can become extremely complex when other factors affecting self-assembly (like temperature, pH, or the presence of multiple amphiphile types) are also taken into account. 

For protocells, the most relevant supra-molecular structures that amphiphiles self-assemble into are micelles and vesicles. As was shown in Fig. \ref{fig:ch1_protocell_architectures}d of Chapter \ref{chapter:1}, micelles are small ordered clusters of relatively few amphiphiles, whereas vesicles represent much larger structures formed of thousands of amphiphiles arranged in a spherical (or quasi-spherical) bilayer. Micelles and vesicles are not static, but very dynamic structures, constantly exchanging lipids with the surrounding solution. As such, one of the key conditions for the self-assembly of micelles and vesicles is the presence of a critical concentration of free amphiphile monomer in the surrounding solution. This concentration is termed the critical micelle concentration (CMC) or critical vesicle concentration (CVC). Figure \ref{fig:ch4_self_assembly}d illustrates the CMC behaviour for micelles. When amphiphiles are initially added to water solvent (x-axis), they exist as free monomers in the solvent. However, steady addition of amphiphiles does not lead to the free amphiphile concentration increasing linearly. Rather, the amphiphiles start becoming clustered into micellar aggregates, and addition of more amphiphiles to the system results in more aggregates being formed instead of an increase in the free monomer concentration (which stays constant at the CMC). The same type of behaviour is observed in the case of vesicles. Typically, the critical concentration depends on the chain length of the amphiphiles forming the aggregates, decreasing as the hydrophobic chain length increases.\footnote{ Amphiphiles with longer hydrophobic chains exchange more slowly between aggregates and solution because of the higher energetic cost of exposing their hydrophobic chain. This translates to a lower critical concentration.} 

An additional factor determining whether micelles or vesicles (or indeed other supra-molecular structures) assemble is the {\em effective shape} of the amphiphiles concerned. The {\em packing parameter} is a useful quantity characterising how bulky an amphiphile head is with respect to its tail, defined as the amphiphile volume v divided by the product of amphiphile head area $a$ and amphiphile chain length $l$ (see Fig. \ref{fig:ch4_self_assembly}e). Amphiphiles with a packing parameter smaller than $\frac{1}{3}$ are conically shaped and are therefore prone to self-assemble into small micelle clusters. Conversely, amphiphiles with a packing parameter close to 1 are cylindrical and form bilayers that can further fold up into vesicles. The packing parameter is influenced by a variety of factors including temperature and pH.\footnote{At the correct pH, adjacent pairs of single chain lipids can develop hydrogen bonds between their head groups and effectively swap their individual cone shapes for a joint cylindrical shape. In this way, by changing the pH, a solution of micelles can be transformed into a solution of vesicles.}

Apart from the core principles reviewed above, many other factors like Van der Waals forces, electrostatic forces, ionic strength, headgroup features, and the presence of co-surfactants and co-solvents (to name but a few) also play subtle but nevertheless important roles in the self-assembly of amphiphiles into micelles and vesicles. In this case, the interested reader is referred to more in-depth treatments by \shortciteA{mouritsen2005,Pohorille2009,Mansy2009,Chen2010,DelBianco2014}.

\subsection{A Lattice Monte Carlo Model of Amphiphilic Self-Assembly}\label{sec:ch4_alice_model}
\label{sec:4_1_2}

To grasp amphiphilic self-assembly more intuitively, a lattice monte carlo (LMC) model of surfactant-water-oil mixtures was developed. The LMC model is technically described in Appendix \ref{appendix:B}, and Fig. \ref{fig:ch4_alice_selected} shows results of using the model to compute selected points on the amphiphile-water-oil phase diagram of Fig. \ref{fig:ch4_self_assembly}c. In Fig. \ref{fig:ch4_alice_selected}, amphiphiles are represented as chains of molecules spanning multiple lattice sites, with the hydrophilic head group coloured red and the hydrophobic tail coloured black. Oil molecules and water molecules each occupy single lattice sites, with oil coloured brown and water coloured blue. 

The LMC model was implemented in just two spatial dimensions and amphiphilic self-assembly was represented in a highly simplified way. Nevertheless, relevant equilibrium self-assembly phenomena could be qualitatively reproduced. As an example, the CMC behaviour of micelles could be observed, with micelles only appearing after a critical surfactant concentration was reached (Fig. \ref{fig:ch4_alice_selected}b,f). Also, familiar structures such as surfactant covered oil droplets in water (Fig. \ref{fig:ch4_alice_selected}c) or surfactant covered water droplets in oil (Fig. \ref{fig:ch4_alice_selected}e) could be produced, along with more complicated `microemulsion' phases where surfactants mediate a complex interface between entangled islands of oil and water (Fig. \ref{fig:ch4_alice_selected}d).

In the literature, the LMC model implemented here has been extended into three dimensions and has notably been used to model the formation of vesicles \shortcite{Bernardes1996,Brindle1992} and to perform detailed calculations of critical micelle concentration \shortcite{Stauffer1994,Brindle1992,Bernardes1994,Larson1992}, including for micelles of two amphiphile types \shortcite{Zaldivar2003}. LMC models are generally a valuable resource for modelling the equilibrium phase behaviour of complex mixtures because, unlike some analytical theories of complex fluids, they don't build in assumptions about self-assembly or restrict the types of structures possible. Instead, self-assembly is modelled from the bottom-up through low-level interactions between individual oil, water and amphiphile chain molecules. Also, LMC models have the benefit of being able to represent the self-assembly of multiple aggregates within the same model, a feat that full-fledged molecular dynamics models are often computationally incapable of.

However, when the intention becomes to go past self-assembly and onto the challenge of modelling aggregates interacting with far-from-equilibrium chemistry, the LMC model described is not the correct formalism to pursue. This is because this formalism is concerned with finding a probable ground state of a closed oil-water-amphiphile mixture at equilibrium (i.e. a state whose energy fluctuations obey the Boltzmann distribution for the final system temperature), rather than being concerned with the kinetic far-from-equilibrium behaviour of an open chemical system (for which no Boltzmann distribution of energy states exists).\footnote{Some studies have tried to make initial attempts at extending the LMC formalism to include basic metabolism alongside the self-assembly of amphiphiles (e.g. see \shortciteNP{McCaskill2007} or \shortciteNP{Ono2005}), pushing the system away from equilibrium. See \shortciteA{Binder1997} Chapter 2 therein, for an advanced discussion of how far lattice monte carlo models are able to be interpreted as dynamical models with a timescale.}

Because the main intention of this thesis is to contribute towards the modelling of autonomous protocells, and because autonomous agents can only operate in far-from-equilibrium conditions \shortcite{Kauffman2000}, the remaining scientific contributions of this thesis are instead based on the semi-empirical vesicle model, as already discussed in Section \ref{sec:3_3_2}. The vesicle model is formulated in terms of kinetic rates, rather than energetic affinities, allowing far-from-equilibrium phenomena such as protocell metabolism to be easily accommodated. The molecular self-assembly of the lipid bilayer membrane in the kinetic vesicle model is not modelled explicitly; modelled instead are the changes to the membrane size and composition in time.

The second scientific contribution, coming next, consolidates the kinetic vesicle model by deriving more realistic kinetic rate equations - based on experimental data - describing how simple fatty acid lipids enter and leave the vesicle membrane.

\begin{figure}
\begin{center}
\includegraphics[width=15.5cm]{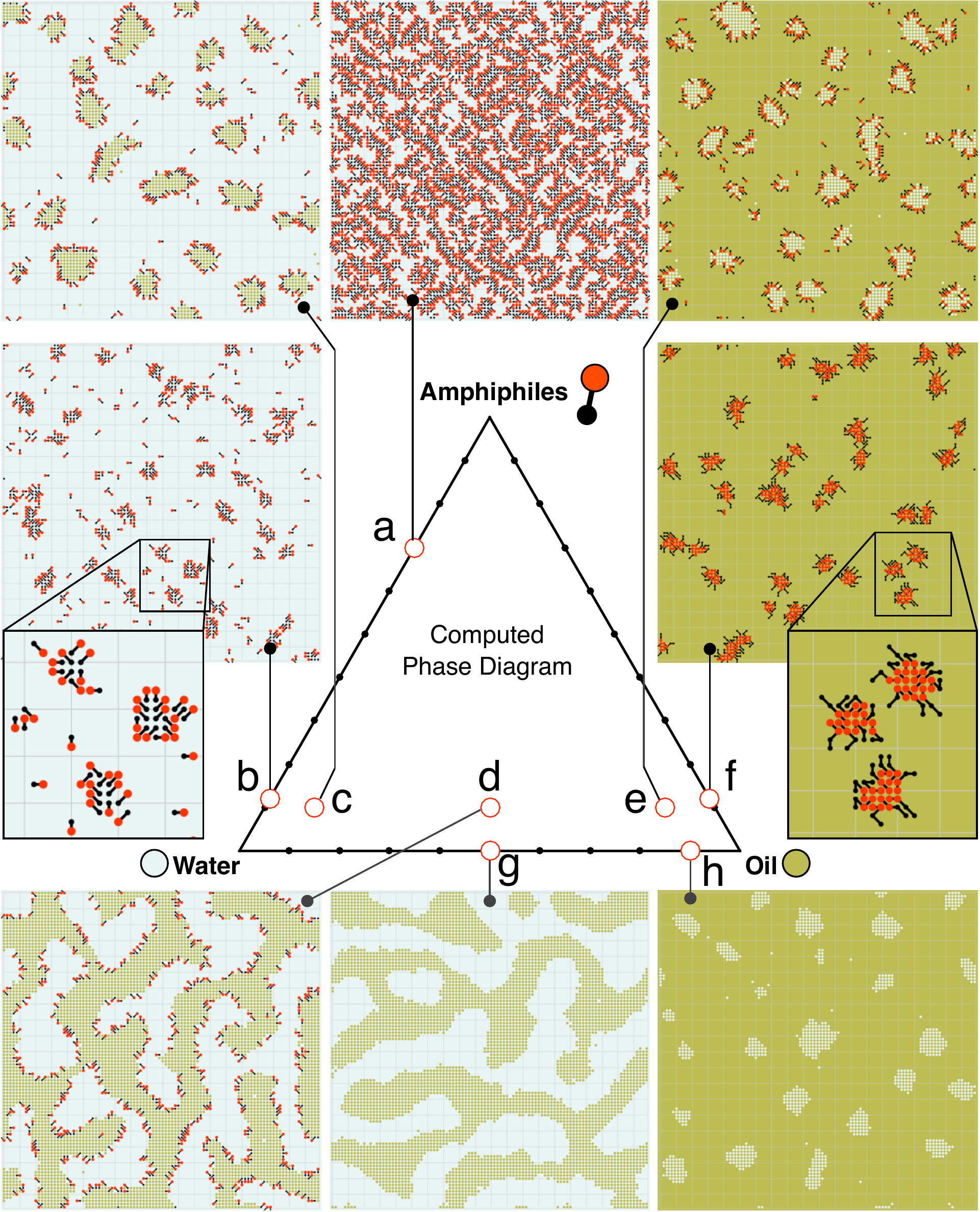}
\end{center}
\caption{
{\textbf{Lattice Model of Amphiphile, Water and Oil Ternary Mixtures.}} Different points on the phase diagram (centre) were computed and found to produce different self-assembled structures, including: stacked bilayers (a), micelles (b), surfactant coated oil/water droplets (c,e), complex microemulsions (d), reverse micelles (f) and oil-water phase separations (g,h). Amphiphiles were modelled as a single head unit connected to a single tail unit except in (f) where amphiphiles had two head units and two tail units. 100x100 two-dimensional lattice.
}
\label{fig:ch4_alice_selected}
\end{figure}
%

%
%
%
\section[Grounding the Semi-Empirical Approach: An Experimentally Validated Kinetics Model for Protocell Membranes]{Grounding the Semi-Empirical Approach: \\ An Experimentally Validated Kinetics Model for Protocell Membranes%
\sectionmark{Membrane Lipid Kinetics Model}}
\sectionmark{Membrane Lipid Kinetics Model}
\label{sec:4_2}

The original Mavelli \& Ruiz-Mirazo semi-empirical vesicle model \shortcite{Mavelli2010,Mavelli2007} introduced in in Section \ref{sec:3_3_2} included a description of the rates at which lipids associate and disassociate with a vesicle membrane. The second scientific contribution, presented below, provides an {\em improved} model of these kinetic equations, extensively validated against experimental data on protocell competition coming from the Szostak lab. These improved membrane lipid kinetic equations allow the semi-empirical model to be used with greater reliability in future studies on the co-evolution of protocell metabolism and membrane. The description of the improved membrane kinetic model is given after the following section, which first explains protocell competition.

%
%
%
%
%
%
%

\subsection{The Phenomena of Protocell Competition}
\label{sec:4_2_1}

\begin{figure}
\begin{center}
\includegraphics[width=16.5cm]{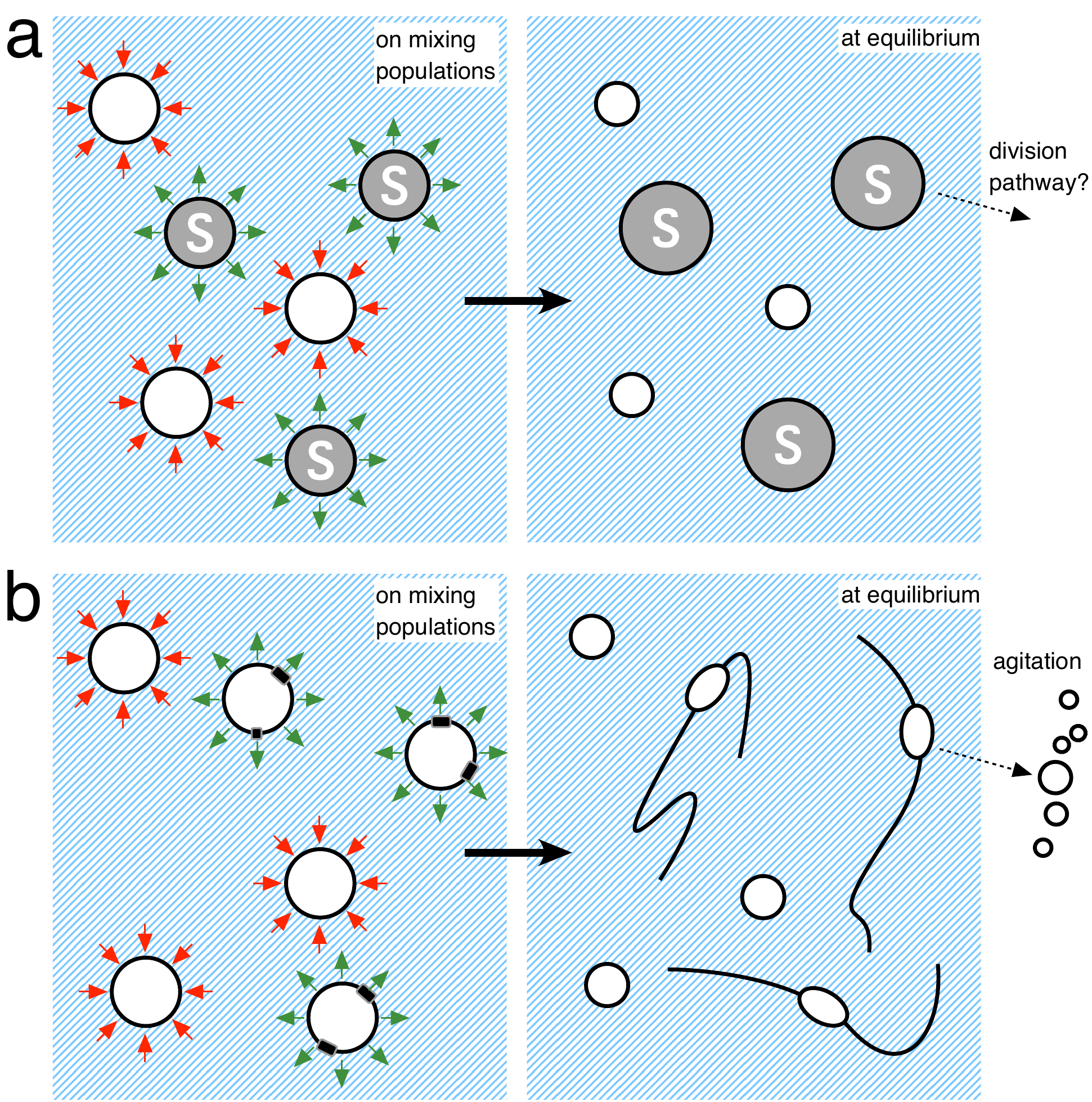}
\end{center}
\caption{
{\textbf{Protocell Competition.}} (a) Osmotic competition. Vesicles marked `S' carry an entrapped cargo of sucrose (or RNA) and grow at the expense of empty vesicles, which shrink.  (b) Phospholipid competition. Vesicles with black segments in the membrane have phospholipids (or other hydrophobic molecules) embedded and grow at the expense of the pure membrane vesicles, which shrink. In this latter case, vesicle growth results in fragile filamentous structures that are prone to divide into smaller vesicles upon agitation. Green and red arrows indicate if vesicles will grow or shrink in surface area respectively from the initial condition.
}
\label{fig:ch4_protocell_competition}
\end{figure}

Associated to the origins of life, a major experimental result over the last 15 years has been the observation that heterogeneous populations of lipid vesicles can compete for shared lipid resources. In a now classic {\em Science} paper, \shortciteA{Chen2004} reported on competition in a population of oleic acid vesicles, wherein vesicles osmotically swollen by an encapsulated cargo of RNA (or sucrose) stole lipids from their empty, osmotically relaxed counterparts by virtue of absorbing lipid monomers more quickly from the solution. As illustrated in Fig. \ref{fig:ch4_protocell_competition}a, this `osmotic competition' resulted in the osmotically tense vesicles growing into larger spheres (relaxing some of their tension), and the empty vesicles shrinking.

Osmotic competition was later supplanted by research into other modes of vesicle competition. This was because it wasn't obvious how the spherical vesicles `winning' osmotic competition could go on to perform the energetically unfavourable feat of division.\footnote{However, if the environment subsequently concentrated, the larger `winner' vesicles would become flaccid (a shape more amenable to division) and the smaller `loser' vesicles would become prone to disintegrate. This possibility does not seem to have been considered.} Phospholipid competition \shortcite{Budin2011} was a different type of competition investigated, resulting from asymmetries in vesicle membrane composition, rather than asymmetries in vesicle membrane tension. In a population of oleic acid vesicles, where some vesicles had a pure oleic acid membrane and others had membranes embedding a small additional fraction of phospholipid, it was observed that the phospholipid laden vesicles would steal lipids and grow at the expense of the pure lipid vesicles. This time, the `winning' vesicles grew larger by following a different pathway, elongating and becoming more prone to division into smaller vesicles (Fig. \ref{fig:ch4_protocell_competition}b).

The phospholipid competition findings corroborate earlier work by \shortciteA{Cheng2003}. Looking into the final size distributions resulting from heterogeneous mixtures of vesicles, the latter authors reported that mixing pure phospholipid vesicles with pure fatty acid vesicles resulted in a single peak size distribution, similar to the initial peak of the phospholipid vesicle sizes. This result suggested that the phospholipid vesicles rapidly `stole' the free fatty acid monomers in the solution, causing the destruction of the fatty acid vesicle population.\footnote{Interestingly, \shortciteA{Cheng2003} also reported competition in a population of pure oleic acid vesicles. Larger vesicles in the population could absorb available micelles faster and thus grow faster than could the smaller vesicles. }

More recently, \shortciteA{Adamala2013} have reported that vesicles containing a membrane fraction of hydrophobic peptides can produce a nearly identical competition effect to vesicles containing a membrane fraction of phospholipid. Therefore, `phospholipid' competition may actually be reflecting a more general effect, initially discovered with phospholipids, but not reliant on their specific molecular properties.

%

%
%
%
%
%
%

\subsection[An Improved Lipid Kinetics Model for Fatty Acid Vesicle Membranes]{An Improved Lipid Kinetics Model for Fatty Acid \\ Vesicle Membranes}
\label{sec:4_2_2}

To arrive at a set of candidate kinetic equations for the uptake and release rates of fatty acids from vesicles, we used experimental data from protocell competition to set up an {\em inverse problem}.\footnote{Inverse problems consist of identifying the probable mechanisms and causes that underlie a given set of macroscopic observations. They are the opposite of (easier) forward problems that involve the generation of macroscopic observations from a known set of causes or mechanisms (e.g. running a simulation). Solving inverse problems is important in many areas of science.} The inverse problem was as follows: given the overall change in vesicle sizes in a competing population, to identify what lipid exchange mechanisms are likely to take place at the micro-scale level of individual vesicles in order to produce the overall population-level effect.\footnote{In fact, a similar inverse problem strategy is used to deduce the elementary reaction steps underlying a chemical reaction, given the concentration dynamics of the reactants and products of the overall reaction.}

\begin{figure}
\begin{center}
\includegraphics[width=15.5cm]{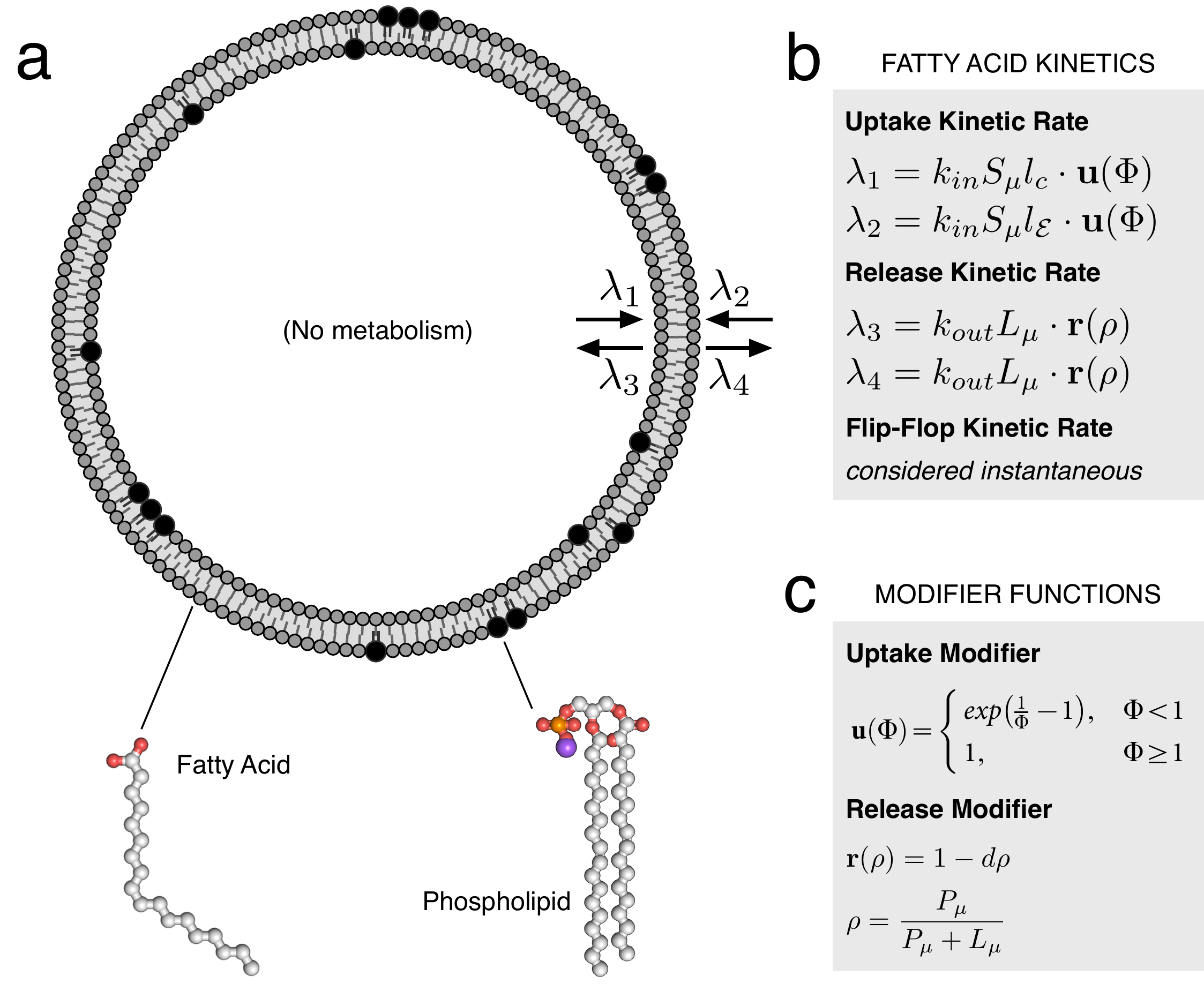}
\end{center}
\caption{
{\textbf{Improved Model of Vesicle Membrane Lipid Exchange Kinetics.}} (a) Empty fatty acid vesicle containing a membrane fraction of phospholipid. (b) Kinetic rates, in molecules per second, at which fatty acid molecules enter an leave the vesicle membrane. Due to their slow exchange rate, phospholipids are considered as static in the membrane. (c) Functions used to modify the uptake and release rates of fatty acid, additional to the original vesicle kinetic model. Refer to Box 2 of Chapter 3 for meaning of constants and symbols. See text for discussion.
}
\label{fig:ch4_competition_model}
\end{figure}

\begin{figure}
\begin{center}
\includegraphics[width=15.5cm]{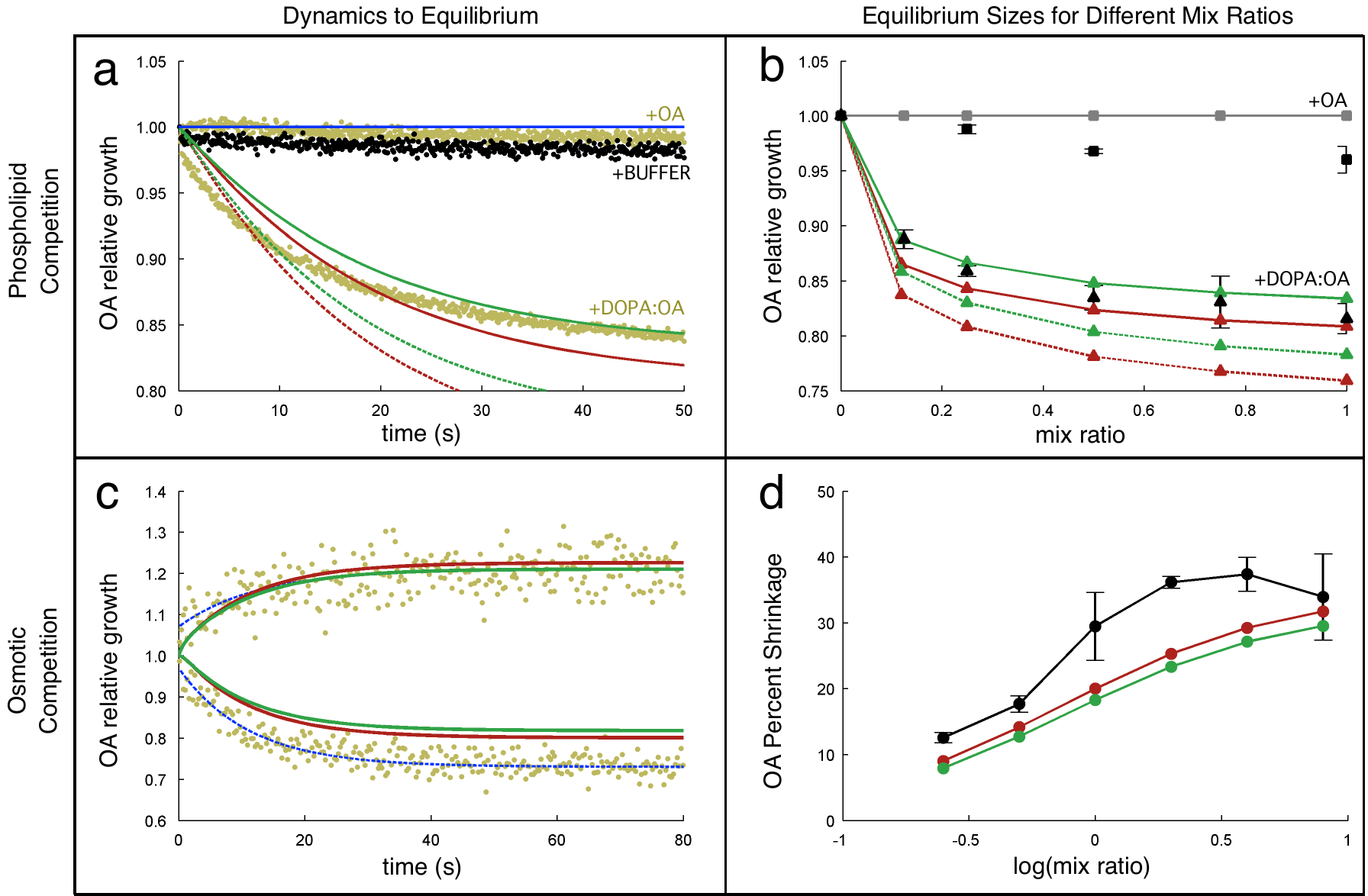}
\end{center}
\caption{
{\textbf{Protocell Competition: Kinetic Vesicle Model vs. Experimental Results.}} Kinetic model outcomes shown as red and green lines. Experimental outcomes shown as background dots, and black squares, circles and triangles. (a,b) Phospholipid competition. (c,d) Osmotic competition. Results adapted from \protect\shortciteNP{shirtediss2014}. See text for discussion.
}
\label{fig:ch4_competition_results}
\end{figure}

Figure \ref{fig:ch4_competition_model} summarises the vesicle model and lipid kinetic rate equations used to re-produce the experimental data on both phospholipid and osmotic competition. In our model of a competing vesicle population, vesicles did not have specific local interactions with other vesicles. Rather, a `meanfield' approach was taken wherein vesicles were assumed well-mixed in solution, affecting each other by consuming or adding to a common pool of free lipid monomer in the environment. Figure \ref{fig:ch4_competition_results} gives a model versus experimental comparison for selected results.

In order to best match the experimental results (discussed shortly), it was crucial to include two additions to the original kinetic vesicle model proposed by Mavelli \& Ruiz-Mirazo \shortcite{Mavelli2010,Mavelli2007}. These additions were the uptake and release modifier functions, detailed in Fig. \ref{fig:ch4_competition_model}c. The release modifier function gave the ability to slow the fatty acid release rate as the phospholipid content of the membrane increased.\footnote{This modelled a mechanism we called the `direct effect' whereby the decreased fluidity brought to the bilayer membrane by the presence of phospholipids slowed down the release rate of fatty acids. It is the main mechanism that \shortciteA{Budin2011} cite as being responsible for phospholipid competition amongst vesicles. In our paper, we actually argued for an `indirect effect' being the more relevant effect in competition. In the indirect effect, the presence of phospholipid in a vesicle membrane drives growth simply through a geometric asymmetry (see \shortciteNP{shirtediss2014}).} More significantly, the uptake modifier function was made into a conditional function that accelerated the rate of lipid uptake only when the vesicle was in osmotically stressed states, not in flaccid ones.\footnote{The conditional function means that all flaccid vesicles uptake lipids at the same rate. Thus a population of vesicles deflated to different degrees can all co-exist in the same solution. This effect is also observed experimentally.} This conditional function turned out to be pivotal for reproducing the experimental competition results -- in particular, for reproducing a `continuous growth' behaviour whereby vesicles with a phospholipid fraction keep growing as more pure fatty acid vesicles are added, even though their phospholipid fraction becomes further and further diluted -- and thus represents a valuable new addition to the Mavelli \& Ruiz-Mirazo vesicle model (already included as Equation 3.5 in Box 2 of Section \ref{sec:3_3_2}). 

It is worth briefly summarising how the kinetic vesicle model could reproduce experimental results. Experimental results on phospholipid and osmotic competition have been presented in the literature in two forms: one form is {\em dynamics}, showing the change in the surface area of vesicles over time from the instant they are mixed, to the time of equilibration; the other form is {\em stoichiometry}, showing how different mix ratios of vesicles lead to different surface area sizes at equilibrium. Figure \ref{fig:ch4_competition_results} demonstrates how the kinetic vesicle model reproduces both dynamic behaviour and stoichiometric behaviour, to a good approximation, in both phospholipid competition and osmotic competition.

For phospholipid competition, Fig. \ref{fig:ch4_competition_results}a shows the relative change in the surface area of a population of pure fatty acid vesicles (oleic acid vesicles, denoted OA), when either more pure fatty acid vesicles are added in an equal ratio (yellow dots marked `+OA') or when fatty acid vesicles containing a membrane phospholipid fraction are added in an equal ratio (yellow dots marked `+DOPA:OA'). The kinetic model reproduces well that the fatty acid vesicles have no surface area change when more fatty acid vesicles are added (blue line) and also quite accurately mimics the shrinkage dynamics of the fatty acid vesicles when vesicles containing a phospholipid fraction are added (red and green lines). Figure \ref{fig:ch4_competition_results}b further shows that the model predicts stoichiometry reasonably well, i.e. how the equilibrium surface area of the fatty acid OA vesicles decreases at equilibrium (black triangles are experimental data) as more and more phospholipid-laden vesicles are added.

Moving onto osmotic competition, \ref{fig:ch4_competition_results}c shows experimental data points (yellow dots) for the growth of swelled and shrinkage of relaxed fatty acid vesicles, when they are mixed in an equal ratio. Again, the kinetic model quantitatively predicts, to a fair accuracy, the dynamic surface area change trends (red and green lines). Stoichiometry for osmotic competition is also approximately reproduced by the model. Figure \ref{fig:ch4_competition_results}d shows that adding more swelled vesicles to a population of initially relaxed (non-swelled) fatty acid vesicles causes the shrinkage of the relaxed vesicles to plateau, rather than to linearly continue.

Other experimental observations reproduced with the model include `winning' vesicles finishing as spherical vesicles in osmotic competition and as flaccid vesicles in phospholipid competition. Also, the model demonstrated competition in mixed vesicle populations, where, for example, {\em all} of the vesicles possessed a differing phospholipid fraction (i.e. none were pure fatty acid vesicles). In fact, the model was able to make predictions about vesicle competition effects in as yet untested scenarios, such as osmotically swelled vesicles versus phospholipid-laden vesicles.

Indeed, vesicle competition is a challenging research sub-field in its own right, and the interested reader is encouraged to read the full paper given in Appendix \ref{appendix:D}. With regards to modelling autonomous protocells, our foray into protocell competition served to provide a more accurate set of lipid exchange rate equations for the membrane of the semi-empirical vesicle model. The vesicle competition scenario was a closed system of empty vesicles that, from a initial non-equilibrium state (after mixing different vesicle populations), settled to a unique equilibrium point. This scenario presented an ideal simplified system from which to deduce membrane fatty acid exchange kinetics.\footnote{One might ask if lipid exchanges rates could be deduced directly from experiments. Experimentally, it is usually difficult to ascertain microscopic rate constants directly. This is because an experiment needs to be performed that effectively isolates a single process (e.g. lipid release rate) and cancels out {\em all other effects}.}

Technically, the use of the word `protocells' in protocell competition is incorrect, since by the definitions given in Chapter \ref{chapter:1}, protocells also require chemical reactions to be linked with self-assembled compartments. In the competition work reviewed above, `protocells' were essentially empty vesicles with a fixed aqueous internal volume (but with variable surface area). The third scientific contribution following below turns the attention away from modelling protocell membranes toward modelling protocell metabolism, and particularly metabolism in changing solvent volume conditions. As a result, a new general paradigm for framing protocell metabolism is introduced.

%
%
%
\section[A New Paradigm for Metabolism in Early Protocells]{A New Paradigm for Metabolism in \\ Early Protocells%
\sectionmark{A New Paradigm for Protocell Metabolism}}
\sectionmark{A New Paradigm for Protocell Metabolism}
\label{sec:4_3}

Modelling protocell membranes has the advantage that empirical data exists for likely prebiotic vesicles. Plausible prebiotic vesicles can be reconstructed in the lab, usually from mixtures of simple fatty acids, and the properties of their membranes can be rigorously investigated. Membranes are the physical structures giving a `handle' into the experimental investigation of protocells.

Modelling protocell metabolism - the other half of the equation - is however a much more difficult feat. The types of prebiotic reactions taking place inside protocells (or in free solution) at the origins of life remain as an extremely speculative matter (see \shortciteNP[Chapter 3]{Dyson1999}). It is known that protocells eventually developed into full-fledged cells with a DNA-RNA-protein metabolism, but the earlier stages leading up to this point admit many different permutations. Some protocell models commit to replicating templates appearing first (e.g. \shortciteNP{Szostak2001}), whereas other experimental models, like \shortciteA{Gardner2009}, have investigated autocatalytic chemical cycles inside vesicles (i.e. the sugar-synthesizing formose reaction). On the other hand, theoretical protocell models have employed various metabolic schemes inspired by the autocatalytic reaction cycles of G{\'a}nti (e.g. \shortciteNP{Ruiz-Mirazo2008,Mavelli2007}) or the collectively catalytic reaction networks of Kauffman (e.g. \shortciteNP{villani2014,hordijkSteel2015}) or Rosen (e.g. \shortciteNP{Piedrafita2012,Piedrafita2013}).

The third scientific contribution presented in this subsection does not commit to, nor passes comment on any one of the specific metabolic schemes mentioned above. Rather, the scientific contribution lies in suggesting a new way that protocell metabolism can be regarded {\em in general}, opening up a whole new line of investigation for modelling metabolism-membrane couplings in future. This new paradigm for protocell metabolism involves taking into serious consideration the possible role(s) of osmotic water flow across the semi-permeable bilayer membrane of a protocell. 

Whereas osmotic water flow is a feature often disregarded completely in models of protocells, its proper consideration actually leads to a unique reactor scenario inside protocells wherein the {\em solvent volume is variable}. As will be more fully developed in Section \ref{sec:4_3_2} and Section \ref{sec:4_4}, regarding the solvent as an active part of the chemical system, rather than just being a passive `backdrop' for the interaction of solutes, has a number of important implications for reaction networks inside protocells. Before that, Section \ref{sec:4_3_1} first explains why osmosis (and variable volume) should be expected to be a significant feature in early protocells.

%
%
%
%
%
%

\subsection{Cells Regulate Their Cytosol Volume, Early Protocells Could Not}
\label{sec:4_3_1}

To survive, cells must keep many variables within homeostatic bounds. One of the fundamental variables that cells require to regulate is their cytosol volume \shortcite{Lang1998,Morris2002,Hoffmann2009}. Because all cells are bound by a semi-permeable lipid membrane, they constantly face an {\em osmotic problem}. 

The cell cytosol contains many trapped macromolecules that carry a charge and thereby attract counter ions into the cell interior. The cytosol also contains numerous metabolites such as sugars, amino acids and nucleotides that also attract further counter ions. The high number density of molecules in the cell interior causes an osmotic water flow across the cell membrane\footnote{An interesting point is how water manages to permeate membranes. Lipid bilayer membranes have an oily hydrophobic core (Section \ref{sec:4_1}), which is actually unfavourable for water molecules to cross. However, water is still able to permeate bilayer membranes because: (i) the hydrophobic domain in the bilayer is only very thin (around 4nm thick), (ii) transient defects caused by lipid flip-flop permit the passage of some water molecules through the bilayer and (iii) there is a huge number of water molecules present (it is the solvent!) on each side of the membrane. Moreover, some cells actively encourage water to flow across their membranes by producing aquaporin channels that permit water molecules to cross the membrane in single file \shortcite{Preston1992}.}, to equalise the osmotic pressure difference between inside and out. Water permeates the lipid membrane of cells at a rate orders of magnitude faster than solutes\footnote{If the opposite was true, i.e. if solutes were able to permeate much faster than water, then solute concentration gradients would tend to equilibrate rapidly before significant cell volume changes could take place. Thus, the osmotic problem would be less relevant.}; this means that variations in the total solute concentration inside or outside the cell can induce rapid and significant changes in cellular volume. Without volume regulation mechanisms, sooner or later, cells would irreversibly `implode' (becoming too small to support vital cell functions leading to apoptosis through excessive crowding and DNA damage), or `explode' (losing critical components of the cytoplasm through a burst cell envelope).

\begin{figure}
\begin{center}
\includegraphics[width=13.2cm]{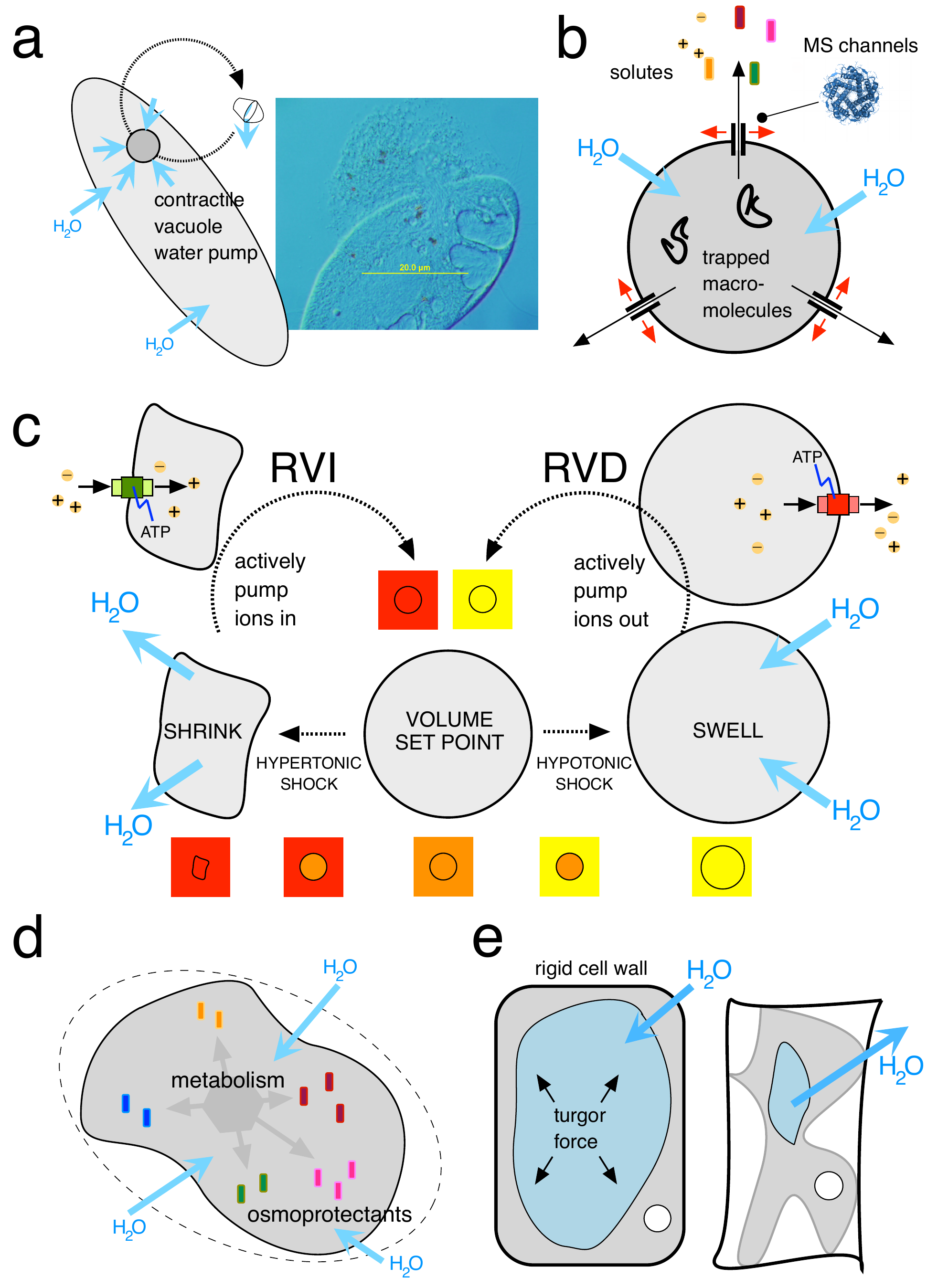}
\end{center}
\caption{
{\textbf{Selected Mechanisms of Volume Regulation in Cells.}} (a) Paramecium water pump. Inset: A paramecium bursts, spilling its cytoplasm when osmoregulation fails. (b) Mechano-sensitive channels forming `emergency release valves'. (c) Regulatory volume increase (RVI) and decrease (RVD). Colours indicate internal / external concentrations: Red = high, orange = medium, yellow = low concentration. The cell eventually returns to its volume set point, but at a different internal concentration. (d) Internal production of osmo-protectants to halt a volume decrease. (e) Rigid cell walls in e.g. plant cells convert excessive cell volume into turgor, and in cases of volume decrease maintain cell shape. See text for discussion.
}
\label{fig:ch4_osmoreg_in_cells}
\end{figure}

Perhaps the most striking example of cell volume regulation is provided by the {\em Paramecium} protozoan. Paramecium is a single-celled eukaryotic organism that prefers to dwell in stagnant pond water. Due to its high internal concentration of macromolecules, metabolites and ions, water constantly flows into the paramecium from the lower concentration (hypotonic) surroundings. To prevent from bursting, the paramecium maintains special internal membrane-bound vacuoles that have a yet higher solute concentration than the cytosol \shortcite{Stock2002}. Water subsequently cascades from the surroundings, into the cytosol and then on into the vacuoles, which swell. When full, the vacuoles contract, mechanically expelling water (and metabolic wastes) out of the paramecium (Fig. \ref{fig:ch4_osmoreg_in_cells}a). In this way, the paramecium is able to actively maintain a concentrated interior whilst inhabiting dilute environments.\footnote{ The paramecium volume regulation mechanism, because it actively pumps water, is able to control volume {\em and} maintain the total concentration inside the cell higher than that of the surroundings. The other volume regulation mechanisms discussed below are also able to control volume, but they always end up with the total concentration of osmolytes inside the cell equal to that of the environment.}

The mechanical pump of the paramecium is an advanced case of osmoregulation, requiring coordinated attachment of the vacuole to the cell plasma membrane and then contraction of the vacuole. More basic mechanical forms of osmoregulation exist, such as mechano-sensitive channels found in {\em E. coli} bacteria and plant and animal cells \shortcite{Kung2005}. Mechano-sensitive channels act as `emergency release valves' for solutes when the environment suddenly dilutes (Fig. \ref{fig:ch4_osmoreg_in_cells}b). Rather than being controlled by complex metabolic processes, mechano-sensitive channels simply open a small water-filled pore (like a camera iris) whenever there is sufficient membrane tension. The open channels cause a rapid re-equilibration of solute gradients and halt the expansion of the cell volume before the cell gets to bursting point. Macromolecules remain safely entrapped inside the cell during the process, due to the small channel diameters. Mechano-sensitive channels are modelled in Section \ref{sec:4_4} to follow.

Other well-documented mechanisms that cells employ to reverse short term changes in volume are {\em regulatory volume increase} (RVI) and {\em regulatory volume decrease} (RVD) \shortcite{Hoffmann2009}. These two metabolic mechanisms shown in Fig. \ref{fig:ch4_osmoreg_in_cells}c involve the cell using an energy carrier (ATP) to actively pump ions across the membrane, thereby approximately restoring the cell volume to its set point. If the cell has experienced a hypertonic shock (i.e. the environment has become more concentrated than the cell interior), then RVI operates and ions are pumped into the shrinking cell to enlarge the water volume. Likewise, if the cell has experienced a hypotonic shock (i.e. the environment has become more dilute with respect to the cell interior), then RVD switches on and ions are pumped out to curb the volume increase of the cell. 

One problem with RVI in particular is that, over prolonged time periods, surplus ions pumped into the cell can start to interfere with the essential functions of enzymes and macromolecules. A longer-term adaptation that cells can make to survive in high concentration environments (such as cells in the kidney, that are continually exposed to high external concentrations of NaCl and Urea) is to start the internal metabolic synthesis of osmo-protectants \shortcite{Burg2007}. Such molecules, also called `compatible solutes', increase the total solute concentration inside the cell without interfering with metabolic functions, even when they are present at high concentrations (Fig. \ref{fig:ch4_osmoreg_in_cells}d). 

Finally, some cells employ rudimentary physical strategies - not regulation mechanisms as such - to cope with osmotic challenges. Plants cells and bacteria, for example, possess an inflexible cell wall which means that cell water volume is only able to expand up until a certain limit, after which water entering the cell instead contributes to internal turgor pressure (Fig. \ref{fig:ch4_osmoreg_in_cells}e). With a cell wall, a cell is able to use energy resources that would have been spent on volume regulation for other means, such as movement. Still other cells, like oocyte cells produced in the ovaries of adult female frogs manage to remain agnostic about osmosis simply by being extremely impermeable to water \shortcite{Preston1992}.

In reality, cell volume regulation is an involved topic and cellular hydration state is closely intertwined with the correct function of cellular metabolism on many levels \shortcite{Haussinger1996}.\footnote{For example, see \shortciteA{Klipp2005} (summarised by \shortciteNP{Dhaeseleer2005}) for a systems biology model that captures some aspects of the response that yeast cells have to a hypertonic shock (a sudden concentration of the environment). This model is notable because it integrates biophysics (i.e. a model of membrane water flow), in addition to standard biochemistry.} Cells may employ one or more of the mechanisms listed above (and others not listed) in order to cope with transient osmotic challenges originating both inside and outside of the cell. The aim here is not to comprehensively review cell volume regulation, but just to give a sense of the relatively advanced mechanisms that cells use for averting immediate osmotic threats and for adapting to persistent environmental changes in the longer term. 

After this brief tour, the essential point to make is this: early protocells could not have possibly possessed such elaborate volume regulation mechanisms. Early protocells were likely to have been instantiated at a much reduced complexity, lacking cell walls and elaborate protein membrane channels able to actively pump ions or act as mechano-sensitive helices. Highly organised water cascading and pumping systems, such as that employed by the paramecium, would have certainly been out of the question. Instead, early protocells were likely to be composed of simple fatty acid based membranes that would have had a significant permeability to water and most solutes \shortcite{Deamer2008}. Hence, it can be reasoned that early protocells were very susceptible to osmotic water flow and volume change. 

In fact, many experimental studies have confirmed that lipid vesicles undergo volume changes readily. As discussed before, \shortcite{Chen2004} reported that fatty acid vesicles became osmotically swelled by an internal cargo of sucrose or RNA, and this lead to them stealing lipids from relaxed vesicles. \shortciteA{Sacerdote2005} have used the shrink-swell volume dynamics of vesicles to calculate membrane permeability to various sugars. \shortciteA{oglecka2012} have shown that complex shape changes can be induced in giant unilamellar vesicles following concentration of the environment, and that vesicles exhibit swell-burst cycles if the environment is instead diluted \shortcite{Oglecka2014}. Finally, in an extreme example of volume change, \shortciteA{Zhu2011} have documented that multilamellar vesicles literally `explode' when intensely illuminated by a metal halide lamp. The lamp light oxides buffer molecules inside the vesicle, quickly making more internal molecules that in turn escalate the internal osmotic pressure to a critical level.

\subsection[Variable Solvent Volume And Emergent Chemical Dynamics in Early Protocells]{Variable Solvent Volume And Emergent Chemical \\ Dynamics in Early Protocells}
\label{sec:4_3_2}

The aqueous interior of simple lipid vesicles actually represents a unique chemical reaction environment. Not only is the solvent volume inside the vesicle variable, but the volume varies as a function of the chemical reactions happening inside the vesicle itself.\footnote{and as a function of concentration fluctuations in the environment.} This provides a very interesting and under-explored scenario for chemical kinetics in early protocells, at a stage before volume regulation mechanisms could have been developed.

\begin{figure}
\begin{center}
\includegraphics[width=15.2cm]{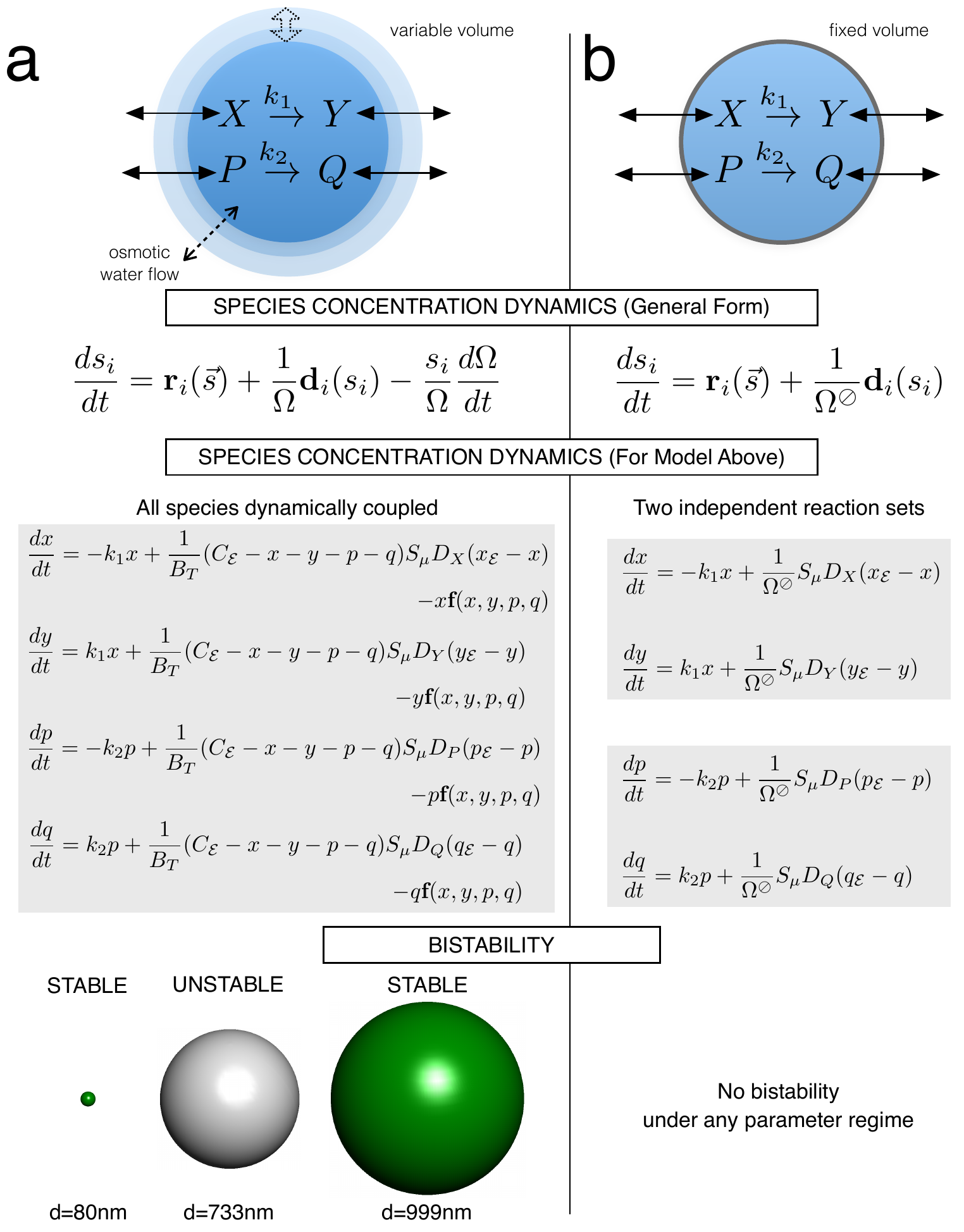}
\end{center}
\caption{
{\textbf{Minimal Model of Chemical Reactions in Protocells with Variable Volume.}} (a) Formulation when solvent volume in vesicle varies as a function of species concentrations. Dynamic equations are derived assuming water flow across the membrane is instantaneous. Function \textbf{f} is a non-linear function of all species concentrations, not included explicitly. (b) Formulation when solvent volume is fixed, for comparison.
}
\label{fig:ch4_varyvol_model}
\end{figure}

For this third scientific contribution, chemical reactions in a variable volume vesicle were modelled by the minimal mathematical formalism given in Fig. \ref{fig:ch4_varyvol_model}a. This is to be contrasted against the (comparatively simple) mathematical formalism for chemical reactions in a fixed volume vesicle given in Fig. \ref{fig:ch4_varyvol_model}b. In both cases, the concentration dynamics for a particular species $s_i$ is determined by the reactions chemically producing that species (function \textbf{r}) and the diffusion of that species into and out of the vesicle (function \textbf{d}). Additionally, in the variable volume formalism, the vesicle volume $\Omega$ is determined by the species concentrations (rather than being a constant), and there is a further `dilution term' that accounts for the effect of volume variation on species concentrations \shortcite{Pawlowski2004}. The negative sign of the dilution term encodes that if the volume increases, chemical concentrations decrease and vice versa.\footnote{Note: Variable volume solvent inside the vesicle does not affect the {\em reaction mechanism} of the internal chemical reactions. The kinetic rate of each reaction remains the same function of the reactant concentrations. However, variable volume does affect the reactant concentrations. Rates of bi-molecular reactions steps (2 reactants turn into $n \ge 1$ products) are directly affected by volume changes whereas rates of unimolecular reaction steps (1 reactant turns into $n \ge 1$ products) are not.} Figure \ref{fig:ch4_varyvol_model} gives an example of two unimolecular reactions $X \rightarrow Y$ and $P \rightarrow Q$ operating in variable and fixed volumes, and lists their respective dynamic equations.

Beyond the technicalities, the most essential points to take away from Fig. \ref{fig:ch4_varyvol_model} are the following. When the solvent volume inside the protocell is variable, (i) the dynamic equations for the species concentrations contain more non-linear terms, and (ii) the concentration dynamics of each chemical species is dependent on the concentrations of {\em all} chemical species inside the vesicle. The very interesting implication is that {\em chemically independent} reaction systems (i.e. that use exclusive sets of chemical species) become {\em indirectly coupled} together via water osmosis in a variable volume vesicle, forming a larger chemical system with potentially very complicated dynamics. We coined this phenomenon `osmotic coupling'.\footnote{This is to be distinguished from `chemiosmotic coupling', as discussed by 
\shortciteA{Harold1991,Mitchell1991}. Chemiosmotic coupling instead refers to the linking of chemical reactions and membrane transporter proteins (`osmotic' in this latter case refers to the membrane itself, rather than water flow through the membrane).} It is an important phenomena, but has received no attention so far in the theoretical or experimental protocell literature.

For a simple example of osmotic coupling, refer again back to Fig. \ref{fig:ch4_varyvol_model} When the vesicle volume is fixed (Fig. \ref{fig:ch4_varyvol_model}b), the dynamics of the unimolecular $X \rightarrow Y$ reaction system are independent from the dynamics of the unimolecular $P \rightarrow Q$ reaction system. Removing either of the reaction systems has no effect on the dynamical behaviour of the other. However, in the variable volume case (Fig. \ref{fig:ch4_varyvol_model}a), the two reaction systems {\em do} mutually influence each other. Their dynamical coupling, rather than being a direct chemical link, instead comes about indirectly by virtue of the fact that both reaction systems are situated in a variable volume solvent (space for molecular collisions) that they jointly determine the size of. In fact, in variable volume solvent, the two unimolecular reactions can produce a unconventional bistable switch, permitting (under the correct parameter regime) two distinct volumes at which flow rates are stable through the vesicle. Conversely, when the vesicle volume is set at a static value (Fig. \ref{fig:ch4_varyvol_model}b), the unimolecular reaction systems $X \rightarrow Y$ and $P \rightarrow Q$ become dynamically independent and are incapable of bistable behaviour.

\begin{figure}
\begin{center}
\includegraphics[width=15.5cm]{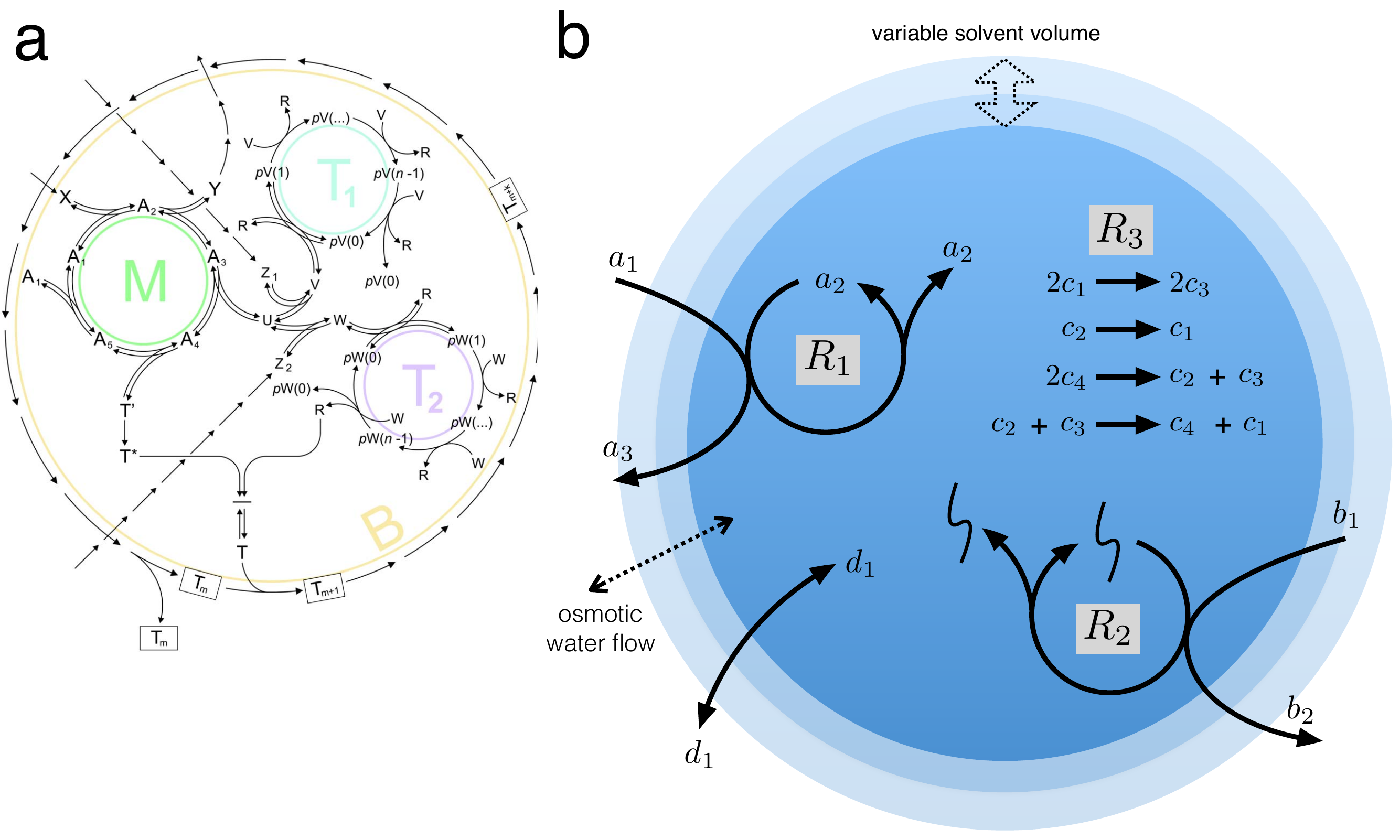}
\end{center}
\caption{
{\textbf{Osmotic Coupling Provides a New Paradigm for Protocell Metabolism.}} (a) The classical `one piece' view of protocell metabolism. A protocell contains a single chemical network responsible for all functions. Pictured is the double template chemoton from \protect\shortciteA{zachar2011}. (b) The unconventional but compelling view of protocell metabolism provided by osmotic coupling. When the aqueous core of a vesicle-based protocell has a variable volume, stoichiometrically independent reaction systems (R1, R2, R3 above) become indirectly coupled. Moreover, entrapped reaction systems with all species membrane impermeable (R3), and also inert solutes that diffuse across the protocell membrane ($d_1$) can both play a role in the transient dynamic behaviour of the protocell system. Osmotic coupling of independent reaction systems provides a new perspective on the dynamic behaviour of protocells.
}
\label{fig:ch4_osmotic_coupling}
\end{figure}

The new paradigm for protocell metabolism brought about by osmotic coupling is illustrated in Fig. \ref{fig:ch4_osmotic_coupling} and can be summarised as follows. Traditional protocell models (e.g. G{\'a}nti's Chemoton, or the Ribocell) blindly follow the assumption of a `one piece' chemical network inside a protocell whereby all chemical species are `stoichiometrically' connected (Fig. \ref{fig:ch4_osmotic_coupling}a). That is to say, a finite sequence of reactions can be traced between one chemical species and another. The dynamical complexity of such a chemical network is tied to the number and organisation of reactions. However, a new possibility presents itself when the physical phenomenon of water osmosis is rigorously considered. In variable volume, a complex chemical network can be created inside a protocell from multiple chemically independent reaction systems that mutually influence each other indirectly through osmotic coupling (Fig. \ref{fig:ch4_osmotic_coupling}b). In this latter case, complex non-linear dynamics can result `for free' from the variable volume coupling of reaction systems.

This work has by no means classified all of the possible dynamic behaviours stemming from osmotically coupled reaction sets; rather, it has opened up a `pandora's box' to be pursued in future works.

%
%
%
\section[Indirect Coupling of Protocell Metabolism and Membrane]{Indirect Coupling of Protocell Metabolism and Membrane%
\sectionmark{Indirect Coupling of Metabolism \& Membrane}}
\sectionmark{Indirect Coupling of Metabolism \& Membrane}
\label{sec:4_4}

Variable solvent volume inside a protocell, as described in the previous contribution, could have had further significance beyond the osmotic coupling of independent aqueous-based reaction systems. Variable volume, through causing effects like internal osmotic pressure, may have additionally served to indirectly couple metabolic (chemical) with membrane (physical/mechanical) processes, promoting further complex protocell dynamics.

One interesting example in this direction involves rudimentary peptide channels which could have embedded into the membranes of early protocells. Osmotic pressure inside a vesicle would act as a chemical-to-mechanical transducer, converting total osmolyte concentration inside the vesicle into membrane tension. Conversely, membrane tension could align and activate the membrane channels, which would increase the passive diffusion rate of solutes and ions across the protocell membrane. Therefore, in the opposite way, the channels would thus work as mechanical-to-chemical signal transducers. Thus, under osmotic stress, a vesicle with membrane channels contains an indirect feedback loop between metabolism and membrane that has potentially interesting consequences for overall protocell dynamics.

Another function cited for rudimentary peptide channels in early protocells has been in the more extreme role as `emergency release valves', guarding against the possibility of the environment suddenly becoming diluted \shortcite{Morris2002}.\footnote{Or, the protocell interior more concentrated. In fact, to achieve a certain osmotic pressure inside a vesicle, an elementary proof can show that the environment requires to be diluted less than the interior of the vesicle requires to be concentrated.} This primitive volume regulation mechanism is suggestive of how early protocells became more robust to environmental change.\footnote{But, channels are only effective against hypotonic shocks in the environment (dilutions). An interesting question is how early protocells countered hypertonic shocks.}

\begin{figure}
\begin{center}
\includegraphics[width=13cm]{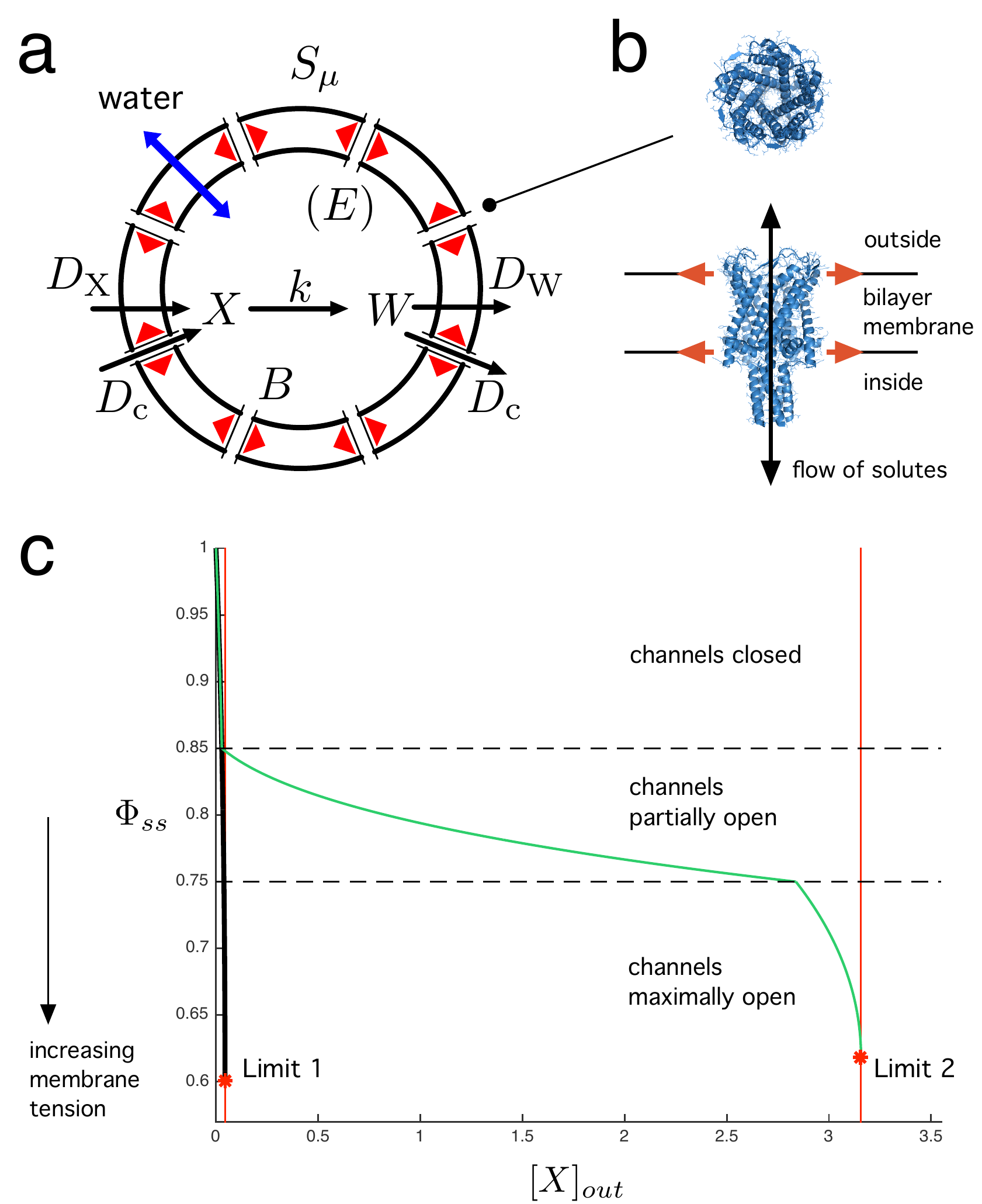}
\end{center}
\caption{
{\textbf{A Protocell Reactor with Mechano-Sensitive (Tension Activated) Channels.}} (a) Schematic showing mechano-sensitive channels open, allowing accelerated passive diffusion of solutes $X$ and $W$. (b) Molecular rendering of a single MscL mechano-sensitive membrane channel, above and side views. (c) Bifurcation diagram detailing how steady states of the reactor (occurring at different membrane tensions $\Phi_{ss}$) depend on the external nutrient concentration $[X]_{out}$. With mechano-sensitive channels, the reactor can tolerate a significantly wider range of external nutrient concentrations (green line, to Limit 2) than without channels (black line, to Limit 1). From \protect\shortciteNP{shirtediss2013}.
}
\label{fig:ch4_ms_channels}
\end{figure}

The final scientific contribution presented in this subsection investigates, using a minimal model, how tension-sensitive membrane channels could improve the robustness of a simple protocell `reactor'. This contribution is directly related to investigating channels in their emergency release role, but can also been seen as generally relevant to the area of how metabolism could have become indirectly coupled, via changing solvent volume, to membrane processes in early protocells. A minimal model of a vesicle was designed, featuring variable volume, fixed surface area, channels that opened progressively with increasing membrane tension (like mechano-sensitive channels) and a simple metabolism (Fig. \ref{fig:ch4_ms_channels}). In order to derive full analytical solutions to the model, the metabolism necessarily consisted of a single irreversible reaction converting nutrient $X$ into waste $W$.

Bifurcation analysis of the model confirmed the expected behaviour that mechano-sensitive channels could significantly extend the range of external $X$ nutrient concentrations at which the protocell could maintain a far-from-equilibrium steady state (Fig. \ref{fig:ch4_ms_channels}c).\footnote{When the membrane diffusion constant for nutrient $X$, was greater than the diffusion constant for waste $W$, leading to an accumulation of waste inside the vesicle and an internal osmotic pressure.} The analysis also revealed some other interesting effects, such as the potential of the protocell reactor to prematurely burst before reaching critical membrane tension, and the possibility of bistable behaviour (and hysteresis) for certain parameters. Additionally, the simplicity of the model allowed an exhaustive search of parameter space. This search revealed that the efficacy of membrane channels, as compared to a vesicle reactor with no channels, was greatest when the metabolism was neither too fast, nor too slow. If the metabolic reaction was too fast, waste would accumulate inside the reactor at a rapid rate, leading to a burst whether the reactor had channels or not. Conversely if the reaction was too slow, $X$ and $W$ inside the vesicle would approximately equilibrate to their external concentrations, and little (if any) membrane tension would be exerted. In this latter case, there was again no stability advantage to be had with channels. For intermediate metabolism speed, however, the vesicle reactor had a wider stability region when channels were present.

\chapter{Discussion}
\label{chapter:5}

%
%
%

Throughout this thesis, the biological autonomy view of living systems has been brought to the fore, and it has been argued how this perspective translates into a very important systems-oriented research program for protocells on the way to full-fledged living cells.

The discussion opens in Section \ref{sec:5_1} by briefly recapitulating how the autonomous systems perspective -- currently heavily marginalised in prebiotic chemistry research in favour of a pure evolutionary view -- demands a radical shake-up of the origins of cellular life problem. A series of major transitions of protocell development in light of an autonomous systems view are suggested, and it is discussed how autonomy represents a general heuristic principle, rather than a well-defined sub-problem for protocell research. Also highlighted is how the investigation and ultimate implementation of autonomous protocells needs to take into account realistic physical and chemical conditions and couplings.

Then, Section \ref{sec:5_2} summarises scientific contributions made by this thesis and where they fit into the reformulation of the major protocell transitions discussed in Section \ref{sec:5_1}. This section also discusses future challenges in advancing a semi-empirical research program toward basic autonomous protocells.

Finally, Section \ref{sec:5_3} concludes by suggesting future directions, including immediate research directions stemming from the scientific contributions made in this thesis, and also longer term research directions beneficial for an autonomous protocells research program.

%
%
%

\section[The Autonomy Perspective: Recapitulation and Implications]{The Autonomy Perspective: Recapitulation and Implications%
\sectionmark{The Autonomy Perspective Recapitulated}}
\sectionmark{The Autonomy Perspective Recapitulated}
\label{sec:5_1}

Earthly life is not embodied as complex chemical networks dispersed in solvents; nor does it take the form of competitive soups of replicating macromolecules. Rather, all known earthly life manifests itself {\em at the cellular level}. Taking seriously this observation, the field of protocells (reviewed in Chapter \ref{chapter:1}) works on the general assumption that {\em compartmentalisation} of chemical systems was essential from an early stage in the origins of life.

The requirement for compartmentalisation narrows down possibilities for the origins of life, but it does not set a definitive direction for protocell research. With no directive, the main trend in protocell research to date has been to follow a {\em default} option: to carry over the evolutionary perspective that existed before in prebiotic chemistry research -- i.e. evolving molecular populations such as the RNA world -- into a compartmentalised setting. Under an evolutionary interpretation, protocells are regarded as a vehicle to assist in the chemical evolution of molecular networks. Compartmentalisation in a protocell permits the segregation of molecule populations and gives rise to the heritable transfer of adaptations from parent to offspring chemical networks when the compartment divides. Various chemical evolution scenarios involving protocells have been proposed, including the evolution of template replicators in protocells (e.g. \shortciteNP{Bianconi2013,Szathmary1987,Szostak2001}), the evolution of autocatalytic networks with multiple `cores' in protocells \shortcite{Vasas2012,hordijkSteel2015}\footnote{In this approach, self-catalysing `cores' in autocatalytic networks -- and not nucleic acids -- are the units envisaged to replicate, propagate and mutate in protocells potentially evolving more complex autocatalytic networks which the authors argue could later be taken over or controlled by template replicators.}, and the evolution of composite protocell aggregates consisting of multiple lipid types \shortcite{Segre2000}.

From the evolutionary perspective, the origins of cellular life problem is reduced to finding the minimal compartmentalised chemical networks (based on template replicators, or otherwise) that could have -- via division and mutation -- started a process of competition, selection and open-ended evolution, to finally culminate in compartmentalised metabolic networks based on informational polymers and protein catalysts (``cells'').

The autonomous systems perspective of protocell development, argued for throughout this thesis, differs quite radically from the above evolutionary view. 

As reviewed at length in Section \ref{sec:2_2}, the autonomy view articulates a much more rigorous, systems-oriented notion of cellularity than the evolutionary view commits to. Autonomy emphasises that cells are systems that can robustly exist far-from-equilibrium because they consist of an integrated network of processes and constraints, and this network is organised such that it can collectively re-fabricate over time the majority of the diverse components of which it is made (by making use of nutrients and utilising energy sources in the surrounding medium). This is the basic idea of organisational closure, or {\em functional integration}, as explained in Section \ref{sec:2_2_2}. Furthermore, under the autonomy view, the cell membrane is not simply a convenient segregator of molecules; rather it serves as an active interface that the cellular system produces and can modify (by adding and modulating various membrane nano-machineries) to control matter and energy flows through itself. Overall, the autonomy view stresses the fundamental organisational scheme of biological cellular systems. The evolutionary approach, on the other hand, simply adopts that cells are reducible to metabolic network diagrams: DNA-RNA-protein metabolism in convenient lipid bags. This latter view is much weaker on the notion of cellular organisation.\footnote{Proteins, for example, under the evolutionary view, are typically seen only in a role as reaction {\em catalysts}. But in cells, proteins play numerous important structural and signalling roles too: they are integral in the cytoskeletal matrix, they can act as motors for internal vesicular transport, and they are vital in the cell membrane as channels and receptors. Cells have an important physical aspect, they are not just an aqueous-based metabolism. }

These diverging views of `cellularity' create diverging protocell research agendas. All include compartmentalisation one way or another, but with differing degrees of rigour. The evolutionary approach sees the grand challenge of protocell research as resolving the problem of how the chemical evolution of molecular networks in compartments -- largely passive and idealised compartments -- was accomplished. Under this approach, metabolism and compartment are quite disjointed and the compartment plays a {\em secondary} role, relevant only for the purpose of containment and division. Conversely, the autonomous systems approach sees the challenge of protocell research as much broader: the problem is not {\em just} to solve how cyclic reaction networks arose, but how {\em integrated} autopoietic cellular systems arose (Section \ref{sec:2_3}). Autonomy brings to the fore that we need to deal with cellular organisation/integration in origins of life from an {\em early stage}. It highlights that it is not enough to assume that the membrane, metabolic and template systems of cells developed independently and then came together almost `by chance' to form functional living units \shortcite{Ruiz-Mirazo2008,DelaEscosura2015}. Rather, the autonomy view highlights that the membrane and metabolism systems of protocells likely co-evolved to become tightly integrated (Section \ref{sec:3_3_1}). This integration allowed protocells to develop increasingly complex chemical networks and overcome the basic problems of cellularity at each stage of their ongoing development, such as the regulation of osmotic pressure and the accessibility of nutrients, and provided a base from which protocells could further increase their levels of molecular/organisational complexity.\footnote{In fact, \shortciteA{DelaEscosura2015} have recently argued that integration of diverse components and processes occurred from almost the beginning in protocell development. They wrote: ``a certain number of different `chemical tasks' (not just catalysis, but also transduction mechanisms, spatial confinement, mediated diffusion or template activity) may need to be jointly performed in order to ensure a minimal level of dynamic stability or robustness, even in the simplest infrabiological systems.'' (p18).} The integration problem stressed by the autonomy approach has severe implications for an evolutionary view of origins, which for example assumes that replicator populations became `functional' protocells simply by synthesising membrane lipids \shortcite{Scheuring2003,Czaran2015}.

It is worth mentioning that the problem of integration of components and processes is actually the main hurdle that synthetic biology approaches are facing in the ongoing challenge to create an artificial cell. Reviewing the state of the art, \shortciteA{Caschera2014} comment ``the integration and the coordination of self-organization, metabolism and information into cell-sized compartments have failed so far.'' (p85). The metabolic, template and compartment subsystems of cells have to be interfaced and coupled together in an effective way. A cell is {\em much more} than just a metabolic network diagram, it is a highly organised and interconnected {\em physico}-chemical system. Simply transplanting transcription/translation machinery from extant cells like {\em E. coli} into lipid vesicles (i.e. how the evolutionary approach conceptualises a ``cell'') does {\em not} lead to functional artificial cell systems that are able to self-maintain far-from-equilibrium, reproduce, and exhibit autonomous behaviour. For example, the bioreactor of \shortciteA{Noireaux2004} demonstrated that an {\em E. coli} transcription/translation system will collapse inside a lipid vesicle after 5 hours. However, these authors found that if the E.Coli machinery could synthesise protein pores for the membrane -- forming a partial integration between compartment and metabolism -- then nutrient permeability restrictions could be overcome, and the internal transcription/translation system could remain active for over 4 days (as evidenced by the expression of green fluorescence protein, see Fig. \ref{fig:ch1_protocell_approaches}d). 

The synthetic protocell study of \shortciteA{Hardy2015} discussed in Section \ref{sec:3_2_2} presents a different example illustrating the necessity of integration. The latter authors embedded a non-trivial phospholipid synthesising metabolism on-board phospholipid vesicles, and they could achieve the growth and division of phospholipid vesicles into functional daughters (also able to grow and divide). However, this was accomplished only by {\em carefully considering} how the entire reaction system (proto-metabolism) was to reproduce itself and transfer to the newly formed daughter vesicles.\footnote{ Nevertheless, the cellular integration problem was only partially solved in this case again. Although the vesicles could continuously grow and divide, they did so as {\em flaccid} structures. The problem of how the protocells could maintain a minimum aqueous volume across generations was not addressed.}

Therefore, how the metabolic and membrane systems in protocells became integrated together is a very important problem. This problem, congruent with a true systems chemistry approach to the origins of cellular life (Section \ref{sec:1_4}), requires the autonomy perspective in order to be fully recognised and properly treated. The following section proposes a set of protocell transitions in the origins of life, as an example of what could be suggested by an autonomy-led approach to the problem.

%
%
%
%
%
%

\subsection[Reformulation of Major Prebiotic Transitions]{Reformulation of Major Prebiotic Transitions}
\label{sec:5_1_1}

Chemical competition and diversification is considered to happen over segregated molecular populations under the evolutionary view. In the autonomous systems view of protocell development, chemical evolution is an equally important driving force, but selection is instead regarded to act over populations of integrated dissipative protocell systems. It could be hypothesised that these protocell systems underwent the following major transitions:

\begin{itemize}

\item \textbf{T1: Self-assembly in heterogenous conditions.} Supramolecular structures with a diverse composition (mixture of  lipids and other small hydrophobic molecules anchored in the membrane), but not yet hosting chemical reactions. In this context, vesicles would have been dynamic but largely passive objects, even if they could have still competed for lipid. In particular, vesicles with mixed lipid membranes would have been able to survive across diverse environments with low lipid monomer concentrations.

\item \textbf{T2: Far-from-equilibrium vesicle reactors.} Vesicles self-assembled from free lipids and other hydrophobic components in the external medium, hosting chemical reactions in their aqueous interior, membrane inner/outer surface or membrane interior (Section \ref{sec:3_4}). The molecular complexity of such systems would be comparable to that of the surrounding medium, but complex chemical and morphological dynamics could still be present in such systems, including adaptive behaviours and directed movement.

\item \textbf{T3: First self-producing protocells.} Protocells hosting chemical reactions that fabricated new molecular species not present in the external medium. These protocells would begin to sequester themselves from the medium, showing a molecular complexity greater than the environment. This would mark the start of an important transition, from {\em self-assembly} to {\em self-production}.

\item \textbf{T4: Intermediate self-producing protocells.} Protocells able to fabricate complex molecular species, such as lipids, peptides, catalysts, possibly nucleotides. Appearance of endergonic-exergonic couplings and energy currencies to enable such syntheses. These protocells would directly control their membrane composition from the inside, thus granting wider possibilities to regulate their coupling with environmental energy/matter flows, increasing their viability range. Catalysts would enable controllable and specific reaction pathways.

\item \textbf{T5: Advanced self-producing protocells with features such as reliable division and heredity.} Self-producing protocells which integrated template biomolecules into metabolism to expand synthetic and catalytic possibilities, together with division coordinated with reliable, heritable transfer of information to offspring. Also, advanced behavioural possibilities such as complex adaptation strategies to diverse environments, and anticipation of future conditions.

\item \textbf{T6: First full-fledged biological cells.} Genetically instructed cellular metabolism. Genotype-phenotype relationship mediated via a code. Open-ended evolution.

\end{itemize}

The above hypothetical transitions have been biased toward lipid vesicles in search of continuity with extant living phenomenon, but it is reasonable to assume that the early stages of protocell development (T1 and T2) could have been based on, or aided by, other types of compartment media, like micelles or coacervates \shortcite{Monnard2015}.

A first significant transition in the above list is that of self-production. At this point, between T2 and T3, protocells started turning from an ``outside-in'', to an ``inside-out'' organisation. That is, protocells went from acquiring their necessary materials from the surrounding medium, to being able to fabricate {\em internally} their own components (e.g. energy rich compounds, lipids, and catalysts), successively replacing the less efficient materials available in the environment. 

For a protocell, the capacity for self-production is intuitively linked with autonomy: a self-producing protocell individuates itself from the environment in a strong sense, for its embodiment is a result of its own metabolic organisation, not just of self-assembly of previously available compounds. Due to an internally synthesised embodiment, a self-producing protocell also has more independence over which environmental regions it can inhabit, i.e. the system can maintain itself even in environments with a low level of molecular diversity/complexity. Last but not least, a protocell with internal component synthesis has a {\em direct and precise} control over its membrane properties, and thus can better manage energy and matter flows to and from the environment in order to remain viable.

A second significant transition in the above list involves the stage at which protocells started using molecular template mechanisms and sequences in order to record (and at the same time, promote) the organised complexity of the protocell system, and pass on this information in a reliable way to offspring (T5). This has a strong evolutionary implication, but this will not be covered in detail in this thesis.

%
%
%
%
%
%

\subsection[Basic Autonomy as a Multidimensional, Heuristic Concept]{Basic Autonomy as a Multidimensional, Heuristic \\Concept}
\label{sec:5_1_2}

In trying to formulate a simplified toy model for the origins of metabolism, Freeman Dyson commented that ``The essential difficulty arises because metabolism is a vague and ill-defined concept. There is no such difficulty with the concept of replication. Replication means exactly what it says….[but] we run immediately into the problem of defining what we mean by metabolism.'' \shortcite[p48]{Dyson1999}. The same type of problem arises for investigations into the origins of autonomy in protocells. `Autonomy' is a multi-faceted and intuitive concept, whose formal expression is not straightforward.

Historically, the approaches to origins of life attracting the most attention have been those which map the grand, unwieldy problem onto some smaller, well-defined sub-problem. Well-defined sub-problems have the advantage of representing a concrete goal to focus investigations, and some even admit analytical formulation and solution. For example, a few popular sub-problems in origins have been as follows: How did replicator molecules manage to replicate rapidly and with high fidelity? How did disordered molecular populations suddenly jump into a state of order? How did compartment division become synchronised with template replication?

The autonomy approach also poses a problem for origins of life: How did protocellular systems become integrated cellular systems, ultimately with a self-producing and self-reproducing autopoietic organisation? However, this is not reducible to a single easy to handle sub-problem, but rather it involves a set of problems, each consisting of many potential strands. A valid general  question is posed, but not at the level where a formal or experimental model could be immediately suggested to resolve it. Therefore, basic autonomy acts more like a {\em heuristic tool} in origins of life protocell research (similar to an ``intuition pump'' in the terminology of \shortciteNP{Dennett1984}). It has value in providing an {\em overall conceptual direction} for protocell research to follow, like a `main trunk' for research activities, useful for pointing out what are the {\em kind} of relevant problems that should be addressed by more concrete protocell studies.

Globally speaking, as discussed in Section \ref{sec:2_2_1}, biological autonomy comprises two dimensions: a behavioural dimension concerned with how a cellular system interacts as an agent with its environment, and a constitutive dimension, concerned with how a cellular system is organised to self-fabricate its embodiment. These dimensions set up multiple sub-problems and questions for the development of protocells, which are typically highly interwoven. These questions include: How did protocells increase in behavioural diversity (dynamical regimes possible)? How did protocells start to show a history dependence and basic learning toward external stimuli? How did protocells start implementing basic mechanisms of motility? How did protocells start to capture and manage energy resources to accomplish endergonic reactions (like nucleotide/peptide polymerisation) and active transport processes? How did protocells develop hypotonic and hypertonic volume regulation mechanisms? How did protocells develop selective membranes to control molecular diffusion? How did supra-molecular structures and/or a heterogeneity of phases start participating in reaction chemistry?

%
%
%
%
%
%

\subsection[Matter Matters: The Value of Semi-Empirical Modelling]{Matter Matters: The Value of Semi-Empirical \\  Modelling}
\label{sec:5_1_3}

The set of example sub-problems above, all stemming from an autonomous systems approach to protocells, cannot be properly tackled without paying attention to the intrinsic activity and constraints inherent in real material systems. Even in a general, universalised conception of the problem of biological organisation, ``matter matters'' \shortcite{Moreno1994, Moreno1999, Ruiz-Mirazo2004BAS}. Therefore, modelling the appearance of autonomy in protocells inherently demands that protocells be represented {\em realistically}, as physicochemical systems, where the way in which chemistry interacts with self-assembly processes is properly captured. For protocell models, it is ultimately not enough to concentrate merely on abstract models of metabolism, nor is it sufficient to construct only toy models of cellular systems (like the computational autopoiesis ``protocell'' models of Fig. \ref{fig:ch3_comp_autopoiesis_models}). In order to get a faithful picture of protocell system dynamics and properties, all of the biophysical effects and consequences of encapsulating chemical processes in dynamic micro-compartments have to be taken into account.

Because of these requirements, a `semi-empirical' approach to modelling becomes essential to pursue \shortcite{Mavelli2007semiempirical,Piedrafita2012}. The semi-empirical approach, discussed in Section \ref{sec:3_3_2}, seeks to capitalise on the advantages of both theoretical {\em in-silico} and experimental {\em in-vitro} approaches to protocells -- namely {\em controllability} and {\em proximity to the real world} -- by proposing a hybrid approach situated roughly in the middle of the two. The semi-empirical approach is itself a theoretical approach, but it proceeds with the explicit aim of constructing protocell models that embed realistic physical processes, empirical parameter values, sizes, molecule numbers and timescales, such that the gap to real in-vitro protocell implementations is closed. Semi-empirical models and protocell experiments work in tandem, bootstrapping each others progress. Semi-empirical models can help to clarify underlying molecular mechanisms and often suggest where new experiments should be focussed. Protocell experiments, on the other hand, provide the hard physical data highlighting where inaccuracies in a semi-empirical model could reside. As mentioned in Section \ref{sec:3_3_2}, semi-empirical models also allows for extrapolation: once a base protocell model has been validated, more complex protocell scenarios can be investigated which are hard to realise with current experimental techniques.

Bridging the gap to in-vitro systems is vital for creating a {\em grounded} protocell modelling approach able to accurately capture the general properties and dynamical behaviour of {\em whole} physicochemical protocell systems. The scientific contributions of this thesis, discussed in the next section, follow precisely this semi-empirical approach and are concerned with modelling autonomous protocells at an early stage of development.

%
%
%

\section[Toward Modelling Autonomous Protocells: Contributions and Challenges]{Toward Modelling Autonomous Protocells: \\ Contributions and Challenges%
\sectionmark{Scientific Contributions and Challenges}}
\sectionmark{Scientific Contributions and Challenges}
\label{sec:5_2}

%
%
%
%
%
%

\subsection{Relevance of Scientific Results}
\label{sec:5_2_1}

The scientific contributions of this thesis are applicable to modelling far-from-equilibrium chemistry in dynamic lipid compartments in a physically realistic way. The contributions have particular value in providing a realistic grounding principles for vesicle models that aim to investigate protocellular development. On the list of major prebiotic transitions toward autonomous cellular systems detailed above, the contributions are specifically related to protocell transition T2, the scenario of `vesicle reactor' (described in Section \ref{sec:3_4}).

The first main contribution of this thesis has been to improve the kinetic model of the bilayer lipid membrane in the Mavelli \& Ruiz-Mirazo semi-empirical vesicle model. This was achieved by validating and refining the membrane lipid exchange equations to match experimental data from two studies on protocell competition.  In additional, this work provides a first theoretical approach to the phenomena of protocell competition from a well-grounded account of its basic dynamics.

A research program into the co-evolution of protocell metabolism and membrane (Section \ref{sec:3_3}) relies heavily on accurate theoretical models for the dynamic behaviour of lipid membranes. There is a general lack of coarse-grain kinetic models for bilayer membranes in the literature, i.e. bilayer models detailing the average rates at which lipids enter, relocate, flip-flop and leave a bilayer, and the corresponding factors affecting these rates. This is mainly because lipid membranes are soft supramolecular structures whose size and shape dynamics are complex, diverse, and influenced by a wide range of competing factors that, unlike chemical reactions, are not trivial to characterise at a coarse-grain level.\footnote{Chemical reactions in solvents lend themselves easily to a kinetic description because they are single chemical transformations whose rates are largely determined by the collision frequency of reactants. Indeed, the Law of Mass Action describing reaction kinetics has existed for 150 years now (see \shortciteNP{Voit2015}).} Nevertheless, the kinetic membrane model developed in Section \ref{sec:4_2} seeks to provide an improved coarse-grain description of a dynamic fatty acid membrane, with the aim of establishing a solid base `chassis' on which to construct future semi-empirical vesicle models that also include metabolism. This first contribution is thus broadly relevant to modelling the development of autonomous protocells.

The second main contribution of this thesis has been to highlight the possible implications that osmotic water flow effects have for the internal chemistries of early protocells. One implicit assumption shared by virtually all current protocell models (theoretical and experimental) is that water solvent in protocells is largely {\em passive}. In general, water has been viewed as a necessary medium for hosting chemical reactions, but as playing no role in protocell dynamics itself. The contribution of this thesis is to challenge this core assumption. It is generally argued that the aqueous volume of early vesicle protocells, through constantly changing in size to equilibrate nett osmotic gradients across the semi-permeable protocell membrane, is likely to have played much more of an {\em active} role in protocell dynamics.

As explained in Section \ref{sec:4_3_2}, one interesting consequence of changing volume would have been {\em osmotic coupling}, i.e. the indirect coupling, via water osmosis, of independent reaction systems (sharing no chemical species) inside the aqueous phase of protocells. Osmotic coupling considerably widens the current view of protocell metabolism — which implicitly assumes `one chemically connected metabolism per protocell' — to a scenario where protocell metabolism can be more fragmented, i.e. consisting of clusters of reaction systems that do not communicate chemically, yet still indirectly couple their dynamics. Osmotic coupling additionally implies that entirely entrapped reaction networks and inert diffusing solutes (i.e. solutes which diffuse across the protocell membrane but don't react further) can damp the dynamical response of a vesicle protocell to environmental perturbations. From a prebiotic chemistry perspective, protocells containing multiple reaction systems each working with a mutually exclusive set of chemical species does not seem unreasonable, given that prebiotic vesicles would have self-assembled in chemical mixtures (possibly including template reactions building `informational' polymers, combined with other non-associated chemical reactions).\footnote{It could be argued that some chemical reaction systems not sharing species may still be weakly coupled by e.g. local pH changes. In this case, osmotic coupling would greatly enhance their degree of coupling.}

As regards to autonomy in protocells, osmotic coupling is a relevant phenomenon because it changes the perception of what could constitute `metabolism' in the membrane-metabolism co-evolution research program. Also, osmotic coupling is one potential route by which protocells could have started developing complex non-linear behaviour without necessarily harbouring complex internal metabolisms (as the bistable example illustrated in Fig. \ref{fig:ch4_varyvol_model}). The principle of osmotic coupling should be applicable from the earliest protocells up until the (relatively advanced) stage when protocells developed regulation mechanisms to actively control water volume, or, alternatively, proto-cell walls to physically resist volume changes. In fact, osmotic coupling is particularly applicable to small vesicle systems with nano-litre water volumes, for it is within this context that small copy numbers of species can lead to large osmolyte concentration changes.

Osmotic coupling and other osmotic effects may be general `systems principles' very relevant to the origins of cellular life. However, it should be noted that osmotic coupling does not apply in all cases. For example, osmotic coupling would cease to operate in vesicles with a high number of entrapped impermeable species (as these molecules would act as a buffer to diminish water volume fluctuations), or in cases where the speed of metabolic processes far exceeded the membrane permeation rate of water. Also, osmotic effects would only be applicable to protocell architectures based on a semi-permeable membrane separating two aqueous phases, or on gel-based systems able to resize their volume (i.e. only protocell architectures a, g and h of Fig. \ref{fig:ch1_protocell_architectures} in Chapter \ref{chapter:1}). Finally, it should be highlighted that osmotic coupling is not a general substitute for stoichiometric coupling: a protocell would still require individual reaction systems that have the appropriate connected chemical transformations to synthesise certain components, or to organise exergonic reactions such that endergonic transformations can take place (as discussed in Section \ref{sec:2_2_3}). 

The final contribution of this thesis, the protocell `bioreactor' of Section \ref{sec:4_4} hinted that, apart from coupling chemical reactions in the internal aqueous phase of a protocell, changing water volume could also provide an indirect coupling between metabolic and membrane processes of a simple protocell. In this case, osmotic pressure inside the vesicle caused membrane tension which opened tension-sensitive membrane channels, but other cases could be imagined whereby osmotic pressure affects membrane properties which then feeds back on metabolic dynamics.

%
%
%
%
%
%

\subsection{Limitations}
\label{sec:5_2_2}

In order to foster further advances, more important than discussing the value of our contributions, one should highlight several limitations of the scientific results presented in this thesis. With regards to the improved kinetic model for fatty acid membranes, it should be emphasised that this theoretical model was validated against outcomes from an {\em artificial} laboratory system of vesicle competition. The original vesicle competition experiments for both osmotic and phospholipid competition were carried out under controlled general conditions (pH and temperature) and also employed controlled populations of vesicles (vesicles were uniform in size and had pure unilamellar bilayers of oleic acid, or oleic acid plus a fraction of a single phospholipid type). Therefore, the lipid kinetics model has been validated, but it has been validated in the context of a {\em constrained} experimental system. Whether the model holds general validity across other scenarios is an open question. For instance, if multiple lipid types were present in a protocell membrane then their combination may change bilayer properties in such a way that the kinetic uptake and release rates of simple fatty acids became substantially altered. Also, the developed lipid kinetics model only strictly gives information about the exchange dynamics of {\em fatty acids} from a membrane. Phospholipids were considered as static amphiphiles in the vesicle membranes and their exchange dynamics were not taken into account.\footnote{It could be considered that phospholipids absorb and desorb from a bilayer membrane with rate equations similar to those of fatty acids (at a much slower rate), but this would be an unverified assumption.}

Our coarse-grain membrane lipid kinetics model also assumed that fatty acid `flip-flop' between the inner and outer leaflet of a vesicle bilayer was very fast, such that any asymmetry in surface area or composition of the two monolayer leaflets could would be instantly equalised. Vesicle shape in our model was solely represented as the ratio between surface area (equal for both leaflets) and volume. However, as already mentioned in Section \ref{sec:3_4}, surface area asymmetries between membrane leaflets do exist and have been demonstrated as a significant factor in determining vesicle geometry \shortcite{Mui1995}.

Moving onto the minimal formalism used to explore changing solvent volume in protocells, at the heart of this model is the assumption that the aqueous core of a vesicle protocell is a {\em well-mixed} water volume, with no localised variations in solute concentrations permitted. This simplification allows to formulate the model in mass action kinetics terms, and means that volume variations are global, affecting the concentrations of all solutes in the protocell core. However, it is sometimes debated whether the well-stirred assumption holds for the aqueous interior of protocells (see below). Additionally, this minimal model of changing volume assumed that the diffusion constants of solutes across the vesicle bilayer membrane were constant and not affected by factors such as osmotic tension, shape, curvature or absolute size of the membrane, amongst others. Moreover, the simple model can be expected to become invalid for vesicles with a gel-like as opposed to watery core, as volume variation dynamics in this former context can assume a very different character \shortcite{Viallat2004}.

Finally, the vesicle reactor model with mechano-sensitive channels, in addition to being built on the  the same the assumptions of the last paragraph, further considered that osmotic pressure was always exerted uniformly across the surface of the protocell membrane. Indeed, the well-stirred and uniform osmotic pressure simplifications made possible an algebraic analysis of the model, but it should be noted that such simplifications may not always hold. Indeed, spatially explicit theoretical protocell models have arrived at relevant results by assuming just the opposite: that internal vesicle chemistry is {\em not} well mixed, and that non-uniform osmotic pressures can exert themselves along the membrane leading to e.g. vesicle division \shortcite{Macia2007,DeAnna2010}. The use of mechano-sensitive channels in the reactor model could also be questioned from a prebiotic perspective, as early protocells would not have been likely to possess advanced channels that function through a highly coordinated -- and likely evolved -- arrangement of protein helices. However, the qualitative outcome of the model analysis is not contingent on the channels specifically being of a mechano-sensitive type: the channels could have been more rudimentary, so long as they permitted some type of solute transport whenever there existed membrane tension.

%
%
%
%
%
%

\subsection[Challenges in Advancing Semi-Empirical Protocell Research]{Challenges in Advancing Semi-Empirical Protocell \\ Research}
\label{sec:5_2_3}

This section documents some modelling challenges that need to be tackled as the semi-empirical research program towards basic autonomous protocells outlined in Section \ref{sec:3_3} is advanced to more complex protocell schemes.

\subsubsection{Analysis of Protocell Models with Many Free Parameters}

A first challenge faced by the semi-empirical protocell research program is that, as protocell models become more complex, a large free parameter space makes a rigorous analysis less forthcoming. It should be emphasised that this is a broad challenge faced by all complex dynamical systems models.

It is straightforward to specify a semi-empirical protocell model of arbitrary complexity. Once all the parameters have been defined, it is also rather straightforward to numerically integrate the dynamical equations of a particular protocell scheme deterministically or stochastically, and observe how the system behaves. These aspects are not so problematic.\footnote{Apart from standard problems arising in numerical simulation, like long simulation times for stochastic simulation when there exists a separation of timescales, or the inaccuracy or instability of deterministic simulation if unsuitable numerical integration algorithms are employed.} Rather, the main challenge comes when trying to draw {\em insightful and general conclusions} about the real dynamical behaviour of the protocell system.

Non-linear (and, worse, high-dimensional) dynamical systems tend not to admit analytical solutions. Often, the only route available to gain information about their behaviour is numerical integration of the equations. However, it is important to note that numerical integration only has leverage as a method of analysis when applied to certain problems. In models where parameters have been fixed beforehand (to values derived experimentally or cited in the literature), numerical integration may be effectively applied to predict outcomes. Alternatively, numerical integration can be used as part of the reverse process of estimating/optimising multiple a set of free parameters in a model, given a target output that the model must reproduce (used in Systems Biology for example, to develop predictive biochemical models, see e.g. \shortciteNP{Ashyraliyev2009}). Numerical integration can also deliver insight when used as a method to compute harder variants of an existing model for which analytical solutions are available. In this case, general solutions from the solvable model can be used as a reference point and compared with computed solutions to the analytically intractable one. Also, for some complex dynamical systems with very small parameter spaces, numerical integration can be used alone as a `brute force' strategy to build a picture of the system bifurcation diagram and exhaustively classify all dynamical possibilities.

The challenge with semi-empirical protocell models like the lipid-peptide or lipid-producing scenario (or even more elaborate models), is that they are complex dynamical systems with a large number of free parameters. Moreover, these models are intended for {\em exploratory} research into the origins of life and so no target behaviour can be defined to estimate free parameters. Indeed, the opposite is required: to characterise all potential behaviours under different parameter settings. These two features mean that numerical integration can be used to solve a protocell model, but that typically, only weak conclusions about model behaviour can be reached.

For an example, consider a theoretical protocell scheme with an autocatalytic metabolism that synthesises a new lipid type. Assume that sustained oscillations are observed in the internal metabolite concentrations, membrane size and membrane composition of the protocell when the new lipid type embeds in the membrane, causing a sharp decrease in membrane permeability. This model forms an interesting {\em existence proof} that sustained oscillations are possible with some parameterisation of this metabolic scheme. However, more penetrating questions would be useful to answer in this context, such as: how does the existence of oscillations depend on the shape of the membrane permeability function? How do oscillations depend on the stoichiometry of the internal metabolism used? In general, what formal requirements must be satisfied by the protocell metabolism and protocell membrane permeability function, for oscillations to result? Decisive answers to these more valuable questions are generally more difficult to produce.

To start addressing this challenge, one tactic could be to interface theoretical models as closely as possible with in-vitro protocell realisations, in order to fix as many of the parameters as possible. Another answer could be to construct more simple `mean field' protocell models of a given protocell scheme which can be analysed, and then use this analysis to shed light on the behaviour of the more complex model (as performed in our protocell competition paper, \shortciteNP{shirtediss2014} , for example). A third option, explained in Section \ref{sec:5_3_2}, could be to start constructing an equivalent of Chemical Reaction Network Theory for protocell systems. Rather than just simulating detailed individual models, this approach would instead seek to identify general constraints on protocell organisation. These constraints could be useful in ruling out certain dynamical behaviours by inspection of a protocell scheme, without the need for simulation.

\subsubsection{Transient Protocell Behaviours are Relevant, in Addition to Long-Term Dynamics}

A second issue that needs to be carefully considered when modelling protocells is how the protocell-environment coupling is represented. In particular, protocell models should not only be tested in the context of {\em invariant environments}, for this strategy does not investigate {\em transient system} responses to stimuli, which may be very relevant.

Both the lipid-peptide and lipid-producing scenarios of Section \ref{sec:3_3_3} considered that the protocell existed in a large expansive `reservoir' environment where all chemical concentrations remained constant for all time. This assumption, also present in all existing kinetic models of the Chemoton, allows bifurcation analysis to be performed on a protocell model, i.e. the investigation into how changing model parameters impacts on properties of some stable long-term state. However, while these studies are undoubtedly valuable, bifurcation analysis does not address the equally (if not more) important {\em transient} aspect of protocell behaviour. Transient aspects are those which are time critical, and depend on the {\em flow features} of the system phase space -- not just the attractors. They include aspects such as how a protocell can temporally manage to avert threatening perturbations in its environment (i.e. by employing regulatory mechanisms to return it to a stable operating regime), how a protocell can show history dependence to a series of environmental stimuli (i.e. to demonstrate a form of primitive learning) or how a protocell can adapt to longer term environmental changes. The conclusion is that, in order to be able to evaluate these equally important transient behaviours, the environment in which the protocell is located is required to be varied and perturbed.\footnote{Incidentally, models of G{\'a}nti's Chemoton have not been studied in the context of response to environmental stimuli. All studies so far have focussed on how factors, like the length of the genetic template, affect the final division regime of the system in a stable environment. }

A simple example of the difference between the bifurcation and transient perspectives can be given by revisiting the protocell reactor with mechano-sensitive channels in Section \ref{sec:4_4}. A bifurcation analysis was performed on this model. This revealed how membrane channels helped the reactor survive in higher external nutrient concentrations. However, this analysis did not characterise the time-response of the channels when there occurred {\em sudden dilutions} of the environment (the main `guarding' function of the channels), nor what intensity/frequency of dilution shock the channels could cope with, nor what effects on protocell metabolic dynamics repeated dilutions would lead to.

In summary, the lipid-peptide and lipid-producing protocell schemes do indeed get closer to the idea of self-production, and controlling of the system-environment interface (i.e. constitutive autonomy). However, in order to start exploring the relevant dimension of adaptive behaviour (i.e. behavioural autonomy), a systemic study and characterisation of the response of such systems to environmental perturbations should be undertaken as well.\footnote{ Incidentally, modelling the environment as invariant also rules out other phenomena. For example, protocell competition as described in Section \ref{sec:4_2_1} cannot be modelled in an invariant environment, because it requires localised lipid `stealing' amongst vesicles. In an invariant environment, protocell competition can only be tackled in a non-spatial way, for example by stating ``protocells which reproduce faster are fitter''. However, this strategy cannot handle the issues of eventual resource limitation and crowding in a rigorous manner.}

It should be noted that introducing environmental changes comes with technical modelling implications. If the protocell environment has constant concentrations, then the protocell model can be simulated stochastically, since the number of reaction and diffusion events are manageable (the only events are lipid fluxes solute fluxes to/from a small vesicle object, and a manageable number of reaction events inside the attolitre vesicle volume). However, if a varying environment is introduced, then the outside concentration dynamics must be modelled deterministically, for there are far too many reaction events in the environment to perform a reaction-by-reaction stochastic computation. Also, in order to properly handle sharp perturbations in environmental concentrations, water flow across the protocell membrane needs to be taken into account explicitly and modelled deterministically (Equation 3.8 of Box 2, Chapter \ref{chapter:3}). Thus, the whole model requires to become deterministic, unless accurate hybrid stochastic-deterministic simulations could be developed.

Another point is relevant. When the protocell environment is regarded as a well-stirred reservoir, even imposing environmental changes and perturbations still does not permit an evaluation of the whole spectrum of protocell behaviour. When environmental conditions are externally imposed, be they fixed or varying concentrations, a model protocell has no causal powers over its environment. The environment affects the protocell, but the protocell cannot directly affect its local environment in return. For example, a protocell cannot excrete local chemicals or change its environmental conditions by movement.\footnote{Modelled in a small closed volume, protocells can influence their environment, but the final long term outcome is always an equilibrium condition (like in protocell competition).} When a protocell cannot change its surrounding environment, the only strategies that a protocell can use to exert control over its viability conditions are to change internal state, or to modify membrane composition, altering the coupling with the environment. The protocell cannot, however, create useful chemical feedback loops via the environment (i.e. secreting chemicals that are later used to cue behaviour), or exhibit taxis toward resources. For these latter behaviours to become possible, a future extension to the semi-empirical protocell model would be required. This extension would perhaps consider the protocell to exist as a point particle in a spatially heterogeneous reaction system, like a reaction-diffusion system, rather than in a well-stirred tank.\footnote{ The position of the protocell point particle would determine local external concentrations that it experienced. The protocell could influence local external concentrations, and could also move (buoyancy being likely the simplest mechanism of movement). This extension would also allow reactions to happen both inside and around the protocell. Indeed, one recurring problem with assuming a fixed, reservoir environment is explaining why chemical reactions can happen {\em inside}, but not {\em outside} of a protocell.}

\subsubsection{Biological Cells are Constructive Dynamical Systems}

Finally, a very important challenge arising as protocell models become more complex toward the complexity level of biological cells, is the issue of {\em constructive dynamical systems}.

As outlined at the end of Section \ref{sec:2_2_3}, biological cells are incredibly complex chemical nano-machines, able to fabricate their own components. Therefore, cells possess the special aptitude of being able to {\em fundamentally reconfigure} their own material structure in response to perturbations and stimuli. As such, at a coarse-grain level of abstraction, the state of a cell does not simply evolve on a fixed $N$-dimensional phase space, but rather the set of relevant variables that describe the cell state also change over time, as do the corresponding set of dynamical equations which couple these relevant  variables together. In fact, the relevant variables characterising cell state will move up and down different levels of abstraction as the cell reconfigures its material structure. Hence, for a cell, it is very difficult -- perhaps impossible -- to reach a fixed coarse-grain dynamical description which is valid in all circumstances.\footnote{ A cell could be modelled at a molecular level. But, molecule-level models don't give any intuition into what are the {\em relevant macro variables} influencing the future evolution direction of the whole system (and molecular-level models also suffer from severe computational restrictions).} This is a deep point and relates back to the discussion of Systems Biology in Section \ref{sec:3_1_2}.

The constructive dynamics problem is still far from being an issue for protocell models like the lipid-peptide scenario. These protocell systems can indeed change their structure, altering their coupling to the environment in order to remain viable (i.e. change their boundary conditions), but these structural changes (like changing membrane composition, permeability, elasticity) are always relatively simple. As such, simple boundary condition changes can be accommodated by adding more state variables to the model, and keeping the description of the overall dynamical system {\em static}. However, as the physicochemical complexity of the cell is approached, it becomes increasingly hard to have a static dynamical system describing its operation: the cell can re-construct itself into very different dynamical systems in its efforts to stay viable. This is an open problem, whose solution or partial solution would bring great benefits, to Systems Biology also.

%
%
%

\section{Future Directions}
\sectionmark{Future Directions}
\label{sec:5_3}

%
%
%
%
%
%

\subsection{Immediate Research Pathways}
\label{sec:5_3_1}


\subsubsection{Osmotic Coupling of Reaction Sets}

The new principle of osmotic coupling of reaction systems inside protocells, introduced in this thesis, gives rise to a number of immediate theoretical research questions. So far in this work, only bistability has been demonstrated in model reaction systems coupled by osmosis. An immediate task would be to find examples of reaction set combinations able to produce other more complex chemical behaviours: e.g. multistability (beyond bistability), oscillations, chaos etc. Multistability is a particularly interesting theme from the point of view of adaptive behaviour, since it would open the possibility for a protocell system to respond in different ways (dependent on its past history), given the same stimulus.

Generally, it could be valuable to investigate {\em motifs} in the context of ensembles of osmotically coupled reaction systems. In standard biochemical circuits, a small number of different reaction motifs have been identified that are abstracted arrangements of positive and negative feedback loops known to be capable of distinct dynamical behaviours, like switching or oscillation (e.g. see \shortciteNP{Tyson2010}). However, these motifs apply to {\em single} reaction networks in {\em reservoir} volume conditions. Can motifs be found in the same way in sets of reaction systems that exist in variable volume, coupled by osmosis? It would be very useful to have a set of rules determining which reaction systems needed to be present together in a variable volume in order for some dynamic regime to be possible, and what were the minimal combinations. Another valuable theoretical contribution would be to calculate an upper limit number of distinct steady states that a set of osmotically coupled reaction systems could have. When dealing with simple individual reaction mechanisms, the limit number of steady states can often be derived algebraically, but this is not an option when sets of multivariate polynomial equations are involved (e.g. as in Fig. \ref{fig:ch4_varyvol_model}a).\footnote{ Conceivably, it could even be the case that osmotic coupling instead places a {\em hard ceiling} on the number of steady states that a variable volume system is able to possess.}

Furthermore, it would be interesting to conduct further analysis on reaction systems that are entirely entrapped in a vesicle. Reaction systems whose species are all membrane impermeable represent a curious case because, although they always inevitably return to an equilibrium, the total number of species that they contain at equilibrium can vary depending on the volume of the protocell when this equilibrium occurs. This feature could potentially change the number of steady states of a variable volume protocell system, although more analysis is needed to confirm this. Another insightful study would be to compare the effect on the stability of steady states when water permeability across the vesicle membrane was considered finite, as opposed to considering water flow as instantaneous. If water permeability was considered finite, additional effects could also be investigated, such as the diffusion rate of water increasing as the absolute external solute concentration increases (reported by \shortciteNP{Sacerdote2005}). 

Another development would be to model osmotically coupled reaction systems inside a dynamic lipid membrane -- merging the two main scientific contributions of this thesis -- to see how having a dynamic surface area affects reaction dynamics. Also, reactions directly consuming or producing water could also be interesting to investigate.

The feasibility of osmotic coupling is also open to be -- and should be -- tested experimentally. A simple experimental scenario to provide a basic proof of the principle, at a more manageable size scale and not employing vesicles, could be the following. Firstly, two chemical reaction systems $R1$ and $R2,$ would be demonstrated to be species independent and non-interfering by an initial experiment. Secondly, an apparatus consisting of two micro-sized water columns connected by a semi-permeable membrane could be introduced. The semi-permeable membrane would facilitate water flow between the columns. Reaction system $R1$ (plus some inert buffer) would be added to one of the micro-sized columns, and the water volume of this column would change as equilibrium was attained. Some or all species in $R1$ may permeate the membrane into the second water column. Concentration profiles for the species in $R1$ on the way to equilibrium in the first water column would be recorded. The same procedure would take place for $R2$. These two experiments would act as the control experiments. Then, the main experiment would introduce reaction systems $R1$ and $R2$ together into one of the micro-sized water columns (without any buffer). The water volume changes promoted by the reaction combination should alter the individual approaches to equilibrium shown by $R1$ and $R2$ in the control experiments.

Finally, it would be valuable to develop knowledge and models of how osmotic effects manifest themselves in vesicles composed of multiple nested compartments.

\subsubsection{Exploring Other Indirect Coupling Mechanisms}

\begin{figure}
\begin{center}
\includegraphics[width=15.5cm]{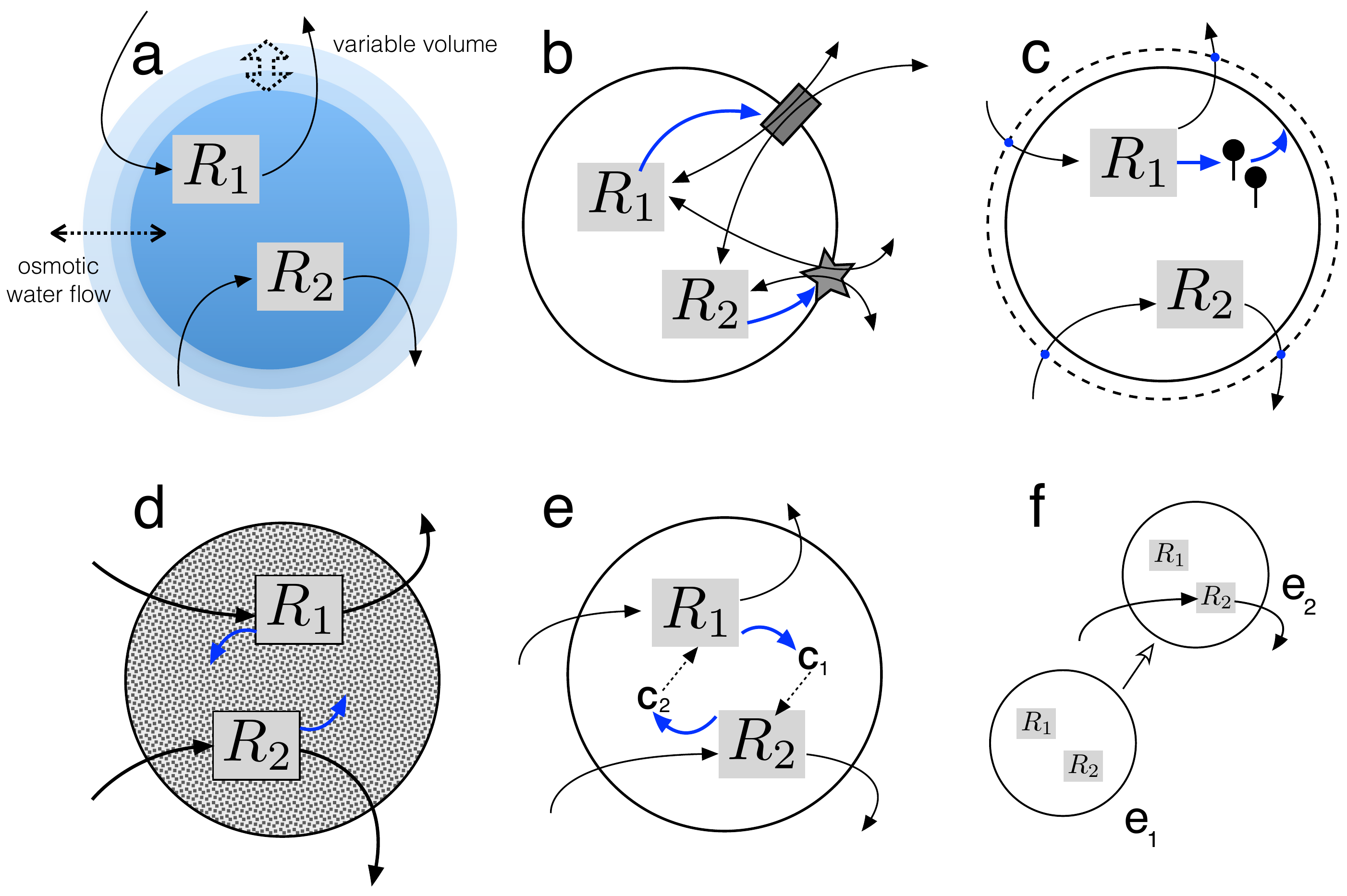}
\end{center}
\caption{
{\textbf{Scenarios of Indirect Coupling Between Mutually Exclusive Reaction Systems in Protocells. }} Multiple scenarios could have operated in parallel. (a) Osmotic coupling as discussed in Section \ref{sec:4_3_2}. Independent reaction systems $R1$ and $R2$ each provoke solvent volume changes that affect the other. (b) Coupling via membrane components. $R1$ and $R2$ produce or modify membrane components (like channels), which in turn affect the diffusion rates of all solutes. (c) Coupling via membrane surface size. $R1$ produces membrane amphiphiles that grow the membrane and/or change its composition and hence permeability, affecting rates of solutes reaching $R2$. (d) Coupling via crowding. $R1$ is able to crowd the medium inside the vesicle, altering the kinetic rate functions of $R2$, and vice versa. (e) Coupling via catalysts. $R1$ produces a set of catalysts that channels reaction pathways in $R2$, and vice versa. (f) Coupling via environment. $R1$ produces chemical species that result in vesicle locomotion from environment $e_1$ to environment $e_2$, changing the external solute species available to $R2$, and vice versa.
}
\label{fig:ch5_indirect_couplings}
\end{figure}

Osmotic coupling can, in fact, be seen as just one of a number of ways in which reaction systems with mutually exclusive species sets could indirectly couple their dynamics inside protocells. Figure \ref{fig:ch5_indirect_couplings} depicts some other possible scenarios. Of these scenarios, osmotic coupling is likely to be the simplest, because in order to become indirectly linked via osmosis, reaction systems are only trivially required to produce osmolytes (species affecting water flow): they are not required to produce specific components, such as membrane amphiphiles, other hydrophobic membrane components, or catalysts. Nevertheless, other scenarios indirectly coupling reactions, like the ones described in the caption of Fig. \ref{fig:ch5_indirect_couplings}, would still be interesting to pursue.

\subsubsection{Incorporation of Statistical Data into Semi-Empirical Vesicle Model}

Vesicle shape transformations, as discussed in Section \ref{sec:3_4}, are important to consider as part of investigations on the interaction between chemical reactions and vesicle compartments. 

One very interesting start in the direction of quantifying the shape transformations of vesicles has recently been performed by \shortciteA{Tsuda2014}. These authors systematically detected and characterised the shapes that vesicles assumed in environments of different osmolarities. Then, from this data, they computed a two dimensional free energy landscape of vesicle shapes. Moreover, they found that the energy landscape was able to accurately predict the shape transformations that vesicles would undergo whilst {\em dynamically} deforming: the transient shapes a vesicle would pass through tended to follow the low energy valleys across the energy landscape. In future, the semi-empirical model would benefit from integrating this type of statistical and empirically-derived data on vesicle shape transformations.

A particularly important transformation is budding and division \shortcite{Svetina2009}. In the current semi-empirical vesicle model, division is deterministic and always occurs as perfect fission. When the vesicle reaches a sufficiently deflated prolate shape such that the surface area is able to perfectly wrap two equal sized spheres made of the aqueous volume, then the vesicle divides into two identical, spherical daughters. 

This division condition (occurring at a certain surface-volume ratio) takes into account vesicle shape and can be considered more realistic than other theoretical protocell models that ignore the vesicle aqueous volume and impose division either when the surface area has doubled from the initial condition (e.g. as in most of the Chemoton models) or when some absolute membrane surface area is achieved (e.g. as in \shortciteNP{villani2014}). However, even though other aspects of the model (like the reaction kinetics or the sharing of metabolic contents between daughter vesicles on division) were considered stochastic, the division condition in the semi-empirical vesicle model has remained deterministic. Namely, the protocell is stable with certainty (probability 1) at any shape, until reaching the surface-volume ratio required for fission into equal daughters; then, division happens, again, with certainty. This deterministic division condition has particularly significant implications for the dynamics of the protocell model across generations. Observed behaviour, like the frequency of a division cycle, is highly contingent on the division condition occurring only at one precise surface-volume ratio, and nowhere else.

One initial way to start relaxing the deterministic assumption could be to give deflated states that are less deflated than the division condition some small probability of division into two unequal spherical daughters. When the vesicle model reached the `perfect fission' condition, division would still happen with certainty, but before that, there would also be some small probability of division. Close to spherical states could be given only a very small chance of division (in this case, one daughter would be a large sphere, and the other a small sphere, as in vesicle budding). It would be interesting to investigate how stochastic division affected population phenomena with the model, and e.g. the conditions necessary to achieve a stationary division cycle (as analysed by \shortciteNP{Mavelli2013}) in this new context. Also, it is a further assumption that vesicle division or budding always leads to identical daughter vesicles {\em outside} of a parent vesicle: another avenue to pursue could be to allow a probability of {\em internal} vesicle budding and/or asymmetric division. 

The idea of stochastic division could be taken further. The assumption of division into two daughters could also be relaxed, and division permitted into multiple approximately spherical progeny (as is observed when filamentous protocells break up into many daughters, see \shortciteNP{Zhu2009}).\footnote{ There are theoretical restrictions on how many spheres a certain vesicle shape can divide into. For example, deflated vesicle shapes where $\phi > \sqrt[3]{2}$ have too much excessive surface area to become two spheres, no matter what sizes these spheres are. They can become 3 spheres, but again, there is a $\phi$ limit for dividing into 3 spheres, and so on.} In this case, statistical data of vesicle division distributions under different conditions would have to be derived experimentally and added into the model.

%
%
%
%
%
%

\subsection{Longer Term Research Pathways}
\label{sec:5_3_2}

\subsubsection{Identifying General Constraints on Protocell Organisation}

In the 1970's, Chemical Reaction Network Theory (CRNT) was born as a response to the difficulties in analysing the dynamical behaviour of complex non-linear chemical systems. Instead of trying to analyse chemical systems on a disparate case-by-case basis, CRNT instead aimed to find general unifying principles in chemical networks of arbitrary complexity. The theory dealt with the static structure of reaction networks, and from mere inspection of structural properties alone, aimed to prove conclusively that certain dynamical regimes could or could not be present in a given chemical network -- sometimes independently of parameter values. An early result was the `Zero Deficiency Theorem' \shortcite{Feinberg1974} which stated that a chemical network with structural properties called `zero deficiency' and `weak reversibility', regardless of its complexity or parameter values, could only ever have a single stable state (see \shortciteNP{craciun2011,Soule2003,Bailey2001} for developments).

Similarly, protocell models could also benefit from more general results in this style. In prebiotic protocells, as stated earlier, it is extremely speculative as to what metabolisms would have been operating (their stoichiometry, if autocatalytic cycles, autocatalytic sets or templates were present etc.). Also speculative is how these metabolisms coupled to, and interacted with, protocell membranes. Therefore, instead of documenting the operation of isolated hypothetical protocell cases, future theoretical studies could be more fruitfully directed toward identifying more general requirements and constraints associated with protocells.

One line of research could be to build a database of ``if, then'' type of system constraints; {\em if} a protocell scheme has a certain feature, {\em then} these implications (or limitations) necessarily follow. Some examples of if-then system constraints might be the following:

\begin{enumerate}
\item \textbf{If} a protocell metabolism produces products in the internal aqueous core that are either impermeable, or have very low permeability, \textbf{then} for the protocell to be capable of a homeostatic steady state regime, these products must be broken down by another reaction pathway. If not, the protocell must also synthesise membrane surfactants, to mitigate the accumulation effects by entering into a growth and division regime. 

\item \textbf{If} molecules are present within the protocell core that are not metabolically synthesised from nutrients, or which do not copy themselves (like template replicators), \textbf{then} these will be diluted upon protocell division, to eventually disappear. As such, division into functional daughters has as a necessary condition that all required molecular components not available in the nutrient medium must be synthesised by the protocell.

\item \textbf{If} a protocell is to possess far from equilibrium dynamical behaviour in general, \textbf{then} metabolic reaction rates must be compatible with membrane permeabilities \shortcite{Piedrafita2012,Piedrafita2013}. A relatively impermeable membrane will force a fast internal metabolism to equilibrate (and infrequent molecular diffusions across the membrane will just perturb this equilibrium). Alternatively, if metabolism is very slow with respect to membrane diffusions, then internal metabolites will equilibrate to external concentrations (and infrequent reaction events will perturb this equilibrium). Far from equilibrium function is only assured between these equilibrium extremes.

\item \textbf{If} there exists a reversible reaction sequence inside a protocell fed by a single nutrient species across the membrane, \textbf{then} there exists no chemical gradient to maintain this reaction sequence far-from-equilibrium, and therefore it will always settle to an equilibrium state. It follows that such reaction sets cannot change the number of steady states already inherent in the protocell system, but they can modify the transient response of the protocell to perturbations.

\item \textbf{If} a protocell does not possess active transport mechanisms (which pump solutes and ions across the membrane against their concentration gradients), \textbf{then} at steady state, all nutrient species inside a protocell cannot exceed their external concentrations. This is because the only process delivering nutrients to the protocell core is a concentration driven, passive diffusion process. During transient dynamics, nutrients may exceed their external concentration, if the protocell volume transiently decreases.

\item \textbf{If} a protocell does not possess water pumping mechanisms, nor hard non-expansive proto-cell walls, \textbf{then} at steady states, the protocell water volume will resize such that the total concentration of solutes and ions inside the protocell will always be equal to the total concentration outside the protocell (isotonic condition). Active pumping mechanisms, or cell walls allowing turgor pressure, are needed to {\em sustain} a disequilibrium between total osmolyte concentration inside and outside of a protocell.

\end{enumerate}

Many more constraints (some much more subtle) could be added to this list. Each constraint could be accompanied by a proof (some proofs would be elementary, others more involved). Used in combination, such constraints could help narrow the space of acceptable chemical architectures for protocells. These system-level constraints could also help deduce if certain dynamical behaviours were impossible in principle, prior to simulation of a protocell scheme.

Two recent theoretical studies serve as examples of more complex constraints that could be added to the above list. These studies tried to formally identify the general type of chemical organisation a protocell must have to solve a key problem. \shortciteA{Mavelli2013}, already mentioned in Section \ref{sec:2_1_1}, focussed on the bottleneck problem of how protocells can grow and divide whilst avoiding the problem of progressive size decrease at each generation, i.e. the general organisational requirement a protocell must have to engage in a `stationary' division regime where daughter protocells were approximately the same size as the original parent. They derived an ``osmotic synchronisation condition'' that described the necessary (but not sufficient) general kinetic relationships that must hold in a protocell system if that protocell was able to engage in a stationary division cycle (neither decreasing nor increasing in size on each division). Their constraint formula, although relying on the deterministic assumption, was broadly applicable to any internal metabolic scheme. In a second study, \shortciteA{Bigan2015b} instead targeted protocell growth. They derived the general conditions that must be necessarily satisfied if an out of equilibrium protocell should be capable of growth while maintaining elevated internal concentrations with respect to its environment. Again, the analysis was broadly applicable to any internal metabolic scheme. An example of a much higher level constraint, also to be included on the list, would be the general protocell organisation required for open-ended evolution \shortcite{VonNeumann1966}: if the organisation does not have these features (i.e. biopolymers able to `fix' and reliably transmit metabolic complexity), then an open-ended increase in complexity cannot take place.

\subsubsection{Metabolic Architectures Suitable for Synthesis of Long and/or High-Energy Molecules}

Another promising longer-term line of research could be to investigate which types of metabolic reaction network motifs are suitable for certain overall tasks, like the fabrication of long and/or high energy molecules. For example, it would be very interesting to devise general constraints on the type of metabolic network architecture that must have been in place in self-producing protocells, allowing them to synthesise macromolecules and energy currency molecules. This work would also start to link to the topic of exergonic-endergonic reaction couplings in protocells.

Prebiotic uncertainty means that reaction networks in protocell models are usually specified in terms of an abstract artificial chemistry. Appendix \ref{appendix:C} provides two tests to help decide whether a set of abstract chemical reactions complies with mass conservation and free energy considerations. These tests prevent physically unrealistic metabolisms from being implemented, but they don't just stop there: they can also be employed in a much more interesting way too. 

If the molecular species participating in an abstract protocell metabolism not only have concentration, but also {\em atom number} and {\em free energy of formation} attributes associated with them, then a metabolism can be checked for {\em additional requirements}.\footnote{Going further, molecular species could also be attributed with how many of each atom type they contained.} These additional requirements may be statements like ``species $X$ must have higher free energy of formation than species $Y$'', or ``species $Z$ must be a molecule larger than 75\% of the molecules in the system, with an free energy of formation exceeding that of $X$ and $Y$ put together''. In this way, protocell reaction schemes can be tested not only for being physically valid, but also for having extra characteristics e.g. the ability to synthesise high energy and/or long macromolecules as mentioned above. For an example, consider G{\'a}nti's autocatalytic reaction cycle. This cycle is certainly valid on physical grounds (i.e. passes the two tests in Appendix \ref{appendix:C}), but when faced with the extra requirement that the nutrient precursor driving the cycle must have lower free energy than all of the other species involved, it would fail. G{\'a}nti's cycle requires a high energy precursor molecule in order to run. Hence, G{\'a}nti's autocatalytic cycle would {\em not} be an acceptable candidate for a protocell that requires to synthesise molecules with higher energy than the nutrient species.

The inverse problem to reaction set validation would also be interesting to pursue. Instead of a specific abstract reaction mechanism being validated to obey a set of constraints, the aim would be instead to develop algorithms capable of returning {\em all} minimal reaction schemes that satisfy a given set of constraints. These constraints might involve the specification of a set of nutrient species to be consumed; a set of species that are required to be produced; a wider pool of total species available for use and the energetic and mass relations that must hold between all the chemical species. Also, the presence or absence of catalytic and autocatalytic loops could be specified. Ideally, such algorithms could be instructed to find, for example, all the minimal (abstract) `autotrophic' metabolisms that converted small lower energy nutrients into a set of higher energy polymers and which contained exactly $N$ autocatalytic cycles.

\subsubsection{Uniting Self-Production with Self-Reproduction}

To conclude this discussion on a more general note, it should be emphasised that a long term goal of a research program into autonomous protocells should also be to investigate how protocells started {\em dividing} in a reliable way (e.g. see \shortciteNP{Mavelli2013}).

To be clear, this thesis has argued for an autonomy perspective on living systems and protocell development, but this has not been for the purpose of replacing evolutionary concepts. Rather, the intent has been to place the evolutionary perspective into a wider context, bringing to the table essential {\em synchronic} problems of cellular systems organisation that a pure {\em diachronic} evolutionary perspective has been blind to. Under the view of protocell development being argued for here, even though the chemical logic of individual protocells has been focussed on, still concepts like protocell division, proliferation, population dynamics (e.g. competition) and chemical evolution remain as extremely relevant to the overall picture of origins.

In autonomous protocells, it is important to understand how diverse components and processes became integrated for robust self-maintenance of single protocells, but also the question of how system-level self-\textbf{\uline{re}}production started is of paramount importance. Division represents a huge organisational challenge for any complex protocell system, as many different components need to be precisely coordinated such that two copies of the original organisational structure can be produced. Moreover, division would have been an inherent property of protocell systems that internally fabricated their own components: not only would these components have granted the ability to precisely control energy and matter flow through the protocell structure, but they would have also likely prompted its growth. Therefore, a research program into autonomous protocells should also aim to investigate, in parallel to the co-evolution of metabolism and membrane, the co-evolution of self-production with self-reproduction.


{  
\chapter*{Conclusions}
\addcontentsline{toc}{chapter}{Conclusions}
\renewcommand{\chaptermark}[1]{\markboth{}{\sffamily #1}}
\chaptermark{Conclusions}

The main conclusions of this thesis are the following:

\begin{enumerate}

\item \textbf{Prebiotic research cannot do without conceptual discussion and clarification.} Origins of life is a field of enquiry that benefits from a joint scientific and conceptual/philosophical approach. Attempts to synthesise life from the bottom up are inevitably immersed in a background conception of what constitutes `life'. Different conceptions of life lead to diverging research agendas in protocell research and in prebiotic chemistry in general. It is important to explicitly recognise and question the general conception of life underlying any research program into origins of life.

\item \textbf{The autonomy perspective brings to the fore system-level challenges related to the emergence of cellular organisation.} A radical reformulation of the origins of cellular life program presents itself when the implications of an autonomous systems view of life are properly developed. In particular, autonomy emphasises that a {\em necessary condition} for functional protocells at each stage of their development would have been a high level of integration between components and processes. Pure evolutionary approaches to protocells miss this necessary condition, for they do not include (nor find important) a rigorously developed concept of cellular organisation. Basic autonomous protocells able to robustly self-maintain far-from-equilibrium would only have been achievable if the chemical metabolism and protocell membrane co-evolved together, integrating their functionality.

\item \textbf{Autonomy is a heuristic concept which can be transformed into a set of concrete research questions able to be pursued through semi-empirical protocell modelling.} A semi-empirical modelling approach can be used to rigorously investigate the co-evolution of metabolism and membrane in protocells. A semi-empirical approach allows the construction and exploration of grounded theoretical protocell models which can be directly related back to {\em in-vitro} protocell experiments, and which can be extrapolated to test more complicated  hypothetical scenarios. It is possible to develop realistic coarse-grain kinetic models of protocells that embed realistic parameters and reproduce various experimental results.

\item \textbf{The synthesis of membrane by metabolism, as advocated in the theory of autopoiesis, is not strictly necessary for protocells to start exhibiting biologically relevant non-linear behaviours.} Self-producing compartmentalised chemical systems with the ability to internally fabricate their own components like membrane lipids, catalysts and peptides, could be considered as a relatively late stage in the evolution of protocellular organisation. Before that, protocells with weaker, more indirect couplings between internal chemistry and compartmentation could have demonstrated emergent dynamical behaviours (as demonstrated by the semi-empirical modelling work of this thesis). A topic of future protocell research should be to rigorously investigate not just the encapsulation of nucleic acids in compartments, but the general indirect coupling effects resulting between basic chemical reactions and dynamic lipid membranes.

\item \textbf{Osmotic coupling is a new and relevant systems principle for protocell metabolism that has received no attention in the origins of life literature.} Early vesicle protocells would have been susceptible to osmosis and, at the same time, poor regulators of internal water volume. Solutes would diffuse across protocell membranes, {\em and so would water solvent}. As such, variable water volume inside early protocells could have been a significant factor in promoting non-linear dynamics in protocell reaction chemistries. In particular, variable water volume can allow effects like {\em osmotic coupling}, where independent reaction systems inside a vesicle couple their dynamics to produce very complicated system-level behaviours.

\item \textbf{Identification of general constraints on the organisation of protocell systems should be given more attention in the field of origins.}  As part of the semi-empirical modelling approach, a constraints-based approach to protocell models would be promising to pursue in the future. This approach would involve identifying general constraints on protocell organisation that must be met in order for protocells to pass through certain bottlenecks, or to possess certain attributes. Such constraints could help ``close the net'' around the problem of how protocells developed into full-fledged autonomous cells.

\end{enumerate}
}


\appendix
\singlespacing

\begin{appendices}

\chapter[\texorpdfstring{Petri Net Framework for Modelling Constructive \\ Dynamical Systems}{}]{Petri Net Framework for Modelling Constructive Dynamical Systems}
\chaptermark{Constructive Dynamical Systems}
\label{appendix:A}

Numerically modelling protocells presents a special challenge. Protocells are  fluid machines which can undergo various morphological transformations. For lipid vesicles, these transformations can include growth, fusion, budding, division, creation of nested internal vesicles (via e.g. invagination), and emission of internal vesicles through the membrane, to name but a few. If protocells are modelled at the level of well-stirred compartments, most of the listed morphological changes present problems for traditional dynamical systems modelling. Excepting growth, mere changes in dynamical state variables are not enough to capture such transformations: they also require modifications to the actual equation structure of the dynamical system. For example, a protocell division event (or the creation of an internal vesicle) requires creating an extra vesicle structure, increasing the degrees of freedom of the system. A protocell burst event conversely requires removing a protocell or turning it into a flat bilayer without an internal aqueous domain, decreasing the degrees of freedom. In an event where two initially separated vesicles coalesce and become enveloped, one inside the other, the degrees of freedom stay constant, but the structure and coupling of the existing evolution equations will change. Therefore, comprehensive protocell modelling demands a {\em constructive dynamical systems} approach (already discussed in Section \ref{ch2_constructive_dynamical_systems}).

Specialised software has been developed to perform such protocell simulations \shortcite{Mavelli2010}, but the purpose of this appendix is to outline a general and flexible architecture, developed as part of this thesis, which can be used for simulating any stochastic system that has changing degrees of freedom and/or boundary conditions\footnote{see also the work by \shortciteA{giavitto2003}.}. Protocells are just one specific application. The architecture is based on the formal object of petri nets.

\section*{A Whistle-Stop Tour of Petri Nets}
\addcontentsline{toc}{section}{A Whistle-Stop Tour of Petri Nets}

Petri nets were originally conceived as a simple yet powerful way to formalise concurrent systems and to decisively prove properties about their operational characteristics (a popular first review is \shortciteNP{Murata1989}). One key feature of petri nets is that they are visually intuitive\footnote{In fact, they were created by Carl Adam Petri at just 13 years of age.}. In their simplest form, petri nets are called `place/transition nets' and are drawn as directed graphs consisting of two types of node. Place nodes (circles) are connected to transition nodes (squares), and transition nodes are connected to place nodes. Each place node is a state variable of the system and holds a certain amount of tokens. As the different transition nodes `fire', the tokens move around the network between the place nodes, updating the system state or `marking'. Tokens are not necessarily conserved and may be spontaneously created or destroyed by the transitions. In order to `fire', a transition must first be enabled, which means that a required number of tokens must exist on each of its input arcs (given by the weighting of those arcs). When enabled, a transition may fire immediately, or may wait for some extra external condition to become true. On firing, a transition takes tokens from all places with arcs coming into it, and gives tokens to all places it has arcs pointing to. The number of tokens transferred depends on the weightings on the individual arcs (by default, they are 1). In place/transition nets, there is no explicit concept of time, only of events which take place in sequences.

Once a system has been converted to a place/transition net representation, a well-developed set of theorems and algorithms can be applied to decide general behavioural possibilities of the network (e.g. see \shortciteNP{heiner2008}). For example, it can be decided whether the network is {\em live} (all transitions are able to contribute to the net behaviour, forever), {\em reversible} (any state can always be returned to) and how the places are {\em bounded} (the maximum number of tokens that each place will hold during execution of the net). Also, invariant features of the network can be identified from its static structure, such as sets of places which tend to conserve total tokens (`P-Invariants') and sets of transitions whose firing does not change the marking (`T-Invariants'). These invariants often have interpretations relevant for the system understudy. For example when metabolic networks are rendered as place/transition nets, P-invariants correspond to pools of conserved substrates, and T-invariants can give insight into steady state behaviour.

\section*{Petri Nets as Structured Descriptions of Model \\ Equations}
\addcontentsline{toc}{section}{Petri Nets as Structured Descriptions of Model Equations}

Since their conception as basic place/transition nets, petri nets have diversified into a `zoo' of many different petri net classes, where extra elements and arcs have been added to increase their modelling power for various different applications. Rather than being applied as a theoretical tool to give insight into network behaviour, one popular application of petri nets is simply as a visual structured description of a dynamical model. A model can be mocked-up by drawing a petri net and then easily converted into either a continuous or stochastic description ready for numerical execution\footnote{The dynamical model usually needs to be expressible as a set of first order derivatives.} \shortcite{heiner2008}.

In particular, a special class of stochastic petri nets (called XSPN) has been developed for turning vague notions of biochemical pathways into exact models, which can then be simulated and investigated quantitatively, allowing comparison to wet lab data \shortcite{heiner2009}. Indeed, this class of petri net can be used to specify the protocell model used in this thesis (see Section \ref{sec:3_3_2}). Then, the protocell model can be executed on a publicly available simulation platform called {\em Snoopy} \shortcite{Rohr2010, heiner2012}. However, the latter simulation platform is not capable of modelling any of the protocell morphological transformations that require modification of the model equation structure (e.g. division) because the petri net simulated is static. Throughout a simulation, the network topology is fixed and the boundary parameters are fixed.

\section*{Extending Petri Nets to Handle Structural Change in Dynamical Systems}
\addcontentsline{toc}{section}{Extending Petri Nets to Handle Structural Change in Dynamical Systems}

Petri nets are actually good candidates for modelling constructive dynamical systems with changing equations and/or boundary conditions. Even though a petri net representation of a dynamical model contains the same amount of information as the differential equations of that model, the visual format of the petri net makes the relationships between processes and state variables much more obvious\footnote{In the same way, a nested LISP program and its tree representation have the same information content, but the tree structure is easier to work with when modifying the program e.g. when combining programs in genetic programming.}. As such, petri nets provide a convenient vehicle for altering the structure of a dynamical model, and then for consistently carrying these changes through to formulate new overall evolution equations.

\begin{figure}
\begin{center}
\includegraphics[width=15.5cm]{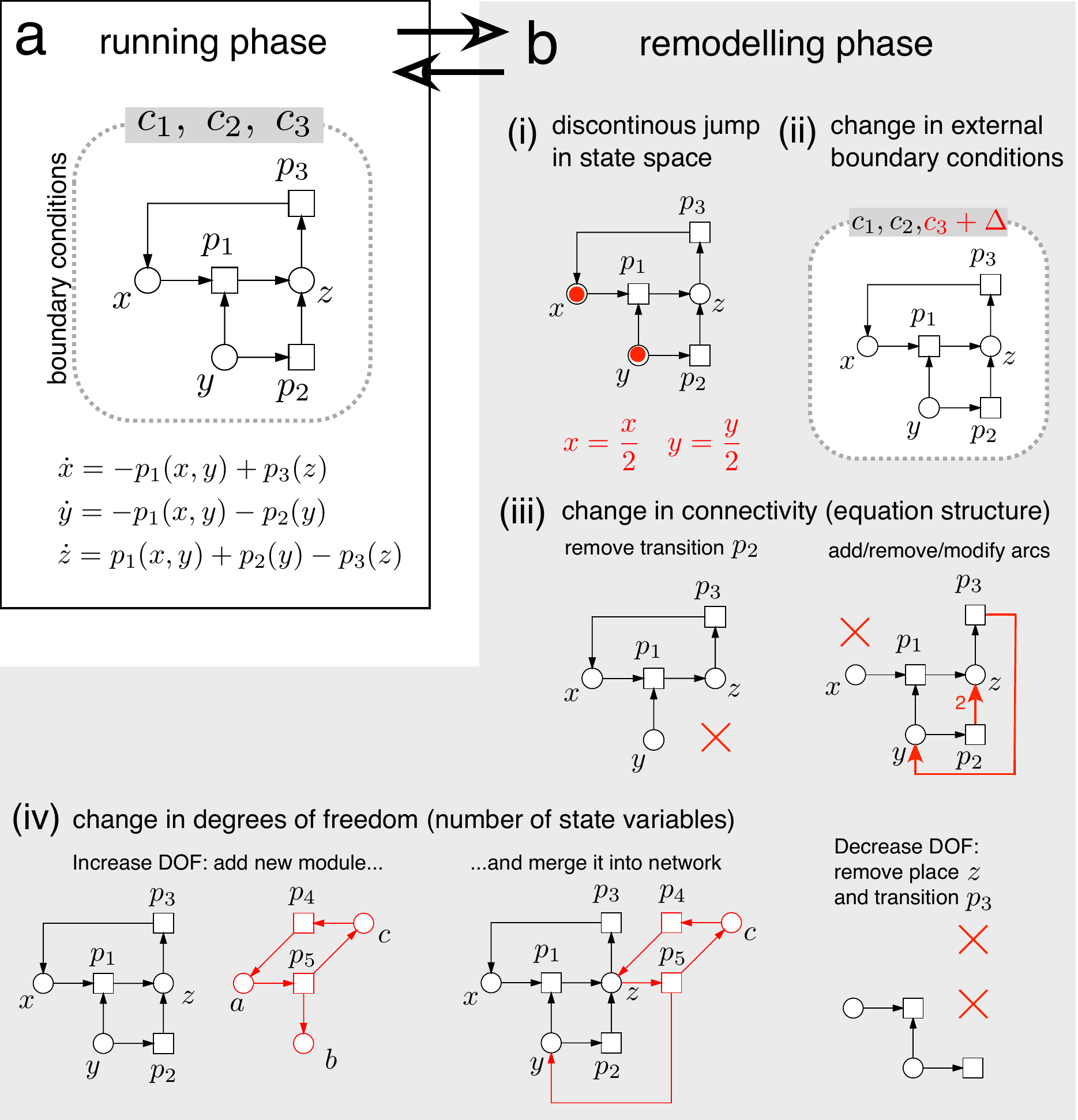}
\end{center}
\caption{
{\textbf{A Two Phase Petri Net Architecture for Modelling Constructive Dynamical Systems.}} See text for discussion.
}
\label{fig:appendixA_twophase}
\end{figure}

Figure \ref{fig:appendixA_twophase} shows how the conventional petri net paradigm can be combined with an unconventional part, in order to obtain a model which can change not only in state, but also in structure, as it executes. The idea of this architecture is that a petri net is simulated in a `running' phase (Fig. \ref{fig:appendixA_twophase}a) where the token marking changes and the structure is fixed, but additionally the petri net may pass transiently into a `remodelling' phase (Fig. \ref{fig:appendixA_twophase}b). In the remodelling phase, four types of more drastic modifications can be made to the network, alone or in combinations: (i) large, discontinuous jumps in the state space, (ii) changes to the boundary parameters of the system, (iii) changes to the connectivity of the system (leaving the degrees of freedom unchanged) or (iv) expansion or contraction of the degrees of freedom of the system.

In order to pass between the running and remodelling phases, a new type of transition can be introduced into a petri net, called a `remodel' transition (not shown in Fig. \ref{fig:appendixA_twophase}). This type of transition fires immediately on becoming enabled, representing that the current state variables have fallen into some critical range or ratios, and the current model structure no longer reflects the underlying system being modelled. Rather than moving tokens, a remodel transition instead executes a sequence of instructions which describe how the petri net network should be remodelled. Then, the modified network is returned to the running phase. Many such remodelling events may happen during the course of a simulation.

Crucially, the instructions executed by the remodel transitions hold information about the {\em underlying objects} the dynamical system is describing (see the discussion of the work of Fontana and Buss in Section \ref{ch2_constructive_dynamical_systems}). When the state of the system gets to the existence limit of the current objects the dynamical system is describing, the remodel transitions fire to modify the underlying objects in the appropriate way, creating a new dynamical system.

\section*{Worked Example: Petri Net Model for Growing and Dividing Protocells}
\addcontentsline{toc}{section}{Worked Example: Petri Net Model for Growing and Dividing Protocells}

As part of this thesis, the novel two phase petri net architecture described above was implemented in C and subsequently used to perform a diversity of numerical protocell simulations. This program (called {\em Flow}) accepts stochastic petri nets in the XSPN format referred to above. Additionally, the petri nets can be specified as modules and connected together, allowing initially complicated or repetitive network designs to be achieved (i.e. grids of connected systems).

\begin{figure}
\begin{center}
\includegraphics[width=15.5cm]{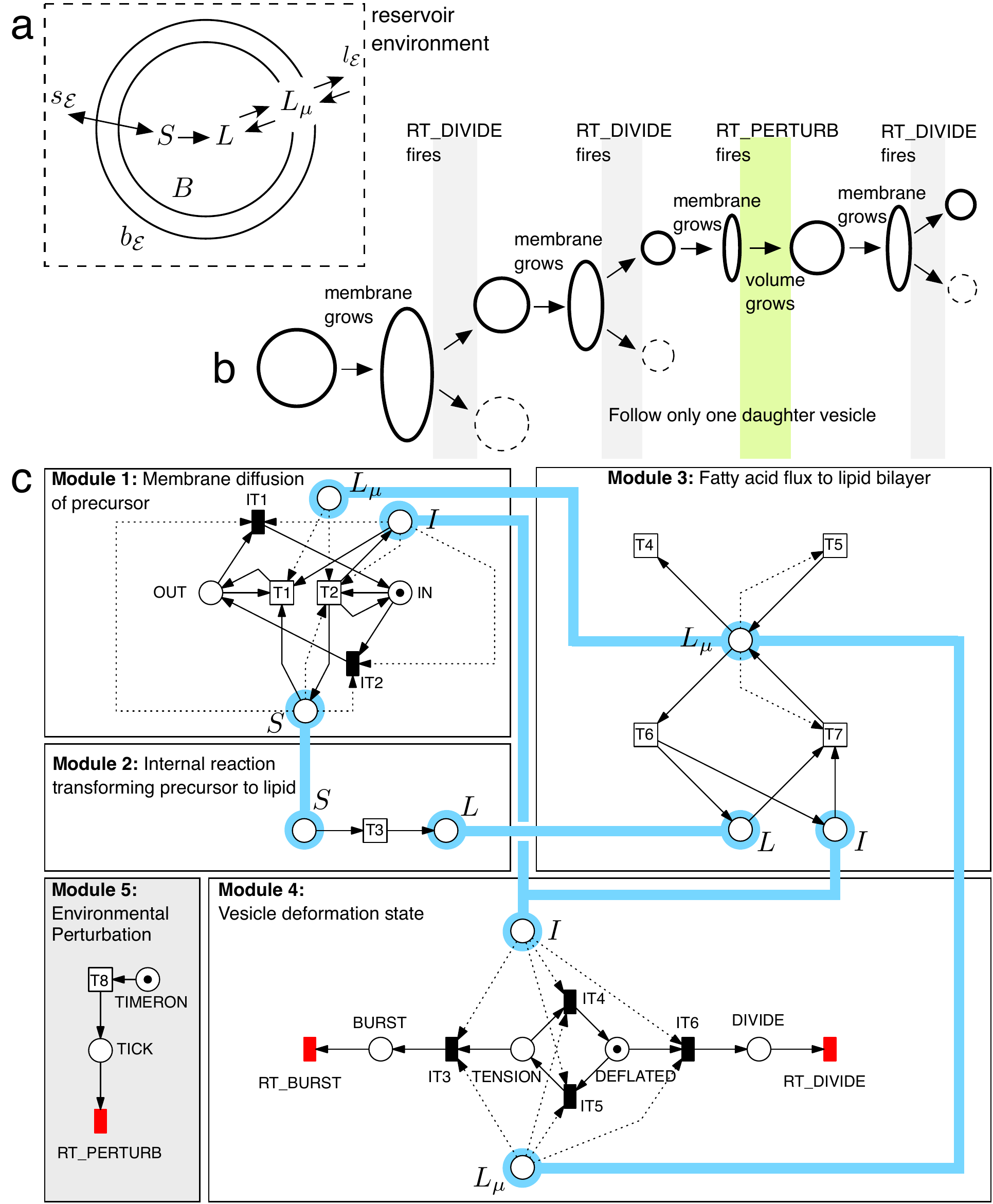}
\end{center}
\caption{
{\textbf{Petri Net Model for a Basic Protocell Division Cycle.}} \textbf{(a)} A vesicle with internal lipid synthesis \textbf{(b)} exhibits the dynamic behaviour of a decreasing division cycle when \textbf{(c)} modelled as an XSPN petri net with added remodel transitions. Table \ref{table:appendixA_divisionmodel} below gives the formula for the petri net transitions.
}
\label{fig:appendixA_divisionmodel}
\end{figure}

For a brief overview, Fig. \ref{fig:appendixA_divisionmodel} shows an example petri net capable of modelling a basic vesicle division cycle. The vesicle grows by internally synthesising membrane lipid from a precursor that diffuses through the membrane (Fig. \ref{fig:appendixA_divisionmodel}a). The petri net consists of five modules, four of which are connected together to make the vesicle (Fig. \ref{fig:appendixA_divisionmodel}c, blue lines). Table \ref{table:appendixA_divisionmodel} details the initial token marking for the petri net, and the rate formula for each of the transitions. The petri net has three types of transition: `normal' transitions (open squares), `immediate' transitions (black rectangles) and remodel transitions (red rectangles), a special type of immediate transition. Normal transitions have a stochastic propensity (rate) and are fired with the standard Gillespie algorithm. Immediate transitions fire as soon as they become enabled\footnote{If many immediate transitions are simultaneously enabled, they are fired at random.}, and they are useful for updating meta-information about the network when certain conditions hold. For example, the immediate transitions \texttt{IT1} and \texttt{IT2} are responsible for monitoring the direction of the concentration gradient of precursor \texttt{S}; similarly, immediate transitions \texttt{IT3}-\texttt{IT6} are responsible for monitoring the surface area to volume ratio of the vesicle. The remodel transitions (red rectangles) execute instructions to make more drastic changes to the petri network. Dotted lines in Fig. \ref{fig:appendixA_divisionmodel}, drawn from places to transitions in the petri net, are `modifier' edges denoting that a place influences the rate of a transition, but the place has no tokens consumed when that transition fires.

In this division example, it does not make sense to follow both daughter vesicles after a fission event because the surrounding environment is a reservoir, and the daughter vesicles will not be able to interact with each other. Therefore, only one daughter is followed, and this means that the degrees of freedom of the system remain the same. The remodel transition handling division \texttt{RT\_DIVIDE} executes the following instructions when enabled by a token in the \texttt{DIVIDE} place:

\begin{enumerate}[label=\textbf{\arabic*})]
\item Calculate how many buffer molecules \texttt{B} are in the vesicle. \texttt{B} = \texttt{I} - \texttt{S} - \texttt{L}
\item Make a \textbf{discontinuous state change}:
	\subitem Half the vesicle contents: tokens on \texttt{S} = \texttt{S} / 2; \texttt{B} = \texttt{B} / 2; \texttt{L} = \texttt{L} / 2
	\subitem Half the membrane surface area: tokens on $L_\mu$ = $L_\mu$ / 2 (some randomness could be added to the division process if desired)
\item Set the total number of solute molecules\footnote{State variable \texttt{I} is not actually a true state variable of the vesicle. It is included however, because it allows the easy addition of processes influencing (and influenced by) the volume of the vesicle. Without this pseudo-state variable, building a protocell from petri net modules would be much more brittle.} inside the vesicle \texttt{I} = \texttt{S} + \texttt{L} + \texttt{B}
\item Calculate $\Phi$ for the new daughter vesicle, and put a token in the \texttt{TENSION} place if $\Phi < 1$, or in the \texttt{DEFLATED} place if $\Phi \ge 1$. 
\item Run the modified petri network (which now represents the followed daughter vesicle).
\end{enumerate}

%

\begin{table}
\centering
\scalebox{0.85}{

\begin{tabular}{|p{7.5cm}|p{7.5cm}|}
\specialrule{.3em}{0em}{0em}
\rowcolor{gray!40}
\multicolumn{2}{| l |}{\textbf{Initial Petri Net Marking}} \\
\hline
\multicolumn{2}{| l |}{Starting with 100nm diameter spherical vesicle} \\
\multicolumn{2}{| l |}{$V_{(sph)}=\frac{4}{3}\pi50^{3}\times10^{-24} \text{litres}$, $S_{\mu(sph)}=4\pi50^{2}\times10^{-16} \text{dm}^{2}$} \\
\multicolumn{2}{| l |}{Buffer $B$ inside vesicle is set to give initial isotonic condition.} \\
\hline
\textbf{State Variable (Place)}	& 	\textbf{Tokens}\\
$S$ 	&	0 \\
$L = l_{\cal E}N_{A}V_{(sph)}$							&	16 		\\
$B = (s_{\cal E}+b_{\cal E})N_{A}V_{(sph)}$				&	66217	\\
$I = S + L + B$											&	66233	\\
$L_\mu=\frac{2S_{\mu(sph)}}{\alpha_{L}}$				&	209440	\\
IN 		& 1		\\
OUT 	& 0		\\
BURST 	& 0		\\
TENSION & 0 	\\
DEFLATED & 1 	\\
DIVIDE 	& 0 	\\
TIMERON & 1 	\\
TICK 	& 0 	\\
\rowcolor{gray!40}
\hline
\multicolumn{2}{| l |}{\textbf{Transition Firing Rates}} \\
\hline
\textbf{Normal Transitions}		&	\textbf{Stochastic Propensity} \\
T1	&	$\text{OUT} \times {\color{blue}D_{S}}S_{\mu}\frac{(S/\Omega)-{\color{blue}s_{\cal E}}}{{\color{blue}\lambda}}$ \\
T2	&	$\text{IN} \times {\color{blue}D_{S}}S_{\mu}\frac{{\color{blue}s_{\cal E}}-(S/\Omega)}{{\color{blue}\lambda}}$ \\
T3	& 	${\color{blue}k_{r}}S$ \\
T4	&	${\color{blue}k_{out}}L_{\mu}$ \\
T5	&	${\color{blue}k_{in}}S_{\mu}{\color{blue}l_{\cal E}}$ \\
T6	&	${\color{blue}k_{out}}L_{\mu}$ \\
T7	&	${\color{blue}k_{in}}S_{\mu}\frac{L}{\Omega}$ \\
T8	&	$\text{TIMERON} \times {\color{blue}k_{tick}}$ \\
\hline
\textbf{Immediate Transitions}	&	\textbf{Fire when True} \\
IT1	&	$\text{OUT} \times ({\color{blue}s_{\cal E}}\geq\frac{S}{\Omega})$ \\
IT2 & 	$\text{IN} \times ({\color{blue}s_{\cal E}}<\frac{S}{\Omega})$ \\
IT3	&	$\Phi<1-{\color{blue}\epsilon}$ \\
IT4	&	$\Phi\geq1$ \\
IT5	&	$\Phi<1$ \\
IT6	&	$\Phi>(1-{\color{blue}\eta})\sqrt[3]{2}$ \\
\hline
\textbf{Remodel Transitions}	&	\textbf{Execute Instructions when True} \\
RT\_PERTURB	& 	$\text{TICK}>{\color{blue}T}$ \\
RT\_DIVIDE	&	$\text{DIVIDE}==1$ 	\\
RT\_BURST	&	$\text{BURST}==1$ 	\\
\specialrule{.3em}{0em}{0em}
\end{tabular}

} 

\caption{\textbf{Initial Marking and Transition Propensities for Petri Net of Fig. \ref{fig:appendixA_divisionmodel}}. Boundary parameters of the system are shown in blue. To prevent the transition formulas being a mess of symbols, `intermediate' variables are used. These are relationships between the true state variables, defined as: total internal molecules $I=S+L+{\color{blue}B}$, scaled volume $\Omega={I}/({{\color{blue}s_{\cal E}+b_{\cal E}}})$, surface area $S_{\mu}=L_{\mu}{\color{blue}\alpha_{L}}/2$, and reduced surface $\Phi={S_{\mu}}/{\sqrt[3]{36\pi\left(\Omega/{N_{A}}\right)^{2}}}$.}
\label{table:appendixA_divisionmodel}
\end{table}
%

In other words, \texttt{RT\_DIVIDE} resets the state of the parent vesicle to that of the first daughter, and then recommences the simulation. Figure \ref{fig:appendixA_divisionmodel}b shows that a decreasing vesicle division cycle will result from this system, because the surface area of the vesicle membrane grows, but the vesicle volume does not. 

Module 5 does not form part of the vesicle petri net, but rather triggers a perturbation to the environment of the vesicle after an approximate time has elapsed. In this case, this perturbation is envisaged to be a decrease in the external buffer concentration $b_{\cal E}$, such that the vesicle experiences an influx of water, growing in volume. After a specific number of tokens have accumulated on place \texttt{TICK}, the remodel transition \texttt{RT\_PERTURB} fires, executing the following instructions:

\begin{enumerate}[label=\textbf{\arabic*})]
\item Make a \textbf{boundary conditions change}: 
	\subitem Set the parameter $b_{\cal E}$, representing the buffer concentration in the environment, to a lower value.
\item Make a \textbf{discontinuous state change}:
	\subitem Reset the tokens on \texttt{TICK} to 0. 
	\subitem Remove the token from \texttt{TIMERON,} to disable any further perturbations to the environment.
\item Run the modified petri network.
\end{enumerate}

The effect of this event is shown as the green bar on Fig. \ref{fig:appendixA_divisionmodel}b: the vesicle instantly increases in volume. Finally, remodel transition \texttt{RT\_BURST} will just contain the instruction to stop the simulation if the vesicle bursts. If the scenario were different, e.g. a population of vesicles existed in a {\em finite} environment, this transition could be used to remove the modules of the vesicle that burst, and add a new module representing a flat bilayer, in their place.

In summary, representing a model protocell system as a flat petri net structure is advantageous for a number of reasons. Firstly, the model is a transparent white box: all of its workings are clear. It is straightforward to publish the precise model used to produce simulation data, so that others can reproduce findings exactly. Secondly, new model features can be added by the user by changing the petri net, and are not dependent on (bug prone) modifications to the underlying software. Finally, a small and  efficient algorithm can be developed for simulating a two phase petri net architecture; an algorithm that will be robust and bug-free regardless of simulating small systems or large, complicated networks.

\chapter{Lattice Model of Self-Assembly in Surfactant-Water-Oil Mixtures}
\chaptermark{Lattice Model of Self-Assembly}
\label{appendix:B}

This appendix contains more technical details on the lattice model of amphiphilic self-assembly reported in Section \ref{sec:ch4_alice_model}. The lattice model was a `spin' model of the Larson type (\shortciteNP{larson1985}; see \shortciteNP{liverpool1996} for a review) and implemented in two dimensions on a 100x100 lattice.

\section*{Spin Models to Capture Universal Properties of \\ Complex Phenomena}
\addcontentsline{toc}{section}{Spin Models to Capture Universal Properties of Complex Phenomena}

Lattice spin models are a class of model typically used in statistical physics to investigate the equilibrium properties of solids and liquids. A famous example of a spin model is the Ising model (\shortciteNP{sole2011}; \shortciteNP{Binder1997}). In this model, a solid piece of magnetic material, say iron, is represented as a regular lattice of spins. Each spin represents an iron atom pointing either upwards (spin of +1) or downwards (spin of -1) and locally adjacent spins have the tendency to align with each other. The lattice is connected to a heat bath which provides a thermal noise, counteracting the alignment of spins. Even though the Ising model is an extremely simplified model of real piece of iron, it is nevertheless able to reproduce the phase transition in the magnetisation of an iron solid when it is heated through a critical temperature (and to quite remarkable accuracy, see \shortciteNP{back1995}). In the Ising model, at low temperatures, there is a global alignment of spins (i.e. the solid is magnetised) up to a critical temperature. At the critical temperature there is the sudden onset of wide fluctuations in the spin alignments, and past the critical temperature all spins are essentially randomised due to excessive thermal noise (i.e. the solid becomes de-magnetised). Therefore, although the Ising model is a brutal abstraction of the atomic structure and interactions of a particular magnetic solid, there is a sense in which the model has captured something \emph{universal} about the way magnetisation in general works.

Spin models of amphiphilic fluids follow a similar formalisation to the Ising model and are also are intended to capture universal features -- this time about the way self-assembly works in ternary fluid mixtures of amphiphiles, water and oil.

\section*{General Description of Larson Model Evolution}
\addcontentsline{toc}{section}{General Description of Larson Model Evolution}

The Larson model of amphiphilic fluids is based on a square lattice of binary spins which have only nearest neighbour and diagonally nearest neighbour interactions.\footnote{That is, the Moore neighbourhood. The four diagonally nearest neighbours are treated as being at the same interaction distance as the four nearest neighbours.} The lattice is typically of two or three dimensions with periodic boundary conditions in all directions. The Larson model uses spin to denote hydrophobicity. In the basic model, a +1 spin signifies the presence of a hydrophilic `water-loving' moiety, and a -1 spin the presence of a hydrophobic `oil-loving' moiety. The lattice is completely filled, with up to three type of molecules: water molecules $W$ are represented as single +1 spins, oil hydrocarbons $O$ as single -1 spins, and amphiphiles, denoted $H_iT_j$, are connected chains of $i$ hydrophilic head $H$ spins (+1) followed by $j$ hydrophobic tail $T$ spins (-1). Amphiphile chains move as self-avoiding walks with the constraint that they cannot be broken, nor cross themselves or other neighbouring chains. Being part of a still liquid, the molecules and amphiphile chains on the lattice are free to move around into new configurations, but the number of molecules of each type is always conserved.\footnote{Conversely in the Ising model (reviewed next), the atoms are fixed in place, and the total number of up and down spins is not conserved.}

The lattice has a total energy in any particular configuration, depending on what spins are next to each other. As \shortciteA{dawson1992} writes, `the lattice model really is simply a way of describing an ensemble of surfaces that can break and tear but that possess an energy of deformation.' (p1592). The hydrophobic effect is modelled by giving all pairs of adjacent spins an appropriate local interaction energy: like spins (hydrophobic-hydrophobic pairs, or hydrophilic-hydrophilic pairs) decrease the lattice energy, whereas unlike spins (hydrophobic-hydrophilic pairs) increase the lattice energy. At equilibrium, closed physical systems always seek their lowest energy configuration: in the case of the lattice, total energy is minimised when there exists a minimal number of unlike spins in contact.

The lattice is started with a random distribution of oil, water and amphiphiles, which represents a high energy non-equilibrium configuration. This initial condition must be transformed into a series of lattice configurations likely at equilibrium. From statistical physics, it is known that when the lattice has equilibrated to some temperature $T$, the fluctuations in the total energy of the configurations appearing should be distributed according to the Boltzmann distribution. A special {\em monte carlo} algorithm called the Metropolis-Hastings algorithm (see \shortciteNP{richey2010} for an excellent pedagogical review) is used to re-arrange the oil, water and amphiphilic chains on the lattice to accomplish this goal. The M-H algorithm drives the system into the region of lattice configuration space where successive lattice configurations have energy distributed according to the Boltzmann distribution for temperature $T$.\footnote{Technically, the Metropolis-Hastings algorithm constructs a markov chain of configurations which is stable on the Boltzmann distribution. The M-H algorithm belongs to the class of Markov Chain Monte Carlo algorithms. Generally, these algorithms are able to find and return samples (termed `important' samples) from regions of multidimensional probability distributions where there is the most `mass' (probability).} When in this region, the M-H algorithm is producing important samples of {\em likely} lattice configurations for temperature $T$. These monte carlo samples can be subsequently used to compute reliable equilibrium averages for certain properties of the amphiphilic fluid on the lattice (e.g. if micelles are formed, their number and size distributions can be calculated).

In practice, to ensure proper lattice equilibration, the lattice temperature should be gradually decreased from a high value to the final temperature $T$. This process is called {\em simulated annealing} \shortcite{kirkpatrick1983} and ensures that the system does not stray too far-from-equilibrium. If the system is not gradually cooled but, rather, is `quenched' to a low temperature rapidly, non-optimal lattice configurations can sometimes become frozen into the system.

\section*{Lattice Energies}
\addcontentsline{toc}{section}{Lattice Energies}

Interaction energies for different pairs of adjacent molecules are given as a matrix and specified in terms of a dimensionless parameter $\omega$.\footnote{Simply put, $\omega$ represents both the temperature and interaction energies of the system. Defined as $\omega = \epsilon / kT$ it corresponds to the lowest common denominator of the interaction energies in Joules, divided by (the Boltzmann constant multiplied by the absolute Kelvin temperature of the system). $\omega$ is used in place of using absolute interaction energies in Joules and absolute temperatures in Kelvin, because it is only the dimensionless {\em ratio} of energy to temperature that is important in the evolution of the model.} For the simple Larson model:

\begin{table}[h]
\centering
\begin{tabular}{c|cc}
 & $W$ or $H$ & $O$ or $T$\tabularnewline
\hline 
$W$ or $H$ & $-\omega$ & $\omega$\tabularnewline
$O$ or $T$ & $\omega$ & $-\omega$\tabularnewline
\end{tabular}
\label{table:InteractionEnergies}
\caption{}
\end{table}

The matrix is symmetric about the diagonal because the order of the adjacent molecules does not matter (WO always has the same energy as OW). The total energy of the lattice is described by the so-called Hamiltonian equation, simply

\begin{equation}
{\cal H}=\sum_{<ij>}E_{ij}N_{ij} = \left(E_{OO}N_{OO} + E_{WW}N_{WW} + E_{WO}N_{WO} \right)
\label{eq:Hamiltonian}
\end{equation}

where $E_{ij}$ is the interaction energy between molecules of type $i$ and $j$ (drawn from the matrix above), and $N_{ij}$ is the number of times that a pair of $i$ and $j$ type molecules are adjacent to each other on the lattice.

\section*{Algorithm for Larson Model Evolution}
\addcontentsline{toc}{section}{Algorithm for Larson Model Evolution}

\begin{enumerate}[label=\textbf{\arabic*})]

\item Initialise a random lattice configuration, completely filling the lattice with water and oil molecules, and amphiphile chains. The concentration of each `species' is defined as the amount of lattice area occupied by that species.

\item Start the system at a high temperature, i.e. with $\omega$ close to 0.

\item Update the lattice to the next configuration:

\subitem (i) \textbf{Propose a move.} Randomly choose to move either a single oil or water molecule, or a whole amphiphile chain. Oil and water molecules move by exchanging with adjacent molecules. Amphiphile chains move through a series a transformations. A common chain movement is a snake-like movement whereby the entire chain slithers one unit along its own path\footnote{Chain transformations generally just need to be able to move chains into varied conformations whilst keeping the system ergodic.}.

\subitem (ii) \textbf{Calculate the energy change associated with the move} ${\triangle{\cal H}}$ from the Hamiltonian \ref{eq:Hamiltonian}.

\subitem (iii) \textbf{Accept the proposed move with probability}: 

\begin{equation}
P(accept)=min\left\{ 1,e^{-\Delta H}\right\}
\end{equation}

Therefore, all moves which lower the lattice energy are automatically accepted, whereas moves which raise the energy are accepted with a probability that exponentially decreases with the increase in energy.

\item Repeat from step 3 until the total lattice energy has stopped changing, signifying that the `burn-in' period is over.

\item Start the next annealing phase. Cool the system temperature a small amount by increasing $\omega$ a small amount toward its target value, and repeat again from step 3. Generally, as the temperature lowers, the more time the lattice needs to equilibrate, and the longer the annealing phase required.

\item When the target temperature is reached, lattice configurations generated by the M-H algorithm can be used for the purposes of estimating desired expected values.

\end{enumerate}

In the model as implemented, the Step 3 above was performed $7 \times 10^7$ times and the lattice temperature was decreased to $\omega = 0.5$ from $\omega = 0.07$ over 7 annealing phases. Each annealing phase was twice the length of the previous one.

\subsection*{A Note on Moving Amphiphile Chains}
\addcontentsline{toc}{subsection}{A Note on Moving Amphiphile Chains}

The problem of moving an arbitrary length amphiphile chain on the lattice without breaking the chain, or crossing other chains, presented a particular challenge (even in two dimensions). In this regard, two checks were helpful to employ:

\begin{enumerate}[label=\textbf{\arabic*})]

\item \textbf{Moore consistency.} After a move, all adjacent beads of a chain must be one Moore distance apart. If not, the chain is broken (illegal).

\item \textbf{Cross consistency.} After a move, all diagonal sections of a chain must be tested and found not to cross another chain segment. This is because on a square lattice, chains can potentially cross each other (or even themselves) when moving diagonally.

\end{enumerate}

\chapter{Two Validity Tests for Abstract Elementary Reaction Sets}
\chaptermark{Validating Abstract Reaction Sets}
\label{appendix:C}

Theoretical protocell models often include a proto-metabolic network specified in abstract terms, with chemical transformations symbolised as letters, e.g. $A + B \rightarrow C$. A useful exercise is to develop validity tests for such abstracted reaction sets, to ensure that they do not negate basic physical laws. Such validity tests also permit the automatic generation of random metabolisms.

This appendix proposes two validity tests which can be applied to abstract sets of elementary reactions, one based on atom number conservation and another on free energy changes. A reaction set passing both tests can be said to be valid on basic conservative mass and energy grounds. The primary use of the tests is to give a way to rule out {\em implausible} reaction mechanisms. The two tests are dependent only on the stoichiometry of the reactions (i.e. the reaction rates are not involved) and are applicable regardless of whether the individual reactions in a reaction set are considered reversible or irreversible, or a combination of both. Table \ref{table:appendixB} gives examples of elementary reaction sets passing zero, one or both tests. 

{\em Related work.} Test 1 presented here seems to accomplish the same as, albeit in a different way, the test for conservative chemical reaction networks presented by \shortciteA{Schuster1991}. Test 2 has also been performed in the literature indirectly by e.g. \shortciteA{Bigan2015b}, but in the latter case, not in the form of an explicit validity test. Moving towards more realistic chemistry, the work of \shortciteA{Nemeth2002} is relevant, as they describe a computer algorithm that generates all the possible multi-step elementary reaction sequences that could underlie a particular overall reaction, given the list of reactants, intermediate and product molecules/ions to be involved in the reaction. Working with real chemistry, these authors are able to use more constraints. The work in this appendix is generally related to the stoichiometric checking of chemical kinetic models, the interested reader is directed to \shortciteA{Wei1962}.

\section*{Validity Test 1: Possibility for Atom Number \\ Conservation}
\addcontentsline{toc}{section}{Validity Test 1: Possibility for Atom Number Conservation}

During the course of a chemical reaction, the law of conservation of mass states that the quantity of each atomic element does not change; the atoms only become re-arranged into new molecules. In the case of real chemical equations, simple elemental balancing or `atom counting' can be used to ensure that no reaction creates nor destroys mass.

In the case of abstracted chemical equations, similar balancing can still be employed if the chemical species have labels which reflect the atomic groups they consist of, e.g. binary polymers $AAB + ABB \rightarrow AABABB$. Here, it is clear how the atom groups have re-arranged in each reaction.

However, if the abstracted chemical equations do not make explicit the atomic constituents of each species, e.g. $A + B \rightarrow C$, then group balancing cannot be performed. Nevertheless, conservation of the total \emph{atom number} in the system can still be verified (a weaker condition) by rewriting the set of reactions as a set of linear simultaneous equations. To do this, every reaction arrow (regardless if irreversible, or bi-directional) is re-written as an equals sign, and the species letters become variables signifying the number of atoms in a molecule of that type. Each linear equation stipulates that the \emph{total number} of atoms on the left and right hand sides has to be equal. 

If at least one positive non-zero solution can be found to the simultaneous equations, then this indicates that all species can be assigned a positive atom number, such that no sequence of reactions will neither create nor destroy atoms. The abstract reaction set has a possibility of being implemented in real chemistry. On the other hand, if a set of elementary reactions fails atom number conservation, then the set is \emph{certainly invalid}: somewhere in the chemical transformations is hidden the destruction or spontaneous creation of atoms.

Usefully, the solutions to the simultaneous equations indicate which are the largest and smallest molecules in the system.

\begin{tcolorbox}[colback=mybrown!50!white,colframe=mybrown!75!black,title=Box C1: Validating conservation of atoms] 

\begin{tabular}{l|l}
Reaction set H in Table \ref{table:appendixB} & Solve simultaneous equations \\
\hline
$2w \rightarrow y$ 			&	$2w - y = 0$ \\
$y + w \rightarrow 2x$ 		&	$y + w - 2x = 0$ \\ 
$x + w \rightarrow y + z$ 	&	$x + w - y - z = 0$ \\
$2z \rightarrow w$ 			&	$2z - w = 0$ \\
& where $x > 0$, $y > 0$, $z > 0$, $w > 0$ \\
\end{tabular}

\end{tcolorbox}

\section*{Validity Test 2: Possibility for All Free Energy \\ Changes as Negative}
\addcontentsline{toc}{section}{Validity Test 2: Possibility for All Free Energy Changes as Negative}

Without having to perform experiments, the standard Gibbs free energy change of a reaction $\Delta G^o$ is a good predictor of whether or not that reaction will take place spontaneously. $\Delta G^o$ is defined as the total standard Gibbs free energy of formation of the products, minus the total standard Gibbs free energy of formation of the reactants: 

\begin{equation}
\Delta G^o = \sum_p^\text{products} \Delta G_f^o(p) - \sum_r^\text{reactants} \Delta G_f^o(r)
\end{equation}

If negative\footnote{i.e. the total standard Gibbs free energy of formation of the products is more negative than that of the reactants}, then the reaction process is spontaneous under standard conditions (1 atm pressure, 298K temperature, 1 M solute concentration). Standard Gibbs free energy of formation is typically listed in tables for many substances. 

The standard Gibbs free energy change can be used as an approximate assessment as to whether a set of abstract elementary reactions is viable from an energetics perspective. The assessment is only approximate, because (i) thermodynamic spontaneity is only assessed relative to standard conditions (whereas reaction conditions could be non-standard, for example), and because (ii) even if a reaction is spontaneous, which in thermodynamic terminology means that it \emph{can} occur, this does not mean that the reaction necessarily does occur at a kinetically appreciable rate. Nevertheless, a basic energetics test is useful to apply to abstract elementary reaction sets, since it rules out perpetual motion cycles like $A \rightarrow B \rightarrow C \rightarrow A$, which have no problems passing validity test 1.

To test for free energy adherence, an elementary reaction set is be transformed into a set of linear inequalities, with each reaction arrow pointing in the spontaneous direction, re-written as a greater-than $>$ symbol, and with each species letter representing the zero or negative standard Gibbs free energy of formation for that species. 

If a set of solutions exists to this set of inequalities (i.e., a solution simplex is defined in the negative orthant of N-dimensional space, where N is the number of species), then this indicates that all species can be assigned a zero or negative standard Gibbs free energy of formation, such that all reactions can proceed spontaneously under standard conditions. The reaction set has a possibility of being implemented in real chemistry.

The solutions to the inequalities, for valid reaction systems, dictate a general energy ordering of molecules in the reaction network, but this ordering is generally flexible and there are many energy combinations which are possible.

$\;$

\begin{tcolorbox}[colback=mybrown!50!white,colframe=mybrown!75!black,title=Box C2: Validating standard free energy change] 

\begin{tabular}{l|l}
Reaction set H in Table \ref{table:appendixB} & Solve inequalities \\
\hline
$2w \rightarrow y$ 			&	$2w - y > 0$ \\
$y + w \rightarrow 2x$ 		&	$y + w - 2x > 0$ \\ 
$x + w \rightarrow y + z$ 	&	$x + w - y - z > 0$ \\
$2z \rightarrow w$ 			&	$2z - w > 0$ \\
							&   $x \le 0$ \\ 
							&   $y \le 0$ \\
							&   $z \le 0$ \\ 
							&   $w \le 0$ \\
\end{tabular}

\end{tcolorbox}

%

\begin{table}
\centering
\scalebox{0.92}{ 

\begin{tabular}{|p{3.1cm}|p{3.1cm}|p{3.1cm}|p{3.1cm}|}
\specialrule{.3em}{0em}{0em}
\rowcolor{gray!40}
Reaction Set A 	& 	Reaction Set B 	& 	Reaction Set C 	& 	Reaction Set D 	\\
\rowcolor{gray!40}
(invalid) 		& 	(invalid) 		& 	(invalid) 		& 	(invalid) 		\\
\hline
1. $2z	\rightarrow x + y$	& 	1. $z \rightarrow y + w$ 	&	1. $2x \rightarrow y + z$ 	& 1. $2w \rightarrow 2x$ 	\\
2. $y 	\rightarrow z + w$	& 	2. $y + z \rightarrow x$	&	2. $y + w \rightarrow 2x$ 	& 2. $2z \rightarrow y$ 	\\
3. $x	\rightarrow y + w$  &	3. $y + w \rightarrow 2x$	&	3. $x + y \rightarrow 2w$ 	& 3. $2x \rightarrow 2w$ 	\\
4. $2w	\rightarrow y + z$  &	4. $y + w \rightarrow z$	&	4. $z \rightarrow x$ 		& 4. $2y \rightarrow w$ 	\\
\hline
\multicolumn{4}{| c |}{valid atom conservation solutions?} \\
\textcolor{red}{NO}					& 	\textcolor{red}{NO}											& 	\textcolor{green}{YES}		&	\textcolor{green}{YES} \\
$w = 0$, $x = 0$, $y = 0$, $z = 0$	& 	$w = \frac{3z}{2}$, $x = \frac{z}{2}$, y = $-\frac{z}{2}$ 	& 	$w = z$, $x = z$, $y = z$ 	&	$w = 4z$, $x = 4z$, $y = 2z$ \\
\hline
\multicolumn{4}{| c |}{valid free energy solutions?} \\
\textcolor{red}{NO}					& 	\textcolor{red}{NO}											& 	\textcolor{red}{NO}			&	\textcolor{red}{NO} \\
\footnotesize{Sub inequality from (3) into (1) gives $2z > (y + w) + y$. Sub inequality from (2) into previous answer gives $2z > 2z + 3w$, which is impossible.}  &
\footnotesize{From (1) $z > y + w$,  but (4) asserts the opposite $y + w > z$, which cannot be true also.} &
\footnotesize{Inequalities for (2) and (3) can be written $y > 2x - w$, $y > 2w - x$ respectively. Sub inequality from (4) into (1) gives $2x > y + x$, or $x > y$. Sub $x$ for $y$ in (2) and (3) gives $y > 2y - w$, $y > 2w - y$ which reduce to the unsolvable pair $w > y$, $y > w$.} &
\footnotesize{From (1) $2w > 2x$, but (3) asserts the opposite $2x > 2w$, which cannot be true also.} \\
\specialrule{.3em}{0em}{0em}
\rowcolor{gray!40}
Reaction Set E 	& 	Reaction Set F 	& 	Reaction Set G 	& 	Reaction Set H 	\\
\rowcolor{gray!40}
(invalid) 		& 	(invalid) 		& 	valid 			& 	valid 			\\
\hline
1. $x + z\rightarrow y + w$			& 	1. $2z\rightarrow y + w$ 		&	1. $2w\rightarrow 2y$ 				& 1. $2w\rightarrow y$ 	\\
2. $x + y\rightarrow z$				& 	2. $x + z\rightarrow y$			&	2. $x\rightarrow w$		 			& 2. $y + w\rightarrow 2x$ 	\\
3. $x + w\rightarrow 2y$			&	3. $z + w\rightarrow x$			&	3. $2z\rightarrow x + y$			& 3. $x + w\rightarrow y + z$   \\
4. $x\rightarrow 2z$  				&	4. $x + z\rightarrow y + w$		&	4. $x + y\rightarrow z + w$ 		& 4. $2z\rightarrow w$ 	\\
\hline
\multicolumn{4}{| c |}{valid atom conservation solutions?} \\
\textcolor{red}{NO}					& 	\textcolor{red}{NO}				& 	\textcolor{green}{YES}				&	\textcolor{green}{YES} \\
$w = 0$, $x = 0$, $y = 0$, $z = 0$	& 	$w = 0, x = z, y = 2z$ 			& 	$w = z$, $x = z$, $y = z$ 			&	$w = 2z$, $x = 3z$, $y = 4z$ \\
\hline
\multicolumn{4}{| c |}{valid free energy solutions?} \\
\textcolor{green}{YES}				& 	\textcolor{green}{YES}			& 	\textcolor{green}{YES}				&	\textcolor{green}{YES} \\
Example: 							&   Example:						&   Example:							& Example: \\
\footnotesize{$w = -305$, $x = -64$, $y = -212$, $z = -375$ kJ/mol} 	& \footnotesize{$w = 0$, $x = -237$, $y = -433$, $z = -105$ kJ/mol}  	& \footnotesize{$w = -374$, $x = -41$, $y = -435$, $z = -189$ kJ/mol}	 & \footnotesize{$w = -72$, $x = -206$, $y = -306$, $z = -20$ kJ/mol}  \\
\footnotesize{1. $(-64 + -375) > (-212 + -305)$}				& \footnotesize{1. $2(-105) > (-433 + 0)$}						& \footnotesize{1. $2(-374)> 2(-435)$}							& \footnotesize{1. $2(-72) > -306$} \\
\footnotesize{2. $(-64 + -212) > -375$}							& \footnotesize{2. $(-237 + -105) > -433$}						& \footnotesize{2. $-41> -374$}									& \footnotesize{2. $(-306 + -72) > 2(-206)$} \\
\footnotesize{3. $(-64 + -305) > 2(-212)$}						& \footnotesize{3. $(-105 + 0) > -237$}							& \footnotesize{3. $2(-189)> (-41 + -435)$}						& \footnotesize{3. $(-206 + -72) > (-306 + -20)$} \\
\footnotesize{4. $-64 > 2(-375)$}								& \footnotesize{4. $(-237 + -105) > (-433 + 0)$}				& \footnotesize{4. $(-41 + -435)> (-189 + -374)$}				 & \footnotesize{4. $2(-20) > -72$} \\
\specialrule{.3em}{0em}{0em}
\end{tabular}

} 

\caption{\textbf{Examples of Valid and Invalid Abstract Reaction Sets.} Each set has 4 species and 4 elementary reactions that are restricted, at most, to be bi-molecular. Solutions to validity tests 1 and 2 computed with Wolfram Alpha.}
\label{table:appendixB}
\end{table}

\section*{Validity Tests are Not Additive}
\addcontentsline{toc}{section}{Validity Tests are Not Additive}

In general, the validity tests proposed above are not additive. This means that if two sets of elementary reactions - both involving some of the same species - each pass both tests individually, then there is no guarantee that the larger chemical system formed by bringing them together will also pass both tests. 

For example, Reaction Set A in Table \ref{table:appendixB} can be separated into two elementary reaction sets: one composed of reactions (1),(3) and (4), and the other composed of reaction (2). Individually, each of these sets passes validity tests 1 and 2, but brought together, the four reaction system fails both tests.

\section*{Reaction Sets with Non-Elementary Steps}
\addcontentsline{toc}{section}{Reaction Sets with Non-Elementary Steps}

Abstract reaction sets found in the literature often contain non-elementary reactions. These are reactions which actually proceed as two or more elementary steps but are written for convenience, or for lack of knowledge, as a single `black box' reaction representing the overall chemical change\footnote{Non-elementary reactions often have kinetics different from the mass action kinetics of their overall chemical equation (e.g. see \shortciteNP{rabai1987}). Reactions catalysed by enzymes are often written in one step e.g. $X + E \rightarrow Y + E$, but actually considered to proceed in two elementary steps involving formation of an enzyme-substrate complex. For this reason, catalysed reactions usually have their rate determined by the Michaelis-Menten approximation (see \shortciteNP{Chen2010}), rather than by mass action kinetics.}. In order to properly apply the two validity measures above, reaction sets with non-elementary steps need to have all such steps expanded to their elementary reactions. Then, the measures should be applied to the entire set of elementary reactions.

Nevertheless, having said the above, abstract reaction systems such as the oscillating Lotka-Volterra model \shortcite{lotka1920}, or minimal bistable systems like Schlogl's \citeyear{schlogl1972} or Wilhelm's \citeyear{schlogl1972} model, often have their kinetics specified as mass action kinetics, despite containing steps which are questionably elementary. This implies that these reactions are treated as if elementary. If the reactions are assumed to be elementary on these grounds, then the validity tests can be applied, and the three examples just cited each pass both tests. The Oregonator (the version presented in \shortciteNP{gillespie1977}) also passes both tests.

\section*{Reservoir Analysis of Reaction Networks Limits Stoichiometric Possibilities}
\addcontentsline{toc}{section}{Reservoir Analysis of Reaction Networks Limits Stoichiometric Possibilities}

Perhaps for mathematical simplicity, chemical reactions are traditionally analysed in `reservoir conditions'. A set of species is designated as nutrient species, a set of species is designated as waste species, and these species sets are given constant reservoir  concentrations.\footnote{The additive free energy of the nutrient species is greater than the additive free energy of the waste species, although individual waste species can have higher free energy than individual nutrient species.} The dynamic behaviour of the remaining {\em intermediate} species is recorded: these are species that are both produced and consumed in reactions. 

However, other reaction sets {\em not} suitable for reservoir conditions, but valid nevertheless, can and do exist. Reaction sets G and H in Table \ref{table:appendixB}, for example, pass validity tests 1 and 2 but are not suitable for reservoir conditions because each one of the species $w$, $x$, $y$ and $z$ is both consumed and produced. Here, designating a species as a constant concentration nutrient has the effect of generally breaking up the reaction set into trivial parts. Instead, the behaviour of reaction sets like G and H can be investigated by also modelling an explicit reactor context in which the reactions take place. Rather than the chemical reaction being sustained by nutrient and waste species at fixed concentrations, it is instead sustained by chemical flows into and out of the reactor. It follows that a greater diversity of reaction mechanisms can be modelled in explicit reactor contexts (including in compartments), than in reservoir conditions.

\chapter{Published Manuscripts}
\chaptermark{Published Manuscripts}
\label{appendix:D}

Manuscripts published as part of this thesis are linked to in this Appendix for reference. All manuscripts were published as open access.

\section*{Journal Papers}

\begin{enumerate}[label=\textbf{\arabic*})]
\item Shirt-Ediss, B., Ruiz-Mirazo, K., Mavelli, F. and Sol{\'e}, R.V. (2014). Modelling Lipid Competition Dynamics in Heterogeneous Protocell Populations. \emph{Scientific Reports}, 4, 5675. \href{http://dx.doi.org/10.1038/srep05675}{doi: 10.1038/srep05675}. 

14 pages of supplementary material is available online (\href{http://dx.doi.org/10.13140/2.1.2170.1448}{doi: 10.13140/2.1.2170.1448}).

\item Shirt-Ediss, B., Sol{\'e}, R.V. and Ruiz-Mirazo, K. (2015). Emergent Chemical Behavior in Variable-Volume Protocells. \emph{Life (Basel)}, 5, 181-211. \href{http://dx.doi.org/10.3390/life5010181}{doi: 10.3390/life5010181}.

16 pages of supplementary material is available online (\href{http://dx.doi.org/10.13140/2.1.4242.7206}{doi: 10.13140/2.1.4242.7206}).

\end{enumerate}

\section*{Conference Paper}

\begin{enumerate}[label=\textbf{\arabic*})]
\item Shirt-Ediss, B., Sol{\'e}, R. and Ruiz-Mirazo, K. (2013). Steady state analysis of a vesicle bioreactor with mechanosensitive channels. In P. Li{\`o}, O. Miglino, G. Nicosia, S. Nolfi, and M. Pavone (Eds.), \emph{Advances in Artificial Life, ECAL 2013: Proceedings of the Twelfth European Conference on the Synthesis and Simulation of Living Systems.} 1162-1169, MIT Press. \href{http://dx.doi.org/10.7551/978-0-262-31709-2-ch178}{doi: 10.7551/978-0-262-31709-2-ch178}.

\end{enumerate}


\end{appendices}

\chapter*{Copyright Permissions for Reproduced Figures}
\addcontentsline{toc}{chapter}{Copyright Permissions for Reproduced Figures}
\renewcommand{\chaptermark}[1]{\markboth{}{\sffamily #1}}
\chaptermark{Copyright Permissions}
\label{chapter:copyright}

%
%

\small

\noindent\textbf{Box 1, Figure (a).} Reproduced from Pohorille, A. and Deamer, D., Research in Microbiology, 160(7), 449-456, 2009 \shortcite{Pohorille2009}. Copyright \copyright 2009 Elsevier Masson SAS. All rights reserved.

$\;$


\noindent\textbf{Figure \ref{fig:ch1_protocell_approaches}(b).} Reproduced from Ichihashi et al., Nature Communications, 4, 2494, 2013 \shortcite{Ichihashi2013}. Reprinted by permission from Macmillan Publishers Ltd: Nature Communications, Copyright 2013.

\noindent\textbf{Figure \ref{fig:ch1_protocell_approaches}(c).} Reproduced from Kurihara et al., Nature Chemistry, 3(10), 775-781, 2011 \shortcite{Kurihara2011}. Reprinted by permission from Macmillan Publishers Ltd: Nature Chemistry, Copyright 2011.

\noindent\textbf{Figure \ref{fig:ch1_protocell_approaches}(d).} Reproduced from Noireaux, V. and Libchaber, A., Proceedings of the National Academy of Sciences of the United States of America, 101(51), 17669-17674, 2004 \shortcite{Noireaux2004}. Copyright 2004 National Academy of Sciences, USA.

\noindent\textbf{Figure \ref{fig:ch1_protocell_approaches}(e).} From Krishna Kumar et al., Angewandte Chemie, 50(40), 9343-9347, 2011 \shortcite{KrishnaKumar2011}. Copyright \copyright 2011 by John Wiley Sons, Inc. Reprinted by permission of John Wiley \& Sons, Inc.

\noindent\textbf{Figure \ref{fig:ch1_protocell_approaches}(f).} Copyright \copyright Hardy et al., Proceedings of the National Academy of Sciences of the United States of America, 112(27) 8187-8192, 2015 \shortcite{Hardy2015}. (The National Academy of Sciences of the United States of America, the publisher of this article, does not require permission to be obtained for reproduction in non-commercial and educational use.)

\noindent\textbf{Figure \ref{fig:ch1_protocell_approaches}(g).} From Luigi Luisi et al., ChemBioChem, 11(14), 1989-1992, 2010 \shortcite{Luisi2010}. Copyright \copyright 2010 by John Wiley Sons, Inc. Reprinted by permission of John Wiley \& Sons, Inc.

\noindent\textbf{Figure \ref{fig:ch1_protocell_approaches}(h).} Reprinted figure with permission from Fallah-Araghi et al., Physical Review Letters, 112, 028301, 2014 \shortcite{Fallah-Araghi2014}. Copyright 2014 by the American Physical Society.

\noindent\textbf{Figure \ref{fig:ch1_protocell_approaches}(i).} Copyright \copyright Budin, I. and Szostak, J., Proceedings of the National Academy of Sciences of the United States of America, 108(13) 5249-5254, 2011 \shortcite{Budin2011}. (The National Academy of Sciences of the United States of America, the publisher of this article, does not require permission to be obtained for reproduction in non-commercial and educational use.)


\noindent\textbf{Figure \ref{fig:ch1_protocell_approaches}(k).} Reprinted (adapted) with permission from Takakura et al., Journal of the American Chemical Society, 125(27), 8134-8140, 2003 \shortcite{Takakura2003}. Copyright (2003) American Chemical Society.

\noindent\textbf{Figure \ref{fig:ch1_protocell_approaches}(l).} Copyright \copyright Zhu et al., Proceedings of the National Academy of Sciences of the United States of America, 109(25), 9828-9832, 2012 \shortcite{Zhu2012}. (The National Academy of Sciences of the United States of America, the publisher of this article, does not require permission to be obtained for reproduction in non-commercial and educational use.)

\noindent\textbf{Figure \ref{fig:ch1_protocell_approaches}(m).} Copyright \copyright Terasawa et al., Proceedings of the National Academy of Sciences of the United States of America, 109(16), 5942-5947, 2012 \shortcite{Terasawa2012}. (The National Academy of Sciences of the United States of America, the publisher of this article, does not require permission to be obtained for reproduction in non-commercial and educational use.)


\noindent\textbf{Figure \ref{fig:ch1_protocell_approaches}(o).} Reprinted  with permission from Szymanski et al., The Journal of Physical Chemistry C, 117, 13080-13086, 2013 \shortcite{szymanski2013}. Copyright (2013) American Chemical Society.

\noindent\textbf{Figure \ref{fig:ch1_protocell_approaches}(p).} Reprinted  with permission from Maselko, J. and Strizhak, P., The Journal of Physical Chemistry B, 108(16), 4937-4939, 2004 \shortcite{maselko2004}. Copyright (2004) American Chemical Society.

$\;$

%
%

\noindent\textbf{Figure \ref{fig:ch2_rube_goldberg}.} Artwork Copyright \textcopyright$\;$and TM Rube Goldberg Inc. All Rights Reserved. RUBE GOLDBERG \textregistered$\;$ is a registered trademark of Rube Goldberg Inc. All materials used with permission. rubegoldberg.com.

$\;$

\noindent\textbf{Figure \ref{fig:ch2_chemoton1997}.} Reprinted from Tibor Ganti, Journal of Theoretical Biology, 187(4), 583-593, 1997 \shortcite{ganti1997}. Copyright 1997, with permission from Elsevier.

$\;$

%
%

\noindent\textbf{Figure \ref{fig:ch3_comp_autopoiesis_models}(b).} Reproduced from McMullin, B. and Varela, F.J., SFI Working Paper 97-02-012 \shortcite{McMullin1997} with permission from Barry McMullin.

\noindent\textbf{Figure \ref{fig:ch3_comp_autopoiesis_models}(f).} Reprinted from Ono, N., BioSystems, 81(3), 223-233, 2005 \shortcite{Ono2005}. Copyright 2005, with permission from Elsevier.

$\;$

\noindent\textbf{Figure \ref{fig:ch3_krm_models}(a).} Reprinted from Ruiz-Mirazo, K. and Mavelli, F., BioSystems, 91(2), 374-387, 2008 \shortcite{Ruiz-Mirazo2008}, Copyright 2008, with permission from Elsevier.

\noindent\textbf{Figure \ref{fig:ch3_krm_models}(b).} Reprinted from Wilson et al., Origins of Life and Evolution of the Biosphere, 44(4), 357-361, 2015 \shortcite{Wilson2014}, with permission from Springer.

\noindent\textbf{Figure \ref{fig:ch3_krm_models}(c).} Reprinted from Ruiz-Mirazo et al., Software Tools and Algorithms for Biological Systems, Chapter 70, 689-696, 2011 \shortcite{Ruiz-Mirazo2011}, with permission from Springer.

$\;$

%
%

\noindent\textbf{Figure \ref{fig:ch4_self_assembly}(c).} Reprinted from Dawson, K. A., Pure and Applied Chemistry, 64(11), 1589-1602, 1992 \shortcite{dawson1992}, with permission from Springer.

\normalsize

%
%
%
%
%
%
%
%
%
%
%
%

\backmatter
\singlespacing

\renewcommand{\bibname}{References}
\bibliography{bib/thesisbib}
\bibliographystyle{apacite}

\end{document}